\documentclass[12pt]{article}
\input epsf.tex
\epsfclipon
\usepackage{epsfig}
\usepackage{epstopdf}
\usepackage{epsf}
\usepackage{graphicx,amsmath,amssymb}
\usepackage{epsfig,multicol}
 \usepackage{epsfig}
\usepackage{amssymb}
\usepackage{jheppub}
\usepackage{mathtools}
\usepackage{framed}
\usepackage{simplewick}
\usepackage[none]{hyphenat}
\usepackage[utf8x]{inputenc}
\def\bg{\begin{eqnarray}}
\def\nd{\end{eqnarray}}

\newcommand{\be}{\begin{equation}}
\newcommand{\ee}{\end{equation}}
\newcommand{\beas}{\begin{eqnarray*}}
\newcommand{\eeas}{\end{eqnarray*}}
\newcommand{\bea}{\begin{eqnarray}}
\newcommand{\eea}{\end{eqnarray}}

\begin{document}
\title{Lectures on Conformal Field Theory }
\author[a]{Joshua D. Qualls}
\affiliation[a]{Department of Physics, National Taiwan University, Taipei, Taiwan}
\emailAdd{joshua.qualls@ntu.edu.tw}

\abstract{
These lectures notes are based on courses given at National Taiwan University, National Chiao-Tung University, and National Tsing Hua University in the spring term of 2015. Although the course was offered primarily for graduate students, these lecture notes have been prepared for a more general audience. They are intended as an introduction to conformal field theories in various dimensions working toward current research topics in conformal field theory. We assume the reader to be familiar with quantum field theory. Familiarity with string theory is not a prerequisite for this lectures, although it can only help. These notes include over 80 homework problems and over 45 longer exercises for students.
}
\maketitle
\newpage

\section{Lecture 1: Introduction and Motivation}

\subsection{Introduction and outline}

This course is about conformal field theory. These lectures notes are based on $8 \times 3$ hours of lectures given for graduate students. Over the last several decades, our understanding of conformal field theories has advanced significantly. Consequently, conformal field theory is a very broad subject. This is not the first set of lecture notes on this topic, nor will it be the last. So why have I bothered making these notes available when there are already so many choices? 

There are two reasons. The first is purely selfish: I have found there is no quicker method of finding mistakes than sharing your results with an audience. It is my hope that I can correct errors if and when they are brought to my attention. Please make me aware of any issues. 

The second reason is more benevolent: I was interested in giving a small course on conformal field theory working toward the conformal bootstrap program. There were already some excellent resources on bootstrapping, so I attempted to cover everything you would need to know before beginning bootstrap research. One thing lead to another, and eventually I had written notes from the basics of conformal field theory all the way to the basics of bootstrapping.

These notes provide only an introduction to the rich field. The course was actually closer to a half-course, and there are portions of the notes that sorely reflect this. Personally, I view these lecture notes as the outline or beginning of a more thorough study of CFTs. While some resources are encylopaedic in their approach, or narrow in their focus, the present volume could serve as introduction to students just beginning their research in string theory or condensed matter. A student of these lectures would not be an expert in gauge/gravity duality, for example, but they would be in a much better position to pursue more focused readings.It is my hope that these notes are general enough that anyone interested in doing research involving conformal field theory could start at the beginning and work through them all, at which point they would be ready to begin a more focused study of whatever applications of CFT techniques are relevant to their interest. In the future, I hope to write lectures that go into more detail about these applications across various fields of physics.

In this lecture, we introduce the motivations for studying conformal field theory. We begin with some examples of classical conformal invariance, before moving on to talk about CFTs in critical phenomena and the renormalization group. We briefly mention some applications of CFTs toward other subjects before finishing the lecture by discussing conformal quantum mechanics---conformal field theory in $d=1$ dimension.

In Lecture 2, we study the basic properies of CFTs in $d > 2$ dimensions. Topics include conformal transformations, their infinitesimal form, a detailed discussion of special conformal transformations, the conformal algebra and group, and representations of the conformal group. We next discuss constraints coming from conformal invariance, followed by the stress-energy tensor and conserved currents. We finish by introducing radial quantization, the state-operator correspondence, and unitarity bounds that come from using both.

In Lecture 3, we shift our focus to CFTs in $d=2$ dimensions. We start again with infinitesimal conformal transformations, before moving on the Witt and Virasoro algebras. We introduce primary fields, and discuss including the  of the conformal group, primary fields, radial quantisation, the operator product expansion, the operator algebra of chrial quasi-primary fields and the representation theory of the Virasoro algebra.

In Lecture 4, we consider simple 2d CFTs. These include the free boson (as well as the periodic boson and the boson on an orbifold), the free fermion, and the $bc$ ghost theory. We then shift our attention to more general CFTs, focusing on descendants, the Ka\u{c} determinant, and constraints on 2d unitarity CFTs.

In Lecture 5, We consider the constraints coming from modular invariance on the torus, bosonic and fermionic theories on the torus, orbifold CFTs, and work toward understanding the Verlinde formula.

In Lecture 6, we will revisit previous topics that are active areas of CFT research. These include the central charge, c-theorems in various dimensions, and whether scale invariance implies conformal invariance.

In Lecture 7, we continue our exploration of CFTs by introducing the conformal bootstrap program. We systematically investigate the operator product expansion and find the constraints imposed upon conformal field theories from crossing symmetry/operator product expansion associativity. 

In Lecture 8, we will attempt to finish all of the topics we have already listed. In theory, this lecture should have introduced boundary conformal field theory. In practice, we finished by talking about the modular bootstrap approach in two-dimensional CFTs and simplifications to the bootstrap program in the limit of large spin.

If you already have experience with conformal field theory, you may find that these notes are lacking several essential topics. We do not get to do justice to Ka\u{c}-Moody algebras, for example. We are not able to present the Sugawara and coset constructions, or the $\mathcal{W}$ algebras. Minimal models corresponding to realized, physical systems do not get nearly enough attention, and we are not able to calculate even one critical exponent. Similarly, there is very little mention of the AdS/CFT correspondence or superconformal symmetry. Every single application we present here should receive at least twice as many lectures as we are able to give, and several interesting applications have been omitted altogether. In the fall, I may have the opportunity to do additional lectures; if this is the case, then I hope to append a variety of topical lectures on more advanced topics/interesting applications.

I will give several references...some of them have been followed closely, some of them are only used in passing. All of them will improve your understanding of this rich field. At the end of each lecture, we give the most relevant references used in preparing the lecture. At the end of the notes are all of the references the author consulted for the entirety of the notes. The first 15 references are the ones that have textual overlap. These include textbooks \cite{difran,bp}, lecture notes \cite{rychkov,naka,ginsparg}, and relevant papers \cite{bigpol,ctheorem,cardymodel,atheorem,rychkovope,rychkovrad,ising,gliozzi1,gliozzi2,hell,mine}. If one of the references is primary but has not been listed first, please let me know. The remaining references are in roughly the order they are relevant to the text. There could be some transposed, however. If I omitted any of these references, please let me know.

Everyone is approaching these lectures from different levels, so I will also provide references to some useful background material. Basic knowledge of quantum field theory is essential at the level of Peskin and Schroeder's ``An Introduction to Quantum Field Theory''. Particularly relevant are chapters 8 (explaining how quantum field theory is relevant for critical phenonmena) and 12.1 (a physical introduction to ideas of the renormalization group). A working knowledge of complex analysis is important, so I recommend familiarity with these methods at the level of Arfken, Weber, and Harris's ``Mathematical Methods for Physicists''.  

This is the second version of these notes available to the public. Based on feedback I have received, as well as several rereadings, I feel I have added appropriate references and corrected unfortunate mistakes. Since I first made these lectures available, there have been several fascinating results and newly discovered directions for research. I have elected not to update the references for these lectures with any of these new results, though I urge you to read as many current papers as possible. I have also started writing additional lectures, though they will not be available until I have tested them on at least one class.

The author wishes to thank the students from NTU, NCTU, and NTHU, with particular thanks to Heng-Yu Chen and C.-J. David Lin. Additionally, the author would like to offer special thanks to Luis Fernando Alday for helpful remarks about the analytic bootstrap and large spin analysis, Michael Duff for helpful remarks about Weyl anomalies, and Slava Rychkov for supportive remarks, as well as his remarkable work that served to interest me initially in this remarkable subject.

\subsection{Conformal invariance: What?}

In this lecture I will give a general introduction to the ideas of conformal field theory (CFT) before moving on to the simplest toy model. This will be a broad introduction, so do not feel discouraged if some of the ideas seem rushed. We will work on filling in details as the lectures progress. Some details are omitted due to time constraints. You should fill them in on your own time. Before telling you what what I'm going to tell you, however, allow me to tell you why it's worth hearing. After all, why should anyone study CFTs? Aren't they a terribly specialized subject? We will argue that conformal field theory is at the very heart of quantum field theory (QFT), the framework describing almost everything we know and experience in nature.

By definition, a conformal field theory is a quantum field theory that is invariant under the \emph{conformal group}. By now you should be familiar with the Poincar\'e group as the symmetry group of relativistic field theory in flat space. That is, Poincar\'e transformations are those that leave the flat space metric $\eta_{\mu\nu}\equiv \mbox{diag}(-,+,+,+)$ invariant. Another way of saying this is that Poincar\'e transformations are \emph{isometries} of flat spacetime. Poincar\'e transformations are transformations of the form
$$
x^\mu \rightarrow \Lambda^\mu_\nu x^\nu + a^\mu
$$
and are a combination of Lorentz transformations parameterized by $\Lambda$ and translations parametrized by $a$.

\begin{figure}
\centering
\includegraphics[scale=.2]{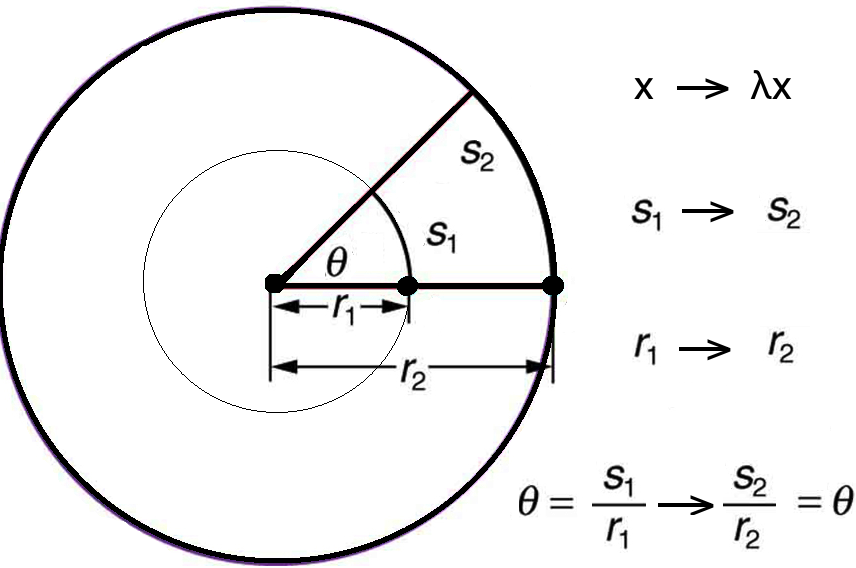} 
\caption{This image illustrates how rescaling distances preserves the angle $\Delta\theta$, even when we have rescaled $r_1\rightarrow r_2$ and $s_1\rightarrow s_2$.}
\label{figure:rescaledcircle}
\end{figure}

In addition to the symmetries of flat spacetime, CFTs have extra spacetime symmetries: the conformal group is the set of transformations of spacetime that preserve angles (but \emph{not} necessarily distances). Conformal transformations obviously include the Poincar\'e transformations. What other transformations should we consider? We will begin with the most intuitive conformal transformation: a scale transformation (as in Figure \ref{figure:rescaledcircle}).  Scale transformations act by rescaling, or zooming in and out of some region of spacetime. If we split the space and time coordinates, then scale transformations act mathematically by taking $x\rightarrow\lambda x$ and $t\rightarrow \lambda^z t$. The quantity $z$ is known as the \emph{dynamical critical exponent} and is an object of great importance in condensed matter physics. In this course, we will mainly be interested in relativistic quantum field theories. This means that space and time coordinates are on equal footing, so that $z=1$ and scale transformations are of the form
$$
x^\mu \rightarrow \lambda x^\mu.
$$
Scale transformations act on momenta in the opposite way: $$p^\mu \rightarrow \lambda^{-1} p^\mu. $$ Mathematically this makes sense: as the product of position and momentum should be dimensionless in natural units. Physically, this scaling behavior reflects the fact that zooming in on a smaller region of spacetime requires higher frequency modes of momentum to probe shorter distances in the system. 

Scale transformations are definitely \emph{not} in the Poincar\'e group. This is obvious from their effect on the flat space metric. Under a scale transformation, we pick up the factor
$$
\eta_{\mu\nu} \rightarrow \lambda^{-2} \eta_{\mu\nu}.
$$
This expression makes it clear that while lengths are rescaled, angles are preserved.
\noindent More generally, a conformal transformation is a generalization of a scale transformation such that under a coordinate transformation $$
x \rightarrow \tilde{x}(x),
$$
the spacetime metric transforms as
$$
\eta \rightarrow f(x) \eta.
$$
Generally speaking, a conformal transformation is a coordinate transformation that is a local rescaling of the metric. We will completely characterize the most general type of this transformation soon. By doing this, we will arrive at the conformal group and investigate the constraints the conformal group imposes on physical quantities.

In the following discussions, we will focus on theories with scale invariance. But we have just claimed that conformal transformations are a generalization of scaling transformations. Is it really enough to restrict our dicussions to scale invariance? What is the distinction between scale invariance and conformal invariance in relativistic quantum field theories? This excellent question will be addressed in detail later in these lectures. To summarize, it can be shown (under some technical assumptions) that scale invariance is enhanced to conformal invariance in $d=2$ dimensions. In $d=4$ dimensions there is a perturbative proof of the enhancement and no known examples of scale-invariant but non-conformal field theories (under some reasonable assumptions); there is also a complementary holographic argument. For now, therefore, we will use the terms interchangeably: a theory without scale invariance will not have conformal invariance and any theory we consider with scale invariance will have conformal invariance.

\subsection{Examples of classical conformal invariance}

Why should we even discuss conformal transformations? For starters, some of the most important equations in physics are conformally invariant. The simplest example of classical conformal symmetry is Maxwell's equations in the absence of sources(/charged particles),
$$
\partial^\mu F_{\mu\nu}=0.
$$
\begin{framed}
\noindent HOMEWORK: Prove the Maxwell action is invariant under scale transformations.
\end{framed}
Another example is the free massless Dirac equation in $d=4$ dimensions,
$$
\gamma^\mu \partial_\mu \psi = 0.
$$

Both of these examples were free massless fields, so the associated Lagrangians have no coupling parameters. But there are also examples of \emph{interacting} theories that have classical conformal invariance. For example, consider classical Yang-Mills theory in $d=4$ dimensions with associated equation of motion
$$
\partial^\mu F^a_{\mu\nu} +g f^{abc}A^{b\mu}F^c_{\mu\nu}= 0.
$$
Yet another familiar theory is the classical $\lambda \phi^4$ theory in $d=4$ dimensions with equation of motion
$$
\partial^2 \phi = \lambda \phi^3 / 3!
$$
Even with interactions, however, the associated Lagrangians describe massless fields. This is because a theory cannot be conformally invariant if the Lagrangian has some mass parameter--the mass introduces a length scale that is not invariant under scale transformations. 

We have specified \emph{classical} conformal invariance, rather than quantum conformal invariance. This is because we know from field theory that even though you write down a Lagrangian with constant couplings, quantum mechanics introduces a dependence on the energy scale-- so-called \emph{running} coupling constants (which we will discuss in more detail shortly). Some theories have classical conformal invariance continue beyond the classical level. Free, massless quantized scalar field theory, for example, has no coupling parameters and is therefore conformally invariant. Similarly, free massless fermions and free Maxwell fields have quantum conformal invariance.  But our other examples have couplings that will become running couplings quantum mechanically. So the couplings are actually functions of some energy scale $\lambda(E), g(E)$. And because they depend on scale, they cannot possible be conformally invariant.

\subsection{Conformal invariance: Why?}

So we expect that interacting quantum field theories can not be conformally invariant quantum mechanically. QED is not scale-invariant, massive scalars are not scale-invariant, Yang-Mills theory is not scale invariant---this is obvious from the associated $\beta$-functions. So the question remains as to why we should bother studying CFTs at all. After all, the interesting theories are clearly not conformally invariant quantum mechanically. I will spend most of the rest of this lecture giving reasons as to why we care about CFTs. The first answer demonstrates how CFTs are relevant in the natural world; the second answer gets to the very heart of our best understanding of quantum field theory.

\subsubsection{CFTs in critical phenomena}

Conformal field theories describe critical points in statistical physics---a \emph{critical point} is the point at the end of a phase equilibrium curve where a continuous phase transition occurs (for example, the liquid-gas transition of water, or at the Curie temperature of a ferromagnet). Mathematically, a phase transition is a point in parameter space where the free energy $F=-T\ln Z$ becomes a nonanalytic function of one of its parameters in the thermodynamic limit\footnote{For a finite system, this can never happen: the partition function $Z$ is a sum over finite, positive terms and thus its derivatives are well-defined and finite}. Phase transitions are often classified by their \emph{order}, which just counts which order derivative of the free energy is discontinuous. The quantity that is different in various phases is known as the \emph{order parameter} and can be used to characterize the phase transition. 

One of the quantities we investigate to determine if we are approaching criticality is the \emph{correlation length}. Roughly speaking, this is a measure of how ``in tune'' different degrees of freedom are. More precisely, the correlation length is the length at which degrees of freedom are still correlated/feel one another's influence strongly. It is computed by the two-point function of basic degrees of freedom (How much does the spin of one atom correlate with the spin of a distant atom? How much is the material's density correlated as you move throughout the sample?)

For a statistical mechanics degree of freedom $\sigma$, we would calculate the correlation function measuring the order of the system via the expression
\begin{equation}
\langle \sigma(x) \sigma(0) \rangle - \langle \sigma(x) \rangle\langle \sigma(0) \rangle
\end{equation}
This expression goes as 
$$
 |x|^{const.} \exp(-|x|/\xi),
$$
where the power-law dependence is dominated by the exponential dependence and $\xi$ is defined as the correlation length. This sort of functional dependence should be familiar: it resembles the Yukawa potential. If you compute the two-point correlation function for a scalar field of mass $m$, it decays exponentially with an associated length scale $1/m$.

In order to approach a critical point, there must be some external parameter we can vary; examples of such a parameter include pressure, temperature, and applied magnetic field. Physicists are usually interested in how various thermodynamic quantities scale as a function of this parameter when we approach a critical point. These scaling behaviors are given by \emph{critical exponents} are are directly related to the dimensions of operators in CFTs that we will study. But why do CFTs enter the picture when we have already mentioned that these theories have a characteristic (correlation) length scale?

For concreteness, consider the case of a ferromagnet placed in an external magnetic field $H$. The degrees of freedom here are individual spins that point either up or down, and the tunable parameter we consider is the temperature $T$. As $T$ approaches some critical temperature $T_c$, thermal fluctuations become large and the material becomes paramagnetic. This is precisely the notion of correlation length that we mentioned; as the temperature increases, the length over which fluctuations have an effect increases such that the correlation length $\xi\rightarrow\infty$. As we approach a critical point, therefore, the corresponding mass scale is vanishing and we have a massless, scale-invariant theory. 

\begin{framed}
\noindent HOMEWORK: This critical temperature is the \emph{Curie temperature}. What is the Curie temperature for various materials? If you do not know, go look it up. Really, go find it. It is never a bad idea to have some idea of relevant physical scales.
\end{framed}

\begin{figure}
\centering
\includegraphics[scale=.3]{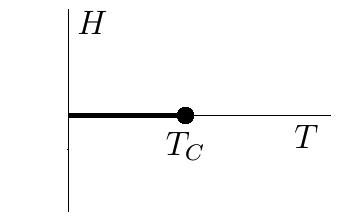} 
\caption{The phase diagram for the ferromagnetic/paramagnetic phase transition. Above the solid line, $M>0$ and below it, $M<0$. A conformal field theory lives at $T_c$. } \label{fig:figure2}
\end{figure}

Let's continue this example. Thermodynamic quantities we may find interesting in this system include the correlation length $\xi$, the free energy $F$, the magnetization $M = -\frac{\partial F }{ \partial H}$, the susceptiblity $\chi = \frac{\partial M}{\partial H}$, and the heat capacity $C = -T \frac{\partial^s F}{\partial T^2}$. The phase diagram for this transition is shown in Figure \ref{fig:figure2}. 
The order parameter for this phase transition is the magnetization, and this transition is second order (the susceptibility diverges near criticality). In this system, we parameterize our proximity to the critical temperature by the dimensionless $\tau \equiv\frac{T-T_c}{T_c}$ and define the critical exponents according to
\begin{gather}
\xi \sim \tau^{-\nu} \\
C \sim \tau^{-\alpha} \\
\chi \sim \tau^{-\gamma} \\
M \sim (-\tau)^\beta \\
M = H^{1/\delta} \\
\langle  \sigma(x)\sigma(0)\rangle =  |x|^{2-d-\eta}  
\end{gather}
These are quantities that we will calculate later\footnote{Notice that I did not specify how much later. Concrete examples of phase transition calculations will have to wait for a later version of this course. If the suspense is unbearable, in four dimensions the values of these exponents for this theory are (in order) $\frac12, 0, 1, \frac12, 3,$ and $0$.}. We will also discover scaling relations between them implying there are actually only two independent exponents (e.g., $\nu$ and $\eta$).

This method of analysis is rather general. A similar phase transition occurs between the liquid and gaseous phases of water. A simplified phase diagram, shown in Figure \ref{fig:figure3}:
\begin{figure}
\centering
\includegraphics[scale=.25]{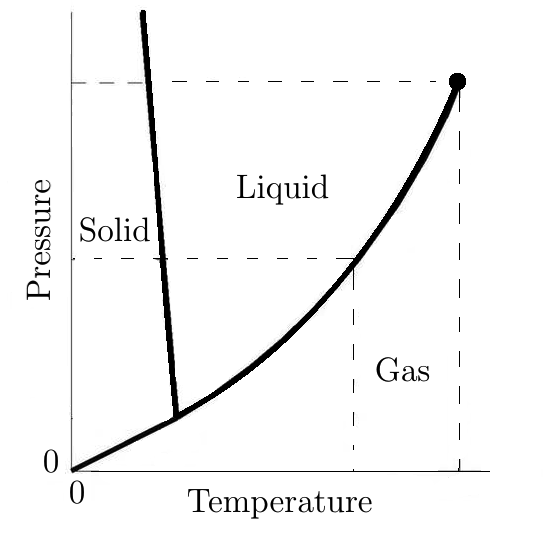}
\caption{A simplified phase diagram for water. The critical point is marked.} \label{fig:figure3}
\end{figure}
\begin{framed}
\noindent HOMEWORK: What are the critical temperature and pressure for this transition? Also, write a few sentences about the phenomenon known as \emph{critical opalescence}. Since you more than likely do not know what it is, go read about it on your own for a few minutes. I can wait.
\end{framed}
Naively, one could expect these two physical systems to be described by completely different Hamiltonians. But one of the basic predictions of the renormalization group (which we will discuss next) is called \emph{universality}. As we shall see imminently, renormalization group flow shows that while we start with different complicated systems at high energies, the behavior of two systems at large distances can be very similar if they have the same low-energy degrees of freedom. 

I will close this section by mentioning \emph{quantum critical points}. The above analysis is not only valid for transitions driven by thermal fluctuations. It turns out there are quantum critical points at $T=0$ where transitions are driven by quantum fluctuations. These phase transitions also exhibit infinite correlation lengths and thus are also describable via CFTs.  A quantum system is characterized by a Hamiltonian with some ground state energy. Typically there is some spectrum of excitations above this ground state; for example, in the quantum harmonic oscillator. The energy gap $\Delta$ between the ground state and the first excited state defines some length scale. Obviously length scales are not allowed with conformal invariance. A useful way of determining when we reach criticality is thus by considering the energy gap and tuning our parameters so that this gap closes to zero.

A quantum critical point described by some two-dimensional conformal field theory means we are considering some quantum mechanical system in one dimension. A good example of such a model is the \emph{Heisenberg spin chain} Hamiltonian.
It consists of some lattice of points (we consider a closed circle). At each point, we have a quantum spin variable. The appropriate Hamiltonian is of the form
\begin{equation}
H \sim J \sum_{j=1}^N  \boldsymbol S_j \cdot \boldsymbol S_{j+1},
\end{equation}
where $J$ is some coupling constant and $S^1,S^2,S^3$ are the Pauli spin matrices.
The sign of the coupling constant $J$ determines whether the system is ferromagnetic or antiferromagnetic, and the model can be generalized to have different couplings in different directions (the Heisenberg XXZ or XYZ models, respectively).The model written here has a gapless spectrum and is described by a $d=2$ CFT for a free, periodic boson. Because this system has an obvious $SU(2)$ symmetry, it turns out the the CFT will also have $SU(2)$ symmetry. This system additionally has a dual description as a $d=2$ $SU(2)$ Wess-Zumino-Witten model at level 1. Hopefully, we will get to these topics\footnote{We did not.}. So in order to try, we will now move forward.

\subsubsection{Renormalization group}
 
We have discussed how conformal field theory is realized in specific physical systems. Now we consider the central role it plays in understanding the space of quantum field theories. I will not take the time to explain why we might care about a deeper understanding of QFT---the main theoretical framework describing most of nature, with applications including elementary particle physics, statistical physics, condensed matter physics, and fluid dynamics\footnote{Sincerely, I hope this is not your first time contemplating why QFT could be important.}. By this point in your education you should have discovered that the description of a physical system very much depends on the energy scale you wish to study, and as I will explain the subject of conformal field theory is essential in studying this question in the realm of quantum field theory.

For example, consider a bucket of water. At the scale of centimeters, the best description for studying the physics of this system is in terms of some Navier-Stokes hydrodynamical equations. But what if I want to probe atomic distances in this system? The previous description is no longer useful, since hydrodynamics is a valid description at wavelengths large compared to water molecules. At some point, we must describe the system in terms of the quantum mechanics of electrons and the nucleus. If we go even smaller, then we must start to consider the constituent quarks in terms of quantum chromodynamics. So any time we study a physical system, we must ask what energy scale we are actually trying to probe.

The situation is similar in quantum field theory. A QFT comes equipped with some ultraviolet cutoff $\Lambda$, the energy scale beyond which new degrees of freedom are necessary. We do not know what's going on past this energy (or equivalently, at distances smaller than $\Lambda^{-1}$). One of the beautiful and remarkable features of physics is that even though we do not have a complete theory of quantum gravity, we can still calculate observable results using low-energy physics. The whole program of the renormalization group in QFT is a way to parameterize this ignorance in terms of interactions or coupling constants that we measure\footnote{Yes, \emph{measure}. We cannot calculate coupling constants from some fundamental principle (...yet?)} between low-energy degrees of freedom. Once we measure these couplings once, quantum field theory is predicted.

Let's discuss the renormalization group (RG) a little more. In the RG framework, we first enumerate the degrees of freedom we wish to study: the field content. So we start with a free field theory action $S_0$\footnote{This is a Gaussian fixed point of the RG. In general we could consider any fixed point, but we will only consider free field theory.}. Next we write down the most general action involving interactions of these degrees of freedom comprised of terms incorporating the symmetries we want to study: global symmetry transformations, for example, or discrete $\mathbb{Z}_2$ transformations. We add local interactions via terms of the form
$$
S_{int} = \int d^d x\sum g_i \mathcal{O}_i(\phi).
$$
These terms consist of operators constructed from low-energy fields and coupling constants describing the relative strength of interactions. To calculate quantities, we use the path integral
\begin{equation}
Z \equiv \int D\phi e^{-S}
\end{equation}

The basic integration variables of the path integral are the Fourier components $\phi_k$ of the field. To impose a cutoff $\Lambda$, we use something like
$$
\int D\phi = \prod_{|k|<\Lambda} \int d\phi_k.
$$
We are interested in relating the coupling constants in a theory having energy cutoff $\Lambda$ to the coupling constants in a theory having energy cutoff $b\Lambda,\;\; b<1$. We redefine $\phi \rightarrow \phi + \phi'$, where $\phi'$ has non-zero Fourier modes in $b\Lambda<|k|<\Lambda$ and $\phi$ has non-zero Fourier modes in $|k|<b\Lambda$. Integrating out the field $\phi'$ (meaning integrating our its Fourier modes) gives us some result written in terms of $\phi$. Whatever the result is, we include it by changing the Lagrangian to a new, \emph{effective} Lagrangian. The explicit disappearance of the highest energy quantum modes is compensated by some change in the Lagrangian. In general, $\mathcal{L}_{eff}$ contains all possible terms involving $\phi$ and its derivatives. This includes terms that were already present in the original Lagrangian. Integrating out these modes thus has the effect of changing the coefficients of terms in the Lagrangian. The effective Lagrangian is parameterized by the coefficients of these terms, and the act of integrating out modes can be considered as moving around inside the space of all possible Lagrangians.

If we let the parameter $b$ be infinitesimally less than $1$, $\mathcal{L}_{eff}$ will be infintesimally close to the original $\mathcal{L}$. Repeatedly integrating out these thin-shells in momentum space corresponds to a smooth motion through this Lagrangian space: this is \emph{renormalization group flow}. A more careful analysis of a particular theory would lead us the beta function $\beta(g)$ describing the dependence of a coupling parameter on some energy scale $\mu$:
\begin{equation}
\beta(g) = \frac{\partial g}{\partial \log(\Lambda)}=\Lambda \frac{\partial g}{\partial \Lambda}.
\end{equation}
We see that this is just picking out the exponent of the energy dependence in the coupling. If you have not had experience calculating $\beta-$functions, well, you are in for a real treat. It is such a pleasure, I will not spoil it by doing any of the calculations here. Enjoy.

At this point, it is clear why some theories with classical conformal invariance do not maintain conformal invariance quantum mechanically. For example, $\phi^4$ theory in $d=4$ dimensions can be shown to have the one-loop $\beta$-function
$$
\beta(g) = \frac{3}{16\pi^2}g^2.
$$
As we will soon see, the positive sign on this expression means the coupling constant increases with energy. Likewise, the (massless) QED one-loop $\beta$-function is
$$
\beta(e)=\frac{e^3}{12\pi^2}.
$$
These contrast with the one-loop QCD $\beta$-function,
$$
\beta(g)=-\left(11-\frac{2N_f}{3} \right)\frac{g^3}{16\pi^2}.
$$
By virtue of the fact that $N_f\leq 16$ in our universe\footnote{At last count.}, this $\beta$-function says the coupling decreases with energy. This is known as \emph{asymptotic freedom}. Each of these theories, although fine classically, have length scales introduced through quantum effects.

The $\beta-$functions controlling RG flow are of gradient type; the topology of RG flow is controlled by fixed points. Fixed points are those points in the coupling parameter space that have vanishing $\beta-$function. If $\beta$ is zero, clearly the coupling is a constant---it is scale invariant and does not change with energy scale. A fixed point $g_*$ of the RG thus corresponds to a scale-invariant (and as far as we are currently concerned, conformally-invariant) QFT. My claim is that these fixed points are crucial to our understanding of \emph{all} QFTs.

\begin{figure}
\centering
\includegraphics[scale=.4]{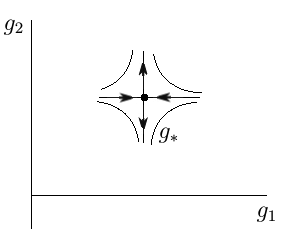}
\caption{Example of RG flow in the space of two couplings. The point $g_*$ is a fixed point, and both stable and unstable flows are visible. The direction of arrows represents the flow from high energies to low energies.} \label{fig:figure4}
\end{figure}

How do these fixed point CFTs control RG flow? Let's consider what RG flow is like in the neighborhood of a fixed point. In the parameter space of QFTs, a particular direction can be \emph{stable} or \emph{unstable}. A stable direction is attractive, in the sense that a flow along this direction will flow toward the fixed point. An unstable direction is repulsive and flows away from fixed points. There are also \emph{marginal} directions corresponding to flows where the coupling does not change. Examples of these types of flows can be seen in Figure \ref{fig:figure4}, where a marginal flow could for example correspond to motion out of the page. Truly marginal flows are  somewhat unusual quantum mechanically, as you have experienced\footnote{Or \emph{should} have experienced.}. A generic point in this diagram corresponds to some general quantum field theory, e.g., QCD at some energy scale described by some set of couplings. The properties of this QFT, however, are dictated largely by the fixed point.

Now that we have this understanding of RG flow in mind, we can characterize the interactions that appear in our theories. Interactions are \emph{relevant} if they are unstable and push you away from the fixed point--relevant operators grow in the infrared. Generally speaking, relevant operators have dimension $\Delta < d$\footnote{This can be seen from a naive counting of powers of energy in each operator. If an operator has $\Delta < d$, then in order to have a dimensionless action the associated coupling must have some scaling dimension. This scaling dimension will determine how the coupling flows---whether the operator contributes more or less at low energies}. Interactions are irrelevant if they are attractive in our RG flow diagram. Irrelevant operators are not important in the infrared and generally have dimension $\Delta > d$. Finally, there are marginal interactions/operators. Marginal operators are invariant under scale transformations. Instead of having isolated fixed points, we could have an entire manifold of conformal invariance. Marginal operators occur when $\Delta=d$, though we can already see why they are unusual---quantum mechanically, scale invariance gets broken and scaling dimensions receive anomalous corrections.

Finally, we understand the importance of conformal field theories to quantum field theory. Given a set of fields, the number of relevant (and marginal) operators is finite and small. Despite starting from general Lagrangians with general couplings, only a few couplings are important at low energies. And since any QFT lives in this ``coupling space'', we can think of any quantum field theory as a perturbation of a conformal field theory by relevant operators. That is to say, any point in our parameter space can be considered as a flow perturbed away from some fixed point CFT.

\begin{framed}
\noindent HOMEWORK: Consider a scalar field $\phi(x)$ with a standard free field kinetic term. For $d=6$, what are the relevant operators? What are the marginal operators? Repeat this for $d=4$ and $d=3$. BONUS: Which of these marginal operators remains marginal quantum mechanically?
\end{framed}

\subsection{A preview for future courses}

Of course, not everyone is interested in conformal field theory for its own sake. In many cases, conformal field theory is a tool used to study other interesting phenomena. One example can be found by considering just how symmetric a quantum field theory can possible be. For a long time, the conventional wisdom (and an important theorem) assured us that the maximal spacetime symmetry for a quantum field theory was conformal field theory. One way of getting around this is \emph{supersymmetry}. In supersymmetric theories, you allow for the existence of anticommuting symmetry generators. Thus the maximal spacetime symmetry is \emph{superconformal} field theory (SCFT). The simplest SCFT in $d=4$ dimensions is $\mathcal{N}=4$ super Yang-Mills (SYM) theory. Some would argue that this theory is the most important toy model of the past three decades, and it certainly deserves its own lecture\footnote{And it will get one! In the sequel to this course. Write to your local politicians.}. One amazing fact about $d=4$ $N=4$ SYM is that it maintains its conformal invariance quantum mechanically-- the $\beta$ function for this theory vanishes to all orders. This is only one example of how the constraints from conformal invariance can combine with additional constraints from supersymmetry.

Another important use for conformal field theories is they can allow us to define a quantum field theory without any reference to a Lagrangian. What we will see in later lectures is that in principle one can solve a CFT without even writing down a Lagrangian---we need only have knowledge of the spectrum and three-point functions of the theory. There are some very interesting theories that simply do not have a Lagrangian description, such as the 6d (2,0) SCFT and its compactifications. Unless you have had some exposure to these theories, their existence may seem a little bizarre. The program of solving a theory using conformal invariance and consistency conditions is the \emph{conformal bootstrap}. We will return to this program in some later lectures.

Along similar lines, conformal field theories provide one of our best understandings of quantum gravity through what is known as the AdS/CFT correspondence. There is a (conjectured) correspondence between a theory of quantum gravity and some dual conformal field theory. The quantum gravity lives in the bulk of a spacetime that behaves asymptotically like anti-de Sitter (AdS) space. AdS spacetimes are maximally symmetric spacetimes that have negative curvature; think of them as the Lorentzian analogues to hyperbolic space. The correspondence tells us that the quantum gravity in the AdS bulk has a formulation in terms of a conformal field theory (CFT) living on the boundary of that space in one fewer dimensions. When this duality holds, an understanding of CFTs can give us profound insights into quantum gravity. This is an extraordinarily active field of research that goes both directions; by studying weakly coupled quantum gravity, we can potentially gain new insights into strongly coupled quantum field theories. The AdS/CFT correspondence could fill an entire course or two; a later version of this course will introduce the basics.

The final tantalizing topic is string theory. String theory is a(/the?) candidate theory for the unification of all interactions. Instead of considering a point particle tracing out a worldline through time, we consider fundamental one-dimensional strings that trace out two-dimensional worldsheets. A CFT lives on this worldsheet moving through some background spacetime. String dynamics are described by a non-linear sigma model. Requiring the worldsheet theory to be a CFT quantum mechanically (that is, demanding the vanishing of the $\beta$ function) gives the string equations of motion---including Einstein's equations of general relativity. The perturbation theory of this sigma model involves an expansion in terms of $\ell_s / R$, where $\ell_s$ is the length of the fundamental string and $R$ is a length scale related to the background geometry (like the curvature).  Using the tools of conformal field theory, we can sum all contributions of world-sheet instantons and solve the theory exactly to all orders in perturbation theory. Because theories in $d=2$ dimensions are the best understood class of CFTs, we will frequently relate our dicussions to string theory. Obviously string theory could fill a few courses; I encourage you to take one if you get the opportunity. 

This is to say nothing of the first two applications discussed in more detail. Studying conformal field theory lets us determine critical exponents describing phase transitions at (quantum) critical points for entire classes of physics theories. And an appropriate unerstanding of CFT and relevant operators give us a powerful means of understanding contemporary renormalization group flow. Each of these topics should be studied, and each will get a lecture or two in a later version of this course.

\subsection{Conformal quantum mechanics}

We will finish this lecture by discussing the simplest type of conformal field theory: one living in $d=1$ dimensions. The one dimension corresponds to time, of course, so we are really talking about conformal quantum mechanics. This theory is simple enough that we can solve it and interesting enough to serve as a nontrivial introduction before continuing to CFTs in higher dimensions. Although we will only consider the theory as a toy model, it has proven useful in contemporary gravitational research. At the end of this lecture, I have provided references for a few randomly chosen works that use conformal quantum mechanics to serve as examples. 

As before, we will consider a conformally-invariant theory by actually considering a scale-invariant theory. Furthermore, we look to our previous examples and decide to start with the Lagrangian for a free particle
$$
L = \frac12 \dot{Q}^2.
$$
What additional terms can we add that will preserve both time-translational (corresponding to the usual Poincar\'e symmetry) and scale invariance? After some effort, you should be able to convince yourself that the most general Lagrangian we can write down is 
\begin{equation}
L=\frac12 \dot{Q}^2 - \frac{g}{2Q^2}.
\end{equation}
At this point, we do not specify whether $g$ is positive or negative (but ultimately, it turns out that $g>0$). 

By construction, we expect this theory to be invariant under time translations and rescalings. But the symmetry is enhanced beyond this. The action is invariant under conformal transformations of the time coordinate: 
\begin{equation}
t\rightarrow t'\equiv \frac{at+b}{ct+d}, \;\;\;Q \rightarrow \frac{Q}{ct+d},\;\;\; \mbox{   with   } ad-bc=1. \label{eq:cqmeqs}
\end{equation}
We wish to remark now (seemingly without motivation, though it is actually because I know what is coming,) that because $d=1$ and $Q$ scales with energy dimension $\Delta=-1/2$, the factor we gain when transforming $Q$ is equivalent to 
$$
\sqrt{\frac{\partial t'}{\partial t}}=\left| \frac{\partial t'}{\partial t}  \right|^{-\Delta/d}.
$$
It is straightforward to see that we can represent the transformation $t\rightarrow t'\equiv \frac{at+b}{ct+d}$ using the matrix
\begin{equation}
\left( \begin{array}{cc}
a & b \\
c & d
\end{array} \right).
\end{equation}
Using this description, successive composition of these transformations amounts to matrix multiplication. Therefore the conformal group for $d=1$ is $SL(2,\mathbb{R})$.
\begin{framed}
\noindent HOMEWORK: Prove the conformal quantum mechanics action is invariant under the transformations (\ref{eq:cqmeqs}). Keep in mind that the action being invariant still allows for the Lagrangian density to change by some total derivative.
\end{framed}

Of course, our physical intuition is somewhat obscured.We would like to understand what these conformal transformations are actually doing. It turns out the the group $SL(2,\mathbb{R})$ is homomorphic\footnote{This is a group homomorphism rather than a group isomorphism. The group isomorphism is between $SO(2,1)\simeq SL(2,\mathbb{R})/\mathbb{Z}_2$. This $\mathbb{Z}_2$ redundancy is apparent from the transformation (\ref{eq:cqmeqs}); we could take the negative of $a,b,c,d$ and it corresponds to the same transformation.} to the group $SO(2,1)$. We will not give a full proof of this fact here, but we will briefly explore this fact. To begin, we consider the algebra of $SL(2,\mathbb{R})$. Every group element $g$ can be parameterized as\footnote{We are not allowing $a=-1$. Work out what this special case is like on your own.}
\begin{equation}
g = \left( \begin{matrix}
1+a & b \\
c & \frac{1+bc}{1+a}
\end{matrix} \right).
\end{equation}
Close to the identity element (meaning for infinitesimal parameters), this element becomes
\begin{equation}
g = \left( \begin{matrix}
1+a & b \\
c & 1-a
\end{matrix} \right).
\end{equation}
From this, we determine the infinitesimal generators
\begin{equation}
X_1 =  \begin{pmatrix}
1 & 0 \\
0 & -1
\end{pmatrix} ,\;\;
X_2 = \left( \begin{matrix}
0 & 1 \\
0 & 0
\end{matrix} \right),\;\;
X_3 = \left( \begin{matrix}
0 & 0 \\
1 & 0
\end{matrix} \right).
\end{equation}
From these expressions, we easily find the algebra
\begin{equation}
[X_1,X_2]=2X_2, \;\;\;\;[X_1,X_3]=-2X_3, \;\;\;\;[X_2,X_3]=X_1.
\end{equation}
\begin{framed}
\noindent HOMEWORK: Explicitly check this easily found algebra.
\end{framed}

My claim is that this Lie algebra is the same as $so(2,1)$. To see this, recall (or go look up) the Lie algebra of $SO(2,1)$:
\begin{equation}
[M_{ab},M_{cd}] = \eta_{bc} M_{ad} - \eta_{bd} M_{ac} + \eta_{ad} M_{bc} - \eta_{ac} M_{bd},
\end{equation}
with $a,b=0,1,2$ and $\eta=\mbox{diag}(-1,+1,-1)$. We now introduce the following generators:
\begin{equation}
P=M_{02}-M_{01}, \;\;K=M_{02}+M_{01},\;\; D = M_{21}.
\end{equation}
It is then straightforward to show in terms of these generators the Lie algebra becomes
\begin{equation}
[D,P]=-P,\;\;\;\;[D,K]=K,\;\;\;\;[P,K]=2D. \label{eq:cqmalg}
\end{equation}
\begin{framed}
\noindent HOMEWORK: Explicitly check this algebra.
\end{framed}
\noindent Under appropriate redefinitions (what are they?),this is exactly the Lie algebra $sl(2,\mathbb{R})$! You may have seen this algebra written in a form obtained by rescaling $P\rightarrow -iP, K\rightarrow -iK, D\rightarrow \frac{i}{2} D$ As we will soon see, $P$ corresponds to time translation and $D$ corresponds to scale transformations. But what about the generator $K$? This is some new, \emph{special} transformation.

We can see the infinitesimal transformations corresponding to these generators in a few ways. We could simply realize that the algebra (\ref{eq:cqmalg}) has the differential realization
\begin{equation}
P=\frac{d}{dt}, \;\;\;\; D = t \frac{d}{dt}, \;\;\;\;K=t^2 \frac{d}{dt}.
\end{equation}
Following the standard procedure (which will be reviewed next lecture), it can be shown the associated finite transformations are
\begin{equation}
P:t\rightarrow t+a, \;\;\;\; D:t\rightarrow ct,\;\;\;\;K:t\rightarrow\frac{t}{1+bt}. 
\end{equation}
Alternatively, we could consider the infinitesimal transformation
\begin{equation}
t\rightarrow\frac{(1+\alpha)t+\beta}{\gamma t+1-\alpha} \approx t + \beta + 2\alpha t - \gamma t^2.
\end{equation}
These infinitesimal transformations again lead to the above finite transformations. Before continuing, we remark upon the curious case of the generator $K$. How are we to understand this special transformation? For now, we only remark that it is equivalent to an inversion, followed by a translation, followed again by an inversion (Check this!).

From here, we could continue studying this theory. For example, we could determine how (some representation of) the infinitesimal generators act on $Q$:
\begin{align}
i[H,Q]&= \frac{d}{dt}Q \\
i[D,Q]&= t\frac{d}{dt}Q-\frac12 Q \\
i[K,Q]&= t^2 \frac{d}{dt}Q-t Q.
\end{align}
We could then use $SO(2,1)$ representation theory (similar to angular momentum results in quantum mechanics) to find eigenstates by constructing ladder operators $L_+,L_-$ and the operator $R$, where
\begin{equation}
R |n\rangle=r_n|n\rangle,\;\;\;\; r_n=r_0+n \;\;(n\in\mathbb{N}),\;\;\;\; \langle m|n\rangle = \delta_{m,n}. \\
\end{equation}
Here the lowest state eigenvalue $r_0>0$ is related to the quadratic Casimir invariant
\begin{equation}
\mathcal{C}\equiv R^2-L_+L_-=\frac{g}{4}-\frac{3}{16},\;\; \;\;\;\;\;\mathcal{C}|n\rangle = r_0(r_0-1)|n\rangle.
\end{equation}
By studying the constraints placed on the theory by conformal symmetry, we could show that the two-point correlator between two fields $Q_h$ and $Q_{h'}$ of scaling dimensions $h$  and $h'$ respectively is fixed to be
\begin{equation}
\langle Q_h(t) Q_{h'}(t') \rangle \sim (t-t')^{-2h}\delta_{h,h'}.
\end{equation}

We could talk about issues with normalizability in this theory, or that we seemingly can not find a normalized vacuum state annihilated by all of the group generators, or study the superconformal extensions of this model, or pursue applications related to two-dimensional gravity via the $AdS/CFT$ correspondence. Instead I will provide some randomly chosen refererences at the end of this lecture to highlight some recent applications of conformal quantum mechanics. If you are interested in these topics, I sincerely recommend reading more about them on your own.
\break
\subsection*{References for this lecture}
\vspace{4mm}
\noindent Main references for this lecture
\\
\begin{list}{}{%
\setlength{\topsep}{0pt}%
\setlength{\leftmargin}{0.7cm}%
\setlength{\listparindent}{-0.7cm}%
\setlength{\itemindent}{-0.7cm}%
\setlength{\parsep}{\parskip}%
}%
\item[]

[1] J. Gomis, \emph{Conformal Field Theory: Lecture 1}, C10035:PHYS609, (Waterloo, Perimeter Institute for Theoretical Physics, 21 November 2011), Video.

[2] Chapter 3 of the textbook: P. Di Francesco, P. Mathieu, and D. Senechal. \emph{Conformal field theory}, Springer, 1997.

[3] Chapters 8 and 12 of the textbook: M. Peskin and D. Schroeder, \emph{An Introduction to Quantum Field Theory}, Westview Press, 1995.

\end{list}

\break

\section{Lecture 2: CFT in $d\geq 3$}

In this lecture, we will study the conformal group for $d\geq3$ dimensions. This is not a typo; we will treat $d=2$ as the particularly interesting case that it is. We will focus on infinitesimal conformal transformations, the conformal algebra and group, representations of the conformal group, radial quantization,the state-operator correspondence, unitarity bounds,  and constraints from conformal invariance imposed on correlation functions. Many of these ideas are also important in $d=2$ dimensions, so this lecture will also serve as an introduction for the richer case of conformal field theory in $d=2$ dimensions.

\subsection{Conformal transformations for $d\geq 3$}

Consider the $d$-dimensional space $\mathbb{R}^{p,q}$ (with $p+q=d$) with flat metric $g_{\mu\nu}=\eta_{\mu\nu}=\text{diag}(-1,\dots,+1,\dots)$ of signature $(p,q)$ and line element $ds^2 = g_{\mu\nu} dx^\mu dx^\nu$. We define a differentiable map $\phi$ as a conformal transformation if $\phi:g_{\mu\nu}(x)\rightarrow  g'_{\mu\nu}(x')=\Lambda(x)g_{\mu\nu}(x)$. Under a coordinate transformation $x\rightarrow x'$, the metric tensor transforms as $g_{\rho\sigma}\rightarrow g'_{\rho\sigma}(x') = \frac{\partial x'^{\mu}}{\partial x^{\rho}} \frac{\partial x'^{\nu}}{\partial x^{\sigma}}  g_{\mu\nu}(x)$ so that conformal transformations of the flat metric therefore obey
\begin{equation}
\eta_{\rho\sigma}  \frac{\partial x'^{\rho}}{\partial x^{\mu}}\frac{\partial x'^{\sigma}}{\partial x^{\nu}} =\Lambda(x) \eta_{\mu\nu} . \label{eq:eq2p1}
\end{equation}
The positive function $\Lambda(x)$ is called the scale factor. The case $\Lambda(x)=1$ clearly corresponds to the Poincar\'e group consisting of translations and Lorentz rotations, and the case where $\Lambda(x)$ is some constant corresponds to global scale transformations. It is also clear from this definition that conformal transformations are coordinate transformations preserving the angle ${u\cdot v / (u\cdot u \,\, v\cdot v)^{1/2}}$ between vectors $u$ and $v$. 

To begin, we consider infinitesimal coordinate transformations to first order in $\epsilon(x)\ll 1$:
\begin{equation}
x'^\mu = x^\mu + \epsilon^\mu(x) + \mathcal{O}(\epsilon^2). \label{eq:eq2p2}
\end{equation}
Under such a transformation, the LHS of eq. (\ref{eq:eq2p1}) becomes
\begin{eqnarray}
\eta_{\rho\sigma} \frac{\partial x'^{\rho}}{\partial x^{\mu}}\frac{\partial x'^{\sigma}}{\partial x^{\nu}}
&=& \eta_{\rho\sigma} \left(\delta^{\rho}_{\mu}+\frac{\partial \epsilon^\rho}{\partial x^{\mu}} + \mathcal{O}(\epsilon^2)  \right)   \left( \delta^{\sigma}_{\nu}+\frac{\partial \epsilon^\sigma}{\partial x^{\nu}} + \mathcal{O}(\epsilon^2) \right) \nonumber \\
&=& \eta_{\mu\nu} + \left( \frac{\partial \epsilon_{\mu}}{\partial x^\nu} + \frac{\partial \epsilon_{\nu}}{\partial x^\mu} \right) + \mathcal{O}(\epsilon^2). \label{eq:eq2p3} 
\end{eqnarray}
Then in order for such an infinitesimal transformation to be conformal, we see that to first order in $\epsilon$ we must have
\begin{equation}
\partial_\mu \epsilon_\nu + \partial_\nu \epsilon_\mu = f(x)\eta_{\mu\nu},
\label{eq:eq2p4}
\end{equation}
where $f(x)$ is some function and we use the notation $\partial_\mu \equiv \frac{\partial}{\partial x^\mu}$. Tracing both sides of eq. (\ref{eq:eq2p4}) with $\eta^{\mu\nu}$, we find that $f(x)=\frac2d \partial^\mu \epsilon_\mu$. Substituting this back into eq. (\ref{eq:eq2p4}) thus gives
\begin{equation}
\partial_\mu \epsilon_\nu + \partial_\nu \epsilon_\mu = \frac{2}{d}(\partial_\rho \epsilon^\rho) \eta_{\mu\nu}.
\label{eq:eq2p5}
\end{equation}
We can also read off at this point that the scale factor for this infinitesimal coordinate transformation is 
$$\Lambda(x) = 1 + \frac2d (\partial_\mu \epsilon^\mu) + \mathcal{O}(\epsilon^2).$$

In order to proceed, we will derive two useful expressions that will soon prove useful\footnote{The author has the benefit of standing on some rather giant shoulders. If these steps seem arbitrary, just have patience that we will use the results we now find. Also, the idea of ``giant'' shoulders is somewhat amusing, given that we are studying conformal field theory.}. Acting on equation (\ref{eq:eq2p5}) with $\partial^\nu$ gives
\begin{equation}
\partial_\mu (\partial \cdot \epsilon) + \Box \epsilon_\mu = \frac2d \partial_\mu (\partial \cdot \epsilon),
\label{eq:eq2p6}
\end{equation}
where $\partial \cdot \epsilon \equiv \partial_\mu \epsilon^\mu$ and $\Box\equiv\partial_\mu \partial^\mu$. Acting on this expression in turn with $\partial_\nu$ gives
\begin{equation}
\partial_\mu \partial_\nu (\partial \cdot \epsilon) + \Box \partial_\nu \epsilon_\mu = \frac2d \partial_\mu \partial_\nu (\partial \cdot \epsilon).
\label{eq:eq2p7}
\end{equation}
By exchanging $\mu \leftrightarrow \nu$ in eq. (\ref{eq:eq2p7}), adding the result back to eq. (\ref{eq:eq2p7}), and using (\ref{eq:eq2p5}), we obtain
\begin{equation}
\left( \eta_{\mu\nu} \Box + (d-2)\partial_\mu \partial_\nu \right) (\partial \cdot \epsilon) = 0.
\label{eq:eq2p8}
\end{equation}
Contracting this equation with $\eta^{\mu\nu}$ finally gives
\begin{equation}
(d-1) \Box (\partial \cdot \epsilon)=0.
\label{eq:eq2p9}
\end{equation}

Before deriving a second expression, we remark upon the dimensional dependence in  equations (\ref{eq:eq2p8}) and (\ref{eq:eq2p9}). So long as $d\geq 3$, eq. (\ref{eq:eq2p9}) takes an identical form---this lecture will focus on this case. In the case $d=2$, however, equation (\ref{eq:eq2p9}) does not follow from equation (\ref{eq:eq2p8})---the second term on the LHS vanishes in two spacetime dimensions. We will consider conformal transformations in two dimensions in the next lecture. For the case of $d=1$, well, we already considered conformal quantum mechanics. We say no more of it here. For the remainder of this lecture, we will only consider conformal transformations in $d\geq 3$ spacetime dimensions.

The second expression we will find useful is also obtained from eq. (\ref{eq:eq2p5}). We act with the derivate $\partial_\rho$ and permute the indices to obtain
\begin{eqnarray}
\partial_{\rho} \partial_{\mu} \epsilon_\nu + \partial_{\rho} \partial_{\nu} \epsilon_\mu &=& \frac2d \eta_{\mu\nu}\partial_\rho (\partial \cdot \epsilon), \label{eq:eq2p1p1} \\
\partial_{\nu} \partial_{\rho} \epsilon_\mu + \partial_{\mu} \partial_{\rho} \epsilon_\nu &=& \frac2d \eta_{\rho\mu}\partial_\nu (\partial \cdot \epsilon), \label{eq:eq2p1p2}\\
\partial_{\mu} \partial_{\nu} \epsilon_\rho + \partial_{\nu} \partial_{\mu} \epsilon_\rho &=& \frac2d \eta_{\nu\rho}\partial_\mu (\partial \cdot \epsilon). \label{eq:eq2p1p3}
\end{eqnarray}
Subtracting (\ref{eq:eq2p1p1}) from the sum of (\ref{eq:eq2p1p2}) and (\ref{eq:eq2p1p3}) gives the expression
\begin{equation}
2 \partial_\mu \partial_\nu \epsilon_\rho = \frac2d \left(  -\eta_{\mu\nu}\partial_{\rho}+\eta_{\rho\mu}\partial_\nu + \eta_{\nu\rho}\partial_\mu \right) (\partial \cdot \epsilon).
\label{eq:eq2p10}
\end{equation}
Now we can continue.

\subsection{Infinitesimal conformal transformations for $d \geq 3$}

Consider again equation (\ref{eq:eq2p9}). This equation implies that $(\partial\cdot\epsilon)$ can be at most linear in $x^\mu$. It follows that $\epsilon_\mu$ is at most quadratic in $x^\nu$ and thus will be of the form
\begin{equation}
\epsilon_\mu = a_\mu + b_{\mu\nu}x^\nu + c_{\mu\nu\rho}x^\nu x^\rho.
\label{eq:eq2p11}
\end{equation}
Here the coefficients $a_\mu,b_{\mu\nu},c_{\mu\nu\rho}\ll 1$ are constants, and the constant $c_{\mu\nu\rho}$ is symmetric in its last two indices. Because the constraints derived here for conformal invariance must be independent of the position $x^\mu$ (this should be a confomal transformation regardless of the value of $x^\mu$), the terms in equation ($\ref{eq:eq2p11}$) can be studied individually.

First, we consider the constant term $a_\mu$. This term corresponds to an infinitesimal translation. The corresponding generator is the momentum operator $P_\mu = -i\partial_\mu$ (this would be a good time to start remembering how infinitesimal transformations relate to their generators, by the way). The term linear in $x$ is more interesting. Inserting a linear term into eq. (\ref{eq:eq2p5}) gives
\begin{equation}
b_{\mu\nu}+b_{\nu\mu}=\frac2d \left( \eta^{\rho\sigma} b_{\rho\sigma} \right) \eta_{\mu \nu}.
\label{eq:eq2p12}
\end{equation}
This equation constrains the symmetric part of $b$ to be proportial to the metric. We therefore divide the $b_{\mu\nu}$ coefficient as
\begin{equation}
b_{\mu\nu}=\alpha \eta_{\mu\nu}+m_{\mu\nu},
\label{eq:eq2p13}
\end{equation}
where $m$ is antisymmetric in its indices and $\alpha$ is some parameter that can be found in terms of eq. (\ref{eq:eq2p12}).  The antisymmetric $m_{\mu\nu}$ corresponds to infinitesimal Lorentz rotations $x'^{\mu}=(\delta^\mu_\nu+m^\mu_\nu)x^\nu$. The generator corresponding to these rotations is the angular momentum momentum operator $L_{\mu\nu} = i(x_\mu\partial_\nu-x_\nu\partial_\mu)$ (Remember generators? Like momentum, these are hopefully familiar to you). The symmetric part of this expression corresponds to infinitesimal scale transformations $x'^\mu=(1+\alpha)x^\mu$ with corresponding generator $D=-ix^\mu \partial_\mu$.

We have skipped the derivations of the momentum and angular momentum operators as generators of translations and Lorentz rotations because they should be familiar. At this point we will pause and consider the generator of scale transformations.  Generic infinitesimal transformations may be written as
\begin{eqnarray}
x'^\mu &=& x^\mu + \epsilon_a \frac{\delta x^\mu}{\delta\epsilon_a} \nonumber \\
 \phi'(x') &=& \phi (x) + \epsilon_a \frac{\delta F}{\delta \epsilon_a(x)}(x), \label{eq:inftrans}
\end{eqnarray}
where $F$ is the function relating the new field $\phi'$ evaluated at the transformed coordinate $x'$ to the old field $\phi$ at $x$
$$
\phi'(x')=F(\phi(x)),
$$
and we are keeping infinitesimal parameters $\{ \epsilon_a \}$ to first order. The convention we follow is that the generator $G_a$ of a transformation action as
\begin{equation}
\phi'(x)-\phi(x) \equiv -i \epsilon_a G_a \phi(x),
\end{equation}
so that 
\begin{equation}
i G_a \phi = \frac{\delta x^\mu}{\delta \epsilon_a} \partial_\mu \phi - \frac{\delta F}{\delta \epsilon_a}. \label{eq:eq1p20}
\end{equation}
If we suppose that the fields are unaffected by the transformation such that $F(\phi)=\phi$ (we will return to this supposition momentarily), then the last term in equation (\ref{eq:eq1p20}) vanishes. Under infinitesimal scale transformations with generator $D$, $x \rightarrow e^\epsilon x \approx (1+\epsilon)x$ so that
\begin{eqnarray}
i D \phi &=&  \frac{\delta x^\mu}{\delta \epsilon}\partial_\mu \phi \nonumber \\
\Rightarrow D \phi &=& -i x^\mu \partial_\mu \phi.
\end{eqnarray}
This is exactly what we previously claimed.

By this point, we have rediscovered the Poincar\'e group supplemented with scale transformations. So far this case is similar to that of conformal quantum mechanics; we expect the remaining quandratic term may correspond to the new, \emph{special} transformation found previously.  What, then, is the transformation corresponding to terms of $\epsilon$ quadratic in $x$? We first insert the quadratic term into expression (\ref{eq:eq2p10}) to see that the parameter $c_{\mu\nu\rho}$ can actually be expressed in the form
\begin{equation}
c_{\mu\nu\rho}=\eta_{\mu\rho} b_\nu + \eta_{\mu\nu} b_{\rho} - \eta_{\nu\rho} b_{\mu}, \;\;\;\;\; b_\mu = \frac{1}{d} c^\nu_{\nu\mu}.
\end{equation}
These transformations are called \emph{special conformal transformations}. Using this expression, we see that they have the infinitesimal form 
\begin{equation}
x'^\mu = x^\mu + 2(x\cdot b)x^\mu - (x^2)b^\mu.
\end{equation}
 After some straightforward calculation, you should be able to convince yourself that the corresponding generator is $K_\mu = -i (2x_\mu x^\nu \partial_\nu - (x^2) \partial_\mu)$. 
\\
\\
\noindent $ \boxed{
  \mbox{HOMEWORK: Do this straightforward calculation. }
 }$
\\
\\
\noindent Combining these new generator expressions with the familiar Poincar\'e generators, we find the generators of the conformal group to be
\begin{align}
&P_\mu = -i \partial_\mu \nonumber \\
&L_{\mu\nu} = i(x_\mu\partial_\nu - x_\nu\partial_\mu) \\
&D = -i x^\mu\partial_\mu \nonumber \\
&K_\mu = -i(2x_\mu x^\nu \partial_\nu - x^2 \partial_\mu) \nonumber
 \end{align} 
 Special conformal transformations are still new and interesting, so it is to them that we now turn our attention.

\subsection{Special conformal transformations and conformal algebra}

The finite conformal tranformations corresponding to most of these infinitesimal conformal transformations are familiar: momentum generates translations, angular momentum generates Lorentz rotations, and $D$ (which we will call the \emph{dilatation} operator) generates scale transformations. But what is a special conformal transformation? We leave it as an exercise to show that the finite transformation associated with the special conformal generator is 
\begin{equation}
x'^\mu = \frac{x^\mu-(x\cdot x) b^\mu}{1-2(b\cdot x)+(b\cdot b)(x \cdot x)}. \label{eq:eq3p20}
\end{equation}
\\
\noindent $ \boxed{
  \mbox{HOMEWORK: Derive the finite special conformal transformation.}
  }$
\\
\\
\noindent The scale factor for special conformal transformations can be shown to be
\begin{equation}
\Lambda(x) = \left( 1-2(b\cdot x) + (b\cdot b)(x \cdot x)  \right)^2.
\end{equation}
\\
$\boxed{
  \mbox{HOMEWORK: Calculate the scale factor for a special conformal transformation.}
 }$
\\
\\
There is a more intuitive understanding of special conformal tranformations motivated in part by our analysis of conformal quantum mechaics. Let us allow ourselves to consider discrete tranformations known as \emph{inversions}:
$$
x^\mu \rightarrow \frac{x^\mu}{x^2}.
$$
Using inversions, we can express finite special conformal transformations in the form
\begin{equation}
\frac{x'^\mu}{x' \cdot x'} = \frac{x^\mu}{x \cdot x} - b^\mu.
\label{eq:eq2p71}
\end{equation}
We see that special conformal tranformations can be thought of as an inversion of $x$, followed by a translation by $b$, followed by another inversion. Note that inversion is a discrete transformation rather than continuous. We are interested only in the continuous transformations associated with the conformal group, and therefore only mention these inversions.

We should also address a potential issue with finite special conformal transformations: they are not globablly defined. From eq. (\ref{eq:eq3p20}), we see that for the point $x^\mu = \frac{1}{b^2} b^\mu$ the denominator vanishes. Even considering the numerator does not resolve this singularity, and we find that $x^\mu$ in this case is mapped to infinity. In order to define finite conformal transformations globablly we should consider the 
$conformal\,\,compactifications$, where additional points are included. We will consider this in more detail in the next lecture for $d=2$ dimensions.

Now that we have discussed the generators, we present the associated algebra. Using the explicit infinitesimal forms, we find
\begin{align}
[D, P_{\mu}] &= \,i P_{\mu} \nonumber \\
[D, K_{\mu}] &= -i K_{\mu} \nonumber \\
[K_{\mu}, P_\nu] &=  2 i ( \eta_{\mu\nu} D - L_{\mu\nu} )  \label{eq:eq228}\\
[K_{\rho}, L_{\mu\nu}] &= i(\eta_{\rho\mu} K_{\nu}-\eta_{\rho\nu} K_{\mu}) \nonumber \\
[P_{\rho}, L_{\mu\nu}] &= i(\eta_{\rho\mu}P_{\nu} - \eta_{\rho\nu}P_{\mu})  \nonumber \\
[L_{\mu\nu}, L_{\rho\sigma}] &= i(\eta_{\nu\rho} L_{\mu\sigma} + \eta_{\mu\sigma}L_{\nu\rho} - \eta_{\mu\rho} L_{\nu\sigma} - \eta_{\nu\sigma} L_{\mu\rho}) \nonumber
\end{align}
These formulas will prove essential in the work that follows. We recover the Poincar\'e algebra by ignoring the commutators with $D$ or $K_\mu$.
\\
\\
\noindent $ \boxed{
  \mbox{HOMEWORK: Explicitly prove at least four of these six equations.}
  }$
\\

\subsection{Conformal group}

Let us now consider the conformal group for $d\geq 3$. As is frequently the case, we will study the group by considering its associated algebra; the conformal algebra is the Lie algebra corresponding to the conformal group. How many generators are in this algebra? We can count the generators explicitly as
\begin{eqnarray}
1 \text{  dilatation} \;\;&+&\;\; d \text{  translations} +\;\; d \text{  special conformal} \nonumber \\
 &+& \frac{d(d-1)}{2} \text{  rotations} = \frac{(d+2)(d+1)}{2} \text{  generators}. \nonumber
\end{eqnarray}
This is precisely the number of generators for an $SO(d+2)$-type algebra (convince yourself of this). This result is not coincidental. Guided by this (and the work of countless others before us), we define alternate generators
\begin{eqnarray}
J_{\mu,\nu} &\equiv& L_{\mu\nu} \nonumber \\
J_{-1,\mu}&\equiv& \frac12 (P_\mu-K_\mu)  \nonumber \\
J_{0,\mu}&\equiv& \frac12 (P_\mu+K_\mu)  \\ 
J_{-1,0}&\equiv& D \nonumber
\end{eqnarray}
These particular generators can be shown to satisfy
\begin{equation}
\left[ J_{mn},J_{pq}   \right] = i \left( \eta_{mq}J_{np} + \eta_{np}J_{mq} - \eta_{mp}J_{nq} - \eta_{nq}J_{mp} \right). \label{eq:129}
\end{equation}
For Euclidean space $\mathbb{R}^{d,0}$, the metric used is diag$(-1,1,\dots,1)$. In this case, these commutation relations correspond to the Lie algebra $so(d+1,1)$. For Minkowski space $\mathbb{R}^{d-1,1}$, the metric used is diag$(-1,-1,1,\cdots 1)$. In this case, these commutation relations correspond to the Lie algebra $so(d,2)$. For $d = p+q$, the conformal algebra is clearly $so(p+1,q+1)$.
\\
\\
\noindent $ \boxed{
  \mbox{HOMEWORK: Explicitly check that our conformal algebra satisfies equation (\ref{eq:129}).}
  }$
\\

The conformal group in $d\geq3$ dimensions is apparently $SO(d,2)$. Although we derived it in a completely different manner, this matches the result we got for conformal quantum mechanics in $d=1$ dimension. Before proceeding to the next topic, we remark that the Poincar\'e and dilatation operators for a subalgebra---a theory could be Poincar\'e and scale invariant without necessarily being invariant under the full conformal group. This point was mentioned earlier, and we will return to it again.

\subsection{Representations of the conformal group}

Earlier we supposed that the infinestimal conformal generators had no effect on fields. We will now consider how classical fields are affected by conformal generators. In general, conformal invariance at the quantum level does not follow from conformal invariance at the classical level. Regularization prescriptions introduce a scale to the theory which breaks the conformal symmetry except at RG fixed points. But we will return to this difficulty later. For now, we seek a matrix representation $T_a$ such that under an infinitesimal conformal transformation parameterized by $\epsilon_a$ a field $\Phi(x)$ transforms as
\begin{equation}
\Phi'(x')=(1-i\epsilon_a T_a) \Phi(x).
\end{equation}

In order to find the allowed forms of these generators, we borrow a trick from the Poincar\'e algebra. We begin by studying the Lorentz group--the subgroup of the Poincar\'e group that leaves the point $x=0$ invariant. We define the action of infinitesimal Lorentz transformations on the field $\Phi(0)$ by introducing the matrix representation $S_{\mu\nu}$,
\begin{equation}
L_{\mu\nu}\Phi(0) = S_{\mu\nu}\Phi(0).
\end{equation}
$S$ is the spin operator associated with the field $\Phi$ (constructed from $\gamma$ matrices, for example). By using the Hausdorff formula
\begin{equation}
e^{-A} B e^{A} = B + [B,A] + \frac{1}{2!} [[B,A], A ] + \frac{1}{3!} [[[B,A], A], A] + \cdots
\end{equation}
we can translate the generator $L_{\mu\nu}$ to nonzero values of $x$ and find
\begin{equation}
e^{i x^\lambda P_\lambda} L_{\mu\nu} e^{-ix^\lambda P_\lambda} = L_{\mu\nu} - x_\mu P_\nu + x_\nu P_\mu.
\end{equation}
Using this fact, we determine
$$
P_\mu \Phi(x) = -i \partial_\mu \Phi(x)
$$
\begin{equation}
L_{\mu\nu} \Phi(x) = i(x_\mu\partial_\nu - x_\nu \partial_\mu)\Phi(x) + S_{\mu\nu}\Phi(x). \label{eq:130}
\end{equation}
\begin{framed}
\noindent HOMEWORK: Using the equations preceding it, derive equation (\ref{eq:130}). 
\end{framed}

Now consider the full conformal group. The derivation is nearly identical: we consider the subgroup that leaves the origin $x=0$ invariant generated by rotations, dilatations, and special conformal transformations. If we denote the values of the generators $L_{\mu\nu}, D$ and $K_\mu$ at $x=0$ by $S_{\mu\nu}$, $\tilde{\Delta},$ and $ \kappa_\mu$, these values must form a matrix representation of the reduced algebra

\begin{align}
[\tilde{\Delta},S_{\mu\nu}] &= 0 \nonumber \\
[\tilde{\Delta},\kappa_\mu] &= -i\kappa_\mu \nonumber  \\
[\kappa_\mu, \kappa_\nu] &= 0 \label{eq:136}\\
[\kappa_\rho,S_{\mu\nu}] &= i(\eta_{\rho\mu}\kappa_\nu - \eta_{\rho\nu} \kappa_\mu  ) \nonumber \\
[S_{\mu\nu}, S_{\rho\sigma}] &= i( \eta_{\nu\rho}S_{\mu\sigma} + \eta_{\mu\sigma}S_{\nu\rho} -\eta_{\mu\rho}S_{\nu\sigma} - \eta_{\nu\sigma}S_{\mu\rho} )\nonumber 
\end{align}
\begin{framed}
\noindent HOMEWORK: Look at eq. (\ref{eq:136}). Compare it to eq. (\ref{eq:eq228}). Look at them again. Convince yourself of the validity of eq. (\ref{eq:136}).
\end{framed}
\noindent Following steps similar to before, we can show
\begin{align}
e^{i x\cdot P} D e^{-ix\cdot P} &= D + x^\nu P_\nu. \nonumber \\
e^{i x\cdot P} K_{\mu} e^{-ix\cdot P} &= K_{\mu}+2x_\mu D -2 x^\nu L_{\mu\nu} + 2x_\mu (x^\nu P_\nu)-x^2 P_\mu. \label{eq:eq138}
\end{align}
Using these in turn, we derive the transformation rules
\begin{align}
D\Phi(x) &= (-ix^\nu\partial_\nu + \tilde{\Delta})  \Phi(x)   \nonumber \\
K_\mu \Phi(x) &= \left[ \kappa_\mu + 2x_\mu\tilde{\Delta}- x^\nu S_{\mu\nu} -2ix_\mu x^\nu\partial_\nu + i x^2 \partial_\mu \right] \Phi(x) \label{eq:eq139}
\end{align}
\begin{framed}
\noindent HOMEWORK: Derive eqs. (\ref{eq:eq138}) and (\ref{eq:eq139}).
\end{framed}

To proceed, we make use of some well-known facts from mathematics that I present without proof. First, we consider a field $\Phi(x)$ that belongs to an irreducible representation of the Lorentz group. According to Schur's lemma, any matrix that commutes with the generators $S_{\mu\nu}$ must be a multiple of the identity. Thus, $\tilde{\Delta}$ is some number. What number? For starters, convince yourself that representations of the dilatation group on classical fields are not unitary\footnote{We are trying to construct a finite-dimensional representation being acted upon my dilatations. But dilatations are not bounded, and a finite dimensional representation of a noncompact Lie algebra is necessarily nonunitary. You have seen this before when considering the boosts of the Lorentz group.}. Thus the generator $\tilde{\Delta}$ is non-Hermitian. The number $\tilde{\Delta}$ equals $-i\Delta$, where $\Delta$ is the \emph{scaling} \emph{dimension} of the field $\Phi$.

We have not really explicitly explained defined yet what we mean by the scaling dimension. The scaling dimension $\Delta$ of a field is defined by the action of a scale transformation on the field $\Phi$ according to
\begin{equation}
\Phi(\lambda x) = \lambda^{-\Delta} \Phi(x).
\end{equation}
For example, consider the action for a free massless scalar field in flat space
\begin{equation}
S = \int d^d x \;\;\partial_\mu \phi(x) \partial^\mu \phi(x).
\end{equation}
In order for the action to be a scale invariant dimensionless quantity, the scaling dimension of the field $\phi$ must be
\begin{equation}
\Delta = \frac12 d - 1.
\end{equation}
\begin{framed}
\noindent HOMEWORK: Verify this is the case. Considering only even $n$ (why?), what terms $\phi^n$ can be added to the Lagrangian that preserve classical scale invariance?
\end{framed}
\noindent We have been finding classical scaling dimensions for awhile now (finding relevant operators, for example); we just have not explicitly noted it.

Finally, the fact that $\tilde{\Delta}$ is proportional to the identity matrix also means that the matrices $\kappa_\mu$ vanish.
This gives us the transformation rules for the field $\Phi(x)$:
\begin{align}
P_\mu \Phi(x) &= -i \partial_\mu \Phi(x)\nonumber \\
L_{\mu\nu} \Phi(x) &= i(x_\mu \partial_\nu - x_\nu \partial_\mu) \Phi(x) + S_{\mu\nu} \Phi(x) \nonumber\\
D \Phi(x) &= -i(x^\mu \partial_\mu + \Delta) \Phi(x) \nonumber\\
K_\mu \Phi(x) &= (-2i\Delta x_\mu - x^\nu S_{\mu\nu} - 2ix_\mu x^\nu \partial_\nu + ix^2 \partial_\mu) \Phi(x)\nonumber
\end{align}
Using these expressions, we can derive the change in $\Phi$ under a finite conformal transformation. In this lecture, I will only give the result for spinless fields; the derivation is left as an exercise. For the scale factor $\Lambda(x)$, the Jacobian of the conformal transformation $x\rightarrow x'$ is given by
\begin{equation*}
\left\lvert \frac{\partial x'}{\partial x} \right\rvert = \frac{1}{\sqrt{\det g'_{\mu\nu}}}=\Lambda(x)^{-d/2}
\end{equation*}
so that the spinless field $\phi$ transforms as
\begin{equation}
\phi(x) \rightarrow \phi'(x') = \left\lvert \frac{\partial x'}{\partial x}\right\rvert ^{-\Delta/d} \phi(x).
\end{equation}
Notice that this is exactly the transformation rule we found when studying conformal quantum mechanic. Fields that have transform according to this expression are called \emph{quasi-primary} fields.

\subsection{Constraints of Conformal Invariance}

We have seen how conformal transformations act on quasi-primary fields. Now we turn our attention to constraints imposed by conformal invariance. We begin by considering the observables of our theory. The quantities of interest in conformal field theories are $N$-point correlation functions of fields. By ``field'', we mean some local quantity having coordinate dependence---in addition to $\phi$, we thus also consider its derivative $\partial_\mu \phi$, the derivative of that, the stress-energy tensor, and so on. This is perhaps more general than your previous experiences with fields as variables in the integral measure, but we find it to be the more useful understanding in this context.

As a concrete example, consider the two-point function
\begin{equation}
\langle\phi_1(x_1)\phi_2(x_2)\rangle = \frac{1}{Z} \int D\Phi_i \phi_1(x_1) \phi_2(x_2) e^{-S[\Phi_i]}.
\end{equation}
Here $\Phi_i$ denotes the set of all fields in the theory, $S$ is the conformally invariant action, and $\phi_1,\phi_2$ are quasi-primary fields. Assuming conformal invariance of the action and integration measure, it can be shown that this correlation function transforms as
\begin{equation}
\langle \phi_1(x_1) \phi_2(x_2) \rangle =  \left\lvert \frac{\partial x'}{\partial x} \right\rvert^{\Delta_1/d}_{x=x_1} \left\lvert \frac{\partial x'}{\partial x} \right\rvert^{\Delta_2/d}_{x=x_2}  \langle \phi_1(x_1') \phi_2(x_2') \rangle
\end{equation}
\begin{framed}
\noindent HOMEWORK: Prove this formula. Begin by proving $\langle\phi(x_1')\cdots\phi(x_n') \rangle=\langle F(\phi(x_1'))\cdots F(\phi(x_n')) \rangle$. Note the assumption that the functional integration measure is conformally invariant is essential; the failure of this to be true is often the reason conformal invariance fails quantum mechanically.
\end{framed}
\noindent For the case of dilatations $x\rightarrow \lambda x$, this becomes
\begin{equation}
\langle \phi_1(x_1) \phi_2(x_2) \rangle = \lambda^{\Delta_1+\Delta_2}\langle \phi_1(\lambda x_1) \phi_2(\lambda x_2) \rangle.
\end{equation}
It is also straightforward to show that Poincar\'{e} invariance implies
\begin{equation}
\langle \phi_1(x_1) \phi_2(x_2) \rangle = f(|x_1-x_2|).
\end{equation}
It immediately follows that
\begin{equation}
f(x)=\lambda^{\Delta_1+\Delta_2} f(\lambda x).
\end{equation}
The symmetries of conformal field theory have therefore constrained the two-point function to be of the form (and make sure you understand this)
\begin{equation}
\langle \phi_1(x_1) \phi_2(x_2) \rangle = \frac{d_{12}}{|x_1-x_2|^{\Delta_1+\Delta_2}},
\end{equation}
where $d_{12}$ is some normalization constant depending on the fields $\phi_1,\phi_2$. This is the only form with the appropriate transformation properties.

We should also examine the consequences of invariance under special conformal transformations. For a special conformal transformation,
\begin{equation}
\left\lvert \frac{\partial x'}{\partial x} \right\rvert =\frac{1}{(1-2b\cdot x + b^2 x^2)^d}.
\end{equation}
The distance between two points transforms as
\begin{equation}
|x_i'-x_j'| = \frac{|x_i-x_j|}{(1-2b\cdot x_i + b^2 x_i^2)^{1/2}(1-2b\cdot x_j + b^2 x_j^2)^{1/2}}.
\end{equation}
Then we have that
\begin{equation}
\frac{d_{12}}{|x_1-x_2|^{\Delta_1+\Delta_2}} = \frac{d_{12}}{\gamma_1^{\Delta_1}\gamma_2^{\Delta_2}} \frac{(\gamma_1 \gamma_2)^{(\Delta_1+\Delta_2)/2}}{|x_1-x_2|^{\Delta_1+\Delta_2}},
\end{equation}
where $\gamma_i \equiv (1-2b\cdot x_i + b^2 x_i^2)$. This constraint is satisfied only if $\Delta_1=\Delta_2$. In summary, 
\begin{equation}
\langle \phi_1(x_1) \phi_2(x_2) \rangle =
\begin{dcases}
\frac{d_{12}}{|x_1-x_2|^{2\Delta_1}}& \text{if } \Delta_1 = \Delta_2\\
0 & \text{if } \Delta_1 \neq \Delta_2.
\end{dcases}
\end{equation}
The constant $d_{12}$ (or more generally, $d_{ij}$) can be further simplified. By redefining our fields, we can always choose a basis of operators so that $d_{ij} = \delta_{ij}$.

We can treat three-point functions in a similar manner. Invariance under rotations, translations, and dilatations force the three-point function to have the form
\begin{equation}
\langle \phi_1(x_1) \phi_2(x_2) \phi_3(x_3) \rangle = \frac{\lambda_{123}^{(abc)}}{x_{12}^a x_{23}^b x_{13}^c},
\end{equation}
where $x_{ij} \equiv |x_i - x_j|$ and
\begin{equation}
a+b+c = \Delta_1 + \Delta_2 +\Delta_3.
\end{equation}
As before, we can further constrain the three-point function by demanding invariance under special conformal transformations. Following similar steps, one can show that 
\begin{align}
a &= \Delta_1 + \Delta_2 - \Delta_3 = \Delta - 2\Delta_3 \nonumber \\
b &= \Delta_2 + \Delta_3 - \Delta_1 = \Delta - 2\Delta_1 \\
c &= \Delta_3 + \Delta_1 - \Delta_2 = \Delta - 2\Delta_2 \nonumber.
\end{align}
where we have defined $\Delta \equiv \sum_i \Delta_i$ for future use. 
\begin{framed}
\noindent HOMEWORK: Derive these values of $a,b,c$. Use the fact that the transformed three-point function being of the same form as the untransformed three-point function gives some conditions on $a,b,c$.
\end{framed}
\noindent The final form of the three-point correlator is therefore
\begin{equation}
\langle \phi_1(x_1) \phi_2(x_2) \phi_3(x_3) \rangle = \frac{\lambda_{123}}{x_{12}^{\Delta - 2\Delta_3} x_{23}^{\Delta - 2\Delta_1} x_{13}^{\Delta - 2\Delta_2}}.
\end{equation}
Unlike the constants $d_{ij}$, the three-point constants cannot be normalized away. They are not determined by conformal invariance and are necessary data to define a particular conformal field theory.

Encouraged by these successes, we might suppose we can continue calculating higher-point correlators. Starting with four points, however, we run into difficulty. Once we have four points $x_1, x_2, x_3, x_4$, we can construct the ratios
\begin{equation}
\left(\frac{x_{12} x_{34}}{x_{13}x_{24}}\right)^2 \equiv {u},\;\;\;\;\;  \left(\frac{x_{12} x_{34}}{x_{23}x_{14}}\right)^2 \equiv {v}.
\end{equation}
These expressions are \emph{anharmonic ratios} or \emph{cross-ratios}; they are invariant under conformal transformations.
\begin{framed}
\noindent HOMEWORK: How many anharmonic ratios can be formed from $N$ points? (Hint: a cute way to do this is by using translational and rotational invariance to describe $N$ coordinates as $N-1$ points in an $N-1$ dimensional subspace. Determine how many independent quantities characterize this subspace, and subtract off the parameters corresponding to the remaining rotational, scale, and special conformal transformations.)
\end{framed} 
\noindent This means the general form of the four-point function is given by
\begin{equation}
\langle \phi_1(x_1) \phi_2(x_2) \phi_3(x_3) \phi_4(x_4) \rangle = f(u,v) \prod_{i<j}^4 x_{ij}^{\Delta/3-\Delta_i-\Delta_j}
\end{equation}
This is the best that we can do at this point, although in later lectures we will see that we can use conformal bootstrapping to extract additional information about theories.

\subsection{Conserved currents and the energy momentum tensor}

Hopefully we are all familiar with Noether's theorem. In short, every continuous symmetry implies the existence of a current. Using the same infinitesimal transformation terminology as before, the conserved current is given by
\begin{equation}
j_a^\mu = \left[\frac{\partial \mathcal{L}}{\partial(\partial_\mu \Phi)}\partial_\nu \Phi - \delta^\mu_\nu \mathcal{L}  \right]
\frac{\delta x^\nu}{\delta \epsilon_a}-\frac{\partial \mathcal{L}}{\partial(\partial_\mu \Phi)}
\frac{\delta F}{\delta \epsilon_a}.
\end{equation}
(If you are unfamiliar with this theorem, I encourage you to complete the appropriate exercise at the end of these lectures.) A conserved current is one such that 
$$
\partial_\mu j^\mu_a = 0.
$$
The conserved charge associated with $j^\mu_a$ is given by
\begin{equation}
Q_a = \int d^{d-1}x j^0_a,
\end{equation}
where we are integrating over all space.
We also remark that this conserved current is ``canonical''. It is straightfoward to see that adding the divergence of an antisymmetric tensor does not affect the conservation of $j$:
\begin{equation}
j^\mu_a \rightarrow j^\mu_a + \partial_\nu B^{\nu\mu}_a, \;\;\;\;B^{\nu\mu}_a = -B^{\mu\nu}_a.
\end{equation}
Thus we have some freedom in redefining our conserved currents.

What are the conserved currents for conformal field theory? The infinitesimal translation $x^\mu \rightarrow x^\mu + \epsilon^\mu$ gives
$$
\frac{\delta x^\mu}{\delta \epsilon^\nu}=\delta^\mu_\nu, \;\;\;\; \frac{\delta F}{\delta \epsilon^\nu} = 0.
$$
The corresponding conserved current is the energy-momentum tensor
\begin{equation}
T^{\mu\nu}_C = -\eta^{\mu\nu}\mathcal{L}+ \frac{\partial \mathcal{L}}{\partial(\partial_\mu \Phi)}\partial^\nu \Phi.
\end{equation}
In general, this quantity is \emph{not} symmetric. This is not good; it means we need consider spin current. We have the freeom, however, to modify this quantity by the divergence of a tensor $B^{\rho\mu\nu}$ antisymmetric in its first two indices. This improved tensor is called the \emph{Belinfante} energy-momentum tensor $T^{\mu\nu}_B$, and it is symmetric. But how do we find the appropriate $B$?
 
One way to do this is to consider infinitesimal Lorentz transformations.  The associated variations are
$$
\frac{\delta x^\rho}{\delta \epsilon_{\mu\nu}}=\frac12 (\eta^{\rho\mu} x^\nu - \eta^{\rho\nu}x^\mu), \;\;\;\; \frac{\delta F}{\delta \epsilon_{\mu\nu}}=\frac{-i}{2}S^{\mu\nu}\Phi,
$$
and the associated conserved current is
\begin{equation}
j^{\mu\nu\rho}=T^{\mu\nu}_Cx^\rho - T^{\mu\rho}_Cx^\nu + i \frac{\partial \mathcal{L}}{\partial(\partial_\mu \Phi)} S^{\nu\rho}\Phi.
\end{equation}
It can be shown that by choosing an appropriate $B$, this current can be expressed as
$$
j^{\mu\nu\rho}= T^{\mu\nu}_Bx^\rho-T^{\mu\rho}_Bx^\nu,
$$
with $T_B$ being symmetric.
The explicit expression for $B$ that does this is
\begin{equation}
B^{\mu\rho\nu}=\frac{i}{2}\left[  \frac{\partial \mathcal{L}}{\partial(\partial_\mu \Phi)} S^{\nu\rho}\Phi
+\frac{\partial \mathcal{L}}{\partial(\partial_\rho \Phi)} S^{\mu\nu}\Phi
+\frac{\partial \mathcal{L}}{\partial(\partial_\nu \Phi)} S^{\mu\rho}\Phi  \right]. \label{eq:Bdef}
\end{equation}
\begin{framed}
\noindent HOMEWORK: Verify this claim.
\end{framed}

We remark that there is an alternate definition of the energy-momentum tensor that is manifestly symmetric (though sometimes requires more complicated calculations). In the derivation of Noether's theorem, it is shown that the variation of the action under an infinitesimal transformation goes as 
\begin{equation}
\delta S = - \int d^dx j^\mu_a \partial_\mu \epsilon_a.
\end{equation}
Under an infinitesimal coordinate-dependent translation $x^\mu \rightarrow x^\mu + \epsilon^\mu(x)$ with the stress-energy tensor as the associated conserved current, the variation of the action is therefore given by
\begin{equation}
\delta S = -\frac12 \int d^dx T^{\mu\nu}(\partial_\mu \epsilon_\nu+\partial_\nu \epsilon_\mu).
\end{equation}
But this diffeomorphism also induces a variation in the metric. The metric tensor varies under this transformation according to
$$
\delta g_{\mu\nu} = - (\partial_\mu \epsilon_\nu+\partial_\nu \epsilon_\mu).
$$
Thus the full variaton of the action is
\begin{equation}
\delta S = -\frac12 \int d^dx \left( T^{\mu\nu}+2\frac{\delta S}{\delta g_{\mu\nu}}  \right) (\partial_\mu \epsilon_\nu+\partial_\nu \epsilon_\mu).
\end{equation}
Demanding the action is invariant under this transformation gives the definition
\begin{equation}
T^{\mu\nu} = -2\frac{\delta S}{\delta g_{\mu\nu}}.
\end{equation}
This expression is manifestly symmetric. This is the stress-energy tensor that appears in general relativity\footnote{There is an interesting interpretation of the Belinfante tensor. The Belinfante tensor includes ``bound'' momentum associated with gradients of the intrinsic angular momentum in analogy with the bound current associated with magnetization density. Just as the sum of bound and free currents acts as a source for the magnetic field, it is the sum of the bound and free energy-momentum that acts as a source of gravity.}. Sometimes this form is easier to derive, so you should be introduced to it.

These conserved currents should be familiar. What about the conserved current associated with scale invariance? An infinitesimal dilatation acts as
$$
x'^\mu = (1+\alpha)x^\mu,\;\;\;\;F(\Phi) = (1-\alpha \Delta)\Phi,
$$
so that by Noether's theorem the conserved current is
\begin{equation}
j^\mu_D = T^\mu_{C\;\nu} x^\nu + \frac{\partial \mathcal{L}}{\partial(\partial_\mu \Phi)}\Delta\Phi.
\end{equation}
Again, we have an additional contribution that ruins what would otherwise be a perfectly lovely current. We were previously able to redefine our energy-momentum tensor; can we do the same thing here without spoiling the other nice features? Specifically, can we kill the second contribution so that the conservation of $j^\mu_D$ corresponds to traclessness of $T_{\mu\nu}$? 

Although not necessarily obvious, it turns out that we \emph{can} add another term to do precisely that. Specifically, we have the freedom to add a term of the form $\frac12 \partial_\lambda \partial_\rho X^{\lambda\rho\mu\nu}$ to our $T_B$ that does not spoil its conservation law or its symmetry (we have left verifying this fact for the exercises). This term is defined so that its trace is given by
\begin{equation}
\frac12 \partial_\lambda \partial_\rho X^{\lambda\rho\mu}_\mu = \partial_\mu V^\mu,
\end{equation}
where the \emph{virial}\footnote{Constructing the $X$ we need depends upon the virial being the divergence of another tensor: $V^\mu=\partial_\alpha \sigma^{\alpha\mu}$. The tensor $X$ is then built out of $\sigma$ so that $\partial_\lambda\partial_\rho X^{\lambda\rho\mu\nu}=2\partial_\mu V^\mu$.This is possible in a large class of physical theories, but it is not necessarily univerally true. When it is true, scale invariance will imply full conformal invariance, as we will see momentarily. For now, we assume we are able to write down the appropriate $X$. We will discuss scale versus conformal invariance in a later lecture.} $V$ is defined as
\begin{equation}
V^\mu =  \frac{\partial \mathcal{L}}{\partial(\partial^\rho \Phi)}(\eta^{\mu\rho} \Delta  +  i S^{\mu\rho})\Phi.
\end{equation}
It follows from these definitions that $T=T_B + \mbox{(new term)}$ satisfies
\begin{equation}
T^\mu_\mu= \partial_\mu j^\mu_D
\end{equation}
so that
\begin{equation}
j^\mu_D = T^\mu_\nu x^\nu.
\end{equation}
Conservation of this current is equivalent to tracelessness of the stress-energy tensor.

And what of the conserved current for special conformal invariance? According to our assumption that ``scale invariance = conformal invariance'', the analysis we have performed so far should be enough to guarantee some current that is trivially conserved. Under an arbitrary change of coordinates $x^\mu \rightarrow x^\mu+\epsilon^\mu$, we know the variation of the action will be
\begin{equation}
\delta S = -\frac12 \int d^dx T^{\mu\nu}(\partial_\mu \epsilon_\nu+\partial_\nu \epsilon_\mu).
\end{equation}
If this infinitesimal transformation is conformal, then using eq. (\ref{eq:eq2p5}) gives
\begin{equation}
\delta S = -\frac{1}{d} \int d^d x T^\mu_\mu \partial_\rho \epsilon^\rho.
\end{equation}
The tracelessness of the energy-momentum tensor, which corresponds to scale invariance\footnote{There is more to this story; the stress-energy tensor can gain a trace quantum mechanically. We will eventually return to this topic.}, implies conformal invariance.

We can write down an expression for the special conformation transformation's associated conserved current. Then we can perform some clever derivations and manipulations to improve/massage it into a form that we find acceptable. In the interest of expediency, we leave this as an exercise. For now, I will quote the result:
\begin{equation}
j^{\mu\nu}_K = T^\mu_\rho (2 x^\rho x^\nu - \eta^{\alpha\nu}x^2).
\end{equation}
Taking the divergence of this quantity, we see that it vanishes due to tracelessness and conservation of $T^{\mu\nu}$.
\begin{framed}
\noindent HOMEWORK: Check this! It is trivial.
\end{framed}

\subsection{Radial quantization and state-operator correspondence}

Before continuing to explore constraints from conformal invariance, we will discuss foliations of spacetime in QFT. By ``foliation'', I mean how we divide our $d$-dimensional spacetime into $d-1$-dimensional regions. For example, for theories with Poincar\'e invariance, we typically choose to foliate our space by surfaces of equal time.  Each surface has its own Hilbert space, and when these surfaces are related by a symmetry transformation then the Hilbert space on each surface is the same. For theories with Poincar\'e invariance, the states that exist on these surfaces are specified by their 4-momenta.

What do we mean by states?  We define \emph{in} states $|\mbox{in}\rangle$ by inserting operators in the past of a given surface and \emph{out} states by inserting operators in the future (think back to QFT scattering amplitudes, if it helps). The overlap of in and out states on the same surface is given by the correlation function of their respective operators
$$
\langle \mbox{out} | \mbox{in} \rangle.
$$
When the in and out states live on different surfaces with no other states happening between then, there exists some unitary operator $U$ connecting the two states. The associated correlation function is
$$
\langle \mbox{out} | U  |\mbox{in} \rangle.
$$
For our Poincar\'e example, the Hamiltonian moves us between surfaces so that the unitary evolution operator is given by
$$
U = e^{iH\Delta t}.
$$

These remarks are very general. This is great news, because we will use a more convenient foliation for CFTs. Thinking back to the conformal algebra, we realize we would prefer some foliation that relates to dilatations (rather than time translations). We will divide spacetime using spheres $S^{d-1}$ of various radii centered at the origin\footnote{We use the origin without loss of generality; the same ``ambiguity'' is present in Poincar\'{e} theories, due to the fact that we have to fix a timelike time vector.} having the corresponding metric
\begin{equation}
ds^2 = dr^2 + r^2 d\boldsymbol{n}^2. \label{eq:flatmet}
\end{equation}
Rather than using the Hamiltonian to move from one foliation to another, we use the dilatation generator $D$. States living on these spheres are classified not by their 4-momenta, as with the Poincar\'e group, but instead by their scaling dimension
\begin{equation}
D | \Delta \rangle = i\Delta | \Delta \rangle
\end{equation}
and their $SO(D)$ spin $\ell$
\begin{equation}
M_{\mu\nu} | \Delta,\ell \rangle = (\Sigma_{\mu\nu}) | \Delta,\ell \rangle.
\end{equation}. 

To express the evolution operator, we define $\tau \equiv \log r$ so that the metric (\ref{eq:flatmet}) becomes
\begin{equation}
ds^2 = e^{2\tau}(d\tau^2+d\boldsymbol{n}^2).
\end{equation}
This metric is conformally equivalent to a cylinder! This coordinate transformation maps from $\mathbb{R}^d$ to $\mathbb{R}\times S^{d-1}$. These $S^{d-1}$ are precisely the spheres from above, and this $\tau$ coordinate is the natural ``time'' coordinate for evolution. The evolution operator is then
\begin{equation}
U = e^{iD \tau}.
\end{equation}
If we act on an eigenstate $|\Delta\rangle$ with this operator, we get
$$
U|\Delta\rangle = e^{-\Delta\tau}|\Delta\rangle = r^{-\Delta}|\Delta\rangle.
$$
This entire discussion and choice of foliation is known as \emph{radial quantization}.

This correspondence between spherical coordinates in Euclidean space and a cylinder with time $\tau$ running along its length is very illuminating. 

We see that moving toward the origin $r\rightarrow0$ in radial quantzation is equivalent to approaching the infinite past $\tau\rightarrow -\infty$ and moving to infinity $r\rightarrow\infty$ is equivalent to approaching the infinite future $\tau\rightarrow\infty$. To create some state at given radius(/time), we would place an object inside the sphere (in the past). Let us consider some examples. If we make no operator insertions, the system should correspond to the vacuum state $|0\rangle$. By the vacuum state, we mean that we assume a unique ground state that that should be invariant under all global conformal transformations. It is therefore labeled by ``0'' because the eigenvalue of the dilatation operator is zero for this state. 

What happens if we insert the operator $\mathcal{O}_\Delta(x=0)$ at the origin? According to the actions of the generator algebra, this creates a state $|\Delta\rangle \equiv \mathcal{O}_\Delta (0)|0\rangle$ with scaling dimension $\Delta$. We like to insert objects in the infinite past; doing so in QFT was how we calculated path integrals by considering the contribution from only the ground state. What happens if we insert the operator $\mathcal{O}_\Delta (x)$ somewhere other than the origin? Using our derived algebras, we see that 
\begin{equation}
|\chi\rangle \equiv \mathcal{O}_\Delta (x)|0\rangle = e^{iPx} \mathcal{O}_\Delta (0) e^{-iPx}  |0\rangle =e^{iPx}|\Delta\rangle
\end{equation}
If we expand this exponential function, we claim that we have a superposition of states with different eigenvalues.
\begin{framed}
\noindent HOMEWORK: Prove this is the case by showing that $P_\mu$  raises the scaling dimension of the state $|\Delta\rangle$ by 1. Similarly, show that $K_\mu$ lowers the dimension by 1.
\end{framed}

This exercise demonstrates that $P$ and $K$ act like ladder operators for dilatation eigenvalues. An operator that is annihilated by the ``lowering'' operator $K$ is called a \emph{primary} operator. The states we get by acting with ``raising operators'' $P$ are called {descendants}.  If we assume that the scaling dimension is bounded from below\footnote{This is the case in unitary theories. We will demonstrate this at the end of the lecture.}, then for some generic state we could always act with $K$ until we hit zero, thus finding a primary operator. A primary operator and all of its descendants form what is known as a \emph{conformal family}.

So: when inserting a primary operator at the origin, we get a state with scaling dimension $\Delta$ that is annihilated by $K$. This procedure can also go the other direction: given some state with scaling dimension $\Delta$ that is annihilated by $K$, we can construct an associated local primary operator. This is the \emph{state-operator correspondence}---states are in a one-to-one correspondence with local operators.The proof of this fact is straightforward. In order to construct an operator, we define its correlators with other operators. We do this definition according to the equation
\begin{equation}
\langle \phi(x_1)\phi(x_2)\cdots\mathcal{O}_\Delta(0)  \rangle = \langle 0| \phi(x_1)\phi(x_2)\cdots|\Delta  \rangle.
\end{equation}
This definition satisfies the usual transformation properties that we expect from conformal invariance; this can be shown, but we do not do it here.

\subsection{Unitarity bounds}

We conclude this lecture by proving that unitarity constrains the scaling dimensions of our conformal field theory. Any theory with operators violating the unitarity bounds would be non-unitary in Lorentzian signature (and non-positive in Euclidean signature). We again consider radial quantization in terms of the cylinder 
$$
ds^2_{cyl}=d\tau^2 + d\boldsymbol{n}^2 = r^{-2}(dr^2+r^2d\boldsymbol{n}^2),
$$ 

To investigate fields in radial quantization, we use the fact that under a conformal transformation scalars change according to
 \begin{equation}
 \langle \phi(x)\cdots  \rangle_{e^{2\sigma(x)}dx^2} = e^{-\sigma(x)\Delta}\langle \phi(x)\cdots \rangle_{dx^2}.
 \end{equation}
In this case, the relation between fields in Euclidean space and fields on the cylinder is 
\begin{equation}
\phi(\tau,\boldsymbol{n})_{cyl}=r^{\Delta}\phi(x)_{\mathbb{R}^d}.
\end{equation}

The cylindrical field is the very same field as in flat space, we are just measuring its correlators in a different geometry. How does Hermitian conjugation work on the cylinder? It is an essential component for finding the norm of a state, after all. Performing the completely standard Wick rotation to get to Euclidean signature, it is straightforward to see for a Hermitian field $\phi$ that
\begin{equation}
\phi(\tau)^{\dagger}_{cyl}=\left( e^{\tau H_{cyl}} \phi(0)  e^{-\tau H_{cyl}}   \right)^\dagger =  e^{-\tau H_{cyl}} \phi(0)  e^{\tau H_{cyl}} = \phi(-\tau)_{cyl}
\end{equation}
(suppressing dependence on $\boldsymbol{n}$). The \emph{reflection positivity} in Euclidean theory corresponds to unitarity in Minkowski theory:
\begin{equation}
\langle   \phi(-\tau)\phi(\tau) \rangle_{cyl} = \langle   \phi(\tau)^\dagger \phi(\tau) \rangle_{cyl} \geq 0.
\end{equation}

Using $\tau=\log r$, this time-reversal transformation becomes a coordinate inversion $R:x\rightarrow x/x^2$ in $\mathbb{R}^d$. Similarly, hermitian conjugation extends to the conformal algebra generators where it corresponds to acting with the inversion operator $R$. This allows us to calculate the extraordinary result that in radial quantization
\begin{equation}
P^\dagger_\mu=RP_\mu R^{-1} = RP_\mu R = K_\mu.
\end{equation}
This seems like quite a claim, given that in flat space know that both $K$ and $P$ are Hermitian. To easily check this result, we can consider the differential operators expressed in terms of cylindrical variables
\begin{eqnarray}
P_\mu &=& -i\partial_\mu\rightarrow -ie^{-\tau}\left[ \boldsymbol{n}_\mu \partial_\tau + (\delta_{\mu\nu} - \boldsymbol{n}_\mu  \boldsymbol{n}_\nu) \partial / \partial \boldsymbol{n}_\nu    \right], \\
K_\mu &=& -i\left[x^2  \partial_\mu-2x_\mu (x\cdot\partial) \right]   \rightarrow -ie^{\tau}\left[ -\boldsymbol{n}_\mu \partial_\tau + (\delta_{\mu\nu}- \boldsymbol{n}_\mu  \boldsymbol{n}_\nu)\partial/\partial  \boldsymbol{n}_\nu    \right].
\end{eqnarray}
From these explicit expressions, we can see these operators are conjugate to one another under time-reversal.
\begin{framed}
\noindent HOMEWORK: Derive the above expressions. This is a straightforward exercise if you remember that $x_\mu=r\boldsymbol{n}_\mu$.
\end{framed}

We can use this fact to extract unitarity bounds in a straightforward way. Here is a simple example. Consider a spinless primary $|\Delta\rangle$ and the quantity
\begin{eqnarray}
\langle\Delta | K_\mu P_\nu|\Delta \rangle &=& \langle\Delta | [K_\mu, P_\nu]  + P_\nu K_\mu  |\Delta \rangle\\
&=& \langle\Delta| 2i(D\delta_{\mu\nu}-M_{\mu\nu}) |\Delta \rangle \\
&=& \Delta\delta_{\mu\nu}\langle\Delta|\Delta\rangle=\Delta\delta_{\mu\nu}.
\end{eqnarray}
Here we have used the fact that $|\Delta\rangle$ is primary and spinless. By setting $\mu=\nu$ and using the fact that norms are positive definite, we thus have our first unitarity bound
\begin{equation}
\Delta \geq 0.
\end{equation}
By considering spinless scalars at level two, where imposing
\begin{equation}
\langle  \Delta|K_\lambda K_\mu P_\nu P_\rho  |\Delta  \rangle \geq 0,
\end{equation}
we can derive the unitarity bound
\begin{equation}
\Delta \geq \frac{d}{2}-1. \label{eq:unitarydel}
\end{equation}
\begin{framed}
\noindent HOMEWORK: Complete the derivation of this bound.
\end{framed}

In theory, we could continue these steps  with more $K$'s and $P$'s to get stronger bounds. It turns out, however, that levels higher than two are not needed for scalars. The constraint (\ref{eq:unitarydel}) is necessary and sufficient to have unitarity at all levels \cite{2-5}. In general, one can derive bounds for fields in any representation of the group of rotations \cite{2-6}.
Similar derivations show that
\begin{equation}
\Delta \geq \frac{d-1}{2}
\end{equation}
for states with $s=\frac12$ and
\begin{equation}
\Delta \geq d+s-2
\end{equation}
for states with $s\geq 1$.
These unitarity bounds were derived using the mapping between Euclidean space and the cylinder, but that was just a convenient way to see the relation between $P$ and $K$ in radial quantization. There are other, more complicated ways to derive these unitarity bounds.

We conclude by pointing out that the free scalar field saturates its unitarity bound. This is also the case for a free massless fermion. In \cite{2-7}, it was shown that for $d=4$ dimensions, a field in the $(s,0)$ or $(0,s)$ representation of the complexified Lorentz group $SL(2,\mathbb{C})\oplus SL(2,\mathbb{C})$ that saturates the unitarity bound is a free field with free field correlation functions. Of course, not all theories of interest are unitary; there are plenty of condensed matter systems described by nonunitary conformal field theories. Even so, the unitary bounds are powerful and useful constraints.

\break

\subsection*{References for this lecture}
\vspace{4mm}
\noindent Main references for this lecture
\\
\begin{list}{}{%
\setlength{\topsep}{0pt}%
\setlength{\leftmargin}{0.7cm}%
\setlength{\listparindent}{-0.7cm}%
\setlength{\itemindent}{-0.7cm}%
\setlength{\parsep}{\parskip}%
}%
\item[]

[1] Chapter 2 of the textbook: R. Blumenhagen, E. Plauschinn, \emph{Introduction to Conformal Field Theory: With Applications to String Theory}, Lect. Notes Phys. 779,  (Springer, Berlin Heidelberg 2009).

[2] Chapter 4 of the textbook: P. Di Francesco, P. Mathieu, and D. Senechal. \emph{Conformal field theory}, Springer, 1997.

[3] S. Rychkov, \emph{EPFL Lectures on Conformal Field Theory in $D\geq 3$ Dimensions: Lecture 1: Physical Foundations of Conformal Symmetry}, (Lausanne, Switzerland, \'Ecole polytechnique f\'ed\'erale de Lausanne, December 2012).

[4] S. Rychkov, \emph{EPFL Lectures on Conformal Field Theory in $D\geq 3$ Dimensions: Lecture 3: Radial quantization and OPE}, (Lausanne, Switzerland, \'Ecole polytechnique f\'ed\'erale de Lausanne, December 2012).

\end{list}

\break

\section{Lecture 3: CFT in $d=2$}

In this lecture, we will consider the conformal group in $d=2$ dimensions. We will find that in this special case, the conformal algebra has infinitely many generators. This additional structure allows for a much richer analysis. We will begin with conformal transformations before discussing the Witt and Virasoro algebras. We briefly discuss primary fields and radial quantization in two dimensions before considering the stress-energy tensor and highest weight states. We finish by considering simple constraints that follow from conformal invariance.

\subsection{Conformal transformations for $d=2$}

During the last lecture, we saw that conformal invariance is special for $d=2$. For  coordinates $(z^0, z^1)$ (with Euclidean metric), a change of coordinates $z^\mu \rightarrow w^\mu (x)$ means the metric transforms as
\begin{equation}
g^{\mu\nu} \rightarrow  \left( \frac{\partial w^\mu}{\partial z^\alpha}\right)\left( \frac{\partial w^\nu}{\partial z^\beta}\right) g^{\alpha\beta}\propto g^{\mu\nu} .
\end{equation}
The condition that makes this a conformal transformation is found to be
\begin{gather}
\left( \frac{\partial w^0}{\partial z^0}\right)^2 + \left( \frac{\partial w^0}{\partial z^1}\right)^2 = \left( \frac{\partial w^1}{\partial z^0}\right)^2 + \left( \frac{\partial w^1}{\partial z^1}\right)^2 \\
\frac{\partial w^0}{\partial z^0} \frac{\partial w^1}{\partial z^0} + \frac{\partial w^0}{\partial z^1} \frac{\partial w^1}{\partial z^1} = 0
\end{gather}
These equations are equivalent to
\begin{equation}
\partial_0 w_1 = \pm\partial_1 w_0, \;\;\;\;\;\;\;\;\;\; \partial_0 w_0 = \mp\partial_1 w_1.
\label{eq:eq3p1}
\end{equation}
As we are diligent students of complex analysis, we recognize these expressions as the holomorphic (and anti-holomorphic) Cauchy-Riemann equations. A complex function $w(z,\bar{z})$ satisfying the Cauchy-Riemann equations is a holomorphic function in some open set. To use this fact, we define
\begin{eqnarray}
\epsilon \equiv \epsilon^0 + i \epsilon^1 \;\;\;\;\;\;  z \equiv x^0 + i x^1, \;\;\;\;\;\; \partial_z \equiv \frac12 \left( \partial_0 - i \partial_1 \right) \nonumber \\ 
\bar{\epsilon} \equiv \epsilon^0 - i \epsilon^1 \;\;\;\;\;\;  \bar{z} \equiv x^0 - i x^1, \;\;\;\;\;\; \partial_{\bar{z}} \equiv \frac12 \left( \partial_0 + i \partial_1 \right).
\end{eqnarray}
We also note that in terms of these coordinates the metric tensor is
\begin{equation}
g_{\mu\nu} = \left(
\begin{array}{cc}
0 & \frac12  \\
\frac12 & 0
\end{array}
\right),
\end{equation}
and we introduce the notation $\partial = \partial_z, \bar{\partial} = \partial_{\bar{z}}$.

Using these coordinates, the holomorphic Cauchy-Riemann equations become
\begin{equation}
\bar{\partial} w(z,\bar{z}) = 0.
\end{equation}
The solution to this equation is any holomorphic mapping
$$
z\rightarrow w(z).
$$
The conformal group for $d=2$ is then the set of all analytic maps! This set is infinite-dimensional, corresponding to the the coefficients of the Laurent series needed to specify functions analytic in some neighborhood. This infinity is what makes conformal symmetry so powerful in two dimensions.

To consider an infinitesimal conformal transformation, we right $f(z)=z+\epsilon(z)$. Because $f(z)$ is a holomorphic function, so too is $f(z)=z+\epsilon(z)$. The same statements hold true for the variable $\bar{z}$. These facts mean that metric tensor transforms as
$$
ds^2 = dz d\bar{z} \rightarrow \frac{\partial {f}}{\partial {z}} \frac{\partial \bar{f}}{\partial \bar{z}} dz d\bar{z}. 
$$
We can also read off the scale factor for these 2d conformal transformations as $\Lambda = \left| \frac{\partial f}{\partial z} \right|^2$.

\subsection{Global Conformal Transformations}

So we have infinitely many infinitesimal conformal transformations. In order to form a group, however, the mappings must be invertible and map the whole plane into itself (including the point at infinity). The transformations that satisfy these conditions are \emph{global} conformal transformations, and the set of them form the \emph{special conformal group}.  Consider such a mapping $f(z)$. Clearly $f$ should not have any branch points or essential singularities (maps are not uniquely defined around a branch point, and the neighborhood of an essential singularity sweeps out the entire plane). The only acceptable singularities are thus poles, so $f$ can be written as
\begin{equation}
f(z) = \frac{P(z)}{Q(z)}.
\end{equation}
If $P(z)$ has several distinct zeros, then the inverse image of zero is not well-defined--$f$ is not invertible. Furthermore, $P(z)$ having a multiple zero means the image of a small neighborhood of the zero is wrapped around $0$: $f$ is not invertible. The same arguments apply for $Q(z)$ when looking at the behavior of $f(z)$ near the point at infinity. Therefore $P(z)$ and $Q(z)$ can be at most linear functions
\begin{equation}
f(z)=\frac{az+b}{cz+d}.
\end{equation}
Futhermore, in order for this mapping to be invertible: the determinant $ad-bc$ must be nonzero. The conventional normalization is that the coefficients have all been scaled so that $ad-bc=1$ (after all, rescaling $a,b,c,d$ by a constant factor does not actually change the transformation).

We have therefore found that the special conformal group is given by the so-called \emph{projective transformations}. We can associate to each transformation a matrix
\begin{equation}
\left( \begin{array}{cc}
a & b \\
c & d
\end{array} \right),\;\;a,b,c,d\in\mathbb{C}.
\end{equation}
The global conformal group in two dimensions is then isomorphic to the group $SL(2,\mathbb{C})$. We also know that this group is isomorphic to the Lorentz group in four dimensions, $SO(3,1)\sim SO(2,2)$. Success! The special conformal group in $d=2$ dimensions matches our expectations from other dimensions.

\subsection{The Witt and Virasoro algebras}

We have found that infinitesimal conformal transformation for $d=2$ must be holomorphic in some open set. It is completely conceivable, however, that $\epsilon(z)$ has isolated singularities outside of this open set; we therefore assume that $\epsilon(z)$ is in general a meromorphic function and perform a Laurent expansion around $z=0$:
\begin{eqnarray}
{z}' = {f(z)} = {z}+{\epsilon}(z) = {z} + \sum_{n\in \mathbb{Z}} {\epsilon}_n \left( -{z}^{n+1}  \right), \nonumber \\ 
\bar{z}' = \bar{f}(\bar{z}) = \bar{z}+\bar{\epsilon}(\bar{z}) = \bar{z} + \sum_{n\in \mathbb{Z}} \bar{\epsilon}_n \left( -\bar{z}^{n+1}  \right).
\label{eq:eq3p3}
\end{eqnarray}
The parameters $\epsilon_n, \bar{\epsilon}_n$ are infinitesimal and constant. Let us consider the $m$-th term in the first sum. What is the generator corresponding to this transformation? The effect of an infinitesimal mapping on a spinless, dimensionless field $\phi(z,\bar{z})$ is
\begin{equation}
\delta \phi = \epsilon(z)\partial\phi + \bar{\epsilon}(\bar{z})\bar{\partial}\phi.
\end{equation}
Thus the generator associated with the $m$-th term in the first sum is
\begin{equation}
\ell_m = - z^{m+1} \partial _z. \label{eq:wittoperator}
\end{equation}
This is true for any $m$, with a corresponding equation for $\bar{\ell}_m$ in terms of $\bar{z}$. Thus we have infinitely many independent infinitesimal conformal transformations for $d=2$.

Alright, we have found the generators; now let us find the conformal algebra. Explicit calculation in terms of $z,\bar{z}$ and $\partial_z, \partial_{\bar{z}}$ gives
\begin{align}
[\ell_m, \ell_n] &=  (m-n)\ell_{m+n}, \nonumber \\
[\bar{\ell}_m, \bar{\ell}_n] &= (m-n)\bar{\ell}_{m+n} , \\
[\ell_m, \bar{\ell}_n] &= 0. \nonumber
\end{align}
\begin{framed}
\noindent HOMEWORK: Derive these commutation relations.
\end{framed}
\noindent The first and second equations are copies of the \emph{Witt algebra}. Because there are two independent copies, we treat $z$ and $\bar{z}$ as independent variables. We will revisit this independence momentarily and only comment here that we are thus considering $\mathbb{C}^2$ rather than $\mathbb{C}$.

Before proceeding, we would like to consider how these $\ell_m$ generators correspond to the earlier generators of conformal transformations. We first notice that each of these infinite-dimensional algebras contains a finite subalgebra generated by $\ell_{-1}$, $\ell_0$, and $\ell_1$.
\begin{framed}
\noindent HOMEWORK: Notice this by actually checking that it is the case.
\end{framed}
\noindent This subalgebra corresponds to the global conformal group. To argue this, we observe that on $\mathbb{R}^2\simeq\mathbb{C}$ the generators are not everywhere defined. Of course, we should probably be working on the Riemann sphere $S^2\sim\mathbb{C}\cup\{\infty\}$, as it is the conformal compactification of $\mathbb{R}$. Even here, however, some generators (\ref{eq:wittoperator}) are not well defined. The generators $\ell_n$ are non-singular at $z=0$ only for $n\geq-1$. By performing the change of variables $z=-1/w$, we can also see that $\ell_n$ are non-singular as $w\rightarrow0$ for $n\leq1$. Therefore globally defined conformal transformations on the Riemann sphere are generated by $\ell_{-1}, \ell_0,$ and $\ell_1$.

That is all well and good, but how does this finite subalgebra correspond to the momentum, rotation, etc. generators? It is clear that $\ell_{-1}$ and $\bar{\ell}_{-1}$ generate translations on the complex plane. Similarly, $\ell_{1}$ and $\bar{\ell}_{1}$ generate special conformal translations. It also follows that $\ell_0$ (with a corresponding statement about $\bar{\ell}_0$) generates the transformation $z\rightarrow az,a\in\mathbb{C}$. To understand this transformation a little better, we can consider complex polar coordinates $z=re^{i\theta}$. In terms of these variables,
\begin{equation}
\ell_0=-\frac12 r\partial_r + \frac{i}{2}\partial_\theta, \;\;\;\;\;\; \bar{\ell}_0=-\frac12 r\partial_r - \frac{i}{2}\partial_\theta.
\end{equation}
Then the useful linear combinations are easily seen to be
\begin{equation}
\ell_0 + \bar{\ell}_0=-r\partial_r, \;\;\;\;\mbox{and}\;\;\;\; i(\ell_0-\bar{\ell}_0)=-\partial_\theta.
\end{equation}
The first corresponds to the generator of dilatations and the second corresponds to the generator of rotations. Together, these operators generate transformations of the form
$$
z\rightarrow \frac{az+b}{cz+d}, \;\;\;\;\;\;a,b,c,d\in \mathbb{C}.
$$
This is precisely the conformal group from our earlier argument, $PSL(2,\mathbb{C})$\footnote{Despite what was stated earlier, there is no infinite-dimensional conformal group for $\mathbb{R}^2$ (as we have just seen). There are two ways to reconcile this claim with the often-claimed ``infinite dimensionality'' in $d=2$ dimensions. to this statement. The first is that frequently physicists consider infinitesimal conformal invariance, so the full Witt algebra is relevant and we do have infinitely many transformations. The second is that the relevant CFT group is usually the conformal group for \emph{Minkowski} (not Euclidean) space $\mathbb{R}^{1,1}$ (or its compactification $S^1\times S^1$). The conformal group is two copies of $\mbox{Diff}^+(S^1)\times \mbox{Diff}^+(S^1)$ and is truly infinite-dimensional, as is shown in the reference \cite{3-5}. }. 

The Witt algebra is not the complete story, however. This algebra admits what is known as a \emph{central extension}. Central extensions are an important subject in infinite-dimensional Lie theory\footnote{Finite-dimensional simple Lie algebras do not have nontrivial central extensions.}. Allowing central extensions into the algebra allows \emph{projective} representations to become true representations. What does all of this mean? A projective representation is a representation up to some scale factor. In quantum field theory, we most often encounter projective representations: the state $|\phi\rangle$ is physically indistinguishable from any nonzero scalar multiple $c|\phi\rangle$. There is, however, an equivalence between projective representations and a true representation with some central extension. So we find it more useful to study this extension so that we may consider true representations of the conformal group\footnote{Of course, if you have familiarity with string theory then you may have seen the central extension arise due to an operator ordering ambiguity when we consider the quantum theory. Normal ordering constants have an important connection to vacuum energy, as we will see in later lectures.}.

For our purposes, the central extension of the Witt algebra expressed in terms of its elements $L,\bar{L}$ is described by
\begin{align}
[L_m, L_n] &=  (m-n)L_{m+n} + c g(m,n), \;\;\;c\in\mathbb{C}, \nonumber \\
[\bar{L}_m, \bar{L}_n] &= (m-n)\bar{L}_{m+n}+ \bar{c} g(m,n) , \\
[L_m, \bar{L}_n] &= 0. \nonumber
\end{align}
Without loss of generality, we will focus only on the case of $L$ (a similar analysis follows for $\bar{L}$). We will only sketch an argument here and leave the detailed calculations as an exercise.

Trivially, the function $g(m,n)$ is antisymmetric in its arguments. We also remark that by redefining $L_n, n\neq0$ and $L_0$, we can arrange for $g(1,-1)=0$ and $g(n,0)=0$. Because $L_n$ is replacing $\ell_n$, and because the generators $\ell_n$ were elements of a Lie algebra, the $L_n$ will also satisfy the Jacobi identity
$$
[A,[B,C]] + [B,[C,A]]+[C,[A,B]]=0.
$$
Using the Jacobi identity with $L_n$, $L_{m}$, and $L_{0}$, we can show that $g(n,m)=0$ for $m+n\neq0$. Finally, we can use the Jacobi identity with $L_n$, $L_{-1}$, and $L_{-n+1}$ to show that
\begin{equation}
g(n,-n)=\frac{1}{12}(n^3-n).
\end{equation}
\begin{framed}
\noindent HOMEWORK: Carry out the steps in this derivation. Use the normalization $g(2,-2)=\frac12$. This normalization is chosen so that $c$ takes a specific value in the case of a free boson theory.
\end{framed}
\noindent The central extension of the Witt algebra is called the \emph{Virasoro algebra}. The constant $c$ is called the \emph{central charge}. In conclusion, 
\begin{equation}
[L_m,L_n]= (m-n)L_{m+n} + \frac{c}{12}(m^3-m)\delta_{m+n,0}
\end{equation}
with a corresponding formula for $\bar{L}$ and $\bar{c}$. Notice that the central extension does not affect our finite subalgebra of conformal transformations. 

\subsection{Primary fields and radial quantization}

We were previously interested in primary fields and their descendants. What are the fields of interest in two-dimensional conformal field theory? We again perform the identification with complex variables, $\mathbb{R}^2\simeq \mathbb{C}$ and thus consider $\phi(x^0,x^1)\rightarrow \phi(z,\bar{z})$. We define a field depending only on $z$ (and not $\bar{z}$) as \emph{chiral} or \emph{holomorphic}, and fields depending only on $\bar{z}$ are anti-chiral, or anti-holomorphic. Previously, we defined quasi-primary fields as fields having a particular transformation rule related to their scaling dimension. To discuss the analogous definition in two dimensions, we define the \emph{holomorphic dimension} $h$ and anti-holomorphic dimension $\bar{h}$. Under the scaling $z\rightarrow\lambda z$, a field $\phi$ that transforms according to
\begin{equation}
\phi(z,\bar{z})\rightarrow\phi'(z,\bar{z})=\lambda^h \bar{\lambda}^{\bar{h}} \phi(\lambda z,\bar{\lambda} \bar{z})
\end{equation}
has holomorphic dimension $h$ and anti-holomorphic dimension $\bar{h}$. Using these quantities (rather than the scaling dimension), we can define quasi-primary fields. Under the global conformal transformation $z\rightarrow f(z)$, a field transforming according to the rule
\begin{equation}
\phi(z,\bar{z})\rightarrow\phi'(z,\bar{z})=\left( \frac{\partial f}{\partial z} \right)^h \left( \frac{\partial f}{\partial z} \right)^{\bar{h}}\phi(\lambda z,\bar{\lambda} \bar{z})
\end{equation}
is a quasi-primary field. If a field transforms according to this rule for \emph{any} conformal transformation, it is a primary field. Clearly every primary field is quasi-primary. There exist quasi-primary fields that are not primary, however, and some fields are neither (as we shall see).

How do primary fields transform infinitesimally? Under the infinitesimal conformal transformation $z\rightarrow f(z)= z+\epsilon(z)$, we know that
\begin{align}
\left( \frac{\partial f}{\partial z} \right)^h = 1 + h \partial_z \epsilon(z) + O(\epsilon^2), \\
\phi(z+\epsilon(z),\bar{z}) = \phi(z) + \epsilon(z)\partial_z \phi(z,\bar{z}) + O(\epsilon^2).
\end{align}
Then from their definition, we see that under an infinitesimal conformal transformation,
\begin{equation}
\delta_\epsilon \phi(z,\bar{z}) = \left( h\partial_z \epsilon + \epsilon \partial_z + \bar{h}\partial_{\bar{z}}\bar{\epsilon} +\bar{\epsilon} \partial_{\bar{z}} \right)  \phi(z,\bar{z}) \label{eq:threetwothree}.
\end{equation}

To continue our investigation, we will compactify the space direction $x^1$ of our Euclidean theory on a circle of radius $R=1$. This CFT lives on an infinite cylinder described by the complex coordinate $w=x^0 + ix^1$. To picture radial quantization in two dimensions, we can map this cylinder to the complex plane via
\begin{equation}
z=e^w=e^{x^0}e^{ix^1}.
\end{equation}
This maps the infinite past to the origin of the complex plane and the infinite future to the point at infinity. Time translations $x^0\rightarrow x^0+a$ are mapped to dilatations $z\rightarrow e^a z$, and space translations $x^1\rightarrow x^1+b$ are mapped to rotations $z\rightarrow e^{ib} z$. Recalling how we expressed dilatations and rotations in terms of Virasoro generators, we thus see that
\begin{align}
H&=L_0+\bar{L}_0 \\
P &= i(L_0 - \bar{L}_0).
\end{align}

In radial quantization, we also needed to discuss Hermitian conjugation and in- and out-states. The discussions are similar to the case of higher dimensions, so we do not provide every detail. We can Laurent expand a field $\phi$ with conformal dimensions $(h,\bar{h})$ as
$$
\phi(z,\bar{z}) = \sum_{m,\bar{n}\in\mathbb{Z}} z^{-m-h}\bar{z}^{-\bar{n}-\bar{h}}\phi_{m,\bar{n}}.
$$
Then we expect we can define an asymptotic in-state as
$$
|\phi\rangle = \lim_{z,\bar{z}\rightarrow0}\phi(z,\bar{z}) |0\rangle.
$$
This could be singular, however, at $z=0$. To avoid this, we require
$$
\phi_{m,\bar{n}}|0\rangle = 0, \;\;\;\;\;\; m > -h, \;\;\bar{n}> -\bar{h}.
$$
Then in-states are defined as
$$
|\phi\rangle = \lim_{z,\bar{z}\rightarrow0}\phi(z,\bar{z}) |0\rangle = \phi_{-h,-\bar{h}}|0\rangle.
$$
Finding conjugate states is similar. Hermitian conjugation acts as $z\rightarrow 1/\bar{z}$\footnote{Why? This is equivalent to time-reversal. Convince yourself.}. Then
$$
\phi^\dagger(z,\bar{z}) = \bar{z}^{-2h}z^{-2\bar{h}}\phi\left( \frac{1}{\bar{z}},\frac{1}{z} \right).
$$
By peforming the Laurent expansion, we see that $\left( \phi_{m,\bar{n}} \right) ^\dagger = \phi_{-m,-\bar{n}}$. Finally, to define out-states we demand there be no singularity at $w,\bar{w}\rightarrow\infty$. A similar discussion results in an out-state being defined as
\begin{equation}
\langle \phi | = \lim_{w,\bar{w}\rightarrow\infty}w^{2h}\bar{w}^{2\bar{h}}\langle 0 | \phi(w,\bar{w}) = \langle 0 | \phi_{h,\bar{h}}.
\end{equation}

\subsection{The stress-energy tensor and an introduction to OPEs}

Recall that  in a field theory, continuous symmetries correspond to conserved currents. For $x^\mu\rightarrow x^\mu + \epsilon^\mu(x)$, the current can be written as $T_{\mu\nu}\epsilon^\nu$. For constant $\epsilon$, conservation of this current implies conservation of the stress-energy tensor $\partial^\mu T_{\mu\nu}$. For general $\epsilon$, using this with conservation of the conserved current gives the tracelessness of $T_{\mu\nu}$:
\begin{equation}
T^\mu_\mu = 0.
\end{equation}
Let us now see what we can learn from a tracless stress-energy tensor for 2d Euclidean CFTs. 

We perform the same complex change of coordinates, under which $T$ transforms as $T_{\mu\nu} \rightarrow \frac{\partial x^\rho}{\partial x^\mu} \frac{\partial x^\sigma}{\partial x^\nu} T_{\rho\sigma}$. Then we find
\begin{align}
T_{z\bar{z}} &= T_{\bar{z}z}=\frac14 T_\mu^{\mu} = 0, \\
T_{zz} &= \frac14(T_{00}-2i T_{10}-T_{11})=\frac12 (T_{00}-iT_{10}), \\
T_{\bar{z}\bar{z}} &= \frac14(T_{00}+2i T_{10}-T_{11})=\frac12 (T_{00}+iT_{10}).
\end{align}
Using the conservation of $T$,
$$
\partial_0T_{00}+\partial_1T_{10} = 0=\partial_0 T_{01}+\partial_1 T_{11},
$$
we are therefore able to show that
\begin{equation}
\partial_{\bar{z}}T_{zz}= 0.
\end{equation}
Similarly, we can show that $\partial_z T_{\bar{z}\bar{z}}=0.$
\begin{framed}
\noindent HOMEWORK: Complete the steps in deriving these equations.
\end{framed}
\noindent Therefore the nonvanishing components of the stress-energy tensor are a chiral and antichiral field, $T(z)$ and $\bar{T}(\bar{z})$.

Because the current $j_\mu=T_{\mu\nu}\epsilon^{\nu}$ associated with conformal symmetry is conserved, the associated conserved charge
\begin{equation}
Q=\int dx^1\; j_0 \label{eq:threethrees}
\end{equation}
is the generator of symmetry transformations for operator $A$
\begin{equation}
\delta A = [Q,A].
\end{equation}
This commutator is evaluated at equal times. In radial quantization, this corresponds to constant $|z|$. The integral over space in real coordinates therefore becomes a contour integral in the complex plane over some circular contour. The appropriate generalization of (\ref{eq:threethrees}) is then
\begin{equation}
Q = \frac{1}{2\pi i}\oint_\mathcal{C} \left( dz T(z) \epsilon(z) + d\bar{z} \bar{T}(\bar{z})\bar{\epsilon}(\bar{z})  \right).
\end{equation}
This allows us to determine the infinitesimal transformation of a field $\phi$:
\begin{equation}
\delta_{\epsilon} \phi(w,\bar{w}) = \frac{1}{2\pi i}\oint_\mathcal{C} dz\; [ T(z) \epsilon(z) ,\phi(w,\bar{w})]+
 \frac{1}{2\pi i}\oint_\mathcal{C} d\bar{z}\; [ \bar{T}(\bar{z})\bar{\epsilon}(\bar{z}),\phi(w,\bar{w}) ].
\end{equation}

\begin{figure}
\centering
\includegraphics[scale=.8]{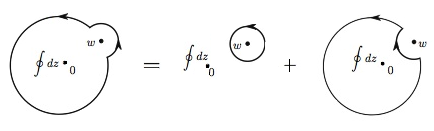}
\caption{Sum of contour integral corresponding to the contour used in the text.} \label{fig:contourcommute}
\end{figure}

There is an ambiguity in this definition, however. Correlation functions in QFT are defined in terms of a time-ordered product. For radial quantization, this becomes a radially-ordered product
\begin{equation}
R A(z)B(w) = \left\{   
\begin{array}{lr}
A(z)B(w)& |z| > |w|,\\
B(w)A(z) & |w| > |z|. 
\end{array}
\right.
\end{equation}
Then using the fact that 
\begin{align}
\oint dz\; [A(z),B(w)] &= \oint_{|z|>|w|} dz\; A(z)B(w) - \oint_{|z|<|w|} dz \;B(w)A(z) \nonumber \\
&= \oint_{\mathcal{C}(w)} dz\; RA(z)B(w), \label{eq:contourcommute}
\end{align}
where the contour of the final integral is taken around the point $w$, then we see
\begin{equation}
\delta_{\epsilon} \phi(w,\bar{w}) = \frac{1}{2\pi i}\oint_{\mathcal{C}(w)} dz\; \epsilon(z) R T(z) \phi(w,\bar{w})
+\mbox{anti-chiral piece}.
\end{equation}
To convince yourself of equation (\ref{eq:contourcommute}), refer to Figure \ref{fig:contourcommute}.

Of course, we have already computed the variation of a field $\phi$ in equation (\ref{eq:threetwothree}). Comparing these expressions gives
\begin{equation}
\delta_\epsilon \phi(w,\bar{w}) = h \partial_w \epsilon(w) \phi(w,\bar{w}) + \epsilon(w) \partial_w \phi(w,\bar{w}) + \mbox{anti-chiral piece}. \label{eq:threefourzero}
\end{equation}
With some work, we can derive
\begin{equation}
RT(z)\phi(w,\bar{w}) = \frac{h}{(z-w)^2}\phi(w,\bar{w}) + \frac{1}{z-w} \partial_w \phi(w,\bar{w})+\cdots, \label{eq:tphiope}
\end{equation}
where the omitted terms are non-singular. 
\begin{framed}
\noindent HOMEWORK: Complete this derivation. To do so, express the quantities $h \partial_w \epsilon(w) \phi(w,\bar{w})$ and $ \epsilon(w) \partial_w \phi(w,\bar{w})$ as contour integrals.
\end{framed}
\noindent Expressions like equation (\ref{eq:tphiope}) are called \emph{operator product expansions} (OPE). In the work that follows, we will omit $R$ and assume radial ordering when we have a product of operators. The OPE is the idea that two local operators inserted at nearby points can be approximated by a string of operators at one of these points. This is an operator equation, and as such we should understand it to hold as an operation insertion inside a radially-ordered correlation function. It is tedious to write the extra characters showing this statement exists inside a correlator, so we typically do not do it. Rather than using our original definitions, we could have defined a primary field as one whose OPE with the stress-energy tensor takes the form (\ref{eq:tphiope}).
\begin{framed}
\noindent HOMEWORK: Show that the OPE of $T(z)$ with $\partial\phi(w)$ (where $\phi$ is a primary field of dimension $h$) has the form
$$
T(z)\partial \phi(w) = \frac{2h\phi(w)}{(z-w)^3}+ \frac{(h+1)\partial \phi(w)}{(z-w)^2} + \frac{\partial^2 \phi(w)}{z-w} +\cdots
$$
Thus as we saw previously, acting with the translation generator $\partial$ increases the holomorphic weight by one.
\end{framed}

Is the stress-energy tensor a primary field? To investigate, we can take the OPE of the stress-energy tensor with itself. To do so, recognize that the Virasoro operators that generate infinitesimal conformal transformations should be the Laurent modes of the stress-energy tensor $T(z)$:
\begin{equation}
T(z) = \sum_{n\in\mathbb{Z}} z^{-n-2}L_n,\;\;\;\; L_n = \frac{1}{2\pi i} \oint dz\; z^{n+1}T(z).
\end{equation}
For a particular conformal transformation $\epsilon(z) = -\epsilon_n z^{n+1}$, then the associated charge is
\begin{equation}
Q_n = \oint \frac{dz}{2\pi i} T(z) \epsilon(z) = -\epsilon_n L_n.
\end{equation}
exactly as we expect. The requirement that these modes obey the Virasoro algebra leads to the $TT$ OPE
\begin{equation}
T(z)T(w) = \frac{c/2}{(z-w)^4} + \frac{2T(w)}{(z-w)^2}+\frac{\partial_w T(w)}{z-w}+\cdots. \label{eq:ttope}
\end{equation}
\begin{framed}
\noindent HOMEWORK: By considering $[L_m,L_n]$ as defined by contour integrals, show that the $TT$ OPE given in (\ref{eq:ttope})  leads to the Virasoro algebra. Thus this is the correct form of the OPE.
\end{framed}
\noindent We also remark here that the condition that $T(z)$ is Hermitian implies $L^\dagger_n = L_{-n}$.

What does this OPE mean? For starters, the stress-energy tensor is generically \emph{not} a primary field. The only way that $T$ can be a primary field is if the central charge vanishes. By performing a calculation similar to this exercise, it can be shown for a chiral primary field $\phi(z)$ that
\begin{equation}
[L_m,\phi_n]=((h-1)m - n)\phi_{m+n}, \;\;\;m,n\in \mathbb{Z}. \label{eq:threefourfive}
\end{equation}
If this relation holds only for $m=0,\pm1$, then $\phi$ is a quasi-primary field.  Then we imediately conclude that although $T(z)$ is not a primary field, it \emph{is} a quasi-primary field of conformal dimension $(h,\emph{h})=(2,0)$. A similar statement holds for $\bar{T}$. To see how $T$ transforms infinitesimally, we can again perform a contour integral calculation to get
\begin{align}
\delta_\epsilon T(z) &= \frac{1}{2\pi i} \oint_{\mathcal{C}(z)}dw \;\epsilon(w)T(w)T(z) \nonumber \\
&= \frac{c}{12}\partial^3_z \epsilon(z) + 2 T(z) \partial_z \epsilon(z) + \epsilon(z) \partial_z T(z). \label{eq:threefoursix}
\end{align}
\begin{framed}
\noindent HOMEWORK: Derive equation (\ref{eq:threefourfive}) or (\ref{eq:threefoursix}). 
\end{framed}

Another important condition on $L_n$ comes from demanding regularity at $z=0$ of 
\begin{equation}
T(z)|0\rangle=\sum_{n\in\mathbb{Z}} L_n z^{-n-2}|0\rangle.
\end{equation}
For terms in the sum with $m> -2$, we find singularities. The only way to ensure that these vanish is by requiring
\begin{equation}
L_n|0\rangle =0, \;\;\;\;\;\; n \geq -1.
\end{equation}
This in turn implies that 
\begin{equation}
\langle 0| L_n = 0, \;\;\;\;\;\; n \leq 1.
\end{equation}
Non-trivial Hilbert space states transforming as part of some Virasoro algebra representation are thus generated by $L_{-n}|0\rangle,\;n\geq 2$. 
The vacuum state is annihilated by the generators of global conformal transformations, exactly as we should expect.

It can be shown (though we will not) that under conformal transformations $f(z)$, the stress-energy tensor transforms as
\begin{equation}
T(z)\rightarrow T'(z) = \left( \frac{\partial f}{\partial z}\right)^2  T(f(z))+\frac{c}{12} S(f(z),z),
\end{equation}
where
\begin{equation}
S(w,z) = \frac{1}{(\partial_z w)^2} \left( (\partial_z w)(\partial_z^3 w)-\frac{3}{2} (\partial^2_z w)^2 \right).
\end{equation}
The quantity $S$ is known as the \emph{Schwartzian} derivative.
If you want, verify this form for the infinitesimal conformal transformation\footnote{Note: by ``if you want,'', I mean ``I want you to''.}.
\begin{framed}
\noindent HOMEWORK: Prove that the Schwartzian vanishes if and only if $w(z)=\frac{az+b}{cz+d}$. This shows that although $T(z)$ is not a primary field, it \emph{is} quasi-primary.
\end{framed}

We conclude this section by claiming that our Virasoro generator constraints and commutation relations imply
\begin{equation}
\langle T(z)T(w) \rangle = \frac{c/2}{(z-w)^4}.
\end{equation}
This gives an easy way to calculate the central charge for some theories. Further, by considering the quantity
\begin{equation}
\langle [L_2,L_{-2}] \rangle,
\end{equation}
we find that for unitary theories the central charge is nonnegative
$$
c \geq 0 \;\;\;\; \mbox{(unitary)}.
$$
\begin{framed}
\noindent HOMEWORK: These derivations are straightforward. Do them both.
\end{framed}

\subsection{Highest weight states and unitarity bounds}

We now turn our attention to the state $|h\rangle = \phi(0)|0\rangle$ created by the chiral field $\phi$ with holomorphic dimension $h$. Using the OPE, we see
\begin{equation}
[L_n, \phi(w)] = \oint \frac{dz}{2\pi i}z^{n+1} T(z) \phi(w) = h(n+1)w^n\phi(w) + w^{n+1}\partial\phi(w).
\end{equation}
This implies that 
\begin{equation}
[L_n,\phi(0)] = 0,\;\; n>0.
\end{equation}
Thus we conclude that the state $|h\rangle$ satisfies
\begin{equation}
L_0 |h\rangle = h|h\rangle, \;\;\;\;\;\; L_n|h\rangle = 0,\;n>0, \label{eq:threefivesix}
\end{equation}
with a straightforward extension for $|h,\bar{h}\rangle$.
\begin{framed}
\noindent HOMEWORK: Prove these statements for either $L_n$ or $\bar{L}_n$. How fun! I'm letting you choose.
\end{framed}
\noindent Recalling that $L_0\pm\bar{L}_0$ are generators of dilatations and rotations, we realize that
\begin{equation}
h = \Delta + s, \;\;\;\;\;\;\; \bar{h} = \Delta - s,
\end{equation}
where $s$ is the Euclidean spin.

A state satisfying (\ref{eq:threefivesix}) is known as a \emph{highest weight state}. Acting with Virasoro generators gives descendant states. From the previous exericise, we have seen that unitarity constrains the allowed value for the central charge. Can we derive additional constraints using descendant states? We evaluate
\begin{align}
\langle L^\dagger_{-n}L_{-n} \rangle &=\langle [L_{n}, L_{-n}] \rangle \\
&=  2n\langle h|L_0|h \rangle+\frac{c}{12}(n^3-n)\langle h|h \rangle\\
&=  \left(2nh + \frac{c}{12}(n^3-n)  \right)\langle h|h \rangle.
\end{align}
For unitarity, this quantity must be nonnegative. For large $n$, the second term dominates and we must therefore have $c>0$. When $n=1$, we have the condition that $h\geq 0$. We also see that $h=0$ only if $L_{-1}|h\rangle=0$, meaning precisely when $|h\rangle$ is the vacuum.

What about the case $c=0$? In this case, the states $L_{-n}|0\rangle$ have zero norm and can therefore be set equal to zero. For arbitrary $h$, we refer you to \cite{3-6}. There, the author considers the matrix of inner products in the basis $L_{-2n}|h\rangle, L^2_{-n}|h\rangle$. The determinant of this matrix is $4n^3 h^2(4h-5n)$. For nonvanishing $h$, we can always choose $n$ large enough to make this quantity negative---but that is impossible. Then for $c=0$, the only unitary representation of the Virasoro algebra has $h=0$ and $L_n=0$.

Of course, we can once again place constraints on two- and three-point functions directly. The arguments rely on using the global conformal symmetry to constrain the allowed forms for correlators of chiral quasi-primary fields. We have already done these derivations for $d\geq 3$, so we will not present every step of the derivation. Invariance under $L_{-1}, L_0$, and $L_1$ fix the two-point function of two chiral quasi-primary fields to be of the form
\begin{equation}
\langle \phi_i(z) \phi_j(w) \rangle = \frac{C_{ij}\delta_{h_i,h_j}}{(z-w)^{2h_i}}.
\end{equation}
This is of precisely the same for as for higher dimensions. We again remark that we can choose a basis for our fields so that the constants $C_{ij}$ go as a Kronecker $\delta_{ij}$. Performing a similar derivation fixes the three-point function to be
\begin{equation}
\langle \phi_1(z_1) \phi_2(z_2) \phi_3(z_3)\rangle = \frac{C_{123}}{z_{12}^{h-2h_3}z_{23}^{h-2h_1} z_{13}^{h-2h_2}},
\end{equation}
where we have defined $h = \sum h_i$. We conclude this brief discussion by considering the effect on the two-point function of a rotation $z\rightarrow e^{2\pi i}z$. In order for this correlator to be single-valued, it must be the case that the conformal dimension of a chiral quasi-primary field is either integer or half-integer. 

\subsection{Ward identities}

\begin{figure}
\centering
\includegraphics[scale=.6]{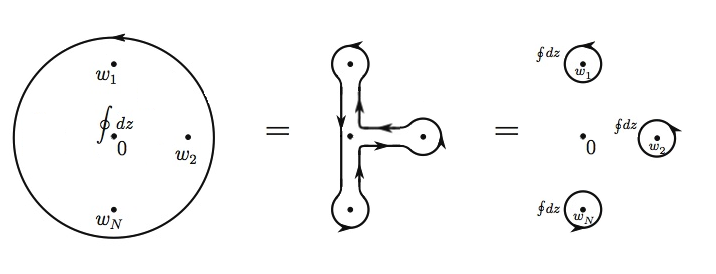}
\caption{Deformation of contours} \label{figure:deformcontour}
\end{figure}

We have a good deal more to say about states in our theory, but we will return to these topics in the next lecture. For now, we will conclude by discussing Ward identities---identities between correlators resulting from symmetries of the theory---and normal ordering.
First, let us begin by considering constraints on $n$-point correlators from global conformal transformations. Recall that global conformal transformations correspond to Virasoro generators $L_k, \bar{L}_k,$ with $k=0,\pm1$. For these generators, we know that the vacuum satisfies
\begin{equation}
\langle 0| L_k = 0, \;\;\;\;\;\; L_k|0\rangle = 0
\end{equation}
(and similarly for $\bar{L}_k$).
Then for quasi-primary fields $\phi_i$, it follows that
\begin{align}
0 &= \epsilon \langle 0 | L_k \phi_1\cdots \phi_n |0\rangle \nonumber \\
&= \epsilon \langle 0| [L_k, \phi_1] \cdots \phi_n |0\rangle + \cdots + \langle 0| \phi_1\cdots [L_k,\phi_n] |0\rangle \label{eq:confward}
\end{align}
where $\epsilon$ is some infinitesimal parameter. These commutators involving Virasoro generators just give the infinitesimal change in the quasi-primary field given by equation (\ref{eq:threefourzero}). Using this expression for the $k=0,\pm1$ modes of $\epsilon$, we find that equation (\ref{eq:confward}) is equivalent to
\begin{align}
\sum_i \partial_{z_i}\langle  \phi_1(z_1)\cdots \phi_n (z_n)   \rangle = 0 \\
\sum_i (z_i\partial_{z_i}+h_i)\langle  \phi_1(z_1)\cdots \phi_n (z_n)   \rangle = 0 \\
\sum_i (z_i^2\partial_{z_i}+2h_i z_i)\langle  \phi_1(z_1)\cdots \phi_n (z_n)   \rangle = 0 .
\end{align}
\begin{framed}
\noindent HOMEWORK: We can investigate consequences of conformal invariance directly from these Ward identifies. (a) Begin with the one-point function. What constraints are placed on this correlator from the Ward identities? (b) Consider the two-point function. What constraints are placed on this correlator from the Ward identities? (c) Consider the three-point function. Speculate what constraints will follow from the Ward identities.
\end{framed}
We also derive the \emph{conformal Ward identity}. To make statements using the local conformal algebra, we consider a collecion of operators located at points $w_i$ surrounded by some $z$ contour (refer to Figure \ref{figure:deformcontour}). To perform a conformal transformation inside this region, we integrate $\epsilon(z)T(z)$ around the contour. This single contour encompassing all of the operators can be deformed to a sum of terms, every term coming from a contour around an individual operator. Then
\begin{align} 
&\langle \oint \frac{dz}{2\pi i} \epsilon(z)T(z) \phi_1(w) \cdots \phi_n (w_n) \rangle \\
&= \sum_{i=1}^n \langle \phi_1(w)\cdots \left( \oint \frac{dz}{2\pi i} \epsilon(z)T(z) \phi_i(w)\right) \cdots \phi_n (w_n) \rangle \\
&=  \sum_{i=1}^n \langle \phi_1(w)\cdots \delta_\epsilon \phi_i(w) \cdots \phi_n (w_n) \rangle ,
\end{align}
where $\delta_\epsilon \phi$ is given by equation (\ref{eq:threetwothree}).

Recalling how we expressed $h \partial_w \epsilon(w) \phi(w,\bar{w})$ and $ \epsilon(w) \partial_w \phi(w,\bar{w})$ as contour integrals (What? You skipped that exercise? Guess you will just have to do it now.), we get
\begin{align}
 0= \oint \frac{dz}{2\pi i} \epsilon(z) \bigg[  \langle T(z) \phi_1(w) \cdots \phi_n (w_n) \rangle 
-  \sum_{i=1}^n  \left(\frac{h_i}{(z-w_i)^2}+\frac{1}{z-w_i}\partial_{w_i}  \right) \langle \phi_1(w)\cdots \phi_n (w_n) \rangle \bigg].
\end{align}
Finally, this expression must be true for all $\epsilon=-z^{n+1}$. Thus the integrand must vanish, and we have derived the conformal ward identity
\begin{equation}
   \langle T(z) \phi_1(w) \cdots \phi_n (w_n) \rangle 
=  \sum_{i=1}^n  \left(\frac{h_i}{(z-w_i)^2}+\frac{1}{z-w_i}\partial_{w_i}  \right) \langle \phi_1(w)\cdots \phi_n (w_n) \rangle.
\end{equation}

\subsection{More about operator product expansions}
The operator product expansion is an incredibly important tool, so we will make a brief aside to perform an example calculation. The OPE is most straightforward in free field theory, where it almost reduces to a Taylor series expansion. For a free, massless scalar $\phi(x)$ in $d=4$ dimensions\footnote{For a free massless scalar, we know $\Delta=1$ and that it receives no quantum corrections; this is truly a quantum conformal field theory.}, we can use Wick's theorem to write
\begin{equation}
\phi(x)\phi(0) = \frac{C}{x^2} + :\phi(x)\phi(0):,
\end{equation}
where : : denotes normal ordering (moving all annihilation operators to the right of creation operators) and $C$ is a numerical normalization constant. The first term reflects the leading singular behavior at short distances. For the normal-ordered term, we can use the Taylor expansion
\begin{equation}
:\phi(x)\phi(0): = \sum_{n=0}^\infty \frac{1}{n!} x^{\mu_1}\cdots x^{\mu_n} :\partial_{\mu_1}\cdots \partial_{\mu_n} \phi(0)\phi(0):.
\end{equation}
The operator appearing in the $n^{\mbox{th}}$ term has dimension $n + 2$, and we understand $\partial_\mu \phi(0) \equiv \partial_\mu \phi(x) \bigg|_{x=0}$. 

The OPE can be easily generalised to composite operators defined by normal ordering. For example, the OPE of $:\phi^2:$ with itself, after we applying Wick’s theorem, becomes
\begin{equation}
:\phi^2(x)::\phi^2(0): = \frac{2C^2}{x^4}  + \frac{4C}{x^2} :\phi(x)\phi(0): + :\phi^2(x)\phi^2(0):.
\end{equation}
We can then apply Taylor series expansions to both normal-ordered operators on the RHS to obtain an infinite sequence of local operators of increasing dimension.

We also note that the expansion in terms of local operators may be reordered. For example, using the equation of motion $\partial^2 \phi = 0$ lets us write
\begin{equation}
\phi(x)\phi(0) = \frac{C}{x^2} + \left(1+\frac12 x^\mu\partial_\mu + \frac14x^\mu x^\nu\partial_\mu\partial_\nu+\frac{1}{16} x^2\partial^2 \right) :\phi^2(0):-\frac12x^\mu x^\nu T_{\mu\nu} + O(x^3).
\end{equation}
Here,
\begin{equation}
T_{\mu\nu} \equiv : \partial_\mu\phi\partial_\nu\phi:−\frac14 \eta_{\mu\nu} : \partial\phi \cdot\partial\phi:
\end{equation}
This demonstrates that many operators appearing in the operator product expansion are expressible in terms of derivatives of lower dimension operators. 
\begin{framed}
\noindent HOMEWORK: Check that this OPE expression is correct.
\end{framed}

For conformal field theories, the generic form of the two- and three-point functions lets us find the general form of the OPE between two chiral quasi-primary fields in terms of other quasi-primaries and their derivatives\footnote{The proof that the OPE involves \emph{only} other quasi-primaries and their derivatives will not be presented here, as it is rather complicated. We refer you to \cite{3-7} and its references}. Our ansatz should thus be of the form
\begin{equation}
\phi_i(z)\phi_j(w) = \sum_{k,m\geq0} C_{ij}^k \frac{a_{ijk}^m}{m!}\frac{1}{(z-w)^{h_i+h_j-h_k-m}}\partial^m\phi_k(w),
\end{equation}
where the $(z-w)$ dependence is fixed by symmetry. By taking $w=1$, we can insert this expression into the three-point function
\begin{equation}
\langle  \left( \phi_i(z)\phi_j(1) \right) \phi_k(0)\rangle = \sum_{l,m\geq0} C_{ij}^l \frac{a_{ijl}^m}{m!}\frac{1}{(z-1)^{h_i+h_j-h_k-m}} \langle \partial^m\phi_l(1) \phi_k(0)\rangle.
\end{equation}
Then using the general form for the two-point function, we know that
\begin{equation}
\langle \partial^m\phi_l(z) \phi_k(0)\rangle\bigg|_{z=1}=\partial^m\left( \frac{d_{lk}\delta_{h_l,h_k}}{z^{2h_k}} \right) \bigg|_{z=1} = (-1)^m m! {2h_k+m-1 \choose m} d_{lk}\delta_{h_l,h_k}.
\end{equation}
We therefore obtain
\begin{equation}
\langle  \phi_i(z)\phi_j(1) \phi_k(0)\rangle = \sum_{l,m\geq0} C_{ij}^l d_{lk} a_{ijk}^m {2h_k+m-1 \choose m} \frac{(-1)^m}{(z-1)^{h_i+h_j-h_k-m}}.
\end{equation}

We can simplify further by comparing this expression to the form of the general three-point function with $z_1=z,z_2=1,z_3=0$. Equating these relations gives
\begin{equation}
 \sum_{l,m\geq0} C_{ij}^l d_{lk} a_{ijk}^m {2h_k+m-1 \choose m} {(-1)^m (z-1)^m} = \frac{C_{ijk}}{(z-1)^{h_i+h_j-h_k}(1+(z-1))^{h_i+h_k-h_j})}.
\end{equation}
The peculiar form of the RHS of this equation is because I know that we should make use of the relation
\begin{equation}
\frac{1}{(1+x)^p}=\sum_{m=0}^\infty (-1)^m  {p+m-1 \choose m}x^m.
\end{equation}
Comparing coefficients then allows us to fix the forms of $C_{ij}^l$ and $a_{ijk}^m$. The final result is that the OPE of two chiral quasi-primary fields has the form
\begin{equation}
\phi_i(z)\phi_j(w) =  \sum_{k,m\geq0} C_{ij}^k \frac{a_{ijk}^m}{m!}  \frac{1}{(z-w)^{h_i+h_j-h_k-m}}\partial^m\phi_k(w),
\end{equation}
where the coefficients are defined as
\begin{align}
a_{ijk}^m &= {2h_k+m-1 \choose m}^{-1}{h_k+h_i-h_j+m-1 \choose m}, \\
C_{ijk} &= C_{ij}^l d_{lk}.
\end{align}
\begin{framed}
\noindent HOMEWORK: Work through this derivation carefully. It is a straightforward (although time-consuming) calculation.
\end{framed}

We can use this result to make statements about the Laurent modes of fields. Recall the Laurent expansion
\begin{equation}
\phi_i(z) = \sum_m z^{-m-h_i}\phi_m^i,
\end{equation}
where $h_i$ for a chiral quasi-primary field are integer or half-integer, $m$ labels the mode, and $i$ labels the field. After expressing the modes as contour integrals, using our expression for the OPE, and performing the actual calculation---an involved sequence of steps that we leave as one of the longer exercises---we finally arrive at the expression
\begin{equation}
[\phi^i_m,\phi^j_n] = \sum_{k} C_{ij}^k p_{ijk}(m,n) \phi_{m+n}^k + d_{ij}\delta_{m,-n} {m+h_i -1 \choose 2h_i -1}. \label{eq:modehw}
\end{equation}
Here we have defined 
\begin{equation}
p_{ijk}(m,n)=\sum_{\substack{r,s\in \mathbb{Z}_0^+ \\  r+s=h_i+h_j-h_k-1}}  C_{r,s}^{ijk} {h_i - m - 1 \choose r} \cdot {h_j-n-1 \choose s}
\end{equation}
and
\begin{equation}
C_{r,s}^{ijk} =(-1)^r \frac{(2h_k-1)!}{(h_i+h_j+h_k-2)!}\prod_{t=0}^{s-1} (2h_i - 2 - r - t) \prod_{u=0}^{r-1}(2h_j-2-s-u)
\end{equation}
Note that only fields with $h_k < h_i+h_j$ can appear in the above mode algebra.

We conclude by using these results to check several earlier calculations. For example, at this point we can show that the norm of the state $|\phi\rangle = \phi_{-h}|0\rangle$ is equal to the structure constant of the two-point function $d_{\phi\phi}$. Similarly, we show that the three-point correlation  function of $\phi_1,\phi_2,\phi_3$ really gives the constant $C_{123}=\lambda_{123}$. A more involved calculation involves the Virasoro algebra. The Virasoro generators are Laurent modes of the energy-momentum tensor; from the above equations, we arrive at
\begin{equation}
[L_m,L_n] = C^L_{LL}p_{222}(m,n)L_{m+n}+d_{LL}\delta_{m,-n} {m+1 \choose 3}.
\end{equation}
We have set $C_{LL}^k = 0$ for $k\neq L$ because we are cheating a little and already know the structure of the Virasoro algebra (and we have started indexing $p$ by the conformal weight of the chiral fields involved). Using the explicit expressions, we find
\begin{align}
p_{222}(m,n) = C_{1,0}^{222} {1-m \choose 1} + C_{0,1}^{222} {1-n \choose1}, \\
C_{1,0}^{222} = -\frac12, \;\;\;\; C_{1,0}^{222} = \frac12.
\end{align}
Combing all of these results gives the Virasoro algebra
\begin{equation}
[L_m,L_n] = C^L_{LL}\frac{m-n}{2}L_{m+n}+d_{LL}\delta_{m,-n} \frac{m^3-m}{6}.
\end{equation}
Comparing this to the previous result, we see the constants are equal to
\begin{equation}
d_{LL}=\frac{c}{2}, \;\;\;\;\;\; C^L_{LL}=2.
\end{equation}
We see that this two-point constant matches our expectations from earlier expressions; we also now have a prediction for the three-point constant.

\break

\subsection*{References for this lecture}
\vspace{4mm}
\noindent Main references for this lecture
\\
\begin{list}{}{%
\setlength{\topsep}{0pt}%
\setlength{\leftmargin}{0.7cm}%
\setlength{\listparindent}{-0.7cm}%
\setlength{\itemindent}{-0.7cm}%
\setlength{\parsep}{\parskip}%
}%
\item[]

[1] Chapter 2 of the textbook: R. Blumenhagen, E. Plauschinn, \emph{Introduction to Conformal Field Theory: With Applications to String Theory}, Lect. Notes Phys. 779,  (Springer, Berlin Heidelberg 2009).

[2] Chapter 4 of the textbook: P. Di Francesco, P. Mathieu, and D. Senechal. \emph{Conformal field theory}, (Springer, 1997).

[3] S. Rychkov, \emph{EPFL Lectures on Conformal Field Theory in $D\geq 3$ Dimensions: Lecture 1: Physical Foundations of Conformal Symmetry}, (Lausanne, Switzerland, \'Ecole polytechnique f\'ed\'erale de Lausanne, December 2012).

[4] S. Rychkov, \emph{EPFL Lectures on Conformal Field Theory in $D\geq 3$ Dimensions: Lecture 3: Radial quantization and OPE}, (Lausanne, Switzerland, \'Ecole polytechnique f\'ed\'erale de Lausanne, December 2012).

\end{list}

\break

\section{Lecture 4: Simple Examples of CFTs}

\subsection{Example: Free boson}
In this lecture, we will present some example conformal field theories and study their properties.
Let us begin our exploration of specific conformal field theories with some simple examples given in terms of a Lagrangian action. We consider first a massless scalar field $\phi(x^0,x^1)$ defined on cylinder of radius $R=1$. The action for this theory is then
\begin{align}
S &= \frac{g}{2} \int dx^0 dx^1\sqrt{|h|}h^{\alpha\beta}\partial_\alpha \phi \partial_\beta \phi \\
&= \frac{g}{2} \int dx^0 dx^1\left( (\partial_0 \phi)^2 + (\partial_1 \phi)^2 \right),
\end{align}
where $h\equiv \mbox{det} \;h_{\alpha\beta}, \;h_{\alpha\beta}=\mbox{diag}(1,1),$ and $g$ is a normalization constant. If you have studied string theory, you recognize this as the Euclidean world-sheet action of a string moving in a flat background in the conformal gauge with coordinate $\phi$. There are no mass/length scales in this action, so we expect conformal invariance. In fact, because this is a free theory we expect that any conformal dimensions will remain unchanged after quantization.

Once again, we perform an exponential map to move from the cylinder to the complex plane. Then the action becomes
\begin{equation}
S = \frac{g}{2} \int dz d\bar{z} \;\partial \phi \cdot \bar{\partial} \phi.
 \end{equation} 
The equation of motion for this action is found to be
\begin{equation}
\partial\bar{\partial}\phi(z,\bar{z})=0.
\end{equation}
\begin{framed}
\noindent HOMEWORK: Follow the steps and derive this equation of motion.
\end{framed}
\noindent 
This equation of motion means that the boson $\phi$ can be split into pieces that are holomorphic and antiholomorphic (or left- and right-moving, from a stringy perspective)
\begin{equation}
\phi(z,\bar{z}) = \phi_L(z) + \phi_R(\bar{z}).
\end{equation}
It is also easy to conclude that $j(z)\equiv i\partial \phi$ is a chiral field and $\bar{j}(\bar{z})\equiv i \bar{\partial}\phi$ is an anti-chiral field. We can also explicitly check this action is invariant under conformal transformations if $\phi$ has conformal dimension equal to zero.
\begin{framed}
\noindent HOMEWORK: Check this.
\end{framed}

Without much effort, we can calculate the propagator $G(z,\bar{z},w,\bar{w})=\langle \phi(z,\bar{z})\phi(w,\bar{w})\rangle$ for this theory. From our experience with quantum field theory\footnote{One way to see this is through canonical quantization. Another way is by demanding the variation of the path integral vanish. Do one or both on your own time.}, we recall that this equation of motion means the propagator must satisfy
\begin{equation}
\partial\bar{\partial}G = -\frac{1}{g} \delta^{(2)}(z-w).
\end{equation}
Using the fact\footnote{
You do not believe this fact? Well, then I suppose
\begin{framed}
\noindent HOMEWORK: Prove this relation. If your complex analysis is a little rusty or you keep misplacing factors of $i$, feel free to switch back to real variables and use Green's theorem to complete the derivation.
\end{framed}} 
that 
\begin{equation}
\bar{\partial}\partial \ln|z-w|^2 =\bar{\partial} \left( \frac{1}{z-w}\right) =  2\pi \delta^{(2)}(z-w), \label{eq:logformula}
\end{equation}
we can show
\begin{equation}
G(z,\bar{z},w,\bar{w}) = \langle \phi(z,\bar{z})\phi(w,\bar{w})\rangle = -\frac{1}{4\pi g} \log |z-w|^2.
\end{equation}
Comparing this to earlier equations, we see that the free boson is \emph{not} a quasi-primary field---we expect pole-type singularities, not logarithms. We can show, however, that
\begin{equation}
\langle j(z) j(w) \rangle = \frac{1}{4\pi g}\frac{1}{(z-w)^2}
\end{equation}
with a similar statement for the anti-chiral fields and with the correlator between chiral and anti-chiral vanishing. Once we define the stress-energy tensor, we can show that the fields $j, \bar{j}$ \emph{are} primary fields with dimensions $(h,\bar{h})=(1,0)$ and $(h,\bar{h})=(0,1)$ respectively.

A chiral (or anti-chiral) field in two dimensions with conformal dimension $h=1$ is called a \emph{current}. In a theory with $N$ quasi-primary currents, we can express them as Laurent series
$$
j_i(z) = \sum_{n\in\mathbb{Z}} z^{-n-1} j_{(i)n}.
$$
Recalling equation (\ref{eq:modehw}), we determine the algebra of the Laurent modes to be
\begin{equation}
[ j_{(i)m},  j_{(j)n}] = \sum_k C^k_{ij}p_{111}(m,n) j_{(k)m+n} + C_{ij} m \delta_{m+n,0}.
\end{equation}
We explictly calculate $p_{111}=1$, so the antisymmetry of the commutator means $C^k_{ij}$ is antisymmetric in its lower indices. By rotating among the fields, we can make $C_{ij}=k\delta_{ij}$ for some constant $k$. Then rewriting $C^k_{ij}$ in the new basis as $f^{ijk}$ and playing index gymnastics, we find that currents satisfy
\begin{equation}
[j^i_m,j^j_n]=\sum_\ell f^{ij\ell} j^\ell_{m+n}+km\delta^{ij}\delta_{m+n,0}.
\end{equation}
The $f$'s are called \emph{structure constants} and $k$ is called the \emph{level}. This algebra is a generalization of a Lie algebra called a \emph{Ka\u{c}-Moody algebra}.

We can simplify these expressions for the case of our current theory. We have only one chiral current, meaning the antisymmetry of $C^k_{ij}$ forces that term to vanish. This means our current algebra becomes
\begin{equation}
[j_m,j_n]=\kappa m \delta_{m+n,0}. \label{eq:modealgebra}
\end{equation}
We will use this expression later when calculating partition functions.

Before moving forward, we should consider an important property of the two- dimensional propagator. In higher dimensions, a free scalar field has an infinite number of ground states determined by the vacuum expectation value. It is possible for the vacuum expectation value not to be invariant under some continuous group of 
transformations even though the group is a symmetry group of the theory, with conserved currents and everything. This is the \emph{Goldstone phenomenon}---Goldstone's theorem says that when this happens, the theory has massless scalar bosons. The massless scalar bosons are the small fluctuations around the vacuum corresponding to the broken translational invariance $\phi\rightarrow \phi+c$. Unlike the cases $d\geq3$, in $d=2$ dimensions (and fewer) the propagator has an infrared divergence. This divergences is telling us that the wavefunction associated with this massless scalar particle wants to spread out. These massless scalar fields are therefore \emph{not} Goldstone bosons; in fact, there are no Goldstone bosons in two dimensions \cite{coleman}. Although our theory looks like it has a translation symmetry, this is not the case at the quantum level due to the logarithmic divergence of the $\phi$ propagator.

To determine the stress-energy tensor, we must first consider \emph{normal ordering}. The building blocks of our CFTs are constructed from operators $\phi_i$, derivatives $\partial\phi_i, \partial^2\phi_i,\cdots$, and products of fields at the same spacetime point. As we know from QFT, we need some sort of rule for operator ordering in such products. When we say ``normal ordering'', you should think ``creation operators on the left''. Let us first consider the stress-energy tensor $T$. 
The stress-energy tensor for a free massless boson is given in real coordinates by
\begin{equation}
T_{\mu\nu}=g (\partial_\mu \phi \partial_\nu \phi-\frac12\eta_{\mu\nu}\partial_\lambda \phi\partial^\lambda \phi).
\end{equation}
The ``quantum'' version of this expression (accounting for normal ordering) written in complex coordinates will be
\begin{equation}
T(z) = -2\pi g :\partial \phi \partial \phi:,
\end{equation}
where in order to match our references we have now chosen the normalization $T=-2\pi T_{zz}$. 

We could go through a careful treatment of various composite fields in terms of their modes, explicitly determining which modes are creation operators and which are annihilation operators. This would involve expressing the stress-energy tensor and currents $j(z)$ in terms of their modes and commuting them according to the algebra (\ref{eq:modealgebra}).  We present only the result
\begin{equation}
L_n \propto \sum_{k>-1} j_{n-k}j_k   \sum_{k\leq -1} j_k j_{n-k}. \label{eq:Lfromjs}
\end{equation}
\begin{framed}
\noindent HOMEWORK: Find this constant of proportionality for the theory we are considering by doing this calculation.
\end{framed}
\noindent In particular, for example, we could normalize our action so that
\begin{equation}
L_0 = \frac12 j_0 j_0 + \sum_{k=1}^\infty j_{-k}j_k. \label{eq:lzerodef}
\end{equation}
In the interest of expediency, we could also just define normal ordering of the stress-energy tensor as
\begin{equation}
T(z) = -2\pi g \lim_{w\rightarrow z} (\partial\phi(z)\partial\phi(w) - \langle \partial\phi(z)\partial\phi(w)\rangle).
\end{equation}
The results are exactly the same.

At this point we may calculate, for example, the OPE of $T(z)$ with $\partial\phi$. To do this, we point out that Wick's theorem still holds for conformal field theories. We must sum over all possible contractions of pairs of operators, where a contraction means we replace the product of fields with the associated propagator. So then
\begin{align}
T(z)\partial\phi(w) &= -2\pi g : \partial\phi(z) \partial\phi(z): \partial\phi(w) \\
 \contraction{-4\pi g : \partial\phi(z)}{\partial\phi(z)}{:}{\partial\phi(w)}
&\sim -4\pi g : \partial\phi(z) \partial\phi(z): \partial\phi(w) \\
&\sim \frac{\partial\phi(z)}{(z-w)^2}.
\end{align}
Expanding $\partial\phi(z)$ around the point $w$, we find the OPE
\begin{equation}
T(z)\partial\phi(w) \sim \frac{\partial\phi(w)}{(z-w)^2} + \frac{\partial^2_w\phi(w)}{(z-w)}.
\end{equation}
This matches our earlier expression (\ref{eq:tphiope})\footnote{If it seems like we missed some contractions in this calculation, it is because we do not ``self-contract'' inside a normal-ordering. If you cannot see why this is the case, try un-normal-ordering before using Wick's theorem.}.

In a similar way, we can calculate the OPE of the stress-energy tensor with itself
\begin{align}
T(z)T(w) &= 4\pi^2 g^2 : \partial\phi(z) \partial\phi(z): : \partial\phi(w) \partial\phi(w): \\
&\sim \frac{1/2}{(z-w)^4}-\frac{4\pi g : \partial\phi(z) \partial\phi(w):}{(z-w)^2}\\
&\sim \frac{1/2}{(z-w)^4} + \frac{2T(w)}{(z-w)^2}+\frac{\partial T(w)}{(z-w).}
\end{align}
Again, we find that $c=1$ for a free boson and that the stress-energy tensor is \emph{not} a primary field.

Using $\phi$,we can define a \emph{vertex operator}
\begin{equation}
V_\alpha(z,\bar{z}) = :e^{i\alpha \phi(z,\bar{z})}:.
\end{equation}
From a completely general point of view, we could be interested in vertex operators due to their very existence; any object in a conformal field theory makes for worthwhile study. To obtain a more physical understanding, we observe that this vertex operator has the same form as a plane wave that we might use when studying in- and out-states for QFT scattering amplitudes. In string theory, vertex operators like this expression (and derivatives of this expression) correspond to initial or final states in string scattering amplitudes. We leave a more thorough discussion as one of the exercises.

To investigate this vertex operator, we first takes its OPE with $\partial \phi$ (using a series expansion):
\begin{equation}
\partial \phi(z) V_\alpha(w,\bar{w})\sim -\frac{i\alpha}{4\pi g}\frac{V_\alpha(w,\bar{w})}{z-w}.
\end{equation}
With this, we can calculate the OPE of $V_\alpha$ with the energy-momentum tensor to find
\begin{equation}
T(z) V_\alpha(w,\bar{w}) = \frac{\alpha^2}{8\pi g}\frac{V_\alpha(w,\bar{w})}{(z-w)^2} + \frac{\partial_w V_\alpha(w,\bar{w})}{z-w}.
\end{equation}
Thus this vertex operator is a primary field with conformal weight $h(\alpha)=\bar{h}(\alpha)=\frac{\alpha^2}{8\pi g}.$
\begin{framed}
\noindent HOMEWORK: Derive these formulas using the series expansion for the exponential function when necessary.
\end{framed}

We can even calculate the OPE of products of vertex operators. It can be shown (although the details are left as an exercise) that
\begin{equation}
:e^{a \phi_1} ::e^{b \phi_2} : =  e^{\langle \phi_1\phi_2  \rangle} :e^{a \phi_1 + b \phi_2}: .
\end{equation}
For vertex operators, this becomes
\begin{equation}
V_\alpha (z,\bar{z}) V_\beta (w,\bar{w}) \sim |z-w|^{2\alpha\beta/4\pi g}V_{\alpha+\beta}(w,\bar{w})+\cdots
\end{equation}
We recall, however, that the two-point functionof primary operators vanishes unless the operators have the same conformal dimension:
$$
\Rightarrow \alpha^2 = \beta^2
$$
Furthermore, the requirement that the correlator between vertex operators does not grow with distance (a physical requirement reflecting how objects correlate over increasing distances) means that $\alpha =-\beta$. Therefore
\begin{equation}
V_\alpha (z,\bar{z}) V_{-\alpha} (w,\bar{w}) \sim |z-w|^{-2\alpha^2 /4\pi g}+\cdots
\end{equation}
In general, the correlator for several vertex operators vanishes unless the sum of their ``charges'' vanishes: $\sum_i \alpha_i=0$. This really \emph{is} a statement about charge conservation, too; in string theory, the charge $\alpha$ is interpreted as the spacetime momentum along the spacetime direction $\phi$.

We will not fill in all of the steps at this point, but we do wish to point out that for special values of $\alpha$ our vertex operators become currents. Setting $g=1/4\pi$ for convenience, we see that choosing $\alpha=\sqrt{2}$ gives us additional currents. We can then study the current algebra of $j(z)$ with the currents
\begin{equation}
j^\pm \equiv :e^{\pm i \sqrt{2} \phi}:.
\end{equation}
The full treatment involves some results about OPEs that we will not prove until a little later (or in the next course). We only say that by defining
\begin{equation}
j^1 = \frac{1}{\sqrt{2}}(j^+ + j^-), \;\;\;\;\;\;j^1 = \frac{1}{\sqrt{2}}(j^+ + j^-),\;\;\;\;\;\;j^3 = j,
\end{equation}
it is possible to show
\begin{equation}
[j^i_m, j^j_n]= i\sqrt{2}\sum_k \epsilon^{ijk} j^k_{m+n} + m \delta^{ij}\delta_{m,-n}.
\end{equation}
This is the $su(2)$ Ka\u{c}-Moody algebra at level $k=1$. Therefore this current algebra is related to the theory of a free boson $\phi$ compactified on a radius $R=1/\sqrt{2}$.

\subsection{Example: Free fermion}

Now we consider the case of a free Majorana fermion in two-dimensional Euclidean space. The action for this theory is
\begin{equation}
S=\frac{g}{2} \int dx^0 dx^1 \bar{\Psi}\gamma^\alpha \partial_\alpha \Psi,
\end{equation}
where here $\eta^{\alpha\beta}=\mbox{diag}(1,1)$, $g$ is a normalization constant, $\bar{\Psi}\equiv \Psi^\dagger\gamma^0$, and the $\gamma^\alpha$ are two-by-two matrices satisfying the Clifford algebra
$$
\{\gamma^\alpha,\gamma^\beta \} = 2 \eta^{\alpha\beta} \boldsymbol{1}_2.
$$
Although many representations of $\gamma^\alpha$ satisfy the Clifford algebra, we will use the representation
\begin{equation}
\gamma^0 = \begin{pmatrix}
0 & 1\\
1 & 0
\end{pmatrix}, \;\;\;\;\;\;\;\; \gamma^1 = \begin{pmatrix}
0 & -i\\
i & 0
\end{pmatrix}.
\end{equation}
In this representation and using the usual definition of $z$
\begin{equation}
\gamma^0(\gamma^0\partial_0+\gamma^1\partial_1) = 2\begin{pmatrix}
\partial_{\bar{z}} & 0\\
0 & \partial_z
\end{pmatrix}.
\end{equation}
Then using $\Psi=(\psi,\bar{\psi})$, the action becomes
\begin{equation}
S = g \int d^2 x (\bar{\psi}\partial \bar{\psi} + \psi \bar{\partial} \psi).
\end{equation}
The associated equations of motion are $\partial\bar{\psi}=\bar{\partial}\psi=0$ whose solutions are the holomorphic $\psi(z)$ and antiholomorphic $\bar{\psi}(\bar{z})$.

Next we need to calculate the propagator for these fields. There are several ways to do this. One way is to write the action as a Gaussian integral; then we know the propagator is a Green's function satisfying a particular differential equation. We can solve this differential equation and write the answer in complex coordinates to finally get
\begin{equation}
\langle \psi(z)\psi(w) \rangle  = \frac{1}{2\pi g} \frac{1}{z-w}.
\end{equation}
There is a similar expression for $\bar{\psi}$, and the two point function between $\psi$ and $\bar{\psi}$ vanishes.
The relevant OPE is thus
\begin{equation}
\langle \psi(z)\psi(w) \rangle  \sim \frac{1}{2\pi g} \frac{1}{z-w}.
\end{equation}
We point out that as in the case of bosonic fields, this OPE reflects the spin nature of the field: exchanging two bosons has no effect on their OPE whereas exchanging two fermions produces an overall negative sign.
\begin{framed}
\noindent HOMEWORK: Check all of the claims made so far in this section.
\end{framed}

Before continuing, we remark upon the boundary conditions for $\psi$. Spinors live in the \emph{spin bundle}, the double cover of the principle frame bundle of the surface. In practice, what this means is that only bilinears in spinors need to transform as single-valued representations of the 2D Euclidean group. As such, there are two different behaviors possible under a rotation by $2\pi$: the different boundary conditions are
\begin{align}
&\psi(e^{2\pi i} z) = + \psi(z) &\mbox{Neveu-Schwarz (NS) sector,} \\
&\psi(e^{2\pi i} z) = - \psi(z) &\mbox{Ramond (R) sector}.
\end{align}
We will return to this point a few more times as the lectures progress.

For now, we consider the stress-energy tensor for the free fermion theory. Using the formula for the stress-energy tensor, we find
\begin{align}
T^{zz} &= 2g\bar{\psi}\bar{\partial}\bar{\psi} \\
T^{\bar{z}\bar{z}} &= 2g\bar{\psi}\bar{\partial}\bar{\psi} \\
T^{z\bar{z}} &= -2g\psi\bar{\partial}\psi.
\end{align}
This is not symmetric, but using the classical equations of motion means the nondiagonal component vanishes. The holomorphic component is then
\begin{equation}
T(z) \equiv -2\pi T_{zz} = -\frac12 \pi T^{\bar{z}\bar{z}}=-\pi g :\psi(z)\partial\psi(z):,
\end{equation}
where we are again promoting this expression to a sensible normal-ordered product. With this expression, we can calculate
\begin{equation}
T(z)\psi(w) \sim \frac{\psi(w)/2}{(z-w)^2}+\frac{\partial \psi(w)}{z-w}.
\end{equation}
We clearly see that the fermion $\psi$ has the conformal dimension $h=\frac12$.
In a similar way, we can show
\begin{equation}
T(z)T(w) \sim \frac{1/4}{(z-w)^4}+\frac{2T(w)}{(z-w)^2}+\frac{\partial T(w)}{z-w}
\end{equation}
Unlike the case of a free boson, the free fermion has central charge $c=1/2$.
\begin{framed}
\noindent HOMEWORK: Do these calculations. Be careful about anticommutation.
\end{framed}

We conclude with another advertisement for the exercises. The fact that a pair of free fermions commutes and has the same central charge as a free boson is suggestive that there may be a relationship between these theories. In an exercise, we show that a boson can be equivalently expressed as a theory of two fermions. This is not surprising. What is remarkable, however, is that in conformal field theories we can also express fermions in terms of the boson. This is the \emph{bosonization} of a complex fermion, and it is recommended you pursue this exercise on your own time.

\subsection{Example: the $bc$ theory}

We now turn our attention to a new theory. We consider fields known as \emph{ghosts} or \emph{reparameterization ghosts}:
\begin{equation}
S = \frac{g}{2} \int d^2x b_{\mu\nu}\partial^\mu c^\nu.
\end{equation}
Both of these fields are anticommuting, and the field $b$ is traceless and symmetric. The physical origin of this theory is not necessary to understand right now. If you are curious, I will say that the $bc$ ghost system arises when studying the bosonic string: this theory represents a Jacobian arising from changing variables in some functional integrals in order to do some type of gauge-fixing.

The equations of motion for this theory are found to be
\begin{equation}
\partial^\mu b_{\mu\nu} = 0, \;\;\;\;\;\; \partial^\mu c^\nu + \partial^\nu c^\mu = 0.
\end{equation}
Switching to complex coordinates, we define $c = c^z$ and $\bar{c}=c^{\bar{z}}$, with the nonvanishing components of $b_{\mu\nu}$ given as $b=b_{zz}$ and $\bar{b}=b_{\bar{z}\bar{z}}$. Then the equations of motion are
\begin{align}
\bar{\partial} b = \partial\bar{b} = 0 \\
\bar{\partial}c = \partial \bar{c} = 0 \\
\partial c = -\bar{\partial}\bar{c}.
\end{align}
\begin{framed}
\noindent HOMEWORK: Derive these equations of motion.
\end{framed}
\noindent We can derive the propagators in a similar way as the case of a free fermion. Solving the associated differential equation gives
\begin{equation}
\langle b(z)c(w) \rangle = \frac{1}{\pi g} \frac{1}{z-w},
\end{equation}
with the relevant OPE being
\begin{equation}
 b(z)c(w)  \sim \frac{1}{\pi g} \frac{1}{z-w}.
\end{equation}
\begin{framed}
\noindent HOMEWORK: Provide a (rough, at least) derivation of this propagator.
\end{framed}

The canonical stress-energy tensor for this system is
\begin{equation}
T^{\mu\nu}_C = \frac{g}{2} (b^{\mu\rho}\partial^\nu c_\rho-\eta^{\mu\nu}b^{\rho\sigma}\partial_\rho c_\sigma).
\end{equation}
Again, this tensor is not symmetric. Recalling our earlier discussions, we add a term of the form $\partial_\rho B^{\rho\mu\nu}$, where
\begin{equation}
B^{\rho\mu\nu}=-\frac{g}{2} (b^{\nu\rho}c^\mu - b^{\nu\mu}c^\rho).
\end{equation}
It can be shown that this gives a symmetric traceless Belinfante tensor
\begin{equation}
T^{\mu\nu}_B=\frac{g}{2} (b^{\mu\rho}\partial^\nu c_\rho    b^{\nu\rho}\partial^\mu c_\rho + \partial_\rho b^{\mu\nu}c^\rho
-\eta^{\mu\nu}b^{\rho\sigma}\partial_\rho c_\sigma).
\end{equation}
\begin{framed}
\noindent HOMEWORK: Prove it.
\end{framed}

To study the holomorphic component (in complex coordinates), we consider $T^{\bar{z}\bar{z}}=4T_{zz}$ and find
\begin{equation}
T(z) = \pi g :(2\partial c b + c \partial b):
\end{equation}
\begin{framed}
\noindent HOMEWORK: Derive this formula.
\end{framed}
Using this, we can calculate the OPE of the stress-energy tensor with both $c$ and $b$. We find that
\begin{align}
T(z)c(w) &\sim -\frac{c(w)}{(z-w)^2}+\frac{\partial_w c(w)}{z-w} \nonumber\\
T(z)b(w) &\sim 2\frac{b(w)}{(z-w)^2} + \frac{\partial_w b(w)}{z-w}.
\end{align}
From these, we see that the conformal dimensions of $c$ and $b$ are $h=-1, 2$ respectively. Finally, we can calculate the OPE of $T$ with itself:
\begin{equation}
T(z)T(w) \sim \frac{-13}{(z-w)^4} + \frac{2T(w)}{(z-w)^2} + \frac{\partial T(w)}{z-w}.
\end{equation}
Again, $T$ has conformal dimension $h=2$ and the central charge for the ghost system is $c=-26$.
\begin{framed}
\noindent HOMEWORK: Derive these equations.
 \end{framed} 
\noindent The negative central charge here means, of course, that this ghost system can not be unitary. This is completely expected, given that these fields have the wrong commutation behavior for their spin.

The fact that this system has $c=-26$ is the reason that we study bosonic string theory in $d=26$ spacetime dimensions. The bosonic fields map the two-dimensional coordinates to a target spacetime, with the number of bosonic fields corresponding to the dimension of the spacetime.  As we will see in an upcoming lecture, we require the total central charge to vanish so that a theory is free of anomalies. Each bosonic field $\phi^i$ contributes $c=1$, so consdering the theory of strings that live in $d=26$ dimensions and the $bc$-ghost theory corresponding to reparameterization invariation of the string gives a theory with total central charge $c=0$. In superstring theory, each boson $\phi^\mu$ has a corresponding fermion $\psi^\mu$. The central charge corresponding to a superstring in $d$-dimensions is then $\frac32 d$. In this case, in addition to the $bc$ ghost system we must include commuting fermionic ghosts $(\beta,\gamma)$ with conformal dimensions $(3/2,-1/2)$. The central charge for this ghost theory is $c=11$. Then the requirement that the complete theory be anomaly free means
$$
\frac32d - 26 + 11 = 0 \Rightarrow d = 10.
 $$ 
In order to have a superstring theory free of anomalies, superstrings must live in $d=10$ spacetime dimensions.

\subsection{Descendant states, Verma modules, and the Hilbert space}

Now that we have dirtied our hands with some explicit theories, let us turn our attention back to generalities. We have already discussed primary fields in some detail; we now turn our attention to descendant fields. The asymptotic state $|h\rangle=\phi(0)|0\rangle$ created by a primary field is the source of an infinite tower of descendant states of higher conformal dimension. Under a conformal transformation, the state and its descendants transform among themselves. Thinking about the state-operator correspondence, each descendant state could actually be viewed as the application of a \emph{descendant field} on the vacuum. For example, consider the descendant
\begin{equation}
L_{-n}|h\rangle = L_{-n}\phi(0)|0\rangle = \frac{1}{2\pi i} \oint_w dz \,\, z^{1-n}T(z)\phi(0)|0\rangle
\end{equation}
This gives a natural definition for a descendant field
\begin{equation}
\phi^{-n}(w) \equiv (L_{-n}\phi)(w) = \frac{1}{2\pi i} \oint dz \,\, \frac{1}{(z-w)^{n-1}}T(z)\phi(w).
\end{equation}
\begin{framed}
\noindent HOMEWORK: Show that the stress-energy tensor is a descendant field. What is its corresponding primary state?
\end{framed}

We will not do the explicit computation\footnote{Which just means that we leave it as one of the exercises.}, but we claim that the correlators of descendant fields can be derived from correlators involving their primary field. To see this, we would consider a correlator $\langle (L_{-n}\phi)(w)X \rangle$, where $X=\phi_1(w_1)\cdots\phi_N(w_N)$ and $\phi_i$ is a primary field with dimension $h_i$. We can then substitute the definition of the descendant field (where the contour encircles only $w$ and none of the $w_i$). Then using the OPE of $T$ with primary fields, we obtain some differential operator acting $\mathcal{L}_{-n}$ acting on $\langle\phi(w)X\rangle$:
\begin{equation}
\langle (L_{-n}\phi)(w)X \rangle \equiv \mathcal{L}_{-n}\langle\phi(w)X\rangle, \;\;n\geq 1.
\end{equation}
\begin{framed}
\noindent HOMEWORK: Find the form of $\mathcal{L}_{-1}$.
\end{framed}
\noindent If a descendant field involves several $L_{-i}$'s, we can define it recursively:
$$
(L_{-m}L_{-n}\phi)(w) \equiv \frac{1}{2\pi i} \oint_w dz \;\; (z-w)^{1-m}T(z)(L_{-n}\phi)(w).
$$
In a similar way it can be shown \cite{4-4} that the three-point function for any three descendant fields (or descendant fields with primary fields) can be determined from the associated primaries. Thus: the information required to completely specify a two-dimensional conformal field theory (meaning to specify the correlators of any collection of fields in the theory) consists of the conformal weights $(h_i, \bar{h}_i)$ of the Virasoro highest weight states and the operator product expansion three-point function constants between the relevant primary fields.

Is there some way to constrain these quantities, or further constrain consistent conformal field theories? We have already seen that unitarity requires $c,\bar{c}, h, \bar{h} \geq 0$. In the next section we will develop even more powerful constrains from unitarity. Further constraints come from demanding our theory be modular invariant on the torus---we will discuss this topic in Lecture 5. Additionally, the conformal bootstrap method provides a powerful way to constrain the conformal weights and three-point function constants; we will see this in Lecture 7. Finally, the addition of supersymmetry to the problem will provide powerful constraints; we do not explore these constraints in this version of the course.

Having considered all of the fields in our theory, let us more deliberately consider the Hilbert space of states. We start with some highest weight state $|h\rangle = \phi(0)|0\rangle$. Acting with $L_n (n<0)$ on the state $|h\rangle$ creates descendant states. The set of all these states is the \emph{Verma module $V_{h,c}$} (where $c$ is the central charge). The lowest few states in the Verma module are
\begin{equation}
L_{-1}|h\rangle, \;\;\;\; L_{-2}|h\rangle,\;\;\;\;L_{-1}L_{-1}|h\rangle,\;\;\;\;L_{-3}|h\rangle, \;\;\;\;\cdots
\end{equation}
We can think about the Verma module as the set of states corresponding to the conformal family of a primary field.
At level $N$, there will be $P(N)$ states, where $P(N)$ is the partition function of $N$---the number of distinct ways of writing $N$ as a sum of positive integers. For example,
\begin{equation}
P(1)=1, \;\;P(2)=2,\;\; P(3)=3, \;\;P(4) = 5, \;\;P(5)=7,\cdots
\end{equation}
The generating function for the number of partitions is given by
\begin{equation}
\sum_{N=0}^{\infty} P(N) q^N = \prod_{n=1}^{\infty} (1-q^n)^{-1}.
\end{equation}

Of course, we do not \emph{actually} have $P(N)$ physical states at level $N$. After all, we have no guarantee that all of the states are linearly independent. A linear combination of states that vanishes is known as a \emph{null state}, and the representation of the Virasoro algebra with highest weight $|h\rangle$ is constructed from a Verma module by removing all of its null states (and their descendants). This is because null states are not physical states. To find null states, we will focus on linear combinations of states that vanish. At level 1, the only way for a state to vanish is
\begin{equation}
L_{-1}|h\rangle=0.
\end{equation}
However, this just implies that $h=0$, meaning $|h\rangle=|0\rangle$---the unique vacuum state. At level 2, on the other hand, we could have some value of $a$ such that
\begin{equation}
(L_{-2}+aL_{-1}^2)|h\rangle=0
\end{equation}
\begin{framed}
\noindent HOMEWORK: By acting with $L_1$ and $L_2$, find $a$ and the relationship between $c$ and $h$ so that this is true.
\end{framed}
\begin{framed}
\noindent HOMEWORK: Find the expression for the null vector at level 3. Determine the corresponding central charge $c$ as a function of $h$.
\end{framed}
\noindent Doing this level by level for infinitely many levels seems somewhat daunting. Let us try to generalize.

\subsection{Ka\u{c}-Determinant and unitary representations}

In order to determine zero-norm states in a Verma module more generally, we first consider an example from linear algebra. Given a vector $|v\rangle$ in a real $n$-dimensional vector space with (not necessarily orthonormal) basis vectors $|a\rangle$, we can express our vector as
\begin{equation}
|v\rangle = \sum_{a=1}^n\lambda_a |a\rangle.
\end{equation}
Then the condition that our vector has a vanishing norm is
\begin{equation}
0 = \sum_{a,b=1}^n \lambda_a \langle a| b\rangle \lambda_b = \vec{\lambda}^T M \vec{\lambda},
\end{equation}
where we have defined the elements of matrix $M$ as $M_{ab}=\langle a|b\rangle$. This expression vanishes when $\vec{\lambda}$ is an eigenvector of $M$ with zero eigenvalue. The number of these eigenvectors (that is, their multiplicity) is given by the number of roots of the equation $\det M = 0$. We can therefore study null states by investigating the determinant of a matrix of inner products.

Let us apply a similar analysis to the null states of the Verma module by computing the \emph{Ka\u{c}-determinant} at level $N$. We define the matrix $M_N(h,c)$, where the entries are defined as
\begin{equation}
\langle h| \prod_i L_{k_{i}} \prod_j L_{-m_{j}}|h \rangle, \;\;\;k_i, m_j \geq 0.
\end{equation}
For $N=1$, the Ka\u{c}-determinant is easily calculated to be
\begin{equation}
\det M_1(h,c) = 2h.
\end{equation}
Requiring this determinant to vanish reproduces the result that we have one possible null state at level $N=1$, when $h=0$. At level $N=2$, there are two states in the Verma module. As such, we must compute four matrix elements for our matrix:
\begin{align}
\langle h| L_{2} L_{-2}  |h \rangle =  4h + \frac{c}{2} \nonumber\\
\langle h| L_{1} L_{1} L_{-2}  |h \rangle = 6h \\
\langle h| L_{2} L_{-1} L_{-1} |h \rangle =  6h\nonumber\\
\langle h| L_{1} L_{1} L_{-1} L_{-1} |h \rangle = 4h+8h^2. \nonumber  
\end{align}
Then we find the Ka\u{c}-determinant
\begin{equation}
\det M_2(h,c) = 32h\left( h^2 - \frac{5h}{8}  + \frac{hc}{8} +\frac{c}{16} \right).
\end{equation}
\begin{framed}
\noindent HOMEWORK: Derive this result.
\end{framed}

We find the roots of this determinant to be 
\begin{align}
h_{1,1}&=0 \nonumber\\
h_{1,2}&= \frac{5-c}{16}-\frac{1}{16}\sqrt{(1-c)(25-c)} \\
h_{2,1}&= \frac{5-c}{16}+\frac{1}{16}\sqrt{(1-c)(25-c)} \nonumber
\end{align}
so that the determinant can be expressed in the form
\begin{equation}
\det M_2(h,c) = 32 (h-h_{1,1}(c))(h-h_{1,2}(c))(h-h_{2,1}(c)).
\end{equation}
Forgive this notation; it will become clearly shortly. We see that we once more have a root at $h=0$. This root is actually due to the null state at level 1. This is a general feature: a root entering for the first time at level $n$ will continue to be a root at higher levels (after all, higher levels just come from acting on a state with Virasoro generators). We can actually figure out the multiplicity of this root at higher levels, since we know that the number of possible operators at a given level is related to the partition function. So a null state state for some value of $h$ appearing at level $n$ implies the determinant at level $N$ will have a $P(N-n)$-th order zero for that value of $h$. In this example, that means we could have expressed the first parenthetical factor as $(h-h_{1,1}(c))^{P(2-1)}$.

So what does this look like in general? V. Ka\u{c} found and proved the general formula for the determinant at arbitrary level $N$:
\begin{align}
\det M_N(h,c) &= \alpha_N \prod_{\substack{p,q\leq N \nonumber \\
                  p,q>0}} \left( h-h_{p,q}(c) \right)^{P(N-pq)}, \label{eq:minmodh}\\
    h_{p,q}(m)&\equiv \frac{((m+1)p-mq)^2-1}{4m(m+1)}, \nonumber \\
    m&=-\frac12 \pm \frac12 \sqrt{\frac{25-c}{1-c}}.
\end{align}
This remarkable formula requires some explanation. The factor $\alpha_N$ is just some positive constant (as we have seen in earlier examples). The variables $p$ and $q$ are positive integers. At its most general, the quantity $m$ is complex. For $c<1$, we typically choose the branch $m\in(0,\infty)$\footnote{This is not necessary, however: $h_{p,q}$ possesses a symmetry between $p$ and $q$ (find it!) so that $\det M$ is independent of the choice of branch in $m$ as it can be compensated by $p\leftrightarrow q$.} Also, we could invert the expression for $m$ to find
$$
c = 1 - \frac{6}{m(m+1)}.
$$
The proof of this formula is not unmanageable, but nothing would be gained by reproducing it here. If you wish to see details, check the references \cite{4-5}.

Having developed a formalism to identify null states, we are now ready to investigate the values of $c$ and $h$ for which the Virasoro algebra has unitary representations. How do null states allow us to study unitarity? In order to consider unitary, we should investigate whether a theory has negative norm states. We claim that this can be seen from the Kac determinant. If the determinant is negative at any given level, it means that the determinant has an odd number of negative eigenvalues---definitely at least one. Then the representation of the Virasoro algebra at those values of $c$ and $h$ include states of negative norm and are thus nonunitary. For QFTs, unitarity relates to the conservation of probability. In statistical mechanics, the corresponding notion is \emph{reflection positivity} and consequently the existence of a Hermitian transfer matrix. In general statistical mechanical systems, unitarity/reflection positivity does not necessarily play an essential role. Models of percolation (the $Q\rightarrow0$ limit of the Q-state Potts model) and the Yang-Lee edge singularity are two incredibly imporant nonunitary conformal field theories. For now, however, we turn our attention to unitary theories.

We already know that unitarity restricts $c,h\geq0$. Referring back to eq. (\ref{eq:minmodh}), we see that we should consider four intervals for the central charge: (1) $c\geq25$, (2) $1 < c < 25$, (3) $c=1$, and (4) $0\leq c < 1$. Again, we provide sketches of proofs and refer the reader to more detailed discussions in the references. For the first interval, we may choose the branch of $m$ such that $-1<m<0$. Doing this, we can convince ourselves that all of the $h_{p,q}<0$---the determinant must therefore be positive. 

A similar thing happens in the second interval where $m$ is not real. In this case, either $h_{p,q}<0$ (which keeps the corresponding factor positive) or $h_{p,q}$ has an imaginary part. From this, we claim that one can show all eigenvalues of the determinant will be positive for this region. In order to see this, we first observe that the highest power of $h$ in the determinant comes from the product of diagonal elements----this is because these elements contain the maximum number of $L_0$ contributions originating in commutators of $L_{k},L_{-k}$. When $h$ is very large, then, the matrix is dominated by its diagonal elements. An explicit computation shows that these matrix elements are all positive, and thus the matrix will have positive eigenvalues for large $h$. But since the determinant never vanishes for $c>1,h\geq 0$ (since the zeros of the determinant are negative or complex), the eigenvalues must therefore stay positive in this whole region. Thus we find no negative norm states and no unitarity constraints for these ranges of our parameters. Success! (Or more accurately, we are successful in considering this case; we have actually failed to find any new constraints.)

The third case is even more straightforward. On the boundary $c=1$, you should be able to see that the determinant vanishes for $h=n^2/4,n\in\mathbb{Z}$. At no point does the determinant become negative; this is trivial to see. Once again, the Ka\u{c}-determinant provides no constraints on having unitary representations of the Virasoro algebra. As a result, two-dimensional conformal field theories with $c>1$ are not well-understood in general. In a later lecture, we will discuss some recent methods that have made some interesting steps in understanding 2d CFTS wih $c>1$. 
\begin{framed}
\noindent HOMEWORK: Make sure that you understand the previous arguments before proceeding.
\end{framed}

\begin{figure}
\centering
\includegraphics[scale=.6]{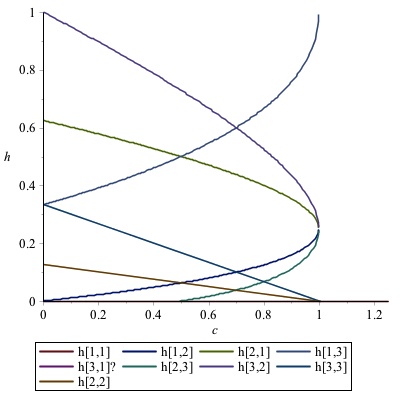}
\caption{First few vanishing curves in the $h,c$ plane.} \label{figure:vanish}
\end{figure}

This leaves only the final case to consider:  the interval $0\leq c < 1$. We start by expressing eq. (\ref{eq:minmodh}) as
\begin{equation}
h_{p,q}(c)=\frac{1-c}{96}\left[ \left( (p+q)\pm(p-q)\sqrt{\frac{25-c}{1-c}} \right)^2 -4\right].
\end{equation}
In the $(c,h)$ plane, the Kac determinant vanishes along the curves $h=h_{p,q}(c)$ (see Figure \ref{figure:vanish}). We will argue that only some points appearing on vanishing curves can correspond to unitary theories, and that other points in the region correspond to nonunitary representations. Consider such a point $P$. Since the determinant does not vanish at $P$, the associated representation does not contain zero-norm states. 
It may, of course, still contain negative norm states. 

Consider what we must demonstrate in order to show that this is the case. We would need to show that for some level $n$, there is a continuous path linking the point $P$ to the region $c>1,h>0$ that crosses a \emph{single} vanishing curve such that $P(n-pq)$ is odd. If this is the case, then the Kac determinant will be negative for the region to the left of the vanishing curve. Referring again to Figure \ref{figure:vanish}, a plot of the first time a zero appears in the Ka\u{c} determinant, we see that we can eliminate large chunks of the region under consideration. 

\begin{figure}
\centering
\includegraphics[scale=.6]{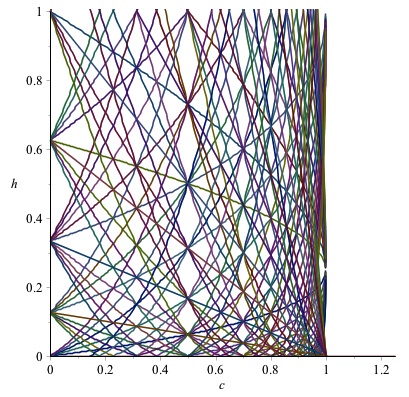}
\caption{Vanishing curves in the $h,c$ plane for $p,q\leq 16$.} \label{figure:lotsvanish}
\end{figure}

We further claim, however, that points not excluded from unitarity at some level are eventually excluded at some higher level. To see this, consider the behavior of the vanishing curves near $c=1$. We can determine the behavior in this region by taking $c=1-6\epsilon$. Then to leading order in $\epsilon$, we find
\begin{gather}
h_{p,q}(1-6\epsilon)\approx\frac14(p-q)^2 +\frac14(p^2-q^2)\sqrt{\epsilon},\nonumber \\
h_{p,p}(1-6\epsilon)\approx\frac14(p^2-1){\epsilon} .
\end{gather}
For a given value of $(p-q)$, the vanishing curve lies closer and closer to the line $c=1$. So each time we increase $pq$ (with fixed $p-q$),  a new set of points is excluded at level $n=pq$ (since $P(n-pq)$ is then one and no other curve lies between the vanishing curve and the line $c=1$). That the vanishing curves actually approach $c=1$ can be seen more clearly in Figure \ref{figure:lotsvanish}.

This argument will exclude all the points in the region $h>0, \;0 < c < 1$---except maybe points lying on the vanishing curves themselves. Verma modules for points on these curves contain null vectors, but it is entirely possible that they also contain negative-norm states. To investigate these points, we must define the \emph{first intersections}. Consider a vanishing curve at a given level; the first intersection associated with that curve (if it exists) is the point intersected by another vanishing curve at the same level lying closest to the line $c=1$.  We can see some first intersections in Figure \ref{figure:vanish}. A more systematic analysis \cite{4-6} of the determinant shows that there is an additional negative norm state everywhere on the vanishing curves except at certain points where they intersect. The central charge at these points is of the form
\begin{equation}
c=1-\frac{6}{m(m+1)}, \;\;\;\; m = 3, 4, \cdots \label{eq:cminmod}
\end{equation}
The case $m=2$ corresponds to the trivial theory with $c=0$. For each value of $c$, there are $m(m-1)/2$ allowed values of $h$ given by
\begin{equation}
h_{p,q}(m) = \frac{[(m+1)p-mq]^2-1}{4m(m+1)}, \label{eq:hminmod}
\end{equation}
where $1\leq p \leq m-1, 1\leq q \leq p$. 
Thus we have \emph{necessary} conditions for unitary highest weight representations of the Virasoro algebra: either $c\geq1, h\geq 0$ or eqs. (\ref{eq:cminmod}) and (\ref{eq:hminmod}) must hold. It turns out that the latter condition is also sufficient \cite{4-7}. This can be shown using a coset space construction, which I intend to present in the future.

This is an incredibly powerful result. Representation theory of the Virasoro algebra for unitary systems with $c\leq 1$ has given us a complete classification of possible two-dimensional critical behavior. The first few members of the series with central charges $c=\frac12, \frac{7}{10}, \frac45,\frac67$, are associated with the Ising model, tricritical Ising model, the 3-state Potts model, and tricritical 3-state Potts model. In general, there may exist more than one model for a given value of $c$ corresponding to different consistent subsets of the full allowed spectrum. The theories we have found are examples of \emph{rational conformal field theories} or RCFTs. These are the \emph{unitary minimal models}. At no point in our discussions did we refer to a concrete realization of a CFT; conformal symmetry and unitarity have given us these constraints for a general theory.

We finish this topic by reiterating that unitarity is not a necessary condition. Weakening this constraint to allow for states with negative norms gives a more general series of minimal models. They have central charges
\begin{equation}
c=1-6\frac{(p-q)^2}{pq},
\end{equation}
where $p,q \geq 2$ and $p,q$ are relatively prime. The conformal weights are given by
\begin{equation}
h_{r,s}(p,q) = \frac{(pr-qs)^2-(p-q)^2}{4pq},
\end{equation}
where $1\leq r \leq q-1$ and $1\leq s \leq p-1$. An example of this type of theory is the Yang-Lee edge singularity, corresponding to $(p,q)=(5,2)$ with central charge $c=-22/5$. To recover unitary models, we choose $p=m+2, q=m+3$. 

\subsection{Virasoro characters}

We conclude by introducing the \emph{character} of a Verma module. To the Verma module $V(c,h)$ generated by Virasoro generators $L_{-n}, n>0$ acting on the highest-weight state $|h\rangle$, we associate the character $\chi_{(c,h)}(\tau)$ defined by
\begin{align}
\chi_{(c,h)}(\tau) &= \mbox{Tr} \;q^{L_0 - c/24} \\
&= \sum_{n=0}^{\infty} \mbox{dim}(h+n) q^{n+h-c/24}. \nonumber
\end{align}
Here we have defined $q\equiv \exp(2\pi i \tau)$, and $\tau$ is a complex variable. The factor $q^{-c/24}$ will make more sense in the next lecture when we discuss modular invariance. The factor dim$(h+n)$ is the number of linearly independent states at level $n$ in the Verma module, a measure of the degeneracy at this level.

We recall the number $P(n)$ of partitions of the integer $n$. The generating function of the partition number is 
\begin{equation}
\sum_{n=0}^\infty P(n) q^n = \prod_{n=1}^\infty \frac{1}{1-q^n} \equiv \frac{1}{\phi(q)}, \label{eq:foureightfour}
\end{equation}
where $\phi(q)$ is the Euler function. Because dim$(h+n)\leq p(n)$, the series (\ref{eq:foureightfour}) will be uniformly convergent if $|q|<1$---this corresponds to $\tau$ in the upper half-plane. Then a generic Virasoro character can thus be expressed as
\begin{equation}
\chi_{(c,h)}(\tau) = \frac{ q^{h - c/24}}{\phi(q)}.
\end{equation}
Alternatively, we can use the Dedekind function
\begin{equation}
\eta(\tau) \equiv q^{1/24}\phi(q) = q^{1/24}\prod_{n=1}^\infty (1-q^n). \label{eq:eta}
\end{equation}
In terms of this function, a Virasoro character can be expressed as
\begin{equation}
\chi_{(c,h)}(\tau) = \frac{ q^{h + (1-c)/24}}{\eta(\tau)}. \label{eq:character}
\end{equation}
We will consider characters in more detail in the next lecture; we have only introduced them here because the next lecture is already overly full.

\break

\subsection*{References for this lecture}
\vspace{4mm}
\noindent Main references for this lecture
\\
\begin{list}{}{%
\setlength{\topsep}{0pt}%
\setlength{\leftmargin}{0.7cm}%
\setlength{\listparindent}{-0.7cm}%
\setlength{\itemindent}{-0.7cm}%
\setlength{\parsep}{\parskip}%
}%
\item[]

[1] Chapter 7 of the textbook: P. Di Francesco, P. Mathieu, and D. Senechal. \emph{Conformal field theory}, Springer, 1997.

[2] Chapter 2 of the textbook: R. Blumenhagen, E. Plauschinn, \emph{Introduction to Conformal Field Theory: With Applications to String Theory}, Lect. Notes Phys. 779,  (Springer, Berlin Heidelberg 2009).

[3] Chapter 4 of : P. Ginsparg, \emph{Applied Conformal Field Theory}, Les Houches, Session XLIX, 1988, \emph{Fields, Strings and Critical Phenomena}, ed. by E. Br\'{e}zin and J. Zinn-Justin, (Elsevier Science Publishers, B.V., 1989), [arxiv:9108028v2 [hep-th]].

\end{list}

\break

\section{Lecture 5: CFT on the Torus}

\subsection{CFT on the torus}

Until now, we have considered conformal field theories defined on the complex plane. On the infinite plane, the holomorphic and antiholomorphic sectors of a CFT can be studied separately. We have done this several times in these lectures, in fact---most frequently when I get tired of writing down two copies of every formula. Because the two sectors do not interfere, they can be considered as different theories. For example, correlation functions factorize into holomorphic and antiholomorphic factors.This situation is unphysical, however. The physical spectrum of the theory should be continuously deformed as we move away from the critical point in parameter space. The coupling between left and right sectors away from the conformal point should therefore lead to constraints between these two sectors of the theory. To impose these constraints while still remaining at the conformal point, we can instead couple these sectors through the \emph{geometry} of the space. 

The infinite plane is topologically equivalent to a sphere, a Riemann surface of genus $g=0$. In general, however, we could study CFTs defined on Riemann surfaces of arbitrary genus $g$. Defining Euclidean field theories on arbitrary genus Riemann surfaces may seem bizarre, particularly when considering critical phenomena. In string theory, of course, higher genus Riemann surfaces correspond to different orders for calculating multiloop scattering amplitudes. It is sensible in the context of critical phenomena to study the simplest nonspherical case: the torus, with $g=1$. This is equivalent to considering a plane with periodic boundary conditions in both directions\footnote{The requirement that a CFT makes sense on a Riemann surface of arbitrary genus adds many constraints to the theory; modular invariance is one of them. We cold consider projections of modularly invariant theories that do not satisfy these constraints. This is the case for boundary conformal field theories, which we will study...one day.}

In this lecture, we will study conformal field theories defined on the torus and extract constraints on the content of these theories. We will start by discussing modular transformations and the partition function. We will then consider the partition function of several simple models, including free bosons, free fermions, and variations of these models. Finally, we discuss the Verlinde formula and fusion rules.

We begin by considering properties of tori. A torus is defined by specifying two linearly independent lattice vectors on a plane and identifying points that differ by integer combinations of these vectors. On the complex plane, these vectors are given by complex numbers $\alpha_1$ and $\alpha_2$, the \emph{periods} of the lattice. As we will soon see, a CFT defined on the torus does not depend on the overall scale of the lattice, or on any absolute orientation of the lattice vectors. The relevant parameter is the ratio of the periods, known as the \emph{modular parameter} $\tau\equiv \alpha_2 / \alpha_1 = \tau_1+i \tau_2$.

In previous lectures, we used radial quantization: curves of constant time were concentric
circles,with time flowing outward from the origin. We defined asymptotic fields at the origin and the point at infinity.
Using an exponential mapping, we saw that this representation was equivalent to a field
theory living on a cylinder; the asymptotic fields correspond to $\pm \infty$ along the
length of the cylinder. To consider the operator formalism on the torus, we just need to impose periodic boundary conditions along this cylinder. The Hamiltonian and the momentum operators propagate states along different directions of the torus, and the spectrum of the theory is embodied in the partition function. 

Recalling its definition, a chiral primary field defined on $\mathbb{C}$ transforms under $z=e^w$ as
\begin{equation}
\phi_{cyl}(w)= \left(\frac{\partial z}{\partial w} \right)^h \phi(z) = z^h \phi(z)
\end{equation}
In terms of the mode expansion, this becomes
\begin{equation}
\phi_{cyl}(w) = z^h \sum_n \phi_n z^{-n-h} = \sum_n \phi_n e^{-nw}.
\end{equation}
If a field is invariant under $z\rightarrow e^{2\pi i}z$ on the complex plane, the same field picks up a phase $e^{2\pi i (h-\bar{h})}$ on the cylinder. If $(h-\bar{h})$ is not an integer, the boundary condition of the field is changed. For example, consider the expansion of a chiral fermion with $(h,\bar{h}) = \left(\frac12,0\right)$ on the cylinder:
\begin{equation}
\psi_{cyl}(w) = \sum_r \psi_r e^{-rw}.
\end{equation}
On the plane, we recall that NS and R boundary conditions were periodic and antiperiodic respectively under $2\pi$ rotations. The opposite is true on the cylinder: the Neveu-Schwarz sector (with $r\in \mathbb{Z}+\frac12$) is antiperiodic under $w\rightarrow w+2\pi i$ while the Ramond seector (with $r\in \mathbb{Z}$) is periodic.

In a similar way, the stress-energy tensor transforms on the cylinder. Because $T(z)$ is not a primary field, we cannot use the above formula. We recall that under transformations $z\rightarrow f(z)$, the stress-energy tensor becomes
\begin{equation}
T'(z) = \left( \frac{\partial f}{\partial z} \right)^2 T(f(z)) + \frac{c}{12} S\left( f(z),z \right), \label{eq:finitettransform}
\end{equation}
where the Schwarzian derivative is defined as
\begin{equation}
S(w,z) = \frac{1}{(\partial_z w)^2} \left((\partial_z w)(\partial_z^3 w)-\frac32(\partial_z^2 w)^2  \right).
\end{equation}
For the exponential map we consider here, then
\begin{equation}
T_{cyl}(w) = z^2 T(z) - \frac{c}{24}.
\end{equation}
The Laurent mode expansion of the stress-energy tensor on the cylinder is therefore
\begin{equation}
T_{cyl}(w) = \sum_{n\in \mathbb{Z}} L_n z^{-n}-\frac{c}{24} = \sum_{n\in \mathbb{Z}} \left( L_n - \frac{c}{24}\delta_{n,0} \right) e^{-nw}.
\end{equation}
The only difference in the Virasoro generators is that now the zero mode is shifted as
\begin{equation}
L_{0,cyl} = L_0-\frac{c}{24}. \label{eq:lzeroshift}
\end{equation}
A similar derivation holds for $\bar{z}$ and $\bar{L}_{0}$.
\begin{framed}
\noindent HOMEWORK: Complete the steps in this derivation.
\end{framed}
\noindent This shift means that the vacuum energy on the cylinder is given by
\begin{equation}
E_0 = -\frac{c+\bar{c}}{24}. \label{eq:casimir}
\end{equation}

Now we are in a position to define the partition function $Z$ in terms of Virasoro generators. This is essentially the same thing for CFTs as in statistical mechanics: a sum over configurations weighted by a Boltzmann factor $\exp(-\beta H).$  It also corresponds to the generating functional in Euclidean QFT due to the fact that the thermodynamic expression can be found by compactifying the time on a circle of radius $R=\beta=1/T$.

\begin{figure}
\centering
\includegraphics[scale=.7]{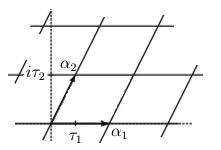}
\caption{A torus generated by $(\alpha_1,\alpha_2)$ chosen as $(1,\tau)$ } \label{figure:torus}
\end{figure}

We choose our coodinate system so that the real and imaginary axes to correspond to the spatial and time directions, respectively, and we consider a torus with modular parameter $\tau = \tau_1 + i \tau_2$. For definiteness, we currently choose $\alpha_1=1,\alpha_2=\tau$ (see Figure \ref{figure:torus}).
\noindent From this picture, it is clear that a time translation of length $\tau_2$ does not come back to where it started. Instead, it is displaced in space by a factor $\tau_1$. A ``closed loop'' in time thus also involves a spatial translation. We are therefore motivated to define the CFT partition function as
\begin{equation}
Z = \mbox{Tr}_\mathcal{H}\left( e^{-2\pi\tau_2 H} e^{2\pi \tau_1 P} \right).
\end{equation}
The Hamiltonian $H$ generates time translations, the momentum operator $P$ generates spatial translations, and the trace is taken over all states in the Hilbert space $\mathcal{H}$ of the theory.

Recalling the relations between $H,P$ and Virasoro generators, we know that 
\begin{equation}
H_{cyl} = L_{0,cyl} + \bar{L}_{0,cyl}, \;\;\;\;\;\;\;\; P_{cyl} = i (L_{0,cyl} - \bar{L}_{0,cyl}).
\end{equation}
Then we can express the partition function as 
\begin{equation}
Z = \mbox{Tr}_\mathcal{H}\left( e^{2\pi i \tau L_{0,cyl}} e^{-2\pi i \bar{\tau} \bar{L}_{0,cyl}} \right).
\end{equation}
Then by defining $q=\exp(2\pi i \tau)$, we conclude that the partition function for a conformal field theory defined on a torus with modular parameter $\tau$ is given by
\begin{equation}
Z = \mbox{Tr}_\mathcal{H}\left( q^{L_{0}-\frac{c}{24}} \bar{q}^{ \bar{L}_{0}-\frac{\bar{c}}{24} } \right).
\end{equation}
Note that this expression for the partition function involves the characters (\ref{eq:character}) defined last lecture. We therefore expect to find the partition function expressible in terms of the characters of irreducible representations
$$
Z(\tau) = \sum_{(h,\bar{h})} \bar{\chi}_{\bar{h}}(\tau) N_{\bar{h}h}\chi_h(\tau),
$$
where the multiplicity $N_{\bar{h}h}$ counts the number of times that the representation $(h,\bar{h})$ occurs in the spectrum.

\subsection{Modular invariance}

The advantage of studying CFTs on a torus is that we get powerful constraints arising from the requirement that the partition function be independent of the choice of periods for a given torus. The pair of complex numbers $(\alpha_1, \alpha_2)$ spans a lattice whose smallest cell is the fundamental domain of the torus. Geometrically, a torus is obtained by idenfitying opposite edges of the fundamental domain. 

There are different choices of this pair of numbers, however, that give the same lattice (and thus the same torus). Let us assume $(\alpha_1, \alpha_2)$ and $(\beta_1, \beta_2)$ describe the same lattice. If we think about this for a minute, that means that we can write the pair $(\beta_1, \beta_2)$ as some integer linear combination of the pair $(\alpha_1, \alpha_2)$:
\begin{equation}
\begin{pmatrix} \beta_1 \\ \beta_2 \end{pmatrix} = \begin{pmatrix} a  &  b\\ c &  d \end{pmatrix} \begin{pmatrix} \alpha_1 \\ \alpha_2 \end{pmatrix}, \;\;\;\; a,b,c,d \in \mathbb{Z}.
\end{equation}
In a similar way, clearly $(\beta_1, \beta_2)$ must be expressible in terms of $(\alpha_1, \alpha_2)$:
\begin{equation}
\begin{pmatrix} \alpha_1 \\ \alpha_2 \end{pmatrix} = \frac{1}{ad-bc}\begin{pmatrix} d  &  -b\\ -c &  a \end{pmatrix} \begin{pmatrix} \beta_1 \\ \beta_2 \end{pmatrix}.
\end{equation}
In order for this inverse matrix to also have integer entries, we need to require that $ad-bc=\pm1$. Furthermore, the lattice spanned by $(\alpha_1, \alpha_2)$ is equal to the one spanned by $(-\alpha_1, -\alpha_2)$. We can therefore divide out by an overall $\mathbb{Z}_2$ action. Matrices with these properties are elements of the group $SL(2,\mathbb{Z})/\mathbb{Z}_2$. By choosing our previous convention, $(\alpha_1, \alpha_2)=(1,\tau)$, we find the \emph{modular group}.
The modular group of the torus is an isometry group acting on the modular paramter $\tau$ as
\begin{equation}
\tau \rightarrow \frac{a\tau + b}{c\tau + d}, \;\;\;\;\;\; \mbox{with } \begin{pmatrix} a & b \\ c & d \end{pmatrix} \in SL(2,\mathbb{Z})/\mathbb{Z}_2.
\end{equation}

There are, of course, infinitely many modular transformations. To get a handle on the modular group, let us try to consider some sort of ``basis'' transformations. First, we consider the modular $T$-transformation, defined by
\begin{equation}
T: \tau\rightarrow \tau+1.
\end{equation}
This transformation can equivalently be expressed by the matrix
\begin{equation}
T = \begin{pmatrix} 1  &  1 \\ 0 & 1 \end{pmatrix}.
\end{equation}
Secondly, we consider the $U$-transformation, defined by
\begin{equation}
U:\tau\rightarrow \frac{\tau}{\tau+1}.
\end{equation}
In a similar way, this can be expressed as
\begin{equation}
U = \begin{pmatrix} 1  &  0 \\ 1 & 1 \end{pmatrix}.
\end{equation}

It turns out, however, that it is more conventient to work not with $U$ but with the modular $S$-transformation. This transformations is defined as
\begin{equation}
S:\tau\rightarrow -\frac{1}{\tau}.
\end{equation}
The corresponding matrix transformation is 
\begin{equation}
T = \begin{pmatrix} -1  &  0 \\ 0 & 1 \end{pmatrix}.
\end{equation}
It is straightforward to show that $S$ can be expressed in terms of $T$ and $U$:
\begin{equation}
S=UT^{-1}U,
\end{equation}
as well as
\begin{equation}
S^2 = (ST)^3 = \boldsymbol{1}.
\end{equation}
Repeated $S-$ and $T-$ transformations generate the entire modular group, though this is somewhat nontrivial to demonstrate. As such, we leave it as exercise at the end of these lectures.
\begin{framed}
\noindent HOMEWORK: Work through the claims made in this paragraph.
\end{framed}

\begin{figure}
\centering
\includegraphics[scale=.4]{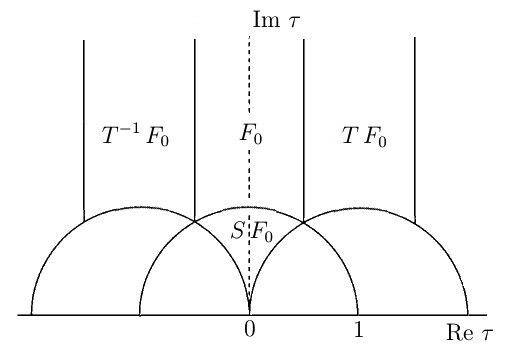}
\caption{The fundamental domain of the modular parameter, as well as images of the fundamental domain under certain modular transformations (adapted from \cite{difran})} \label{figure:fundomain}
\end{figure}

The action of the modular group on the $\tau$ upper half-plane is nontrivial. We consider a \emph{fundamental doman} of the modular group such that no points inside the domain are related by a modular transformation and any point outside the domain can be reached from a unique point in the domain. We will choose the conventional fundamental domain $F_0$:
\begin{equation}
F_0 = \{z\}, \;\;\;\mbox{such that} \;\; \left\{  \begin{array}{ccc} \operatorname{Im} z > 0, & -\frac12 \leq \operatorname{Re}  z \leq 0, & |z|\geq 1 \\
\; \\
  \mbox{or} \\
\; \\
    \operatorname{Im}  z>0, & 0 < \operatorname{Re} z <\frac12, & |z|> 1
  \end{array}
  \right.
\end{equation}
The fundamental domain $F_0$ is shown in Figure \ref{figure:fundomain}, along with domains obtained from simple modular transformations.

\subsection{Construction partition functions on the torus}
\subsubsection{Free boson on the torus}

Having introduced the partition function on the torus and the modular group, we now turn our attention to constraints in specific models. As always, we start with the simplest model: a single free boson. We recall that we have already found an expression for $L_0$ in this theory, given by equation (\ref{eq:lzerodef})
\begin{equation}
L_0 = \frac12 j_0 j_0 + \sum_{k=1}^\infty j_{-k}j_k.  \label{eq:lzerodef2}
\end{equation}
We also know that because the current $j(z)$ has conformal dimension one, $j_n|0\rangle=0$ for $n>-1$. Generic states in the Hilbert space come from acting with creation modes $j_{-k}$ so that states are of the form
\begin{equation}
|n_1,n_2,n_3,\cdots\rangle = j_{-1}^{n_1}j_{-2}^{n_2}\cdots|0\rangle, \;\;\;\; \mbox{with } n_i \geq 0, n_i\in \mathbb{Z}.
\end{equation}
In order to proceed, we also recall the current algebra for the Laurent modes
$$
[j_m, j_n]m \delta_{m,-n},
$$
and make the claim that
$$
[j_{-k}j_k,j_{-k}^{n_k}]=n_ k j_{-k}^{n_k}.
$$
\begin{framed}
\noindent HOMEWORK: Prove this formula via induction.
\end{framed}

Using these formulas, we can show
\begin{align}
L_0 |n_1,n_2,n_3,\cdots\rangle &= \sum_{k\geq 1} j_{-1}^{n_1}j_{-2}^{n_2}\cdots (j_{-k}j_k) j_{-k}^{n_k} \cdots|0\rangle \nonumber \\
&= \sum_{k\geq 1} k n_k |n_1,n_2,n_3,\cdots\rangle
\end{align}
Once we have this expression, we can calculate the partiton function. 
\begin{align}
&\mbox{Tr}(q^{L_0-\frac{c}{24}})\\
&=q^{-\frac{1}{24}} \sum_{n_1=0}^\infty \sum_{n_2=0}^\infty \cdots \left\langle n_1,n_2,\cdots\left| \sum_{m=0}^\infty \frac{1}{p!} (2\pi i \tau)^p (L_0)^p \right| n_1,n_2,\cdots \right\rangle  \nonumber\\
&\;\;\;\;\;\;\;\;\;\;\;\;\vdots\nonumber \\
&=q^{-\frac{1}{24}} \prod_{k=1}^\infty \sum_{n_k=0}^\infty q^{k n_k}.
\end{align}
We have omitted some of the intermediate steps because 
\begin{framed}
\noindent HOMEWORK: Fill in the missing steps in this derivation.
\end{framed}
\noindent Using the definition of the Dedekind $\eta$-function (\ref{eq:eta}) and including the anti-holomorphic contribution, we have therefore found the partition function
\begin{equation}
Z_{bos.}'(\tau,\bar{\tau}) = \frac{1}{|\eta(\tau)|^2}.
\end{equation}

Let us make some important points. First of all, we hae constructed this free bosonic theory on the torus. As such, we expect that the partition function should have the property of modular invariance. To check if this is the case, we need to see how the $\eta$-function changes under the modular $S-$ and $T$- transformations. The effect of the $T$-transformation is trivial to calculate. We leave the derivation of the effect under the $S$-transformation as an exercise and only give the results
\begin{align}
\eta(\tau+1) = e^{\frac{2\pi i}{24}}\eta(\tau), \;\;\;\;\;\; \eta\left(-\frac{1}{\tau}\right) = \sqrt{-i\tau} \eta(\tau).
\end{align}
Disaster! This partition function is \emph{not} invariant under the $S$-transformation (as you can straighforwardly check). This is a problem.

But there is another problem: namely, we cheated in our derivation of the partition function. By performing the series expansion of the exponential, we would have the zero-mode piece $j_0 j_0$ to the $L_0$ operator giving no contribution. But this piece actually appears in an exponential. The vanishing of this piece corresponds to a factor of $e^0=1$, which means the zero mode actually contributes an infinite amount to $Z$. This is definitely cheating. Our method ignores this zero-mode, which is the origin of the primed notation on $Z$---this means we are omitting the zero-mode contribution. To obtain this contribution honestly, we must turn to the path-integral formalism. I'm not going to do this\footnote{You are! We leave it as one of the exercises.} Instead, I will quote the result:
\begin{equation}
Z_{bos.}(\tau,\bar{\tau}) = \frac{1}{\sqrt{\tau_2}|\eta(\tau)|^2}.
\end{equation}
\begin{framed}
\noindent HOMEWORK: Check that \emph{this} expression is modularly invariant. 
\end{framed}
\noindent This additional factor has a natural origin in string theory; it comes from integrating over the center of mass momentum of the string.

\subsubsection{Compactified free boson}

As another example, we consider the free boson $\phi$ compactified on a circle of radius $R$; this means we identify the field like
\begin{equation}
\phi(z,\bar{z}) \sim \phi(z,\bar{z}) + 2\pi R n, \;\; n\in \mathbb{Z}. \label{eq:circboson}
\end{equation}
We could interpret $\phi$ as an angular variable\footnote{This is a confusing point for some students: this identification has nothing to do with the torus periodicity. The torus is the surface on which the theory is defined by variables $z, \bar{z}$ and is periodic.}. To see how this compactification changes the partition function, we must consider the mode expansion of the bosonic field $\phi$:
\begin{equation}
\phi(z,\bar{z}) = x_0 - i\left(j_0 \ln z + \bar{j}_0 \ln \bar{z}  \right) + i \sum_{m\neq0} \frac{1}{n} \left( j_n z^{-n} + \bar{j}_n \bar{z}^{-n} \right).
\end{equation}
\begin{framed}
\noindent HOMEWORK: Derive this expression by integrating the mode expansion for the currents.
\end{framed}
\noindent To find the interesting new constraints, we require that the field $\phi$ is invariant up to identifications (\ref{eq:circboson}) under rotations $z\rightarrow e^{2\pi i}z$
\begin{equation}
\phi\left(e^{2\pi i}z,e^{-2\pi i }\bar{z} \right) = \phi(z,\bar{z}) + 2\pi R n.
\end{equation}
Using this relation with the mode expansion gives
\begin{equation}
j_0 - \bar{j}_0 = R n. 
\end{equation}

If we performed this calculation for the original free boson, we would find that $j_0 = \bar{j}_0$. Thus we see that the ground state has a non-trivial charge under these zero modes. We express this fact as
\begin{align}
j_0|\Delta,n\rangle = \Delta |\Delta,n\rangle, \nonumber \\
\bar{j}_0|\Delta,n\rangle = (\Delta-Rn)|\Delta,n\rangle.
\end{align}
Thinking back to the case of the free boson, we calculate the partition function to be
\begin{align}
Z_{R} &= Z_{bos.}' \cdot \sum_{\Delta,n} \langle \Delta,n | q^{\frac12 j^2_0} \bar{q}^{\frac12 \bar{j}^2_0}  |\Delta, n\rangle \\
&=  \frac{1}{|\eta(\tau)|^2}\sum_{\Delta, n} q^{\frac12 \Delta^2} \bar{q}^{\frac12 (\Delta-Rn)^2}.
\end{align}
This (albeit sloppy) notation should be understood to mean we should perform a sum for discrete values of $\Delta$ or an integral for continuous values\footnote{We have focused thus far, and will continue to focus on, the former case.} Once again, we must check for invariance under modular transformations. Under the modular $T$-transformation, the argument of this sum picks up an additional factor of 
$$
\exp\left(2\pi i n \left(\Delta R - \frac{R^2 n}{2}\right)\right).
$$
Thus demanding modular invariance means
$$
\Delta=\frac{m}{R}+\frac{Rn}{2}, \;\;\;\;m\in\mathbb{Z}.
$$

This clarifies the action of $j_0$ and $\bar{h}_0$ on the ground state:
$$
j_0|m,n\rangle = \left( \frac{m}{R}+\frac{Rn}{2} \right) |m,n\rangle, \;\;\;\;\;\; \bar{j}_0 |m,n\rangle =\left( \frac{m}{R}-\frac{Rn}{2} \right) |m,n\rangle.
$$
In string theory, states with $n\neq0$ are called \emph{winding states}. They correspond to strings winding $n$ times around the circle. States with $m\neq 0$ are called momentum or \emph{Kaluza-Klein} states. This is because the sum of $j_0$ and $\bar{j}_0$ corresponds to the center of mass momentum of the string. With these expressions for our currents, we have
\begin{equation}
Z_{R}(\tau,\bar{\tau}) = \frac{1}{|\eta(\tau)|^2}\sum_{m,n} q^{\frac12 (\frac{m}{R}+\frac{Rn}{2})^2} \bar{q}^{\frac12 (\frac{m}{R}-\frac{Rn}{2})^2}. \label{eq:bosonZ}
\end{equation}
But wait: what about invariance under the modular $S$-transformation? Proving this invariance requires the \emph{Poisson resummation formula}
\begin{equation}
\sum_{n\in\mathbb{Z}}\exp\left( -\pi a n^2 + bn \right) = \frac{1}{\sqrt{a}} \sum_{k\in\mathbb{Z}}\exp\left( -\frac{\pi}{a}\left( k  + \frac{b}{2\pi i} \right)^2\right). \label{eq:poisson}
\end{equation}
The derivation of this expression and its application to deriving invariance of the partition function under the modular $S$-transformation are left as an exercise.

Before moving on, we remark upon two things. First, we mention $T$-\emph{duality}:
\begin{equation}
Z_{2/R}(\tau,\bar{\tau} ) = Z_{R}(\tau, \bar{\tau}).
\end{equation}
In string theory, this is a statement about how closed strings propagating around a circle cannot distinguish whether the size of the circle is $R$ or $2/R$. The self-dual radius $R=\sqrt{2}$ can be interprested as a minimal length scale that this string can resolve. Finally, we can investigate what vertex operators are allowed for this theory. To respect the symmetry of the theory, we find the condition that
\begin{equation}
\alpha = \frac{m}{R}, \;\;\;\; m\in \mathbb{Z}.
\end{equation}
This makes sense, of course; if we are interpreting $\alpha$ as the spacetime momentum, this condition says that the momentum along the compactified direction must be quantized.

\subsubsection{An aside about important modular functions}

We will now consider the previous theory at the radius $R=\sqrt{2k}$. This theory will help us investigate some important modular functions. We begin by considering chiral states in this theory:
\begin{equation}
\bar{L}_0 |m,n\rangle = 0 \Rightarrow m=kn.
\end{equation}
Then the sum in our partition function (\ref{eq:bosonZ}) becomes
\begin{equation}
\sum_{n\in \mathbb{Z}} q^{kn^2} \equiv \Theta_{0,k}(\tau).
\end{equation}
As we shall soon see, the modular $S$-transformation takes this $\Theta$ function into a finite sum of the more general functions
\begin{equation}
\Theta_{m,k}(\tau) \equiv \sum_{n\in \mathbb{Z}+\frac{m}{2k}} q^{kn^2}, \;\;\;\;\;\;-k+1 \leq m \leq k.
\end{equation}
Using these functions, we see that we can express the partition function $Z_{R}$ in the form
\begin{equation}
Z_{R} = \frac{1}{|\eta(\tau)|^2} \sum_{m=-k+1}^{k} \left| \Theta_{m,k}(q) \right|^2.
\end{equation}
\begin{framed}
\noindent HOMEWORK: Verify this claim. This is more involved than you might expect. You may find it easier to work from both directions.
\end{framed}

For the modular $T$-transformation, it is straightforward to compute
\begin{equation}
\Theta_{m,k}(\tau+1) = e^{\pi i \frac{m^2}{2k}} \Theta_{m,k}(\tau).
\end{equation}
The effect of the modular $S$-transformation is, unsurprisingly, more complicated to derive. We again leave this as one of the exercises, and quote only the result. The modular $S$-transformation of the $\Theta$-functions is of the form
\begin{equation}
\Theta_{m,k}\left( -\frac{1}{\tau}\right) = \sqrt{-i\tau} \sum_{m=-k+1}^k S_{m,n} \Theta_{n,k}(\tau),
\end{equation}
where the \emph{modular S-matrix} is defined as
\begin{equation}
S_{m,n} \equiv \frac{1}{\sqrt{2k}} \exp\left(  -\pi i \frac{mn}{k} \right). 
\end{equation}
Thinking back to Virasoro characters (\ref{eq:character}), we see that can use these functions to rewrite the character of an irreducible representation $|h_i\rangle$ with highest weight $h_i$
\begin{equation}
\chi_m^{(k)} = \frac{\Theta_{m,k}(\tau)}{\eta(\tau)},
\end{equation}
Then the partition function $Z_{circl.}$ can be written in the form
\begin{equation}
Z_{R=\sqrt{2k}} = \sum_{m=-k+1}^k \left|\chi_m^{(k)}\right|^2.
\end{equation}
In particular, at the self-dual radius $k=1$ we have
\begin{equation}
Z_{R=\sqrt{2}} = \left|\chi_0^{(1)}\right|^2 + \left|\chi_1^{(1)}\right|^2
\end{equation}

Before moving on, we will introduce additional functions that will prove useful. We define
\begin{equation}
\vartheta\left[ \begin{smallmatrix} \alpha \\ \beta \end{smallmatrix} \right](\tau,z) = \sum_{n\in \mathbb{Z}} q^{\frac12 (n+\alpha)^2  }e^{2\pi i (n+\alpha)(z+\beta)}
\end{equation}
We can use this general formula to study the \emph{Jacobi theta functions}
\begin{align}
\vartheta_1(\tau) &= \vartheta \left[ \begin{smallmatrix} 1/2 \\ 1/2 \end{smallmatrix} \right](\tau,0) = 0 \nonumber \\
\vartheta_2(\tau) &= \vartheta\left[ \begin{smallmatrix} 1/2 \\ 0 \end{smallmatrix} \right](\tau,0)  = \sum_{n\in \mathbb{Z}}q^{\frac12 (n+\frac12)^2} = 2\eta(\tau) q^{\frac{1}{12}}\prod_{r=1}^\infty (1+q^r)^2, \nonumber \\
\vartheta_3(\tau) &= \vartheta\left[ \begin{smallmatrix} 0 \\ 0 \end{smallmatrix} \right](\tau,0)  = \sum_{n\in \mathbb{Z}}q^{\frac{n^2}{2} } = \eta(\tau)q^{-\frac{1}{24}}\prod_{r=0}^\infty (1+q^{r+\frac12})^2, \\
\vartheta_4(\tau) &= \vartheta\left[ \begin{smallmatrix} 0 \\ 1/2 \end{smallmatrix} \right](\tau,0)  = \sum_{n\in \mathbb{Z}}(-1)^n q^{\frac{n^2}{2}} = \eta(\tau) q^{-\frac{1}{24}}\prod_{r=0}^\infty (1-q^{r+\frac12})^2. \nonumber
\end{align}
To simplify the expressions as we have done, we have used something called the \emph{Jacobi triple product identity}
\begin{equation}
q^{-\frac{1}{24}} \prod_{r\geq 0} \left( 1+q^{r+\frac12}w \right) \left( 1+q^{r+\frac12}w^{-1} \right) = \frac{1}{\eta(\tau)} \sum_{n\in\mathbb{Z}} q^{\frac{n^2}{2}}w^n.
\end{equation}
We leave the derivation of this identity as an advanced exercise.

From these explicit formulas, we can derive the actions of the modular $S$- and $T$-transformations on the Jacob theta functions:
\begin{align}
\vartheta_1(\tau+1) &= e^{\frac{\pi i}{4}} \vartheta_1(\tau) , \;\;\;\;\;\; &\vartheta_1\left(-\frac{1}{\tau}\right) = e^{\frac{\pi i}{2}}\sqrt{-i\tau} \vartheta_1(\tau) ,\nonumber \\
\vartheta_2(\tau+1) &= e^{\frac{\pi i}{4}} \vartheta_2(\tau) , \;\;\;\;\;\; &\vartheta_2\left(-\frac{1}{\tau}\right) =\sqrt{-i\tau}\vartheta_4(\tau) , \nonumber \\
\vartheta_3(\tau+1) &= \vartheta_4(\tau) , \;\;\;\;\;\; &\vartheta_3\left(-\frac{1}{\tau}\right) = \sqrt{-i\tau}\vartheta_3(\tau) ,   \\
\vartheta_4(\tau+1) &= \vartheta_3(\tau) , \;\;\;\;\;\; &\vartheta_4\left(-\frac{1}{\tau}\right) = \sqrt{-i\tau}\vartheta_2(\tau) . \nonumber
\end{align}
These Jacobi theta functions will be used when studying the fermionic theory on the torus.

\subsubsection{Free fermions on the torus}

The subject of fermionic conformal field theories could fill an entire lecture. In the interest of completion, however, we should say something about these theories---even if it is hurried. Most expressions follow in a straightforward manner, and so for now we encourage the reader to check the claims made here on their own. 

The mode expansion for a free fermion $\psi(z)$ with Neveu-Schwarz boundary conditions is
\begin{equation}
\psi(z) = \sum_{r\in\mathbb{Z}} \psi_r z^{-r-\frac12}.
\end{equation}
Recall our discussion from earlier: on the torus with variable $w$, this expansion corresponds a field with anti-periodic boundary conditions.
States in the Fock space $\mathcal{F}$ of this theory are obtained by acting with creation operators $\psi_{-s}$ on the vacuum $|0\rangle$
\begin{equation}
|n_{\frac12}, n_{\frac32}\cdots\rangle =\left(\psi_{-\frac12} \right)^{n_{\frac12}} \left(\psi_{-\frac32} \right)^{n_{\frac32}} |0\rangle, \;\;\;\; n_s = 0,1.
\end{equation}
These occupation numbers reflect the fermionic nature of this field. 

We will also need the mode expansion for the stress-energy tensor. The relevant formula for calculating the partition function is
\begin{equation}
L_0 = \sum_{s=\frac12}^\infty s \psi_{-s}\psi_s.
\end{equation}
Then we can use the anti-commutation relation $\{ \psi_r, \psi_s \}=\delta_{r,-s}$ to investigate the action of $L_0$ on a general state
\begin{align}
L_0 |n_{\frac12}, n_{\frac32}\cdots\rangle &= L_0 \left(\psi_{-\frac12} \right)^{n_{\frac12}} \left(\psi_{-\frac32} \right)^{n_{\frac32}} |0\rangle \nonumber \\
&= \sum_{s=\frac12}^\infty s \left(\psi_{-\frac12} \right)^{n_{\frac12}} \cdots n_s (\psi_{-s}\psi_s\psi_{-s}) \cdots |0\rangle \\
&= \sum_{s=\frac12}^\infty s n_s |n_{\frac12}, n_{\frac32}\cdots\rangle   \nonumber.
\end{align}

Using this expression, it is straightforward to compute the character
\begin{align}
\chi_{NS,+}(\tau) &= \mbox{Tr}_\mathcal{F}\left( q^{L_0-\frac{c}{24}}  \right) \nonumber \\
&= q^{-\frac{1}{48}}  \sum_{n_{\frac12}=0}^1 \sum_{n_{\frac32}=0}^1 \cdots \langle n_{\frac12}, n_{\frac32}\cdots| q^{L_0} | n_{\frac12}, n_{\frac32}\cdots\rangle    \nonumber\\
&\;\;\;\;\;\;\;\;\;\;\;\;\vdots \nonumber \\
& = q^{-\frac{1}{48}}\prod_{r=0}^\infty \left( 1+q^{r+\frac12} \right) = \sqrt{\frac{\vartheta_3(\tau)}{\eta(\tau)}}
\end{align}
\begin{framed}
\noindent HOMEWORK: Complete this derivation.
\end{framed}
\noindent The (NS,+) notation will become clear momentarily.

The character we have computed is part of the partition function, but we want to construct a partition function that is invariant under modular transformations. Because we have already discussed the properties of $\eta$ and $\vartheta$ in detail, it immediately follows that 
$$
S(\chi_{NS,+}(\tau)) = \chi_{NS,+}.
$$
This time, it is the modular $T$-transformation that gives us trouble. We imediately see that
$$
T\left(\sqrt{\frac{\vartheta_3(\tau)}{\eta(\tau)}} \right) = e^{-\frac{i\pi}{24}}\sqrt{\frac{\vartheta_4(\tau)}{\eta(\tau)}}.
$$
The phase factor will cancel when we include the antiholomorphic contribution, but we still have a different $\vartheta$-function. In order to construct a modular invariant partition function, it looks like we must include additional sectors.
To do this, we introduce the fermion number operator $F$ such that
$$
\{(-1)^F, \psi_r \}=0.
$$
Then we can define a new character $\chi_{NS,-}(\tau)$ as
\begin{equation}
\chi_{NS,-}  = \mbox{Tr}_\mathcal{F}\left( (-1)^F q^{L_0-\frac{c}{24}}  \right)  = \sqrt{\frac{\vartheta_4(\tau)}{\eta(\tau)}}.
\end{equation}
\begin{framed}
\noindent HOMEWORK: Derive this fact by performing a computation along the same lines as the previous one.
\end{framed}
\noindent So both of these sectors correspond to anti-periodic boundary conditions; the additional term in the argument ofthis trace is a way to implement different periodicity conditions in the time direction (see the exercises for more details).

So now we have two sectors, but still no guarantee that we can construct a modular invariant partition function. We must check the modular transformation properties of this new character. It is straightforward to check that the modular $T$-transformation takes this sector back to the (NS,+) sector. This time it is the modular $S$-transformation that has a new effect. We see from our earlier calculations that
$$
\vartheta_4\left( -\frac{1}{\tau} \right) = \sqrt{-i\tau}\vartheta_2(\tau)
$$
so that
\begin{equation}
S(\chi_{NS,-}(\tau)) = \sqrt{2}q^{\frac{1}{24}}\prod_{r\geq1}(1+q^r) = \sqrt{\frac{\vartheta_2(\tau)}{\eta(\tau)}}.
\end{equation}
The exponent of $q$ takes integer values $r$ which indicates that this is a partition function for fermions $\psi_r$ with $r\in\mathbb{Z}$---fermions in the Ramond sector. As such, we label this new sector $\chi_{R+}$. At this point, we might worry that this pattern will continue indefinitely. Investigating the modular transformation properties of this character, however, we find closure:
\begin{equation}
T(\chi_{R+}(\tau)) = e^{\frac{i\pi}{12}}\chi_{R+}(\tau), \;\;\;\;\;\; S(\chi_{R+}(\tau)) = \chi_{NS-}(\tau)
\end{equation}

We are now in a position to construct the modular invariant partition function. In particular, starting from a free fermion in the NS sector, we have seen that modular invariance requires us to also consider the R sector as well as the operator $(-1)^F$. We write the partition function
\begin{equation}
Z_{ferm.}(\tau,\bar{\tau}) = \frac12 \left(  \left| \frac{\vartheta_3}{\eta} \right| + \left| \frac{\vartheta_4}{\eta} \right| + \left| \frac{\vartheta_2}{\eta} \right|   \right).
\end{equation}
The overall factor of $1/2$ is necessary to ensure the NS ground state only appears once; otherwise, we are overcounting states. Previously we found it convenient to express partition functions in terms of characters. We define
\begin{gather}
\chi_{0} = \frac12 \left( \sqrt{\frac{\vartheta_3}{\eta}} + \sqrt{\frac{\vartheta_4}{\eta}} \right) = \mbox{Tr}_{NS}\left( \frac{1+(-1)^F}{2}q^{L_0-\frac{c}{24}} \right), \nonumber \\
\chi_{\frac12} = \frac12 \left( \sqrt{\frac{\vartheta_3}{\eta}}-\sqrt{\frac{\vartheta_4}{\eta}} \right) = \mbox{Tr}_{NS}\left( \frac{1-(-1)^F}{2}q^{L_0-\frac{c}{24}} \right), \\
\chi_{\frac{1}{16}} = \frac{1}{\sqrt{2}} \sqrt{\frac{\vartheta_2}{\eta}} = \mbox{Tr}_{R}\left(q^{L_0-\frac{c}{24}} \right). \nonumber
\end{gather}
The subscripts label the conformal weight of the highest weight representations.
\begin{framed}
\noindent HOMEWORK: Check that these weights are correct. An easy way to do this is perform a series expansion of the LHS to find the exponent on the leading power of $q$.
\end{framed}
\noindent Using these expressions, we can write the partition function for a single free fermion as
\begin{equation}
Z_{ferm.}(\tau, \bar{\tau}) = \chi_0\bar{\chi}_0 + \chi_{\frac{1}{2}}\bar{\chi}_{\frac{1}{2}} + \chi_{\frac{1}{16}}\bar{\chi}_{\frac{1}{16}}.
\end{equation}
The structure of this partition function also appears when studying superstrings in flat backgrounds. The projection given by the operator $\frac12 (1+(-1)^F)$ is known as the Gliozzi-Scherk-Olive (GSO) projection.

\subsubsection{Free boson orbifold}

Often in string theory, we are interested in describing strings moving in a compact background manifold. We have already considered compatification on a circle; we now turn our attention to \emph{orbifold} models. Although this is only a quotient of a torus, this simple model will capture some of the features of more general compactifications on highly curved background geometries. 

We will therefore study the $\mathbb{Z}_2$-orbifold of the free boson on the circle. What this means is that we are not only performing the identification $\phi \sim \phi+2\pi R$, we are also imposing a $\mathbb{Z}_2$ symmetry $\mathcal{R}$ that acts as
\begin{equation}
\mathcal{R}: \phi(z,\bar{z}) \rightarrow -\phi(z,\bar{z}).
\end{equation}
Identifying the fields $\phi(z,\bar{z})$ and $-\phi(z,\bar{z})$ means the circle we had previously considered now becomes a line with a fixed point on each end.

The Hilbert space of CFTs on orbifolds will only contain states that are invariant under the orbiold action. To calculate the partition functin, we must therefore project onto invariant states. We use the projector $\frac12 (1+\mathcal{R})$ so that the partition function is
\begin{align}
Z(\tau,\bar{\tau}) &= \mbox{Tr}_{\mathcal{H}} \left( \frac{1+\mathcal{R}}{2} q^{L_0-\frac{c}{24}}\bar{q}^{\bar{L}_0-\frac{\bar{c}}{24}} \right) \nonumber \\
&= \frac12 Z_{R} +\frac12 \mbox{Tr}_{\mathcal{H}} \left( \mathcal{R} q^{L_0-\frac{c}{24}}\bar{q}^{\bar{L}_0-\frac{\bar{c}}{24}} \right).
\end{align}
Only the second term gives us a new contribution, we we will focus on it.

By the definition of our current $j(z)$, we easily find the action of $\mathcal{R}$ on the modes $j_n$:
\begin{equation}
\mathcal{R}j_n\mathcal{R} = -j_n,
\end{equation}
with a similar statement for the antiholomorphic current. The action on a general state is also straightforward to find:
\begin{equation}
\mathcal{R}|n_1,n_2,\cdots\rangle = (-1)^{n_1+n_2+\cdots}|n_1,n_2,\cdots\rangle, \label{eq:rstates}
\end{equation}
where we have chosen the action of $\mathcal{R}$ so that the vacuum $|0\rangle$ is left invariant. We also need to discuss the action of $\mathcal{R}$ on the momentum and winding states $|m,n\rangle$. To this end, we calculate
\begin{equation}
j_0 \mathcal{R}|m,n\rangle = \mathcal{R}(\mathcal{R}j_0\mathcal{R})|m,n\rangle = -\left(\frac{m}{R}+\frac{Rn}{2} \right) \mathcal{R}|m,n\rangle,
\end{equation}
with a similar calculation for $\bar{j}_0$. We have therefore found that
$$
\mathcal{R}|m,n\rangle = |-m,-n\rangle,
$$
so that only states with $|m=0,n=0\rangle$ can contribute.

Taking into account the effect of $\mathcal{R}$ on states (\ref{eq:rstates}), we follow steps similar to before, ultimately differing from the calculation of the free boson result as
\begin{equation} 
q^{-\frac{1}{24}} \prod_n \frac{1}{1-q^n} \rightarrow q^{-\frac{1}{24}}\prod_n \frac{1}{1-(-q)^n} = \sqrt{2}\sqrt{\frac{\eta(\tau)}{\vartheta_2(\tau)}}.
\end{equation}
We therefore arrive at the partition function
\begin{equation}
Z(\tau,\bar{\tau}) = \frac12 Z_{circ.}(\tau,\bar{\tau}) + \left| \frac{\eta(\tau)}{\vartheta_2(\tau)} \right| \label{eq:zorbpart}
\end{equation}
\begin{framed}
\noindent HOMEWORK: Derive equation (\ref{eq:zorbpart}) by carefully working through the steps in the last few paragraphs.
\end{framed}

Of course, we are now experts on modular invariance. From our work on the free fermion theory, we recognize that this partition function cannot be modular invariant. By performing the appropriate modular $S$- and $T$-transformations, we find that the modular invariant partition function of the $\mathbb{Z}_2$-orbifold of the free boson on the circle is
\begin{equation}
Z_{orb.}(\tau,\bar{\tau}) = \frac12Z_{circ.}(\tau,\bar{\tau}) + \left| \frac{\eta(\tau)}{\vartheta_2(\tau)} \right| + \left| \frac{\eta(\tau)}{\vartheta_4(\tau)} \right| + \left| \frac{\eta(\tau)}{\vartheta_3(\tau)} \right|.
\end{equation}
We have had to add the contribution from the \emph{twisted} sector. For the fermion, the additional contributions come from sectors with different boundary conditins and ground state charges. What is the origin of these contributions for the orbifold partition function? Consider the explicit form
\begin{equation}
\sqrt{\frac{\eta(\tau)}{\vartheta_4(\tau)}} = q^{\frac{1}{16}-\frac{1}{24}} \prod_{n=0}^\infty \frac{1}{1-q^{n+\frac12}}.
\end{equation}
This can be interpreted as the partition function in a sector with ground state energy $L_0|0\rangle = \frac{1}{16}|0\rangle$ and half-integer modes $j_{n+\frac12}$:
$$
j(z) = i \partial \phi(z,\bar{z}) = \sum_{n\in\mathbb{Z}}j_{n+\frac12}z^{-\left(n+\frac12\right)-1}.
$$
This mode expansion respects the symmetry
\begin{equation}
j(e^{2\pi i}z) = -j(z) = \mathcal{R}j(z)\mathcal{R},
\end{equation}
so that the free boson $\phi(z,\bar{z})$ is invariant under rotations inthe complex plane up to the action of the discrete symmetry $\mathcal{R}$. In general, for an orbifold with abelian symmetry group $G$, the partition function is of the form
\begin{equation}
Z(\tau,\bar{\tau}) = \frac{1}{|G|} \sum_{g,h\in G} \mbox{Tr}_h\left(g q^{L_0-\frac{c}{24}} \bar{q}^{\bar{L}_0 - \frac{\bar{c}}{24}}  \right),
\end{equation}
where the trace is over all twisted sectors for which the fields $\phi$ obey
$$
\phi(e^{2\pi i}z, e^{-2\pi i}\bar{z}) = h \phi (z, \bar{z}) h^{-1}.
$$

We conclude with a few remarks about this result. For starters, this example demonstrates the amazing relationship between conformal field theory on the world-sheet of a string and the background geometry through which the string propagates. Also note that the twisted sector has an overall two-fold degeneracy. The origin of this fact is that the twisted sectors are localized at the fixed points of the orbifold action; in this case, there are two fixed points corresponding to the end points of our line segment/identified circled. Finally, note that only the first term in $Z_{orb.}$ depends on the radius of the circle. As such, the orbifold partition function is also invariant under $T$-duality. Moreover, it can be shown that $Z_{orb.}\bigg|_{R=\sqrt{2}} = Z_{circ.}\bigg|_{R=2\sqrt{2}}$. The moduli spaces of these partition functions intersect. In fact, the moduli space of conformal field theories with $c=1$ has been classified; refer to the references for more information.

\subsection{Fusion rules and the Verlinde formula}
 
 We finish this lecture by discussing a powerful result known as the \emph{Verlinde formula}. Before discussing this, however, we need to introduce \emph{fusion rules}. Recall\footnote{It was a HOMEWORK.} that the null state at level $N=2$ satisfies
 \begin{equation}
 \left( L_{-2}-\frac{3}{2(2h+1)}  \right) |h\rangle = 0.
 \end{equation}
 for a theory with central charge $c=\frac{2h}{2h+1}(5-8h)$. The corresponding descendant field
 \begin{equation}
 \hat{L}_{-2}\phi(z) -\frac{3}{2(2h+1)}\hat{L}^2_{-1}\phi(z)
 \end{equation}
 is thus a null field. This relation implies an expression for the differential operators acting on the correlation functions involving $\phi(z)$
 \begin{equation}
0 = \left(  \mathcal{L}_{-2}\phi(z) - \frac{3}{2(2h+1)}\mathcal{L}^2_{-1}\phi(z) \right) \langle \phi(w) \phi_1(w_1)\cdots\phi_n(w_n)\rangle. 
 \end{equation}
 Working out this differential equation for two-point function, we see that it is trivially satisfied.
 \begin{framed}
 \noindent HOMEWORK: See that it is trivially satisfied.
 \end{framed}
 
 A more interesting constraint comes from acting with this differential operator on the three-point correlator $\langle \phi(w)\phi_1(w_1)\phi_2(w_2)\rangle$. Using the known form of the three-point function, we obtain the constraint on the conformal weights $\{ h,h_1,h_2 \}$
 \begin{equation}
 2(2h+1)(h+2h_2-h_1)=3(h-h_1+h_2)(h-h_1+h_2+1).
 \end{equation}
 Solving this expression for $h_2$ gives
 \begin{equation}
 h_2=\frac16 + \frac{h}{3} + h_1 \pm \frac23 \sqrt{h^2 + 3h h_1 -\frac12 h + \frac32 h_1 + \frac{1}{16}}. \label{eq:h2}
 \end{equation}
 
This is all well and good, but what does this have to do with modular invariance? First, let us apply equation (\ref{eq:h2}) to the primary fields $\phi_{(p,q)}$. In particular, choosing $h=h_{2,1}$ and $h_1=h_{p,q}$, then the two solutions for $h_2$ are precisely $\{ h_{p-1,q}, h_{p+1,q}   \}$. At most, two of the coefficients $C_{\phi\phi_1\phi_2}$ will be non-zero. The OPE of $\phi_2=\phi_{(2,1)}$ with any other primary field in a unitary minimal model is then restricted to be of the form
 \begin{equation}
 [ \phi_{(2,1)}]\times[\phi_{(p,q)}] = [\phi_{(p+1,q)}] + [\phi_{(p-1,q)}],
 \end{equation}
 where $[\phi_{(p,q)}]$ denotes the conformal family descending from $\phi_{(p,q)}$. This equation means that the OPE between a field in the first conformal family and a field in the second conformal family involves only fields belonging to one of the conformal families on the RHS. The coefficients could still be zero, actually, but no more than these families can contribute. This is an example of a \emph{fusion rule}.
 
We could express more general fusion rules for the unitary minimal models of the Virasoro generators; seeing their form is not helpful at the moment. We could generalize to arbitrary RCFTs. We will only use that the OPE between conformal families $[\phi_i]$ and $[\phi_j]$ gives rise to the concept of a \emph{fusion algebra}
\begin{equation}
[\phi_i]\times[\phi_j] = \sum_k N_{ij}^k [\phi_k].
\end{equation}
Here $N_{ij}^k \in \mathbb{Z}_0^+$, and $N_{ij}^k=0$ if and only if $C_{ijk}=0$. This algebra is commutative, meaning
\begin{equation}
N^k_{ij} = N^k_{ji},
\end{equation}
and it is associative.
\begin{framed}
\noindent HOMEWORK: To see consequences of associativity, consider $[\phi_i]\times[\phi_j]\times[\phi_k]$  two different ways to conclude
\begin{equation}
\sum_l N^l_{kj} N^m_{il} = \sum_l N^l_{ij} N^m_{ik} .
\end{equation}
\end{framed}
\noindent The vacuum representation, $[0]$, contains the stress-energy tensor and its descendants. We label it in this way because it is the unit element
\begin{equation}
N^k_{i1}=\delta_{ik}.
\end{equation}
 
Again, this is an interesting line of inquiry. But \emph{what} does it have to do with modular invariance? One of the most incredible results in CFT is that there does exist a relation between the fusion algebra for the OPE on the sphere (which is at tree-level) and the modular $S$-matrix (related to the torus partition function). We previously studied considered $S_{mm'}$ for the $\Theta_{mk}$-functions. But we can consider a more general RCFT with central charge $c$ and a finite number of highest weight representations $\phi_i$ having characters $\chi_i$. Then there exists a representation of the modular group on that space of characters; in particular, there is a matrix $S_{ij}$ such that
\begin{equation}
\chi_i\left( -\frac{1}{\tau}\right) = \sum_{j=0}^{N-1} S_{ij}\chi_j(\tau). \label{eq:smatrix}
\end{equation}
 In all known cases, the $S$-matrix is unitary and symmetric
 \begin{equation}
 SS^\dagger = S^\dagger S = \boldsymbol{1}, \;\;\;\;\;\;\;\; S = S^T.
 \end{equation}
 The Verlinde formula gives us a way to calculate the fusion coefficients from the $S$-matrix:
 \begin{equation}
 N_{ij}^k = \sum_{m=0}^{N-1}\frac{S_{im}S_{jm}S^*_{mk}}{S_{0m}}.
 \end{equation}
 In this formula, $S^*$ denotes the complex conjugate of $S$ and the subindex 0 labels the identity representation.
 
We will not give a full proof of the Verlinde formula; at this time, we will not even give a very detailed overview of the proof. The proof relies on something called the \emph{pentagon identity} for fusing matrices and monodromy transformations on the space of conformal blocks. In a later course, these lectures will be structured so that this can be detailed. For now, we refer you to the references. We are in an excellent place to push this formalism further; we can calculate fusion coefficients for different theories and construct entire classes of modular invariant partition functions. Alas, we must bring this discussion to an end. We leave a fusion coefficient calculation as an exercise. We finish by mentioning that similar to equation (\ref{eq:smatrix}), there is a matrix $T_{ij}$ that gives a similar relation for the modular $T$-transformation:
 \begin{equation}
 \chi_i(\tau+1) = \sum_{j=0}^{N-1} T_{ij}\chi_j(\tau). \label{eq:tmatrix}
 \end{equation}
 Again, we skip a detailed derivation and claim that we can choose a basis such that 
 \begin{equation}
 T_{ij} = \delta_{ij}e^{h_i-\frac{c}{24}},
 \end{equation}
 where $h_i$ denotes the conformal weight of the heighest weight representation for character $\chi_i(\tau)$.
 
 \break
 
\subsection*{References for this lecture}
\vspace{4mm}
\noindent Main references for this lecture
\\
\begin{list}{}{%
\setlength{\topsep}{0pt}%
\setlength{\leftmargin}{0.7cm}%
\setlength{\listparindent}{-0.7cm}%
\setlength{\itemindent}{-0.7cm}%
\setlength{\parsep}{\parskip}%
}%
\item[]

[1] Chapter 4 of the textbook: R. Blumenhagen, E. Plauschinn, \emph{Introduction to Conformal Field Theory: With Applications to String Theory}, Lect. Notes Phys. 779,  (Springer, Berlin Heidelberg 2009).

[2] Chapter 10 of the textbook: P. Di Francesco, P. Mathieu, and D. Senechal. \emph{Conformal field theory}, Springer, 1997.

[3] Chapters 7,8 of : P. Ginsparg, \emph{Applied Conformal Field Theory}, Les Houches, Session XLIX, 1988, \emph{Fields, Strings and Critical Phenomena}, ed. by E. Br\'{e}zin and J. Zinn-Justin, (Elsevier Science Publishers, B.V., 1989), [arxiv:9108028v2 [hep-th]].

\end{list}

\break

\section{Lecture 6: Central Charge and Scale vs. Conformal}

We have made some significant strides toward a general understanding of conformal field theory. We have studied theories in various dimensions, found conformal algebras and groups, and constructed representations of these conformal groups; we have studied conserved currents and constraints coming from conformal invariance; in the case of two dimensions, we were able to completely classify the unitary representations of the Virasoro algebra for a particular range of the central charge. Yet our work is built upon a bed of lies.

Well, that is an exaggeration. But there are important topics and significant issues that we have been ignoring for several lectures. In this lecture, we will go back to some of these earlier topics in order to clarify some points, flesh out additional details, and touch base with active areas of conformal field theory research.

\subsection{The central charge}

We begin by studying the central charge. If I asked you to explain in a couple of sentences what we mean we talk about the central charge, what would you say? They are perhaps the most important numbers characterizing the CFT, and thus far we have only said that they are somehow measuring the number of degrees of freedom in the CFT. Can we make this understanding more explicit? Let us find out!

Recall that under a finite conformal transformation $z\rightarrow f(z)$, the stress-energy tensor transforms according to equation (\ref{eq:finitettransform})
\begin{equation}
T'(z) = \left( \frac{\partial f}{\partial z} \right)^2 T(f(z)) + \frac{c}{12} S\left( f(z),z \right),
\end{equation}
where $S$ is the Schwartzian derivative. Note that this term is the same evaluated on all states; it only affects the constant term/zero mode in the energy. When studying conformal field theory on the cylinder, we calculated this contribution in equation (\ref{eq:lzeroshift})
\begin{equation}
L_{0,cyl} = L_0-\frac{c}{24},
\end{equation}
with a corresponding change in $\bar{L}_0$. Considering both of these terms, we found (\ref{eq:casimir}) the ground state energy on the cylinder to be
\begin{equation}
E_0 = -\frac{c+\bar{c}}{24}.
\end{equation}
For a free scalar field having $c=\bar{c}=1$, the energy density is $-1/12$. This is the infamous vacuum energy in bosonic string theory that can be found by adding together all of the positive integers\footnote{This is not a typo.}.

If we wanted to compare this a physical system, the cylinder would have some radius $L$. Then the Casimir energy becomes
$$
E = -\frac{c+\bar{c}}{24 L}.
$$
In your studies of quantum field theory, you may have considered the Casimir force between two parallel plates. In the case of this cylinder, there is a similar calculation for QCD-like theories. We can consider two quarks in a confining theory separated by some distance $L$. If the tension of the confining flux tube is $T$, then this string will be stable so long as $TL\lesssim m$, the mass of the lightest quark. The energy of the stretched string as a function of $L$ is given by
\begin{equation}
E(L)=TL + a - \frac{\pi c}{24 L}+\cdots
\end{equation}
Here $a$ is some undetermined constant and $c$ counts the number of degrees of freedom of the flux tube\footnote{Two important points: first, there is no analog of $\bar{c}$ here because of the reflecting boundary conditions at the end of the string; second, there is a factor of $2\pi$ that is different between the two cylindrical energies we have expressed. This factor is related to our earlier definition of the holomorphic stress-energy tensor $T(z)\sim 2\pi T_{zz}$.}. This contribution to the string energy is known as the \emph{Lüscher term}.

Of course, there is another important manner in which the central charge affects the stress-energy tensor. Recall that one of the defining features of a CFT was the vanishing of the trace of the stress-energy tensor
$$
T^\mu_\mu = 0.
$$
Of course, this result was derived at the classical level. When we consider the full quantum theory, the quantity $\langle T^\mu_\mu \rangle$ may not vanish. On a curved background, there will be a \emph{trace anomaly}. We will now argue that
\begin{equation}
\langle T^\mu_\mu \rangle = - \frac{c}{12} R. \label{eq:canomaly}
\end{equation}
Before doing this derivation, we make a few general statements related to this claim. 

First of all, why does this only involve the left-moving central charge? Is there something special about the left-moving sector? Of course this is not the case; we could also write
$$
\langle T^\mu_\mu \rangle = - \frac{\bar{c}}{12} R.
$$
In flat space, CFTs are perfectly fine with different $c$ and $\bar{c}$. If we want these theories to be consistent in fixed, curved backgrounds, we must require $c=\bar{c}$. We also remark that this trace anomaly exists in higher dimensions, although the specific terms that appear depend on the dimension of the spacetime. For example, 4d CFTs are characterized by two numbers $a$ and $c$. The trace anomaly in four dimensions is
\begin{equation}
\langle T^\mu_\mu \rangle_{4d} = \frac{c}{16\pi^2}C_{\kappa\lambda\rho\sigma}C^{\kappa\lambda\rho\sigma} - \frac{a}{16\pi^2} \tilde{R}_{\kappa\lambda\rho\sigma} \tilde{R}^{\kappa\lambda\rho\sigma},
\end{equation}
where $C$ is the Weyl tensor (built from the Riemann tensor and Ricci tensor and scalar) and $\tilde{R}$ is the dual of the Riemann tensor \cite{duff1}. We will return to the $a$ ``central charge'' later.

We also remark that the result (\ref{eq:canomaly}) is not just true for the vacuum; it holds for any state. This can be seen as a reflection of the fact that this anomaly comes from regulating short distance divergences; at short distances, all finite energy states look basically the same and so the expression will be the same as for the vacuum expectation value. Because this expectation value is the same for any state, regardless of the states in our theory, we expect that it must equal something depending on the background metric (the object that will be appearing in our CFT coupled to gravity regardless of the other fields present in the theory). This something should be local, and by dimensional analysis we see that it should be dimension $2$. The natural candidate is the Ricci scalar $R$. Through an appropriate choice of coordinates, we can always put a 2d spacetime metric in the form $g_{\mu\nu}=e^{2\omega(x)}\delta_{\mu\nu}.$ The Ricci scalar is then given by
\begin{equation}
R = -2e^{-2\omega}\partial^2 \omega.
\end{equation}
Thus according to (\ref{eq:canomaly}), any CFT with $c\neq 0$ has a physical observable taking different values on backgrounds related by a Weyl transformation $\omega$. This is why this anomaly is also referred to as the \emph{Weyl anomaly}.

Alright, let us actually derive the Weyl anomaly. Our starting point is the equation for energy conservation\footnote{This expression follows from definitions and steps taken back in Lecture 3. Work through the steps if at any point the equations seem too unfamiliar.}
\begin{equation}
\partial T_{z\bar{z}}= -\bar{\partial}T_{zz}.
\end{equation}
Using this expression, we can write the OPE
\begin{equation}
\partial_z T_{z\bar{z}} \partial_w T_{w\bar{w}} = \bar{\partial}_{\bar{z}} T_{zz} \bar{\partial}_{\bar{w}} T_{ww} = \bar{\partial}_{\bar{z}}\bar{\partial}_{\bar{w}} \left( \frac{c/2}{(z-w)^4}+\cdots\right). \label{eq:sixnine}
\end{equation}
Naively, we could expect this quantity to vanish. After all, we are taking the antiholomorphic derivative of a holomorphic quantity. There is a singularity, however, at $z=w$ that could affect this result. Recall our derivation of the free bosonic propagator; we had a similar situation happening in (\ref{eq:logformula}). Using that result, we find
\begin{equation}
\bar{\partial}_{\bar{z}}\bar{\partial}_{\bar{w}} \frac{1}{(z-w)^4} = \frac16 \bar{\partial}_{\bar{z}}\bar{\partial}_{\bar{w}} \left(\partial_z^2\partial_w \frac{1}{z-w} \right) = \frac{\pi}{3}\partial_z^2\partial_w\bar{\partial}_{\bar{w}}\delta^{(2)}(z-w).
\end{equation}
Comparing this expression to (\ref{eq:sixnine}), we find the OPE
\begin{equation}
T_{z\bar{z}}(z,\bar{z})T_{w\bar{w}}(w,\bar{w}) = \frac{\pi c}{6} \partial_z \bar{\partial}_{\bar{w}}\delta^{(2)}(z-w).
\end{equation}
We find that this expression does \emph{not} vanish, as we might have naively expected, but instead has a contact term. 

We assume that $\langle T^\mu_\mu \rangle=0$ in flat space (as we have found to be the case), and derive an expression for the Weyl anomaly for some background infinitesimally close to flat space. First, we know that under a general shift of the metric $\delta g_{\alpha\beta}$ we get the variation
\begin{align}
\delta \langle T^\mu_\mu(\sigma) \rangle &= \delta \int D\phi \; e^{-S} T^\mu_\mu(\sigma) \\
&= \frac{1}{4\pi} \int D\phi \; e^{-S}\left( T^\mu_\mu(\sigma) \int d^2\sigma'\;\sqrt{g}\delta g^{\alpha\beta}T_{\alpha\beta}(\sigma') \right).
\end{align}
If we consider a Weyl transformation, then $\delta g_{\alpha\beta}=2\omega\delta_{\alpha\beta}$ so that $\delta g^{\alpha\beta}=-2\omega\delta^{\alpha\beta}$. This gives
\begin{equation}
\delta \langle T^\mu_\mu(\sigma) \rangle = -\frac{1}{2\pi}  \int D\phi \; e^{-S}\left( T^\mu_\mu(\sigma) \int d^2\sigma'\;\ \omega(\sigma') T^\nu_\nu(\sigma') \right).
\end{equation}

Now to calculate the Weyl anomaly, we change between complex coordinates and Cartesian coordinates. We find\footnote{Again, this follows from definitions of complex coordinates. The formulas necessary are expressions like $T_{z\bar{z}}=\frac14(T_{00}+T_{11})$.}
\begin{equation}
T^\mu_\mu(\sigma) T^\nu_\nu(\sigma')=16 T_{z\bar{z}}(z,\bar{z}) T_{w\bar{w}}(w,\bar{w}).
\end{equation}
We also use the fact\footnote{\emph{Again}, this follows from all of our conventions. Convince yourself of this fact if you need.} that 
\begin{equation}
8\partial_z\bar{\partial}_{\bar{w}}\delta^{(2)}(z-w)=-\partial^2 \delta^{(2)}(\sigma-\sigma').
\end{equation}
Substituting these expressions, we obtain
\begin{equation}
T^\mu_\mu(\sigma) T^\nu_\nu(\sigma')= -\frac{c\pi}{3}\partial^2 \delta(\sigma-\sigma').
\end{equation}
Then plugging this into the expression for $\delta \langle T^\mu_\mu(\sigma) \rangle$ and integrating by parts, we are left with
\begin{equation}
\delta \langle T^\mu_\mu(\sigma) \rangle = \frac{c}{6}\partial^2\omega.
\end{equation}
To do the final step, we use the fact that we are working infinitesimally to replace $e^{-2\omega}=1$, so that $R=-2\partial^2\omega$. Then
\begin{equation}
\langle T^\mu_\mu(\sigma) \rangle = -\frac{c}{12}R. 
\end{equation}
Thus we have completed the proof for spaces infinitesimally close to flat space. Without providing the proof, I claim that $R$ remains on the RHS for general 2d surfaces. This fact follows from the fact that we need the expression to be reparameterization invariant.

In both of these examples, the central charge has provided an extra contribution to the energy. But we will now argue that it also tells us the density of high energy states. To do this, we consider a CFT on a Euclidean torus (as in Lecture 5). Of course, the key idea we discussed when considering CFTs on the torus was modular invariance. In particular, we expect the partition function of our theory to be invariant under the modular $S$-transformation $\tau\rightarrow -1/\tau$. We will normalize the spatial direction so that $\sigma\in[0,2\pi)$. The partition function for a theory with periodic Euclidean time can be related to the free energy of the theory at temperature $T=1/\beta = 1/2\pi\mbox{Im}(\tau)$. Invariance of the partition function under the modular $S$-transformation thus means
\begin{equation}
Z[4\pi^2/\beta]=Z[\beta].
\end{equation}
We thus have a simple way to study the very high temperature behavior of the partition function. But this high temperature limit is sampling all states in the theory, and on entropic grounds this sampling should be dominated by the high energy states. Thus this computation is really telling us how many high energy states there are. 

The partition function is generically given by
\begin{equation}
Z[\tau,\bar{\tau}]=\mbox{Tr} \;e^{2\pi i(\tau L_0 - \bar{\tau} \bar{L}_0)} = \langle 0 |e^{2\pi i(\tau L_0 - \bar{\tau} \bar{L}_0)} | 0 \rangle + (\mbox{excited states}).
\end{equation}
At low temperatures, corresponding to $T=1/\mbox{Im}(\tau)\ll 1$, the trace is well-approximated by the vacuum contribution. We therefore have
\begin{equation}
Z_{{low}}[\tau,\bar{\tau}] = e^{2\pi i\frac{c}{24}(-\tau + \bar{\tau} )} + O\left(e^{-\mbox{\rm Im}(\tau)}\right).
\end{equation}

Now we need to discuss the partition function at \emph{high} temperature. We will denote the eigenvalues of $L_0,\bar{L}_0$ at high temperatures by $\ell_0,\bar{\ell}_0$, and we introduce the density of states $\rho(E)=e^{S(E)}$, where $S(E)$ is the entropy. Then the partition function can be expressed as 
$$
Z = \int dE\,\, e^{S(E)+2\pi i (\tau \ell_0-\bar{\tau}\bar{\ell}_0)}.
$$
We can find the leading-order behavior via a saddle-point approximation:
\begin{equation}
\log Z_{{high}}[\tau,\bar{\tau}] \sim S(\ell_0,\bar{\ell}_0) + 2\pi i (\tau \ell_0 - \bar{\tau}\bar{\ell}_0),
\end{equation}
where $\ell_0$ and $\bar{\ell}_0$ are functions of $\tau$ and $\bar{\tau}$ respectively that extremize the right-hand side.
\begin{framed}
\noindent HOMEWORK: Perform this saddle-point approximation.
\end{framed}
\noindent So we have expressions for the partition function at high and low temperatures. Equating the logarithms of these expressions gives
\begin{equation}
S(\ell_0,\bar{\ell}_0) \simeq 2\pi i \frac{c}{24}\left( \frac{1}{\tau}-\frac{1}{\bar{\tau}} \right) - 2\pi i (\tau \ell_0 - \bar{\tau}\bar{\ell}_0).
\end{equation}

In this formula, $\tau$ and $\bar{\tau}$ are functions of $\ell_0$ and $\bar{\ell}_0$ that extremize the right-hand side. We find the extremal values for $\tau$ and $\bar{\tau}$ to be
\begin{equation}
\tau(\ell_0) = i \sqrt{\frac{c}{24\ell_0}},\;\;\;\;\;\;\;\; \bar{\tau}(\bar{\ell}_0) = -i \sqrt{\frac{c}{24\bar{\ell}_0}}.
\end{equation}
The signs for these roots have been chosen so that the temperature is positive. Substituting these values back into the above expression, we arrive at \emph{Cardy's formula} \cite{bigcardy}
\begin{equation}
S \simeq 2\pi \sqrt{\frac{c\ell_0}{6}} + 2\pi \sqrt{\frac{c\bar{\ell}_0}{6}}.
\end{equation}
The eigenvalue $\ell_0$ was for the Virasoro generator on the cylinder. Switching to the Virasoro generators on the plane, we pick up the Casimir energy contribution to get 
\begin{equation}
S \simeq 2\pi \sqrt{\frac{c\left(L_0-\frac{c}{24}\right)}{6}} + 2\pi \sqrt{\frac{c\left(\bar{L}_0-\frac{\bar{c}}{24}\right)}{6}}. \label{eq:cardy}
\end{equation}

In a paper by Verlinde \cite{6-8}, a generalization of equation (\ref{eq:cardy}) was proposed for CFTs in arbitrary dimensions. Consider a conformal field theory in $(n+1)$-dimensional spacetime described by the metric
\begin{equation}
ds^2 = -dt^2 + R^2 d\Omega_n^2,
\end{equation}
where $R$ is the radius of an $n$-dimensional sphere. The entropy of this CFT can be given by the \emph{Cardy-Verlinde formula}
\begin{equation}
S = \frac{2\pi R}{\sqrt{ab}}\sqrt{E_c(2E-E_c)}, \label{eq:cardyv}
\end{equation}
where $E_c$ represents the Casimir energy, and $a$ and $b$ are two positive coefficients which are independent of $R$ and $S$.

In this version of the course, we will not be able to discuss the AdS/CFT correspondence in detail. This is obviously a terrible shame; the conjectured correspondence between conformal field theories and anti-de Sitter spaces is arguably the most important advance in our understanding of quantum gravity in the last couple of decades. The correspondence relates a stringy theory of quantum gravity on an AdS spacetimes with a conformal field theory without gravity living on the boundary of that spacetime. In the current context, Strominger \cite{6-9} used the correspondence between AdS space in three-dimensions and the two-dimensional CFT living on the boundary. He showed that the Cardy formula (\ref{eq:cardy}) gives an entropy that is exactly the same as the Bekenstein-Hawking entropy--a calculation of the entropy of a three-dimensional black hole from a purely gravitational perspective. These results are obtained in vastly different ways, but in light of the AdS/CFT correspondence their equality make sense. Do similar statements hold in higher dimensions? It was argued by Witten \cite{6-10} that the thermodynamics of a CFT at high temperature can be identified with the thermodynamics of black holes in AdS space even in higher dimensions. Verlinde checked the formula (\ref{eq:cardyv}) for AdS Schwarzschild black holes using the AdS/CFT correspondence and found it holds exactly. Some of the recent work in this topic is provided at the end of the lecture.

\subsection{The $c$-theorem and $d=2$ scale invariance}

After that lengthy discussion about the central charge, it is time to return to a statement we have been taking for granted: does scale invariance imply conformal invariance? In this section, we will show a proof by Zamolodchikov and Polchinski that global scale invariance does imply local scale invariance in two dimensions under broad conditions. We will also discuss the status of this question in higher dimensions.

To begin, recall scale transformations
\begin{equation}
\delta x^\mu = \epsilon x^\mu
\end{equation}
Following the Noether procedure as in Lecture 2, we find that the scale current for the dilatation symmetry will be of the general form
\begin{equation}
S^\mu(x) = x^\nu T_\nu^{\mu}(x) + K^\mu(x).
\end{equation}
Here $T_{\mu\nu}$ is the symmetric stress-energy tensor and $K$ is a local operator without explicit dependence on the coordinates. 
The conservation of this scale current implies
\begin{equation}
T_\mu^\mu(x) = -\partial_\mu K^\mu (x).
\end{equation}
Given any stress-energy tensor, the necessary and sufficient condition for existence of a conserved scale curent is that its trace be the divergence of a local operator.

We also recall conformal transformations
\begin{equation}
\delta x^\mu = \epsilon b^\mu(x),
\end{equation}
such that
\begin{equation}
\partial_\mu b_\nu(x) + \partial_\nu b_\mu(x) = \frac{2}{d} g_{\mu\nu}\partial\cdot b(x).
\end{equation}
We did not explicitly calculate it earlier, but the Noether procedure shows us that a conformal current must be of the form
\begin{equation}
j^\mu_b(x) = b^\nu(x)T_\nu^\mu(x) + \partial \cdot b(x) K'^\mu + \partial_\nu \partial \cdot b(x) L^{\nu\mu}(x). \label{eq:sccurrent}
\end{equation}
Here $K'$ is the same as $K$ up to possibly some conserved current, and $L$ is a local operator.
\begin{framed}
\noindent HOMEWORK: Derive this form for the conformal current. You can also use general reasoning to determine where each term originates (e.g., the first term is determined by the spacetime nature of the transformation, etc.).
\end{framed}
\noindent For $d\geq 3$, we know $\partial\cdot b$ is a linear function of $x^\mu$. By taking the divergence of (\ref{eq:sccurrent}), we find that conformal invariance is equivalent to
\begin{equation}
T_\mu^\mu(x) = -\partial_\mu K'^\mu(x), \;\;\;\;, K'^\mu = -\partial_\nu L^{\nu\mu}(x).
\end{equation}
\begin{framed}
\noindent HOMEWORK: Derive these conditions.
\end{framed}
\noindent For $d=2$, $\partial\cdot b$ is a general harmonic function and conservation also implies $L^{\nu\mu}(x) = g^{\nu\mu}L(x)$. Thus we have the conditions
\begin{equation}
T_\mu^\mu(x)= \left\{ 
\begin{array}{ll}
\partial_\nu\partial_\mu L^{\nu\mu}(x), & \;\; d\geq 3 \\
\partial^2 L(x), & d=2
\end{array}\right.
\end{equation}
The trace of the stress-energy tensor being of this form means that our theory will have the full conformal invariance. This also makes it clear that conformal invariance implies scale invariance.

We now see that a system will be scale invariant without being conformally invariant if the trace of the stress-energy tensor is the divergence of a local operator $-K^\mu$ which is \emph{not} itself a conserved current plus a divergence (or gradient, for $d=2$). This matches well with our earlier understanding of the relationship between scale and conformal invariance, where the virial being the divergence of another tensor naively let us promote scale invariance to full conformal invariance. When this is the case, we can also define the improved stress-energy tensor (for $d>2$ dimensions)
\begin{equation}
\Theta_{\mu\nu} = T_{\mu\nu}+\frac{1}{d-2}(\partial_\mu\partial_\lambda L^\lambda_\nu + \partial_\nu\partial_\lambda L^\lambda_\mu - \partial^2 L_{\mu\nu}-\eta_{\mu\nu}\partial_\lambda\partial_\rho L^{\lambda\rho})
\end{equation}
with a similar definition for $d=2$ dimensions. This improved tensor is traceless, symmetric, and conserved.
\begin{framed}
\noindent HOMEWORK: Show that this is the case.
\end{framed}
\noindent So a traceless stress-energy tensor really does imply conformal invariance.

This more detailed understanding of scale and conformal invariance gives us an obvious condition under which scale invariance with conformal invariance: if there is no suitable candidate for $K^\mu$. For example, consider perturbative $\phi^4$ theory in $d=4$ dimensions. The only possible vector with the correct dimension is $K^\mu \sim \partial^\mu (\phi^2)$ (check this fact). Therefore scale invariance implies conformal invariance for the nontrivial fixed points in $4-\epsilon$ dimensions\footnote{We have not derived these conformal field theories; we refer you to a proper course on the renormalization group for more details.}. The same will be true for $\phi^3$ theory in $d=6$ dimensions and $\phi^6$ theory in $d=3$ dimensions. What  about gauge theories? In both abelian and non-abelian gauge theories coupled to fermions, BRST invariance of the stress-energy tensor means the only candidate is $A_\mu\partial_\nu A^\nu + \alpha \bar{c}D_\mu c$ (with gauge parameter $\alpha$)\footnote{We leave this as an exercise.}. The perturative fixed point for $SU(N_c)$ when $0<1-2N_f/11N_c\ll 1$ is therefore a conformally invariant theory\footnote{This condition comes from demanding the $\beta$ function for non-abelian gauge theory with $SU(N_C)$ gauge group and $N_f$ fermions inthe fundamental representation be very small---close to zero. Look it up in a QFT textbook.}. There are also many statistical mechanical systems that have a small number of low dimension operators and thus no candidate for $K^\mu$.

Of course, this depends on knowing the spectrum of a theory with only a small number of low dimension operators. If we restrict ourselves to two dimensions, we can provide a proof of the fact that scale invariance implies conformal invariance.
Consider the two-point function of the stress-energy tensor $T_{\mu\nu}$ in complex coordinates. We define
$T\equiv T_{zz}$ and $\Theta\equiv T^\mu_\mu$. Following [2], we also define
\begin{gather}
F(|z|^2) =  z^4 \langle T(z,\bar{z}) T(0) \rangle, \\
G(|z|^2) =  z^3 \bar{z} \langle T(z,\bar{z}) \Theta(0) \rangle, \\
H(|z|^2) =   z^2\bar{z}^2 \langle \Theta(z,\bar{z}) \Theta(0) \rangle.
\end{gather}
By Poincar\'{e} invariance,we know that $T_{\mu\nu}$ is conserved
\begin{equation}
\bar{\partial} T + 4\partial\Theta= 0.
\end{equation}
Now by taking the correlation function between this equation of motion and either $T$ or $\Theta$, one can derive the equations
\begin{gather}
\dot{F} + \frac14 (\dot{G}-3G) = 0 \\
\dot{G} - G + \frac14(\dot{H}-2H)=0.
\end{gather}
Here we have defined $\dot{X}\equiv z\bar{z} X'(z\bar{z})$.
\begin{framed}
\noindent HOMEWORK: Derive these equations. Really, do it. They are not difficult, and the result is worth it.
\end{framed}

Now we can define the function $C$ as
\begin{equation}
C \equiv 2F - G - \frac38 H.
\end{equation}
Using the above equations, we arrive at the conclusion that
\begin{equation}
\dot{C} = -\frac34 H.
\end{equation}
By unitarity/reflection positivity, we know the quantity $H\geq 0$. Therefore the function $C$ is a decreasing function of $R\equiv \sqrt{z\bar{z}}$:
\begin{equation}
\dot{C} \leq 0.
\end{equation}
In a theory with coupling constants $g_i$, we can write the renormalization group equation for $C$ as
\begin{equation}
\left[R\frac{\partial}{\partial R} + \beta_i(g) \frac{\partial}{\partial g_i}  \right] C(g,R) = 0.
\end{equation}
Here the $\beta_i$ are the renormalization group beta-functions. At a fixed point corresponding to a conformal field theory, $\beta_i=0$. We can also find that for a conformal field theory, $G = H = 0$ and $F = c/2$. Thus for a CFT, the function $C$ equals the central charge $c$. This is \emph{Zamolodchikov's c-theorem}: if renormalization flows connect different conformal field theories, then $C$ decreases from the ultraviolet to the infrared with $C=c$ at criticality.

This is an amazing result, but we have gotten sidetracked. At a scale-invariant fixed point, we will assume the stress-energy tensor scales canonically so that $T_{\mu\nu}$ has a scaling dimension $\Delta=2$ and $C$ is constant. Then
\begin{equation}
\langle \Theta(z,\bar{z})\Theta(0)\rangle = 0
\end{equation}
which means from unitarity and causality (according to the Reeh-Schlieder theorem \cite{6-7}), $\Theta(z,\bar{z}) =0$ as an operator identity. Because $\Theta$ is the trace of the stress-energy tensor, the scale invariance implies conformal invariance. Success!

Before continuing, we need to make a few remarks. First, we can expand $\Theta$ with respect to  operators in our theory via something like
\begin{equation}
\Theta = \mathcal{B}^I O_I
\end{equation}
where the $\mathcal{B}$ are related to the $\beta$-functions.The $c$-theorem can then be expressed as
\begin{gather}
\frac{dc}{d \log \mu} = \mathcal{B}^I \chi_{IJ} \mathcal{B}^J \geq 0, \\
\chi_{IJ} \equiv \frac32 |z|^4 \langle O_I(z,\bar{z}) O_J(0)\rangle\bigg|_{|z|=\mu^{-1}}.
\end{gather}
The positive definite metrix $\chi_{IJ}$ is known as the \emph{Zamolodchikov metric}. It is not immediately obvious that the $C$ function is a function of the running coupling constants alone and does not depend on the energy scale $\mu$ explicitly. A local renormalization group analysis tells us that this is precisely the case; we refer the reader to the references for more details.

We must address one final technicality. This derivation tacitly assumed that the stress-energy tensor had a canonical scaling dimension. We can prove this is the case in $d=2$ dimensions when we also make the assumption of the discreteness of scaling dimensions of operators in our theory. The violatation of canonical scaling of the stress-energy tensor means that $T_{\mu\nu}$ is not an eigenoperator under dilatations
\begin{equation}
i[D,T_{\mu\nu}] = x^\lambda \partial_\lambda T_{\mu\nu}+dT_{\mu\nu} + y_a \partial^\rho\partial^\sigma Y^a_{\mu\rho\nu\sigma}.
\end{equation}
Here $Y$ is the complete set of tensor operators that have the symmetry of the Riemann tensor and the scaling properties
\begin{equation}
i[D,Y^a_{\mu\rho\nu\sigma}] = x^\lambda\partial_\lambda Y^a_{\mu\rho\nu\sigma}+\gamma^a_b Y^b_{\mu\rho\nu\sigma}.
\end{equation}
Polchinski [3] argued that we can improve the stress-energy tensor so it has a canonical scaling dimension so long as there is no dimension zero operator other than the identity operator. He introduces the improved
\begin{equation}
T'_{\mu\nu} = T_{\mu\nu}+y^a(d-2-\gamma)^{-1}_{ab}\partial^\rho\partial^\sigma Y^b_{\mu\rho\nu\sigma}.
\end{equation}
There are subtleties in other dimensions, but as of now we are only considering $d=2$ dimensions.

\subsection{Example of scale without conformal invariance}

It would be easy to assume that theories that have scale invariance without having conformal invariance are bizarre or nonphysical in some fundamental way. In this section, we consider a simple example that illustrates this is not always the case: the theory of elasticity in two dimensions:
\begin{equation}
S = \frac12 \int d^2x [2 g u_{\mu\nu}u^{\mu\nu}+k(u_\rho^\rho)^2], \label{eq:elasticaction}
\end{equation}
where $u_{\mu\nu}=\frac12(\partial_\mu u_\nu + \partial_\nu u_\mu)$ is the strain tensor built from displacement fields $u_\mu$, and the coefficients $g$ and $k+g$ represent the shear modular and bulk modulus of the material respectively.This is certainly a well-defined physical theory; let us investigate the properties of this theory. We omit several of the details and leave the verification of some claims as one of the detailed exercises.

What are the symmetries of this theory? It is straightforward to see that this action is invariant under translations. This action is also invariant under rotations if the fields $u_\mu$ transform as vectors
\begin{equation}
u'_\mu(x') = \Lambda_\mu^\nu u_\nu(x).
\end{equation}
Knowing how the measure and metric transform under dilatations, we can find what conformal dimension for $u_\mu$ will leave the action invariant under a scale transformation.
\begin{framed}
\noindent HOMEWORK: Find this conformal dimension for $u_\mu$.
\end{framed}

Rather than considering special conformal transformations directly, let us focus on the stress-energy tensor. The canonical stress-energy tensor
$$
T_C^{\mu\nu} = \frac{\partial \mathcal{L}}{\partial(\partial_\mu u_\rho)} \partial^\nu u\rho - g^{\mu\nu} \mathcal{L}
$$
is not symmetric for this theory. We can add an improvement term via the Belinfante procedure
$$
T_B^{\mu\nu} = T_C^{\mu\nu} + \partial_\rho B^{\rho\mu\nu}, 
$$
where $B^{\rho\mu\nu}$ is defined in equation (\ref{eq:Bdef}). The field $u_\mu$ transforms as a vector, and the only non-vanishing components of $S_{\mu\nu}$ act as
\begin{gather}
S_{12}u_1 = i u_2, \;\;\;\;\;\;\;\;S_{12}u_2 = -iu_1.
\end{gather}
Given this fact, it follows that the trace of the stress-energy tensor is of the form
\begin{equation}
T^\mu_\mu = -\partial^\mu V_\mu, \;\;\;\;\;\;\mbox{with}\;\;\;\;\;\; V_\mu = -B_{\mu\rho}^\rho.
\end{equation}
This is in agreement with the scale invariance of this theory. 

To investigate whether this theory has full conformal invariance, we explicitly write $V_\mu$ in coordinates to get
\begin{gather}
V_1 = \partial_1 \left( -\frac{k}{2}u_1^2 - \frac{g}{2} u_2^2 \right) - (k+2g)u_1\partial_2 u_2 + g u_2\partial_2 u_1, \\
V_2 = \partial_2 \left( \frac{g}{2}u_1^2 - \frac{k}{2} u_2^2 \right) - (k+2g)u_2\partial_1u_1 + gu_1\partial_1u_2.
\end{gather}
\begin{framed}
\noindent HOMEWORK: Find these expressions.
\end{framed}
\noindent Playing with these equation for a bit, we see that $V_\mu$ cannot be expressed as a gradient. Therefore conformal invariance does not hold for this theory; this Belinfante stress-energy tensor cannot be improved to be traceless. But we have a two-dimensional CFT that has scale and Poincar\'{e} invariance. What went wrong?  

Let us push farther using our normal approach. We again write the action in complex coordinates $z=x^1+ix^2$ to obtain
\begin{equation}
S = \frac12 \int d^2z \; \left[ (k+g) (\partial \bar{u} + \bar{\partial} u )^2 + 4 g (\partial u) (\bar{\partial}\bar{u})\right). \label{eq:selastic}
\end{equation}
We know from the transformation properties we discussed earlier that the fields $u$ and $\bar{u}$ must have spins $s=1$ and $\bar{s}=-1$ respectively. We also know that both of their scaling dimensions must vanish in order to ensure scale invariance (go back and do that exercise if you skipped it). We obtain these properties with the conformal weights
\begin{equation}
h_u = \bar{h}_{\bar{u}} = \frac12, \;\;\;\;\;\;\;\; \bar{h}_u = h_{\bar{u}}= -\frac12.
\end{equation}
Then we can investigate the effect of a generic conformal transformation $z\rightarrow w = f(z)$, under which the fields transform as $\phi\rightarrow (\partial f)^{-h} (\bar{\partial}\bar{f})^{-\bar{h}} \phi$. We find that this action is not invariant under this transformation.
\begin{framed}
\noindent HOMEWORK: Explicitly check that the action (\ref{eq:selastic}) is not invariant under a conformal transformation.
\end{framed}

With the proof from earlier in mind, let us see exactly where this theory fails to be conformally invariant. We can express the trace of the stress-energy tensor at the quantum level as
\begin{equation}
T^\mu_\mu = (k+g) (:\partial \bar{u}\partial\bar{u}: +:\bar{\partial}u\bar{\partial}u:+2:\partial\bar{u}\bar{\partial}u:) -g (:\partial u \bar{\partial}\bar{u}: - :u\partial\bar{\partial}\bar{u}: - :\bar{u}\partial \bar{\partial}u:).
\end{equation}
By using the explicit expressions for the two-point correlators\footnote{Their calculation is involved enough to be left as an exercise.}
\begin{gather}
\langle {u}(z) {u}(w) \rangle = \frac{k+g}{4\pi g (k+2g} \frac{\bar{z}-\bar{w}}{{z}-{w}} ,\\
\langle \bar{u}(z) \bar{u}(w) \rangle = \frac{k+g}{4\pi g (k+2g} \frac{{z}-{w}}{\bar{z}-\bar{w}}  \\
\langle {u}(z) \bar{u}(w) \rangle = \frac{k+g - (k+3g)\log(z-w)(\bar{z}-\bar{w})}{4\pi g (k+2g)},
\end{gather}
and Wick's theorem, we can find the two-point correlator 
\begin{equation}
\langle T^\mu_\mu(z) T^\nu_\nu(0) \rangle = \frac{-2 (k+g)(k+3g)}{\pi^2 (k+2g)^2} \frac{1}{z^2\bar{z}^2}.
\end{equation}
This expression does not vanish.

To investigate further, we first see that the operator $V_\mu$ expressed in complex coordinates takes the form
\begin{gather}
V_z = \partial(g u \bar{u})-\frac{k+g}{2} u \bar{\partial} u - \frac{k+3g}{2} u \partial \bar{u},\\
V_{\bar{z}} = \bar{\partial} (g u \bar{u}) - \frac{k+g}{2} \bar{u}\partial\bar{u} - \frac{k+3g}{2} \bar{u} \bar{\partial}u.
\end{gather}
This operator has some contribution going as a gradient. We therefore choose $L=-gu\bar{u}$ and naturally define $T'_{\mu\nu} = T_{\mu\nu}+\partial_\mu\partial_\nu L(x) - g_{\mu\nu} \partial_\rho \partial^\rho L(x)$.Then we \emph{do} find that
\begin{equation}
\langle T'^\mu_\mu(z) T'^\nu_\nu(0) \rangle = 0. \label{eq:sixseventyone}
\end{equation}
Yet the trace itself does \emph{not} vanish! It is given by
\begin{equation}
T'^\mu_\mu =(k+g)[:\partial \bar{u}\partial \bar{u}: + :\bar{\partial}u\bar{\partial}u:]+ 2(k+3g):\partial\bar{u}\bar{\partial}u:.
\label{eq:sixseventytwo}
\end{equation}
Seriously, what is happening here?

This suggests the theory of elasticity lacks reflection positivity\footnote{We can see other evidence that this theory lacks reflection positivity. If this theory had reflection positivity, any two-point function involving the trace $T'^\mu_\mu$ should vanish. We can show several instances where this is not the case; for example
$$
\langle T'^\mu_\mu(z) :\partial u \partial u:(0) \rangle = - \frac{k+g}{2\pi^2 g(k+2g)}\frac{1}{z^4}.
$$}. The lack of reflection positivity is equivalent to non-unitarity in Minkowski coordinates. If we express the Hamiltonian associated to (\ref{eq:elasticaction}) we find
\begin{equation}
H = \frac12 \int dx\; \left[ \frac{1}{k+2g}\pi^2_t + g (\partial_x u_t)^2 -\frac{1}{g}(\pi_x-(k+g)\partial_x u_t)^2 - (k+2g)(\partial_x u_x)^2   \right],
\end{equation}
where the conjugate momenta are given by
\begin{equation}
\pi_t = (k+2g)\partial_t u_t, \;\;\;\;\;\;\;\; \pi_x = g\partial_t u_x + (k+g)\partial_x u_t.
\end{equation}
Here the nonunitarity is explicit in the form of negative signs. These negative signs originate from the signature of the Minkowski metric. The question of when and if scale invariance implies conformal invariance can be complicated.

\subsection{Generalizations for $d>2$ scale invariance}

Given our successful proof in $d=2$ dimension, we conjecture that any scale invariant quantum field theory in $d>2$ dimensions is conformally invariant under the same assumptions as before: unitarity, Poincar\'{e} invariance, unbroken scale invariance, the existence of a scale current, and a discrete scaling dimension spectrum. In terms of the stress-energy tensor, our conjecture is that given these assumptions whenever the trace of the stress-energy tensor is the divergence of the virial current
$$
T^\mu_\mu = \partial^\mu V_\mu,
$$
the virial current can be removed by an improvement (or equivalently, it is itself the derivative of a local scalar operator
$$
K_\mu = \partial_\mu L
$$
as discussed earlier). For most of this section we will focus on $d=4$ dimensions. Near the end of the lecture we will mention possibilities in other dimensions. 

In $d=2$ dimensions, the proof of the enhancement from scale invariance to conformal invariance was almost identical to the proof of the $c$-theorem. It is therefore natural to consider a generalization of the $c$-theorem to higher dimensions. In $d=4$ dimensions, the most generic possibility for the Weyl anomaly is given by
\begin{equation}
\langle T^\mu_\mu \rangle = c C^2 - a E + b R^2 + \tilde{b} D^\mu D_\mu R + d \epsilon^{\mu\nu\rho\sigma}R_{\mu\nu}^{\alpha\beta}R_{\alpha\beta\rho\sigma}. \label{eq:4danomaly}
\end{equation}
Here, $C$ is the Weyl tensor with $C^2 = R^2_{\mu\nu\rho\sigma}-2R^2_{\mu\nu}+\frac13 R^2$ and $E = R^2_{\mu\nu\rho\sigma} - 4 R^2_{\mu\nu} + R^2$ is the Euler scalar. The term $\tilde{b}D^2 R$ can be removed by adding a local counterterm proportional to 
$$
\int d^4 x\sqrt{|g|}\tilde{b}R^2,
$$
so it is not an anomaly in the traditional sense\cite{duff1}. In addition, it is possible to show that $b=0$ in order to satisfy the Wess-Zumino consistency condition\cite{duff2}\footnote{These are consistency conditions for how the partition function must behave under gauge transformations. In the current context, The W-Z consistency condition is a statement about Weyl variations of terms that could possibly appear in the anomaly.}. Finally, the Pontryagin $d$ term is consistent. It does, however, break invariance under the CP transformation. There is no known unitarity field theory model that gives the Pontryagin term as a Weyl anomaly\footnote{Although the Pontryagin term shows up in the Euclidean formulation for (A,B) representations, with $A>B$ (or with the opposite sign for $A<B$)\cite{duff3}.}. We will not be considering this term in the work to follow.

This leaves us with the result we quoted earlier in the lecture: the Weyl anomaly will be of the form
\begin{equation}
\langle T^\mu_\mu \rangle = c C^2 - a E.
\end{equation}
It is not immediately clear which combination of $a$ and $c$ will count the degrees of freedom like $c$ did in $d=2$ dimensions. One can show \cite{duff2,birrell} that the $a$ term for a real scalar, a Dirac fermion, and a real vector are given by $\frac{1}{90(8\pi)^2}$, $\frac{11}{90(8\pi)^2}$, and $\frac{62}{90(8\pi)^2}$ respectively. Similarly, the $c$ term for a real scalar, a Dirac fermion, and a real vector are given by $\frac{1}{90(8\pi)^2}$, $\frac{6}{90(8\pi)^2}$, and $\frac{12}{90(8\pi)^2}$. There are known examples where $c$ does \emph{not} show monotonicity along renormalization group flow \cite{nonct}, so the remaining possibility is $a$. Cardy formulated the \emph{a-theorem}: the quantity
\begin{equation}
a = -\frac{1}{64\pi^2}\int_{\mathbb{S}^4}d^4x\; \sqrt{|g|}\langle T^\mu_\mu \rangle
\end{equation}
will behave in a similar manner in $d=4$ dimensions as the central charge $c$ in $d=2$ dimensions\cite{acardy}.

The conjectured $a$-theorem can be formulated as different statements. Some different formulations are
\begin{align}
(1&) \;\;a_{IR}\geq a_{UV} \mbox{ between the flow of two CFTs.} \\
(2&)\;\;\frac{d a}{d\log\mu}\geq 0 \mbox{ along renormalization group flow.} \\
(3&) \;\;\mbox{gradient formula}: \mathcal{B}^I = \chi^{IJ}\partial_J a, \mbox{ so that} \frac{d a}{d\log\mu} = \mathcal{B}^I \chi_{IJ}\mathcal{B}^J.
\end{align}
In $d=2$ dimensions, we were able to prove the analogue of each of these statements using the fact that the Weyl anomaly $c$ was related to the two-point function of the stress-energy tensor (although we omitted the explicit renormalization group proof of statement (3)). Specifically,it appears in the contact terms of two-point functions involving the trace, and we could use conservation of $T_{\mu\nu}$ to relate the trace to the stress-energy tensor $T_{zz}$. In $d=4$ dimensions, the situation is more complicated; $\langle T^\mu_\mu\rangle$ contains quartic and quadratic divergences that must be subtracted. These steps spoil naive positivity arguments, so that a similar approach does not work. 

We strongly refer the reader to \cite{naka}; it is a fantastic resource with more details than I could hope to adequately cover. The recent status of this problem is as follows\footnote{All efforts were made to keep this information up-to-date as of when I began writing these notes. Any mistakes or missing information will happily be corrected; please contact me.}.
In $d=4$ dimensions, there is a recent nonperturbative proof of (1); we will return to this proof momentarily. Under some technical assumptions, scale invariant fixed points can be shown to be conformal invariant perturbatively. Beyond perturbation theory, the proof is not complete. There is also a perturbative proof of the strong version as well as the gradient formula \cite{158,159}. The subsequent results shows that subject to our assumptions, scale invariance implies conformal invariance perturbatively in $d=4$ dimensions. The general idea is to use the local renormalization group to generalize the Wess-Zumino consistency condition for the Weyl anomaly not only in the non-trivial metric background but with spacetime dependent coupling constants. This is natural, as the Weyl transformation acts on coupling constants non-trivially so that they must be treated in a spacetime dependent way even if we started with a constant background. The full argument is too involved for this lecture; we strongly encourage the reader to check the references for this lecture to read a complete discussion.

As previously remarked, Cardy's conjecture has a natural generalization in even dimensions: the coefficient in front of the Euler density in the Weyl anomaly must be monotonically decreasing along the renormalization group flow. In $d = 6$ dimensions, there has not yet been success in using the dilaton-scattering argument that we will discuss in $d=4$ dimensions to prove (1)---it is difficult to show the positivity of the dilaton scattering amplitudes in $d = 6$ dimensions\cite{6d}. On the other hand, there is no counterexample known, and there are no known theories which have scale invariance and not conformal invariance (with a gauge invariant scale current! See the additional exercises.) Within perturbation theory, an argument similar to the one in $d=4$ dimensions can be found in \cite{6-11}.

There has also been important work done from a holographic perspective (along the same lines as the already mentioned AdS/CFT correspondence). Investigating RG flows in a holographic framework means the results are readily extended to arbitrary dimensions. By studying holographic models with higher curvature gravity in the $(d+1)$-dimensional bulk, \cite{6-6} was able to distinguish the various ``central charges'' appearing in the Weyl anomaly of the $d$-dimensional boundary CFT. They found that the coefficient $a$ of the Euler scalar has a natural monotonic flow in various dimensions.  In fact, they found a quantity $a^*_d$ that satisfies (1) for \emph{any} $d$. Given that there is no Weyl anomaly in odd dimensions, a new interpretation for this quantity must be found. 

In $d = 3$ dimensions, the candidate for the $a$-function is the finite part of the $\mathbb{S}^3$ partition function $F = − \log Z_{\mathbb{S}^3}\bigg|_{reg.}$ \cite{6-6,3d1}. This is equivalent to the finite part of the \emph{entanglement entropy}\footnote{The connection between entanglement entropy and CFTs is interesting, and I hope to address it in a later version of this course. If you came down to this footnote to learn about entanglment entropy, I can only say that it is an entanglement measure for a state divided into two partitions $A$ and $B$. Specifically, $S(\rho_A) = -\mbox{Tr}[\rho_A\log \rho_A]$, where $\rho_A=\mbox{Tr}_B(\rho_{AB})$ is the reduced density matrix for a pure state $\rho_{AB}=|\psi\rangle\langle\psi|_{AB}$.  } of the half $\mathbb{S}^3$ when the theory is at the conformal fixed point \cite{3d2}. It is currently an active area of research as to whether there is a strong version of the F-theorem that would imply enhancement from scale invariance to conformal invariance in d = 3 dimensions \cite{3d3,3d4}. We refer the reader to references cited.

In $d = 1$ dimension, we cannot use the Reeh-Schlieder theorem due to the lack of Poincar\'{e} invariance. If we assume its validity regardless, then scale invariance implies conformal invariance. On the other hand, $d=1$ QFTs are equivalent to simple quantum mechanical systems. There are certainly examples of cyclic renormalization group flow realized in non-relativistic field theories, as well as systems having scale invariance without conformal invariance. In these cases, the Reeh-Schlieder theorem does not hold no matter how much we may wish to assume its validity. 

Finally, for $d \geq 7$ dimensions it is likely that there is no interacting unitary conformal field theory \cite{7d}. There are no classically scale invariant Lagrangians having two-derivative kinetic terms other than free field theories. Higher-dimensional free Maxwell theory cannot be conformally invariant for $d\geq 7$; this makes sense when you consider the fact that there is no superconformal algebra for $d\geq 7$.

But wait! We became sidetracked considering $c$, $F$, and $a$ theorems in higher dimensions. Does scale invariance imply conformal invariance in dimensons $d>4$? It turns out that scale invariance does \emph{not} imply conformal invariance in higher dimensions. There is a very simple counterexample using the ideas we have already discussed. We leave it as an exercise (but its definitely worth it; I highly recommend it).

\subsection{Overview of nonperturbative proof of the $a$-theorem}

We will conclude this lecture by giving a (terribly) brief overview of the nonperturbative proof of the $a$-theorem. Some of the mathematical details will be left as additional exercises. And of course, the original reference covers this topic in much greater detail.

We consider a UV CFT perturbed by relevant operators. In flat space, this looks like
\begin{equation}
S = S_{CFT, UV} + \sum_j \lambda_j \int \Phi_j(x) d^4 x,
\end{equation}
where $\Phi_j$ has dimension $\Delta_j < 4$. By defining the dimensionless coupling $g_j \equiv \lambda_j \ell^{4-\Delta_j}$, we can write 
$$
\beta_j = (\Delta_j - 4) g_j.
$$
Under renormalization group flow, $g_j\rightarrow \infty$ and the theory flows as $S\rightarrow S_{CFT,IR}$. The a-theorem concerns whether $a_{UV} > a_{IR}$.

To address this question, we consider the theory coupled to an additional scalar $\tau$. This scalar is known as the \emph{dilaton}, and it is related to broken scale symmetry. In flat space, the modified theory is
\begin{equation}
S = S_{CFT, UV} + \sum_j \lambda_j\int \Phi_j(x) e^{\Delta_j - 4}\tau d^4 x + f^2 \int e^{-2\tau} (\partial \tau)^2 d^4 x \end{equation}
\begin{framed}
\noindent HOMEWORK: Under a scale transformation, $x^\mu \rightarrow e^b x^\mu$ and $\Phi_j \rightarrow e^{-b\Delta_j}\Phi_j$. How much $\tau$ transform in order to maintain scale invariance?
\end{framed}
\noindent It can be shown that this action is conformally invariant, with $T^\mu_\mu=0$.
\begin{framed}
\noindent HOMEWORK: This last term might look a little strange; we claim it is actually the action for a free scalar in disguise. Find the relation between $\phi$ and $\tau$ that brings this term into canonical form. How does the scale transformation rule for $\tau$ affect $\phi$?
\end{framed}

The coefficient $f$ has dimensions of mass. By taking $f\rightarrow\infty$, we select a vacuum expectation value for $\tau$. Without loss of generality, we could say that we pick out $\tau=0$, and thus we end up back at the original theory. In practice, it is sufficient to take $f$ to be much larger than any other mass scale in the theory in order to see the change from ultraviolet to infrared behavior. As the crossover from the UV to the IR happens, some UV CFT degrees of freedom will become massive. By integrating out these degrees of freedom, we are left with the IR CFT plus some effective low-energy dilaton theory $S_{dil}$; this effective action decouples for large $f$. Since the total theory is conformally invariant, we know that 
\begin{equation}
a_{CFT, UV} = a^{tot.}_{UV} = a^{tot.}_{IR} = a_{CFT, IR} + a_{dil}.
\end{equation}
Thus what we must argue is that $a_{dil} > 0$.

Of course, we have been quoting formulas in flat space. In curved space, the coupling to the dilaton is of the form
\begin{equation}
\sum_j \lambda_j \int \Phi_j (x) e^{(\Delta_j-4)\tau} \sqrt{g} d^4x.
\end{equation}
The scale invariance of flat space will now show up as invariance under Weyl transformations
\begin{equation}
g_{\mu\nu} \rightarrow e^{2\sigma}g_{\mu\nu}, \;\;\;\;\;\;\tau\rightarrow\tau+\sigma.
\end{equation}
The effection action should respect this symmetry, up to the anomaly term. The authors Komargodski and Schwimmer determined the effective action $S_{dil}$ up to four derivatives. 

So we need to construct an action $S_{anomaly}$ such that its Weyl variation produces the trace anomaly terms we expect:
\begin{equation}
\delta S_{anomaly} / \delta \sigma = c_{dil} C^2 - a_{dil} E.
\end{equation}
In the exercises, we argue that the result up to four derivatives is
\begin{align}
S_{anomaly} &= \int d^4 x \,  \sqrt{g} \tau (c_{dil} C^2 - a_{dil}E)   \label{eq:4deriv}    \\
&- a_{dil} \int d^4 x\,\sqrt{g} \left[ 4\left( R^{\mu\nu}-\frac12 g^{\mu\nu} R \right) \partial_\mu\tau \partial_\nu \tau - 4 (\partial\tau)^2 \partial^2 \tau + 2(\partial\tau)^4  \right]. \nonumber
\end{align}
It is interesting that even in the flat-space limit there are terms involving $a_{dil}$ that survive. In the exercises, we show that the terms proportional to $a_{dil}$ that survive in flat space after using the equation of motion for $\tau$ are
\begin{equation}
S_{anomaly} \rightarrow 2 a_{dil} \int(\partial\tau)^4 d^4 x.
\end{equation}

We therefore see that $a_{dil}$ determines the on-shell low-energy elastic dilaton-dilaton scattering amplitude. The amplitude is given by
\begin{equation}
A(s,t,u) = \frac{a_{dil}}{f^4} (s^2+t^2+u^2) + \cdots
\end{equation}
Additional terms are suppressed. By considering the forward direction ($t=0, u=-s$, this scattering amplitude becomes
\begin{equation}
A(s) = \frac{2 a_{dil}}{f^4} s^2.
\end{equation}
We also know that for forward scattering we can use the optical theorem
\begin{equation}
\mbox{Im} A(s) = s \sigma_{tot}(s). 
\end{equation}

The final steps in the proof require some facts about dispersion relations\footnote{At some later date, we might include some proofs as additional exercises.}. We want to consider the amplitude $A/s^3$ and write a dispersion relation for it. This requires knowledge of the singular behavior. There are branch cuts both at positive and negative $s$. Negative $s$ cuts just correspond to physical states in the $u$ channel, so the $s\leftrightarrow u$ symmetry means these contributions will be equivalent to ones for positive $s$. In addition, $A/s^3$ has a pole at the origin that gives the coefficient $a_{dil}$. By closing the contour, then we find the dispersion relation
\begin{equation}
a_{dil} = \frac{f^4}{\pi} \int_{s' > 0} ds' \frac{\mbox{Im} A(s')}{s'^3}.
\end{equation}
Using the optical theorem, this becomes
\begin{equation}
a_{dil} = \frac{f^4}{\pi} \int_{s' > 0} ds' \frac{\sigma_{tot}(s')}{s'^2}.
\end{equation}
This discontinuity will therefore be positive, and we have thus argued that $a_{dil} >0$. Thus we have therefore proven that $a_{UV} > a_{IR}$.

Of course, there are important details we have omitted and open questions that need addressing. Relating the difference in $a$-charges between the UV and IR to a physical quantity like dilaton scattering avoids the previously mentioned issues with subtractions. This proof is similar to earlier work in two-dimensions using dispersion techniques. In $d=2$ dimensions, however, we also have the cleaner Zamolodchikov proof. Maybe improvements can be made by considering the flat space $\langle TTTT \rangle$ four-point function. At the time of these lectures, I do not know of any additions or improvements to this nonperturbative proof. And so it is here that we pause; I refer you to the original notes for more details \cite{naka}.

\break

\subsection*{References for this lecture}
\vspace{4mm}
\noindent Main references for this lecture
\\
\begin{list}{}{%
\setlength{\topsep}{0pt}%
\setlength{\leftmargin}{0.7cm}%
\setlength{\listparindent}{-0.7cm}%
\setlength{\itemindent}{-0.7cm}%
\setlength{\parsep}{\parskip}%
}%
\item[]

[1] Y. Nakayama, \emph{Scale invariance vs conformal invariance,} Physics Reports 569 (2015): 1-93.
[arxiv:1302.0884 [hep-th]].

[2] D. Tong, \emph{Lectures on String Theory, 4. Introducing Conformal Field Theory,} 2009.  www.damtp.cam.ac.uk/user/tong/string/four.pdf

[3] A. B. Zamolodchikov, \emph{Irreversibility of the Flux of the Renormalization Group in a 2D Field Theory,} JETP lett 43.12 (1986): 730-732.

[4] J. Polchinski, \emph{Scale and conformal invariance in quantum field theory,} Nuclear Physics B 303.2 (1988): 226-236.

[5] V. Riva and J. Cardy, \emph{Scale and conformal invariance in field theory: a physical counterexample,} Physics Letters B, 622.3 (2005): 339-343 [arxiv:0504197 [hep-th]] .

[6] Z. Komargodski and A. Schwimmer, \emph{On renormalization group flows in four dimensions.} Journal of High Energy Physics 2011.12 (2011): 1-20. [arxiv:1107.3987 [hep-th]].

\end{list}

\break

\section{Lecture 7: Conformal Bootstrap}

\subsection{A brief recap}

By now, we have seen that any CFT is characterized by the spectrum of local primary operators; by this, we mean the pairs $\{ \Delta, \mathcal{R} \}$, where $\Delta$ is an operator's scaling dimension and $\mathcal{R}$ is the representation of the $SO(D)$ under which it transforms. We have seen that all other operators are obtained by differentiating primary operators to get descendant operators. We also showed that there is a one-to-one correspondence between operators $\mathcal{O}_\Delta$ and the states of a radially quantized theory. This correspondence is obtained by inserting the operator at the origin $|\Delta\rangle = \mathcal{O}_\Delta |0\rangle$. We even showed that there exist unitarity bounds for operator dimensions 
$$
\Delta \geq \Delta_{min}(\mathcal{R}),
$$
where $\Delta_{min}(\mathcal{R})$ is the lowest allowed value for an operator in the representation $\mathcal{R}$.

By using constraints from conformal invariance, we were able to completely fix the form of two-point functions of primaries (and descendants, though that was more complicated). In the case of identical scalars, for example, we found
$$
\langle \phi(x) \phi(y) \rangle = \frac{d_{\phi\phi}}{|x-y|^2\Delta}
$$
where the normalization is usually chosen so that $d_{\phi\phi}=1$. We also found that the three-point functions of primary operators are fixed up to a constant. For three scalars, we found
$$
\langle \phi_1(x_1)\phi_2(x_2)\phi_3(x_3)  \rangle = \frac{\lambda_{123}}{x_{12}^{\Delta-2\Delta_3}x_{13}^{\Delta-2\Delta_2}x_{23}^{\Delta-2\Delta_1}},
$$
where $\Delta = \Delta_1+\Delta_2+\Delta_3$ and $x_{ij}=|x_i-x_j|$. The constant $\lambda_{123}$ is a physical parameter that cannot be rescaled away once the two-point function normalization has been fixed. Analogously, we can compute the most general three-point function of three spin-$\ell$ operators. It turns out that there are a finite number of tensor structures that are consistent with conformal symmetry and thus a finite number of constants multiplying these tensors.

Finally, we have studied the operator product expansion (OPE). We found that the three-point correlator constant $\lambda_{123}$ appears in the OPE
\begin{equation}
\phi_1(x)\phi_2(0) = \sum_{\rm primaries\;\mathcal{O}} \lambda_{12\mathcal{O}}C_\mathcal{O}(x,\partial_y)\mathcal{O}(y)\bigg|_{y=0}.
\end{equation}
We are using new notation for the OPE that we will find more convenient, but nothing has actually changed. Note that while the operator $\mathcal{O}$ and its derivatives are being computed at $y=0$, we could actually calculate them at any point between $0$ and $x$ (although the coefficient functions $C_\mathcal{O}$ will then be changed appropriately). 

With knowledge of the \emph{CFT data}, the spectrum and OPE coefficients for a particular theory, we can compute \emph{any} $n$-point correlation function of the theory. By using the OPE, we can recursively reduce an $n$-point function to $(n-1),(n-2),\cdots$ and finally to some combination of three-point functions. Schematically, this looks like
\begin{equation}
 \langle \phi(x_1)\phi(x_2) \prod \psi_i(y_i)\rangle  = \sum_\mathcal{O} \lambda_{\phi\phi\mathcal{O}} C_\mathcal{O}(x_1-x_2,\partial_{x_2})\langle  \mathcal{O}(x_2)\prod\psi_i(y_i) \rangle. \label{eq:opereduce}
\end{equation}
The first correlator that we haven't completely utilized is the four-point correlation function. In the majority of this lecture, we will restrict our attention almost exclusively to four-point correlators.

This brings us to a very important fact about the conformal OPE that has yet to be adequately emphasized: the OPE is a convergent expansion. It is precisely this convergence that allows us to compute correlation functions of arbitrarly high order. And it is precisely this convergence that will allow us to actually constrain the CFT data itself using the \emph{conformal bootstrap}. We claim that the OPE $\phi_1(x_1)\phi_2(x_2)$ will converge as long as $x_1$ is closer to $x_2$ than any other operators inserted at $y_i$
\begin{equation}
|x_1-x_2| < \mbox{min}_i |y_i - x_2|.  \label{eq:opemin}
\end{equation}
A good discussion of this proof can be found in Section 2.9 of reference \cite{polbook}. We will provide only a rough outline.

We radially quantize the theory with $x_2$ as the origin. If equation (\ref{eq:opemin}) is satisfied, there exists a sphere separating the points $x_1,x_2$ from the other operators. The LHS of (\ref{eq:opereduce}) can then be understood as an overlap function $\langle \Psi | \Phi \rangle$ between two states living on the sphere; these states are produced by acting with $\phi$'s and $\psi$'s on the \emph{in} and \emph{out} vacua
\begin{equation}
|\Phi\rangle = \phi(x_1)\phi(x_2) |0\rangle, \;\;\;\;\;\;\;\;\langle\Psi| =\langle 0 | \prod \psi_i(y_i).
\end{equation}
Furthermore, we can expand the state $|\Phi\rangle$ into a complete basis of energy eigenstates\footnote{
Since the radial quantization Hamiltonian is the dilatation generator $D$, these states are generated by acting on the vacuum with local operators of definite scaling dmension $\Delta_n = E_n$. Moreover, there is a one-to-one correspondence between this expansion and the OPE; for every primary $\mathcal{O}$, the eigenstate expansion will contain a series of states produced by $\mathcal{O}(x_2)$ and its descendants
\begin{equation}
|E_n\rangle = (\partial_{x_2})^n\mathcal{O}(x_2)|0\rangle, \;\;\;\; E_n = \Delta_\mathcal{O} + n.
\end{equation}
The coefficients $C_n$ are found by picking up the coefficient of $(\partial_{x_2})^n$ in the OPE.
}
\begin{equation}
|\Phi\rangle = \sum_n C_n (x_1-x_2) |E_n\rangle.
\end{equation}
Convergence of the OPE then follows from a basic theorem about Hilbert spaces: the scalar product of two states converges when one of the two states is expanded into an orthonormal basis. We also refer the reader to \cite{rychkovope}.

\subsection{Conformal bootstrap: the general picture}

We are finally in a position to discuss the conformal bootstrap technique. Given that we can compute all the correlators in a theory given CFT data, it is natural to ask if a random set of CFT data defines a consistent theory. The answer is \emph{no}; in imposing consistency conditions on CFT data, we can rule out certain candidate theories. In order to impose a consistency condition, we will study the four-point function. Consider a scalar four-point function. To compute it via the OPE, we surround two of the operators, say $\phi_1$ and $\phi_2$, by a sphere; we then expand into radial quantiation states on this sphere. This means we are writing
\begin{gather}
\phi_1(x_1)\phi_2(x_2) = \sum_\mathcal{O} \lambda_{12\mathcal{O}}C_{\mathcal{O}}(x_{12},\partial_y)\mathcal{O}(y)\bigg|_{y=\frac{x_1+x_2}{2}}, \\
\phi_3(x_3)\phi_4(x_4) = \sum_\mathcal{O} \lambda_{34\mathcal{O}}C_{\mathcal{O}}(x_{34},\partial_z)\mathcal{O}(z)\bigg|_{y=\frac{x_3+x_4}{2}}.
\end{gather}
Substituting these expressions, we find
\begin{gather}
\langle \phi_1(x_1) \phi_2(x_2) \phi_3(x_3) \phi_4(x_4) \rangle \label{eq:bigblock}
= \\  \mbox{ }\sum_\mathcal{O} \lambda_{12\mathcal{O}} \lambda_{34\mathcal{O}} \left[   
C_\mathcal{O}(x_{12},\partial_y) C_\mathcal{O}(x_{34},\partial_z) \langle \mathcal{O}(y)\mathcal{O}(z) \rangle 
\right]. \nonumber
\end{gather}
\begin{figure}
\centering
\includegraphics[scale=.5]{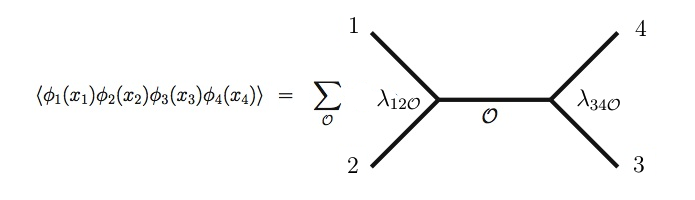}
\caption{A diagrammatic representation of the conformal partial wave expansion. Connected lines do \emph{not} a Feynman diagram make.} \label{figure:cpw1}
\end{figure}
The quantity in square brackets is completely fixed by conformal symmetry in terms of the dimensions of $\phi_i$ and of the dimension and spin of $\mathcal{O}$\footnote{We rushed a little here, and after receiving questions I am clarifying: generically, the operators appearing in the $\phi_1\phi_2$ and $\phi_3\phi_4$ are different and we have a double sum. But we can choose a basis for our fields so that the two-point functions go as Kronecker $\delta$'s thus collapsing the double sum into a single sum.}. These functions are called \emph{conformal partial waves}. We can express the expansion into conformal partial waves diagramatically as in Figure (\ref{figure:cpw1}). We emphasize that this diagram is \emph{not} a Feynman diagram; it is a separate concept.
\begin{figure}
\centering
\includegraphics[scale=.4]{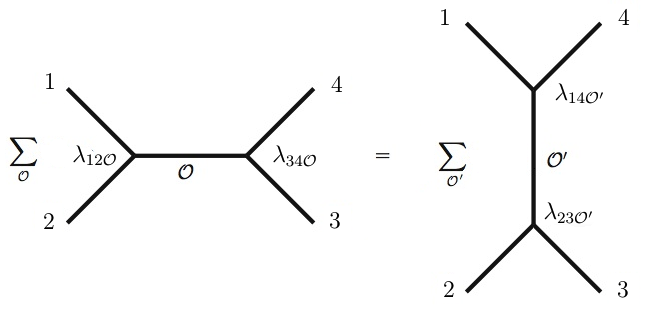}
\caption{A diagrammatic representation of the OPE associativity of the four-point correlation function of four fields. Connected lines do \emph{not} a Feynman diagram make.} \label{figure:cpw2}
\end{figure}

\begin{figure}
\centering
\includegraphics[scale=.5]{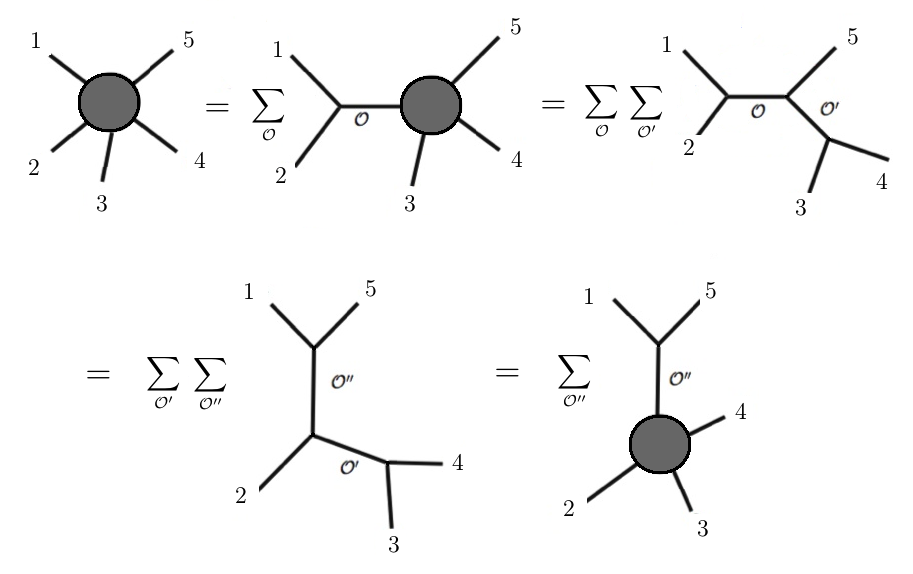}
\caption{A diagrammatic explanation of why crossing symmetry of five-point correlation functions does not give new constraints (adapted from \cite{rychkov}). This particular example uses OPEs in the (12) and (15) channels. (1) The first equality comes from performing the (12) OPE. (2) The second equality comes from expressing the remaining four-point functions using OPE expansions. (3) The third equality comes from using the four-point function crossing symmetry constraint. (4) The final equality is simplifying the sum over $\mathcal{O}'$ operators into a four-point correlation function. Thus we get an equality between expansions in the (12) channel and (15) channels.} \label{figure:cpw3}
\end{figure}

Now we realize a powerful fact: we just as easily could have chosen to compute the same four-point correlation function by choosing a sphere enclosing $\phi_1$ and $\phi_4$. We mean that we could have chosen a different OPE ``channel'', calculating the OPEs (14)(23) instead of (12)(34). This would give a different conformal partial wave expansion, but the end result should be the same. It must be the same. This leads to a non-trivial consistency relation, diagrammatically expressed in Figure (\ref{figure:cpw3}). This condition is called the conformal bootstrap condition, or \emph{OPE associativity}, or \emph{crossing symmetry} (also this final name belongs to an unrelated concept in field theory and we will try to avoid it in order to avoid unncessary confusion).

Before continuing, we claim that considering the conformal bootstrap condition for four-point functions is sufficient for our purposes. By imposing OPE associativity on all four-point functions, no new constraints appear at higher $n$-point functions. This can be seen diagrammatically in Figure (\ref{figure:cpw3}).

\subsection{Conformal bootstrap in $d=2$ dimensions}

Before considering the conformal bootstrap in higher dimensions, we will consider the conformal bootstrap in two-dimensional CFTs (first applied in \cite{3-7}). As previously discussed, the two-dimensional conformal algebra has an infinite-dimensional extension called the Virasoro algebra. The generators $L_{-1}, L_0,$ and $L_1$ (and the corresponding antiholomorphic generators) correspond to the finite-dimensional subalgebra of global conformal transformations. The generators $L_2, L_3, \cdots$ correspond to extra raising operators and the generators $L_{-2}, L_{-3}, \cdots$ correspond to extra lowering oeprators. By this, we mean the generator $L_n$ raises the scaling dimension by $n$ units (with corresponding statements for $\bar{L}_n$ and the lowering operators). A Virasoro primary field then satisfies
$$
L_{-n}|\Delta\rangle = 0, \forall n\geq 1.
$$

We additionally found strong conditions related to unitarity of CFTs in $d=2$ dimensions depending on the central charge $c$. For $c\geq 1$, the unitarity conditions are more or less the same as in higher dimensions. But for $0 < c < 1$, requiring unitarity is quite restrictive. Only a discrete sequence of values for $c$ is allowed
$$
c=1-\frac{6}{m(m+1)}, \,\, \mbox{ with } \, m=3,4,\cdots.
$$
Moreover, we found that only a finite discrete set of operator dimensions is allowed to appear
$$
\Delta_{r,s} = \frac{(r+m(r-s))^2-1}{2m(m-1)}, \,\, \mbox{with} \, 1 \leq s \leq r \leq m-1\,\, \mbox{are integers}.
$$
The conformal bootstrap approach is perfect for this problem; we have finitely many primaries and we know all of the operator dimensions. The OPE associativity equations then reduce to a problem of finite-dimensional linear algebra.

The simplest minimal model has $c=1/2$ and corresponds to the two-dimensional Ising model at the critical temperature. The Virasoro primary field content includes the identity operator/vacuum $\mathbf{1}$, the spin $\sigma$ (which is $\mathbb{Z}_2$ odd), and the energy density $\epsilon$ (which is $\mathbb{Z}_2$ even). These fields have dimensions $\Delta_{\mathbf{1}}=0, \Delta_\sigma = \frac18, \Delta_\epsilon = 1$. The nontrivial OPEs are
\begin{gather}
\sigma\times\sigma = \mathbf{1} + \lambda_{\sigma\sigma\epsilon}\epsilon \\
\epsilon\times\epsilon = \mathbf{1}+\lambda_{\epsilon\epsilon\epsilon}\epsilon \\
\sigma\times\epsilon = \lambda_{\sigma\sigma\epsilon}\sigma.
\end{gather}
Here, $\lambda_{\sigma\sigma\epsilon}$ is determined by solving the bootstrap equation, while $\lambda_{\epsilon\epsilon\epsilon}$ is due to the Kramers-Wannier duality. We leave the detailed computations as an additional exercise, but already we can see how simplied the case is for this class of 2d CFTs. For $c\geq 1$, the conformal bootstrap becomes difficult to solve even in two dimensions. There are notable exceptions\footnote{One notable example is the Liouville theory.}, but in general conformal bootstrap techniques will be similar from two to higher dimensions. It is thus to $d\geq 3$ we now turn our attention.

\subsection{Conformal bootstrap in $d\geq 3$ dimensions}

We have argued that we can express the operator product expansion in terms of conformal partial waves. If we think for a moment, we realize that each conformal partial wave will have the same transformation properties under the conformal group as the four-point function itself. With this in mind,  we can rewrite eq. (\ref{eq:bigblock}) for four fields with the same scaling dimension as
\begin{equation}
\langle \phi_1\phi_2\phi_3\phi_4 \rangle = \frac{g(u,v)}{x_{12}^{2\Delta_\phi} x_{34}^{2\Delta_\phi}},
\end{equation}
the variables $u$ and $v$ are the anharmonic ratios previously defined 
$$
u = \frac{x_{12}^2 x_{34}^2}{x_{13}^2 x_{24}^2}, \;\;\;\;\;\;\;\; v = \frac{x_{14}^2 x_{23}^2}{x_{13}^2 x_{24}^2}.
$$
The \emph{conformal block} $g(u,v)$ is the interesting part of the conformal partial wave. In the case of four identical fields $\phi$, we can express
\begin{equation}
g(u,v) = 1 +  \sum_{\mathcal{O}} \lambda_{\mathcal{O}}^2 {G_{\mathcal{O}}(u,v)}
\end{equation}
where we are slightly changing our notation (though in a way that should be straightforward to follow)\footnote{The first term is always the number one; refer to \cite{dolan1} for details.}.

Of course, OPE associativity tells us that we could have expressed the four-point correlation function by calculating different OPEs. 
If we exchange $(2\leftrightarrow4)$, our expression for the four-point function becomes
\begin{equation}
\langle \phi(x_1)\phi(x_2)\phi(x_3)\phi(x_4) \rangle = \frac{g(u', v')}{x_{14}^{2\Delta_\phi} x_{23}^{2\Delta_\phi}}.
\end{equation} 
The variables $u'$ and $v'$ are the conformally invariant cross sections calculated with exchanged indices. For $(2\leftrightarrow4)$, this means
$$
u' = v, \;\;\;\;\;\; v' = u.
$$
Notice that the function $g$ is the same for both of these expressions; this is because the four-point correlation function is totally symmetric under permutations. OPE associativity then tells us
\begin{equation}
\frac{g(u,v)}{x_{12}^{2\Delta_\phi} x_{34}^{2\Delta_\phi}} = \frac{g(v,u)}{x_{14}^{2\Delta_\phi} x_{23}^{2\Delta_\phi}}.
\end{equation}
Multiplying through by $x_{14}^{2\Delta_\phi}x_{23}^{2\Delta_\phi}$, we find that the conformal blocks must satisfy the \emph{bootstrap equation}
\begin{equation}
\left( \frac{v}{u}  \right)^{\Delta_\phi}  g(u,v) = g(v,u) \label{eq:bootstrapeq}.
\end{equation}
For the case of identical fields, this equation further simplifies to
\begin{equation}
(v^{\Delta_\phi}-u^{\Delta_\phi})+\sum_\mathcal{O} \lambda_\mathcal{O}^2 \left[ v^{\Delta_\phi}G_\mathcal{O}(u,v) - u^{\Delta_\phi}G_\mathcal{O}(v,u) \right] =0.
\end{equation}
Sometimes this is written as a \emph{sum rule}
\begin{equation} 
1 = \sum_\mathcal{O} \lambda_\mathcal{O}^2 \left[ 
\frac{
v^{\Delta_\phi}G_\mathcal{O}(u,v) - u^{\Delta_\phi}G_\mathcal{O}(v,u) }
{u^{\Delta_\phi} - v^{\Delta_\phi}}.
\right] 
\end{equation}

This is quite a non-trivial equation; it is not satisfied term by term but only in the sum. We can already see hints, however, of how the bootstrap equation could further constrain CFT data. We are free to adjust the spectrum and $\lambda$'s. For what spectra can we find $\lambda's$ such that the crossing requirement is satisfied? Presumably we cannot do this for just any spectrum. In the next few sections, we will investigate this question using a variety of methods.

\subsection{An analytic example}

Before proceeding to detailed numerical computations, let us gain some intuition about conformal blocks. In order to do this, let us use a conformal transformation to map our coordinates to convenient values. We first map $x_4\rightarrow\infty$, and then shift $x_1\rightarrow 0$. A combination of a rotation and then a dilatation maps $x_3\rightarrow (1,0,\cdots, 0)$ \footnote{These transformations all leave the one $\infty$ point invariant.}. Finally, we rotate about the $x_1-x_3$ axis to put $x_2$ in the plane of the page. We will use the complex coordinate $z$ in this plane. Choosing this configuration, we find the cross-rations are given by
\begin{equation}
u = |z|^2, \;\;\;\; v=|1-z|^2.
\end{equation}
We will be interested in the neighborhood of the special point $z=1/2$ (corresponding to $u=v=1/4$) since this configuration treats the OPE channels symmetrically. 

Immediately, however, we will consider a different configuration (obtainable by some conformal transformation) \cite{rychkovrad}. This new configuration is given in the $z$-plane by
\begin{equation}
x_1 = \rho \equiv re^{i\alpha}, \;\;\; x_2 = -\rho, \;\;\;, x_3 = -1, \;\;\;, x_4 = 1.
\end{equation}
\begin{framed}
\noindent HOMEWORK: Find the correspondence between $\rho$ and $z$ by making $u$ and $v$ the same. What value of $\rho$ corresponds to $z=1/2$?
\end{framed}
\noindent Clearly this configuration puts the points $x_1$ and $x_2$ (or $x_3$ and $x_4$) symmetrically with respect to the origin.

\begin{figure}
\centering
\includegraphics[scale=.5]{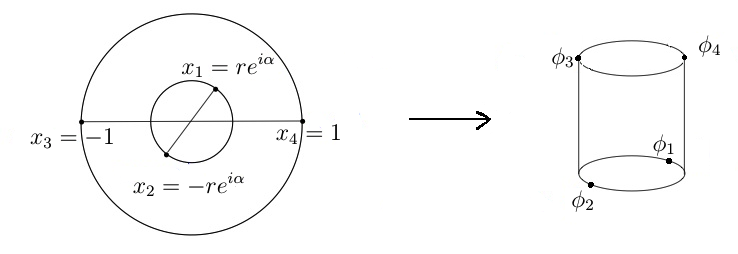}
\caption{The configuration described in the text and the configuration on $S^{d-1}\times\mathbb{R}$ obtained by a Weyl transformation (adapted from \cite{rychkov}). The pairs $\phi_{3,4}$ and $\phi_{1,2}$ are in antipodal positions on the spheres at cylindrical time 0 and $\log r$. Their positions on the respective spheres are rotated from one another by angle $\alpha$.} \label{figure:cbweyl}
\end{figure}

As we have seen in previous lectures, we can use a Weyl transformation to map flat space to the cylinder (Figure \ref{figure:cbweyl}). 
We can compute the conformal block on the cylinder using the expression
\begin{equation}
C.B. = \sum \langle 0 | \phi_1\phi_2| n \rangle e^{-E_n\tau}\langle n | \phi_3 \phi_4 | 0 \rangle,
\end{equation}
where the sum is over all the descendants of $|\Delta,\ell\rangle, E_n = \Delta+n,$ and $\tau = -\log r$ is the cylinder time interval over which we propagate exchanged states.

Referring to the diagrams, we realize the product of the matrix elements depends only on $\alpha$. We conclude that the conformal blocks must have the form
\begin{equation}
C.B. = \sum_{n=0}^\infty A_n(\alpha) r^{\Delta+n}.
\end{equation}
The coefficients $A_n$ are completely fixed by conformal symmetry; their exact values can be found. We will not do that now. We will instead argue the leading coefficient $A_0$ on physical grounds. The states $\phi_1\phi_2|0\rangle$ and $\phi_3\phi_4|0\rangle$ differ by a rotation of angle $\alpha$. Thus $A_0(\alpha)$ measures how the matrix elements with a spin $\ell$ state change under rotation by an angle $\alpha$. 

Let us parametrize the state on the cylinder by the unit vector $\boldsymbol{n}$ pointing to where $\phi_1$ is inserted on the sphere. The state $|\Delta,\ell\rangle$ has internal indices $|\Delta,\ell\rangle_{\mu_1,\mu_2,\cdots}$ that form a symmetric traceless spin $\ell$ tensor. Then the individual matrix elements are
\begin{equation}
\langle 0 | \phi_1\phi_2 |\Delta,\ell \rangle_{\mu_1,\mu_2} \propto (\boldsymbol{n}^{\mu_1}_1\cdots \boldsymbol{n}^{\mu_\ell} - \mbox{traces})
\end{equation}
since there is only the one traceless and symmetric spin $\ell$ tensor constructible from a single vector $\boldsymbol{n}_1$. Then up to some normalization the leading coefficient will be
\begin{equation}
A_0(\alpha) = (\boldsymbol{n}^{\mu_1}_1\cdots \boldsymbol{n}^{\mu_\ell}_1 - \mbox{traces} )(\boldsymbol{n}^{\mu_1}_2\cdots \boldsymbol{n}^{\mu_\ell}_2 - \mbox{traces} ) = \mathcal{P}(\boldsymbol{n}_1\cdot \boldsymbol{n}_2) = \mathcal{P}(\cos \alpha).
\end{equation}

Here $\mathcal{P}$ is a polynomial whose coefficients can only depend on  the spin $\ell$ and the number of dimensions $d$. For $d=2$, symmetric traceless tensors mean that $A_0$ is of the form
$$
A_0(\alpha) = (n^z_1 n^{\bar{z}}_2)^\ell + c.c = \cos(\ell \alpha).
$$
For $d=3$, the answer is the Legendre polynomials
$$
A_0(\alpha) = P_\ell(\cos \alpha).
$$
For $d=4$, the answer is related to the Chebyshev polynomials
$$
A_0(\alpha) = \frac{1}{\ell+1}\frac{\sin((\ell+1)\alpha)}{\sin \alpha}.
$$
In general $d$, the answer is related to the Gegenbauer polynomials
$$
A_0(\alpha) = C^{(d/2-1)}_\ell(\cos \alpha).
$$
The appearance of Gegenbauer polynomials is not surprising, as they arise in a similar situation in the theory of angular momentum in quantum mechanics. When two spinless particles scatter through a spin-$\ell$ resonance, it is known that the amplitude is given by the Legendre polynomial of the scattering angle.

Using our correspondences between $u,v,z$, and $\rho$, we can express the structure of $G_\mathcal{O}$ as
\begin{equation}
G_\mathcal{O}(u,v) = C_\ell(\cos \alpha) r^\Delta \left[1+\mathcal{O}(r^2)  \right].
\end{equation}
Since the bootstrap equations must be satisifed for any $u$ and $v$, we can consider the points having $0<z<1$ real so that $\rho$ is real. Then
\begin{equation}
\left[(1-z)^{2\Delta_\phi}-z^{2\Delta_\phi} \right] + \sum_\mathcal{O} \lambda_\mathcal{O}^2 \left[ 
(1-z)^{2\Delta_\phi}\rho(z)^{\Delta}-z^{2\Delta_\phi}\rho(1-z)^\Delta
  \right] =0.
\end{equation}
Is this equation with the approximate conformal blocks even valid? We can trust it near $z=1/2$ where both $\rho(z),\rho(1-z)\sim 0.17$. The omitted terms are then suppressed by approximately 0.0289. When we Taylor expand near $z=1/2$, the first term gives
\begin{equation}
\left[(1-z)^{2\Delta_\phi}-z^{2\Delta_\phi} \right] \sim -C_{\Delta_\phi}\left(x+\frac43(\Delta_\phi-1)(2\Delta_\phi-1)x^3 + \mathcal{O}(x^5)
\right),
\end{equation}
with $x=z-1/2$ and $C_{\Delta_\phi}>0$ a positive constant.

We will now consider the case where all operators have $\Delta \gg \Delta_\phi$ and show that this is inconsistent. In this limit, the conformal block terms go as
\begin{equation}
\rho(z)^{\Delta} - \rho(1-z)^\Delta \sim B_\Delta \left(
x+\frac43\Delta^2 x^3 +\cdots
\right),
\end{equation}
where $B_\Delta>0$ is another positive constant and we neglect the $z^{2\Delta_\phi}$ factors since $\Delta \gg \Delta_\phi$. We can normalize away this positive constant by swallowing it into the $\lambda_\mathcal{O}^2$ constant. By requiring the bootstrap equation be satisfied term by term around $z=1/2$, we get the first two conditions as
\begin{gather}
-C_{\Delta_\phi} + \sum_\mathcal{O} \lambda_\mathcal{O}^2 = 0 \\
-\frac43 C_{\Delta_\phi} ({\Delta_\phi}-1)(2{\Delta_\phi}-1) + \frac43 \sum_\mathcal{O} \lambda_\mathcal{O}^2 \Delta^2 = 0.
\end{gather}
It trivially follows that
\begin{equation}
\Delta_{min}^2 C_{\Delta_\phi} = \Delta_{min}^2 \sum_\mathcal{O} \lambda_\mathcal{O}^2 \leq \sum_\mathcal{O} \lambda_\mathcal{O}^2 \Delta^2 = ({\Delta_\phi}-1)(2{\Delta_\phi}-1)C_{\Delta_\phi}.
\end{equation}
We therefore conclude
\begin{equation}
\Delta_{min}\leq \sqrt{({\Delta_\phi}-1)(2{\Delta_\phi}-1)}\sim \mathcal{O}({\Delta_\phi}).
\end{equation}
This is a contradiction; we assumed $\Delta_{min}\gg {\Delta_\phi}$. Thus we arrive at our first conclusion:
\begin{equation}
\Delta_{min.}\leq f({\Delta_\phi}).
\end{equation}

\subsection{Numerical bootstrapping}

Of course, the previous result is only the simplest conclusion we can draw from the conformal bootstrap program. There are obviously many ways to improve this analysis: we could use more exact expressions for the conformal blocks; we could consider values of $z$ off of the real line (allowing us to distinguish between operators of different spins); we could expand to higher order in $x$. Depending on the particular model we may be interested in studying, there is also the possibility that we will have additional information to help constrain our problem: the presence of supersymmetry relates OPE coefficients of components of SUSY multiples in some correlators, fixes dimensions of protected operators, and imposes stronger unitarity bounds in terms of $R$-charge\footnote{If this terminology is completely alien to you, then you are reading an earlier version of this course without superconformal field theory. Check back in a few months/years.}; global symmetries (e.g., the $O(N)$ vector model) can provide additional input into our bootstrapping program; similarly, considering things like $\mathbb{Z}_2$ symmetric models allow us to constrain properties of $\mathbb{Z}_2$-even and -odd operators.

First: what expressions should we actually be using for conformal blocks? In [nb DO], the authors found recursion relations for the conformal blocks and in specific cases even solved for their exact explicit form. For example, in $d=4$ dimensions, we can write the conformal block \cite{dolan1} appearing in the bootstrap equation for four identical scalars in a very symmetrical form via
\begin{gather}
G_{\Delta,\ell}(u,v) = F_{h\bar{h}}(z,\bar{z}),\;\;\;\;\Delta=h+\bar{h}, \ell = h-\bar{h}\in \mathbb{Z},\\
F_{h\bar{h}}(z,\bar{z}) = \frac{1}{\ell+1}\frac{z\bar{z}}{z-\bar{z}}\left[k_{2h}(z)k_{2\bar{h}-2}(\bar{z}) - (z\leftrightarrow\bar{z})  \right],\\
k_\beta(z)= z^{\beta/2}{}_2F_1\left(\beta/2, \beta/2; \beta, z \right).
\end{gather}
The function ${}_2F_1$ is a hypergeometric function.

Given these explicit expressions (or working from the recursion relation in $d=3$ dimensions), we can recast the problem we are interest in solving into the form
\begin{equation}
\sum_{\Delta,\ell} p_{\Delta,\ell} F_{\Delta,\ell}(z,\bar{z}) = 0.
\end{equation}
By Taylor expanding around $z=\bar{z}=1/2$ and requiring each order to vanish, we see we are trying to solve the matrix equation
\begin{equation}
\begin{pmatrix}
  F^{(0,0)}_1 & F^{(0,0)}_2& F^{(0,0)}_3 & \xrightarrow{\Delta} \\
  F^{(2,0)}_1 & F^{(2,0)}_2& F^{(2,0)}_3 & \cdots \\
  F^{(0,2)}_1 & F^{(0,2)}_2& F^{(0,2)}_3 & \cdots \\
  \downarrow \partial  & \vdots  & \vdots & \ddots  
 \end{pmatrix}
 \begin{pmatrix}
  p_1  \\
  p_2 \\
  p_3 \\
  \vdots 
   \end{pmatrix}
 =
 \begin{pmatrix}
 0  \\
 0 \\
 0 \\
  \vdots 
   \end{pmatrix}
\end{equation}
The rows of this matrix are Taylor coefficients labeled by derivatives $\partial^m \bar{\partial}^n$\footnote{In truth, we need to consider every single order. In practice, we truncate at some large number of derivatives and argue that higher orders do not change the results toward which the bootstrap converges.}. The columns are operators $\mathcal{O}_k$ allowed in the spectrum\footnote{In truth, there should be a continuum of allowed values here. But that is not very easy to do on a computer, so in practive we discretize the conformal dimension and increment it over a small step size. The hope is that for a small enough step size we get a convergent result.}. Finally, we seek to solve this matrix equation subject to the constraint that $p_i \geq 0$ (which is just a statement about unitarity).

Thus at its heart, the numerical bootstrap program is like a linear programming problem. Several authors have developed a (free-to-use) modified simplex algorithm for semi-continuous variables \cite{cbsolver}. The general routine goes like this: consider a CFT living in $d=3$ dimensions. We want to study OPE associativity using a single scalar correlator $\langle \sigma\sigma\sigma\sigma \rangle$, where the OPE for this lowest-lying scalar $\sigma$ goes schematically as $\sigma \times \sigma \sim 1 + \epsilon+ \cdots$. First, we would suppose a trial spectrum. For example, we would fix $\Delta_\sigma=0.6$ and $\Delta_\epsilon=\Delta_1=1.8$. Our trial spectrum is thus that all $\Delta \geq \Delta_{unitarity}$ and $\Delta_{\ell=0}\geq \Delta_1$. If our linear programming techniques find a $p$ vector satisying the bootstrap equation for this CFT data, then we have learned nothing. This is an important point: we are never proving that a CFT exists. What we can prove, however, is that a CFT cannot exist having certain properties. For when our linear programming techniques find no such $p$ vector, then \emph{no} CFT exists with $\Delta_\epsilon \geq \Delta_1$. We have excluded this CFT due to its inconsistency with the bootstrap equation.That is precisely what happens with the trial spectrum we have stated. At this point, we select a new trial spectrum and begin again.

\begin{figure}
\centering
\includegraphics[scale=.5]{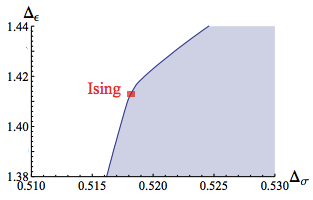}
\caption{This figure has been adapted from \cite{ising}. The shaded portion represents the spectra allowed by the OPE associativity constraint. The white region has been excluded as being inconsistent with the bootstrap equation. They have also marked a kink on the boundary that seems suspiciously close to the Ising model in $d=3$ dimensions.} \label{figure:3dising}
\end{figure}

This argument generalizes to conformal dimensions of higher-spin operators. And in the case that a bound is saturated, we can actually compute the full OPE. At this point, though, we will content ourselves with referring to Figure (\ref{figure:3dising}).
We remark that we have excluded several conformal field theories that were previously permitted from conformal invariance and unitarity alone. 

\begin{figure}
\centering
\includegraphics[scale=.5]{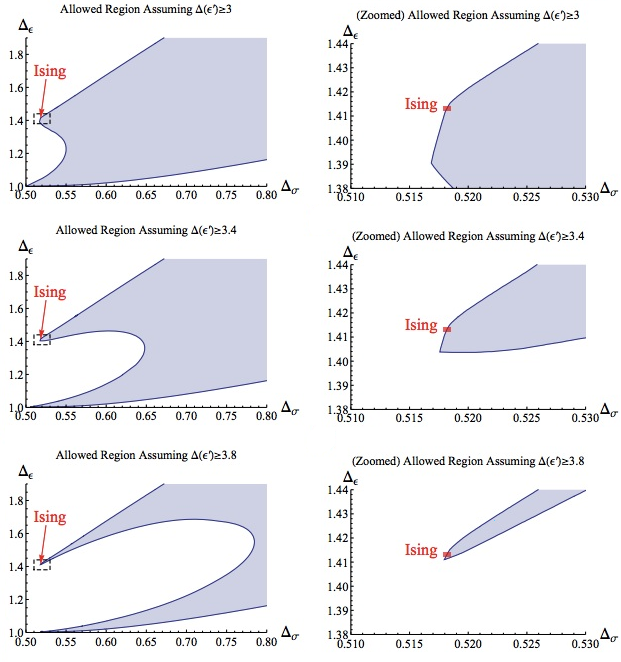}
\caption{This figure has been adapted from \cite{ising}. It shows excluded CFTs from imposing the extra constraints $\Delta_{\epsilon'} \geq \{3,3.4,3.8\}$. Best estimates from other methods give the value $\Delta_{\epsilon'}=3.832(6)$.} \label{figure:3disingmore}
\end{figure}

\begin{figure}
\centering
\includegraphics[scale=.5]{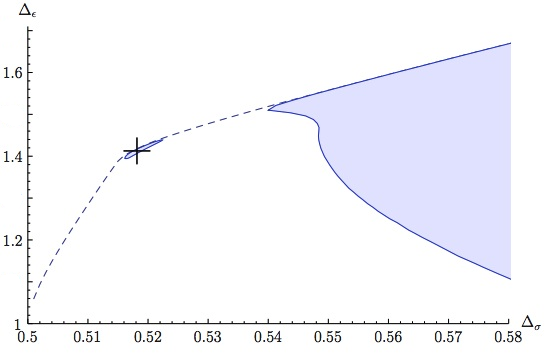}
\caption{Adapted from \cite{mixed}. Region of parameter space allowed in a $\mathbb{Z}_2$-symmetric CFT$_3$ with only one $\mathbb{Z}_2$-odd scalar. Unlike the previous plots, this plot uses multiple correlators to tightly constrain the 3d Ising model.} \label{figure:3disingmost}
\end{figure}

We also stress once again that we can not make any existence claims about CFTs in the shaded region. At best, we can try to find known CFTs in this parameter space and see what that tells us. For example, there is a kink in Figure (\ref{figure:3dising}) suspiciously close to the 3d Ising model. Using additional inputs from the Ising model, we can push farther with the bootstrap equation. In addition to scalars $\sigma$ and $\epsilon$, the 3d Ising model has a scalar $\epsilon'$. By considering trial spectra with restricted conformal dimensions for an additional scalar, we can exclude even more regions of our parameter space---and the kink just become more and more interesting (see Figure (\ref{figure:3disingmore}).
And it is beyond the scope of these lectures, but recent work [poly] has considered constraints coming from OPE associativity of multi-field correlators. By considering a $3d$ $\mathbb{Z}_2$-symmetric CFT having only one relevant $(\Delta<3) \mathbb{Z}_2$-odd scalar, the authors used unitarity and crossing symmmetry of the $\langle \sigma\sigma\sigma\sigma \rangle$, $\langle \sigma\sigma\epsilon\epsilon\rangle$, and $\langle\epsilon\epsilon\epsilon\epsilon\rangle$ correlators to arrive at Figure (\ref{figure:3disingmost}). This gap in the odd sector has created a small closed region around the point corresponding to the 3d Ising model. While work is still being done, it seems as though the conformal bootstrap method is truly solving the 3d Ising model.

\subsection{Future directions}

The methods we have presented so far have several obvious benefits. They allow us to make rigorous statements about the nonexistence of conformal field theories having various trial spectra, and the detailed analysis presented above gives us a great deal of control over the sources of error in our calculations. Yet there are also issues with the numerical bootstrap program. First, this type of analysis is computational intensive. To calculate Figure \ref{figure:3disingmost}, additional correlators (such as $\langle \sigma\epsilon\sigma\epsilon \rangle$) were added; this adds hours and hours of computation time. Furthermore, this method has no means on deciding which theory we can study. And if the theory we are attemping to study doesn't saturate an exclusion boundary, then we can't uniquely solve for the spectrum. Finally, the positivity required for all of these arguments means that numerical bootstrap methods only applie to unitary conformal field theories. White this covers many interesting CFTs, there are also important nonunitary CFTs we might wish to consider.

Recently Gliozzi \cite{gliozzi1,gliozzi2} proposed an alternative formulation of the conformal bootstrap. His method involves truncating the operator spectrum to some finite number of primary operators---like we would find for $c<1$ in $d=2$ dimensions. By doing this, he can look for approximate solutions to the crossing equation. We will follow the terminology of this reference for this discussion; it should be clear how the definitions differ.

Recall our sum rule from before; by truncating our CFT we are considering 
\begin{equation}
\sum_{\Delta,\ell=2n}^k p_{\Delta,\ell} \frac{v^{\Delta_\phi}G_{\Delta,\ell}(u,v)-u^{\Delta_\phi}G_{\Delta,\ell}(v,u)}{u^{\Delta_\phi}-v^{\Delta_\phi}}=1. \label{eq:gliozzisum}
\end{equation}
We before the change of variables
$$
u=z\bar{z}, \;\;\;v=(1-z)(1-\bar{z}); \;\;\;\;\;\;\;2z=a+\sqrt{b}, \;\;\;2\bar{z} = a-\sqrt{b}.
$$
Then we define
\begin{equation}
f^{(m,n)_{\Delta_\phi,\Delta,\ell}}=\left( \partial^m_a \partial^n_b  \frac{v^{\Delta_\phi}G_{\Delta,\ell}(u,v)-u^{\Delta_\phi}G_{\Delta,\ell}(v,u)}{u^{\Delta_\phi}-v^{\Delta_\phi}}   \right)  \bigg|_{a=1,b=0}.
\end{equation}
Then we can express eq. (\ref{eq:gliozzisum}) as
$$
\sum_{\Delta,\ell} f^{(0,0)_{\Delta_\phi,\Delta,\ell}} p_{\Delta,\ell} = 1
$$
$$
\sum_{\Delta,\ell} f^{(2m,n)_{\Delta_\phi,\Delta,\ell}} p_{\Delta,\ell} = 0,\;\;\;\;\;\; m+n\neq 0.
$$

Consider the homogeneous equations with $M$ different sets of derivates and a set of $N$ operators, such that $M\geq N$. Then we consider a finite system of equations:
\begin{equation}
\begin{pmatrix}
  f^{1}_{\Delta_\phi} & \cdots & f^{1}_{\Delta_\phi,N} \\
  \vdots &\ddots & \vdots \\
  f^{M}_{\Delta_\phi} & \cdots& f^{M}_{\Delta_\phi,N}
 \end{pmatrix}
 \begin{pmatrix}
  p_0  \\
  \vdots \\
  p_N  
   \end{pmatrix}
 =
 \begin{pmatrix}
 0  \\
 \vdots \\
 0  
   \end{pmatrix}
\end{equation}
 A system of $M$ linear homogeous equations with $N$ unknowns admits a non-vanishing solution if and only if all the minors of order $N$ are vanishing. Thus the probem has been translated into the search of the zeros of a system of non-linear equations. By plotting curves representing the location of zeros of minors, the mutual intersections should accumulate around the expected exact value. Once we have a solution, we can calculate the squared three-point correlator constants $p_i$; they are just the values of the eigenvector.
  
We will not pursue this topic further currently; even an understanding at this level is enough to see the advantages and disadvantages of this alternate method. Firstly, these computations are simple enough that they can be done on a laptop\footnote{Very much like the one from which you are mostly likely reading these notes.}. Additional correlators also doesn't overly complicate the aforementioned techniques. Furthermore, this method allows us to calculate the spectrum and OPE for theories even when they do not saturate the full bootstrap equations. This includes the values of $\lambda^2$, regardless of their sign. That means that this method will also hold for nonunitarity theories. 

Yet this technique is clearly not without its problems. Firstly, this method is not exactly systematic. When reading the original resources, some of the steps they take seem almost as much art as they are science. These techniques also offer very little control over the error that arises from approximations. We skipped over some of the details, but another issue with this method is that it requires some input from other methods---like the conformal dimension of some additional operator in the CFT spectrum. The CFT must also be truncable so that the system of equations can be found that has a solution. Nevertheless, this promising method is worth additional study.

While the numerical conformal bootstrap program is incredibly powerful and will only continue to give new and exciting results, there is also something to be said for analytic results. Recently, analytic bootstrap methods have been used to study the four point function of four identical scalar operators. It was shown that there must exist towers of operators at large spins with twists $2\Delta_\phi+2n$, where $\Delta_\phi$ is the dimension of the scalar and $n\geq \mathbb{Z}^+$. When there is a single tower of such operators and there is a twist gap between these and any other operator, one can calculate the anomalous dimensions \cite{sen1,sen2}. Large spin simplifies conformal blocks and thus the conformal bootstrap equation. Refer to \cite{oldspin,bigspin,otherspin,kaplan,spin3,spin4,spin5,sen1,sen2,spin6} for an introduction to this exciting new direction.

\subsection*{References for this lecture}
\vspace{4mm}
\noindent Main references for this lecture
\\
\begin{list}{}{%
\setlength{\topsep}{0pt}%
\setlength{\leftmargin}{0.7cm}%
\setlength{\listparindent}{-0.7cm}%
\setlength{\itemindent}{-0.7cm}%
\setlength{\parsep}{\parskip}%
}%
\item[]

[1] S. Rychkov, \emph{EPFL Lectures on Conformal Field Theory in $D\geq 3$ Dimensions: Lecture 4: Conformal Bootstrap}, (Lausanne, Switzerland, \'Ecole polytechnique f\'ed\'erale de Lausanne, December 2012).

[2] D. Pappadopulo, S. Rychkov, J. Espin, and R. Rattazzi, \emph{OPE Convergence in Conformal Field Theory}, Phys. Rev. D 86.10 (2012): 105043. [arxiv:1208.6449 [hep-th]].

[3] M. Hogervorst and S. Rychkov, \emph{Radial Coordinates for Conformal Blocks}, Phys. Rev. D 87.10 (2013): 106004. [arxiv:1303.1111 [hep-th]].

[4] S. El-Showk, M. F. Paulos, D. Poland, S. Rychkov, D. Simmons-Duffins and A. Vichi, Phys.
Rev. D 86(2012) 025022 [arXiv: 1203.6064 [hep-th]].

[5]  F. Gliozzi, \emph{More constraining conformal bootstrap}, Phys. Rev. Lett. 111, 161602 (2013)[arxiv:1307:3111 [hep-th]].

[6] F. Gliozzi and A. Rago, JHEP 1410 (2014) 42 [arXiv: 1403.6003[hep-th]].

\end{list}

\break

\section{Lecture 8: Misc.}

The original lecture was cancelled due to lack of time and boundaries. Specifically, the boundaries dividing lectures were not strictly enforced: we therefore ran out of time. So this lecture was devoted to finishing up earlier topics, discussing exercises/results, and answering questions. 

We did present one topic that has not yet been covered: the \emph{modular bootstrap}. I did not prepare many lecture notes for this topic, instead working directly from the primary sources. Really, we were completely out of time. I have chosen to put a few remarks here, though the reader is referred to \cite{hell,mine,mine3,fried,mine2,hell2,hart,mine4} for details.

\subsection{Modular Bootstrap}

As already mentioned, in $d=2$ dimensions we could allow our theory to live on an arbirary Riemann surface with some number of handles; we could calculate $n$-point correlation functions by giving this surface $n$ punctions. The OPE associativity we have been studying in this lecture corresponds to a symmetry relating the exchange of these punctures. In a similar/complementary way, we could investigate constraints coming from a symmetry relating the exchange of cycles corresponding to these handles. 

In Lecture 5 we considered precisely this type of symmetry by investigating CFTs living on the torus. We found that the partition function of such a CFT must be modular invariant; specifically, it must be invariant under the modular $S$-transformation (relating to the exchange of our space and time directions, or, the cycles of the single handle for this Riemann surface). We found that the partition function
\begin{equation}
Z(\tau) = \mbox{Tr} \left(   q^{L_0 - c/24} \bar{q}^{\bar{L}_0 - \bar{c}/24} \right). \label{eq:ztorus}
\end{equation}
must be invariant under $\tau\rightarrow-1/\tau$. In Lecture 6, we showed how this invariance lead to the Cardy formula (\ref{eq:cardy}). Cardy's formula alone does not make a statement about a CFT spectrum that can be tested at finite energies or temperatures; we considered only leading terms and the formula only applies for $h\gg c, \bar{h} \gg \bar{c}$. 
Cardy's formula can be used, however, to show that the partition function and all its derivatives converge and are continuous in the upper half plane. We can use this fact to study fixed points of the partition function, such as the modular $S$-transformation fixed point $\tau=i$. 

We can parameterize the neighborhood of this fixed point conveniently using $\tau\equiv i \mbox{ exp}(s)$. Then invariance of the partition function $Z(\tau,\bar{\tau})$ under the modular $S$-transformation $\tau \rightarrow -\frac{1}{\tau}$ can be expressed as 
\begin{equation}
Z\left(i e^s, - i e^{\bar{s}}\right) =Z(i e^{-s},-i e^{-\bar{s}})
\end{equation}
By taking derivatives of this expression with respect to $s, \bar{s}$, one obtains an infinite set of equations
\begin{equation}\left(\tau\frac{\partial}{\partial \tau}\right)^{N_{L}}  \left(\bar{\tau}\frac{\partial}{\partial \bar{\tau}}\right)^{N_{R}} Z(\tau,\tilde{\tau})\bigg|_{\tau=i}=0, \;\;N_{L} + N_{R} \text{ odd} \end{equation}
For purely imaginary complex structure $\tau = i\beta/2\pi$, this condition implies 
\begin{equation}\left(\beta\frac{\partial}{\partial \beta}\right)^{N}   Z(\beta)\bigg|_{\beta=2\pi}=0, \;\;N \text{ odd} \label{eq:twothreeish} \end{equation}

We will assume a unique vacuum and a discrete spectrum. By further assuming cluster decomposition and no chiral operators other than the stress-energy tensor\footnote{This latter assumption is a little restrictive, but it ultimately just serves to simplify the calculation. The restriction is removed in \cite{mine3}.}, the Virasoro structure theorem implies that the partition function  $Z(\beta)$ can be expressed as a sum over conformal families:
\begin{equation}
Z(\beta)=Z_{id}(\beta) + \sum_A Z_{A}(\beta).
\end{equation}
Here $Z_{id}(\beta)$ is the sum over states in the conformal family of the identity; $Z_{A}(\beta)$ is the sum over all states in the conformal family of the $A^{th}$ primary operator, which has conformal weights $h_A, \tilde{h}_A$ and conformal dimension $\Delta_A =  h_A +\tilde{h}_A $. 

For CFTs with $c,\tilde{c}>1$ (since theories with smaller central charge in two dimensions are classified), we have that:
\begin{equation}
Z_{id}(\tau)=q^{-\frac{c}{24}}\bar{q}^{-\frac{\tilde{c}}{24}}\prod_{m=2}^{\infty}{ (1-{q}^{m})^{-1}}\prod_{n=2}^{\infty}{(1-\bar{q}^{n})^{-1}}
\end{equation}
\begin{equation}
Z_{A}(\tau)=q^{h_A-\frac{c}{24}}\bar{q}^{\bar{h}_A-\frac{\tilde{c}}{24}}\prod_{m=1}^{\infty}{ (1-{q}^{m})^{-1}}\prod_{n=1}^{\infty}{(1-\bar{q}^{n})^{-1}}
\end{equation}
where $q=\exp(2\pi i \tau)$.
The full partition function with $\tau = i\beta/2\pi$ is then given by the expression
\begin{gather}
Z(\beta)= M(\beta)Y(\beta)+B(\beta) \label{eq:twofiveish}, \\
M(\beta)\equiv \frac{\exp(-\beta\hat{E}_0)}{\eta(i\beta/2\pi)^2}, \\
B(\beta)\equiv M(\beta) \left(  1-\exp(-\beta) \right)^{2},
\end{gather}
where $\hat{E}_0 \equiv E_0 + \frac{1}{12} = \frac{1}{12}-\frac{c+\tilde{c}}{24}$ and $\eta$ is the Dedekind eta function. For real $\beta$, the partition function over primaries $Y(\beta)$ is
\begin{equation}
Y(\beta)=\sum_{A=1}^{\infty}e^{-\beta \Delta_{A}}.
\end{equation}
\begin{framed}
\noindent HOMEWORK: Work through these steps to find this form of the partition function.
\end{framed}
By applying the differential constraints (\ref{eq:twothreeish}) to the partition function (\ref{eq:twofiveish}). To simplify the analysis, we introduce polynomials $f_p(z)$ defined by
\begin{equation}
(\beta \partial_\beta)^{p}M(\beta)Y(\beta)\bigg|_{\beta=2\pi} = (-1)^{p} \eta(i)^{-2}\text{exp}(-2\pi \hat{E}_0) \sum_{A=1}^{\infty}\text{exp}(-2\pi \Delta_A)f_p(\Delta_A+\hat{E}_0).
\end{equation}
The polynomials relevant to us are
\begin{equation} f_1(z)=(2\pi z)-\frac12 \label{eq:fs} \end{equation}
$$f_3(z)=(2\pi z)^3-\frac{9}{2}\left(2\pi z\right)^2+\left(\frac{41}{8}+6r_{20}\right)(2\pi z)-\left( \frac{17}{16}+3r_{20}\right),$$
where 
\begin{equation*}
r_{20} \equiv \frac{\eta''(i)}{\eta(i)}\approx0.0120...
\end{equation*}
We also define the polynomials $b_p(z)$ by
\begin{equation}
(\beta \partial_\beta)^{p}B(\beta)\bigg|_{\beta=2\pi} = (-1)^{p} \eta(i)^{-2}\text{exp}(-2\pi \hat{E}_0)b_p(\hat{E}_0),
\end{equation}
Explicitly,
\begin{equation}
b_p(z)= f_p(z) - 2e^{-2\pi} f_p(z+1) + e^{-4\pi}f_p(z+2) .
\end{equation}
Using these polynomials, the equations (\ref{eq:twothreeish}) for modular invariance of $Z(\beta)$ for odd $p$ become
\begin{equation}
\sum_{A=1}^{\infty}f_p(\Delta_A+\hat{E}_0)\text{exp}(-2\pi\Delta_A)=-b_p(\hat{E}_0)  \label{eq:modeq}
\end{equation}
 
It is this expression that is used to derive an upper bound on the conformal dimension $\Delta_1$. In \cite{hell}, Hellerman takes the ratio of the $p=3$ and $p=1$ expressions to get
\begin{equation}
\frac{\sum_{A=1}^{\infty}f_3(\Delta_A+\hat{E}_0)\text{exp}(-2\pi\Delta_A)}{\sum_{B=1}^{\infty}f_1(\Delta_B+\hat{E}_0)\text{exp}(-2\pi\Delta_B)}     =    \frac{b_3(\hat{E}_0)}{b_1(\hat{E}_0)}\equiv F_1.
\end{equation}
Or, upon rearrangement,
\begin{equation}
\frac{\sum_{A=1}^{\infty}   \left[ f_3(\Delta_A+\hat{E}_0)  - F_1(\hat{E}_0) f_1(\Delta_A+\hat{E}_0)    \right]        \text{exp}(-2\pi\Delta_A)}{\sum_{B=1}^{\infty}f_1(\Delta_B+\hat{E}_0)\text{exp}(-2\pi\Delta_B)}     =   0. \label{eq:hellratioone}
\end{equation}

Next assume that $\Delta_1 > \Delta_1^+$, where $\Delta_1^+$ is defined as the largest root of the numerator, and proceeds to obtain a contradiction. Because $\Delta_A \geq \Delta_1$, this assumption implies that every term in both the numerator and denominator is strictly positive. Then equation (\ref{eq:hellratioone})  says that a positive number equals zero --- an impossibility. Therefore
$$
\Delta_1 \leq \Delta_1^+.
$$ 
Finally, by analyzing $\Delta_1^+$ as a function of $c_{\rm tot}$ Hellerman proves that for the given assumptions, $\Delta_1^+ \leq \frac{c_{\rm tot}}{12}+   \frac{(12-\pi) + (13\pi -12)e^{-2\pi}}{6\pi(1-e^{-2\pi})}$, implying the bound
\begin{equation}
 \Delta_1 \leq \frac{c_{\rm tot}}{12}+ 0.4736...
\end{equation}
\begin{framed}
\noindent HOMEWORK: Make sure you understand the preceding argument.
\end{framed}

\subsection{More modular bootstrapping}

Can we use modular bootstrapping to learn more about conformal field theories? Building from these techniques, the work \cite{fried} applied the next several higher-order differential constraints following from $S$-invariance. The work \cite{mine2} considered additional invariance of the partition function under $ST$-transformation in CFTs with only even spin primary operators. In \cite{hell2}, the authors use modular bootstrapping results with some additional assumptions to give bounds on the entropy and an upper bound the number of marginal operators in some theories. In \cite{hart}, the authors used modular bootstrapping results and assumed a sparse light spectrum to derive upper bounds on the number of operators. In \cite{mine4}, the assumption of a light spectrum was removed, and several earlier results were checked for a large class of CFTs\footnote{It is more difficult than you might think to generate CFTs in $d=2$ dimensions with small central charge and/or large conformal dimensions.}. The only extension I will discuss today\footnote{because it was easy to modify what I had already presented on the board.} was found in \cite{mine} and involves larger conformal dimensions and a lower bound on the number of primary operators.

We can extend these methods to derive bounds on primary operators of second and third-lowest dimension. In order to bound the conformal dimension $\Delta_2 (\Delta_3)$, we move the $\Delta_1$(and $\Delta_2$) term(s) of equation (\ref{eq:modeq}) to the RHS. We then form the ratio of the $p=3$ and $p=1$ equations to get (for the case of $\Delta_2$)
\begin{eqnarray}
\frac{\sum_{A=2}^{\infty}f_3(\Delta_A+\hat{E}_0)e^{-2\pi\Delta_A}}{\sum_{B=2}^{\infty}f_1(\Delta_B+\hat{E}_0)e^{-2\pi\Delta_B}} = \frac{f_3(\Delta_1+\hat{E}_0)e^{-2\pi\Delta_1}+b_3(\hat{E}_0)}{f_1(\Delta_1+\hat{E}_0) e^{-2\pi\Delta_1}+b_1(\hat{E}_0)} \equiv F_2(\Delta_1,c_{\rm tot}).  \label{eq:fdef}
\end{eqnarray}
Moving $F_2$ to the left side,  we get
\begin{equation}
\frac{\sum_{A=2}^{\infty} \left[f_3(\Delta_A+\hat{E}_0)-  f_1(\Delta_A+\hat{E}_0  ) F_2\right]\text{exp}(-2\pi\Delta_A)}{\sum_{B=2}^{\infty}f_1(\Delta_B+\hat{E}_0)\text{exp}(-2\pi\Delta_B)} =0  \label{eq:mine}
\end{equation}

Before proceeding, we make some definitions. Define $\Delta^+_{f_p}$ to be the largest root of $f_p(\Delta+\hat{E}_0)$ viewed as a polynomial in $\Delta.$ The bracketed expression in the numerator is a polynomial cubic in $\Delta_2$; we denote it by $P_2(\Delta_2)$, and define the largest root of $P_2$ to be $\Delta^+_2(c_{\rm tot},\Delta_1)$, where $\hat{E}_0$ dependence has been replaced by $c_{\rm tot}$. We now assume that $\Delta_2>\text{ max}(\Delta^+_{f_1}, \Delta^+_2)$ and work to obtain a contradiction.
\begin{framed}
\noindent HOMEWORK: Complete this proof by contradiction.
\end{framed}
\noindent We have thus derived a bound on the conformal dimension $\Delta_{2(3)}$:
\begin{equation}
\Delta_{2(3)}\leq \text{ max}(\Delta^+_{f_1},\Delta^+_{2(3)}). \label{eq:bbound}
\end{equation}
From the explicit form of $f_1(\Delta+\hat{E}_0$) in (2.12), we see that 
\begin{equation}
\Delta^+_{f_1} = \frac{c_{\rm tot}}{24}+\frac{(3-\pi)}{12\pi}. \label{eq:delfp}
\end{equation}
Knowing the explicit expression for $\Delta^+_{2(3)}$, we can find its least upper linear bound such that 
$$
\Delta^+_{2(3)} \leq \frac{c_{\rm tot}}{12} + \mbox{const}_{2(3)}.
$$
Doing this gives the bounds
\begin{gather}
\Delta_2 \leq \frac{c_{\rm tot}}{12} + 0.5338... \label{eq:chiraldelta2bound} \\
\Delta_3 \leq \frac{c_{\rm tot}}{12}+ 0.8795...
\end{gather}

There are issues if we try to extend the proof to larger conformal dimensions. Starting with $\Delta_4$, we can develop singularities that ruin this analysis. It can shown that requiring
\begin{equation}
\log{n} \lesssim \frac{\pi c_{\rm tot}}{12} + O(1) \label{eq:nbound},
\end{equation}
results in the bound
\begin{equation}
\Delta_n \leq \frac{c_{\rm tot}}{12} + O(1)\label{eq:deltanbound}.
\end{equation}
For large $c$, this $O(1)$ term never contributes to leading order. We can also invert this statement to give a lower bound. If we think for a moment, we realize that there must be \emph{at least} $n$ primary operators obeying the bound (\ref{eq:deltanbound}). There could be more. We thus know that the number $N$ of primary states with conformal dimension satisfying (\ref{eq:deltanbound}) the lower bound
\begin{equation}
\log{N} \gtrsim \frac{\pi c_{\rm tot}}{12} + O(1).
\end{equation}

There are many topics mentioned at the beginning of this lecture that we could discuss; we must omit these due to time. In principle, the most powerful constraints on CFTs in $d=2$ dimensions should come from combining earlier results. 
Crossing symmetry of four-point functions on the sphere (see Lecture 7) and modular invariance of the partition function and one-point functions on the torus (see this lecture) are necessary and sufficient to define a conformal field theory on \emph{all} Riemann surfaces \cite{moore}. This work will have to wait for future papers, however. It's time to look through some exercises.

\break

\subsection*{References for this lecture}
\vspace{4mm}
\noindent Main references for this lecture
\\
\begin{list}{}{%
\setlength{\topsep}{0pt}%
\setlength{\leftmargin}{0.7cm}%
\setlength{\listparindent}{-0.7cm}%
\setlength{\itemindent}{-0.7cm}%
\setlength{\parsep}{\parskip}%
}%
\item[]

[1] S. Hellerman, \emph{A universal inequality for CFT and quantum gravity}, JHEP 08 (2011) 130 [arXiv:0902.2790v2  [hep-th]].

[2] J. D. Qualls and A. D. Shapere, \emph{Bounds on Operator Dimensions in 2D conformal field theories}, JHEP 05 (2014) 091 [arXiv:1312.0038 [hep-th]].

\end{list}

\break

\section{Exercises}
\noindent Here we have collected additional exercises that are either (1) more labor-intensive/challenging and thus requiring more explanation/attention, or (2) outside of the immediate focus of our course. In addition to original exercises, we have modified some exercises from the references \cite{difran,bp,cardybook} or adapted work from the other sources cited.

\vspace{4mm}
\begin{list}{}{%
\setlength{\topsep}{0pt}%
\setlength{\leftmargin}{0.7cm}%
\setlength{\listparindent}{-0.7cm}%
\setlength{\itemindent}{-0.7cm}%
\setlength{\parsep}{\parskip}%
}%
\item[]

(1) {\bf One-dimensional Ising model renormalization group} \cite{cardybook}:  Consider the Ising model in $d=1$ dimension having Hamiltonian
$$
H=-J \sum_i s_i s_{i+m}-h\sum_i s_i.
$$
The coefficients $J,h$ are some coupling constants/parameters, and $s_i = \pm 1$. We can study the renormalization group for this system. $\;\;\;\;\;\;\;\;\;\;$
\linebreak (a) Perform a renormalization group transformation by summing over every other spin (one way to do this is by splitting the partition function into a sum over even and odd spin sites). $\;\;\;\;\;\;\;\;\;\;\;\;\;\;\;\;\;\;\;\;\;\;\;\;\;\;\;\;\;\;\;\;$
\linebreak(b) Investigate the renormalization group flows in the $(e^{-2J}, e^h)$ plane. $\;\;\;\;\;\;\;\;\;\;\;$
\linebreak

(2) {\bf One-dimensional three states Potts model } \cite{cardybook}:
(a) Consider the Hamiltonian for the one-dimensional three states Potts model
$$
H = -J \sum_i \delta_{t_i,t_{i+1}},
$$
\linebreak where the label $t_i=\{1,2,3\}$ is a spin at the $i^{\mbox{th}}$ site and $J$ is some coupling. 
\linebreak Perform the transformation from Exercise 1 and derive the flow equation. $\;\;\;\;\;\;\;\;$
\linebreak (b) Show that there are no non-trivial fixed points.     $\;\;\;\;\;\;\;\;\;\;\;\;\;\;\;\;\;\;\;$
\linebreak

(3) {\bf Special conformal transformation}: Prove that the action for massless $\phi^4$-theory in $d=4$ dimensions is invariant under infinitesimal special conformal transformations. Check if this is true for finite special conformal transformations. $\;\;\;\;\;\;\;\;\;\;\;\;\;\;$
\linebreak

(4) {\bf Masses in conformal field theory }: We have stressed that particles in our example CFTs are massless; after all, a mass scale would introduce a corresponding length scale and thus break conformal invariance. The details are actually more interesting that that. Consider the commutator of $D$ and $P$:
$$
[D,P^\mu] = iP^\mu.
$$
(a) Use this to calculate the quantity $e^{i\alpha D} P^2 e^{-i\alpha D}$, $\alpha\in\mathbb{R}^+$. $\;\;\;\;\;\;\;\;\;\;\;\;\;\;\;\;\;\;\;\;\;$
\linebreak (b) Relate the masses of states $|P\rangle$ and $e^{i\alpha D}|P\rangle$. What are the possible masses allowed in a conformal field theory having a state with mass $m^2>0$? $\;\;\;\;\;\;\;\;\;\;\;\;$
\linebreak

(5) { \bf Scale invariance in momentum space} \cite{difran}: In many situations, we prefer to work in momentum space rather than position space. Rather than working with correlation functions in position space, we therefore consider the Fourier transform
$$
\langle \phi_1(x_1)\cdots\phi_n(x_n) \rangle = 
\int \frac{d\boldsymbol{k}_1}{(2\pi)^d} \cdots \frac{d\boldsymbol{k}_{n-1}}{ (2\pi)^d} \Gamma(\boldsymbol{k}_1,\cdots,\boldsymbol{k}_n)
e^{i(\boldsymbol{k}_1\cdot x_1+\cdots +\boldsymbol{k}_n\cdot x_n)}.
$$
We also know $\sum_i \boldsymbol{k}_i = 0$ by translation invariance/momentum conservation.
\linebreak (a) Show that scale invariance implies
$$
\langle \phi_1(\boldsymbol{k}_1) \cdots \phi_n(\boldsymbol{k}_n)\rangle =\lambda^{(n-1)d - \Delta_1 -\cdots - \Delta_n} 
\Gamma(\lambda\boldsymbol{k}_1,\cdots,\lambda\boldsymbol{k}_n).
$$
\linebreak (b) Prove that the two-point function of a scale invariant theory is of the form
$$
\langle \phi_1(\boldsymbol{k}) \cdots \phi_2(-\boldsymbol{k})\rangle \sim \frac{1}{|k|^{2-\eta}}.
$$
\linebreak (c) Now consider the case of $d=2$ dimensions. Show that the two-point function in coordinate space must be
$$
G(r) =\int_{1/L}^{\infty} \frac{dk}{k^{1-\eta}}J_0(kr),
$$
where $L^{-1}$ is some infrared cutoff\footnote{We mention short-distance divergences in $d=2$ dimensions in Lecture 4.}.  $\;\;\;\;\;\;\;\;\;\;\;\;\;\;\;\;\;\;\;\;\;\;\;$
\linebreak (d) We previously saw that conformal invariance fixes the form of the two-point correlator. Expain how this form is compatible with the result found in part (c).
\linebreak

(6) {\bf Nonrelativistic CFTs, Galilean group} \cite{ncft}: We will spend several exercises developing a formalism for a nonrelativistic analogue of relativistic conformal field theory. In nonrelativistic theories, we scale space and time differently:
$$
t\rightarrow \lambda^z t, \;\;\;\;\;\; x^i\rightarrow \lambda x^i.
$$
As such, the previous conformal algebra no longer holds. In fact, we do not even start with the Poincar\'{e} algebra. Instead, we split Lorentz rotations into spatial rotations and \emph{Galilean boosts}
$$
x\rightarrow x' = x-vt
$$
in order to consider the \emph{Galilean group}.
Derive the commutation relations for the Galilean algebra, with infiniesimal generators corresponding to the Hamiltonian $H$, momentum $P_i$, angular momentum $L_{ij}$, and boosts $G_i$. $\;\;\;\;\;\;\;\;\;\;\;\;\;\;\;\;\;\;$
\linebreak

(7) {\bf Nonrelativistic CFTs, Part 2} \cite{ncft}: We have found the Lie algebra of the Galilean group. As mentioned in lectures, we would like to promote projective representations of this group to unitary representations of the central extension of the group. We claim that the operator that does this is the mass operator $M$. To further investigate this algebra, we now turn our attention to a 
$d$-dimensional nonrelativistic theory (in units $\hbar=m=1$ described by some quantized field $\psi_\alpha(\bar{x})$ (with spin index $\alpha$). For now, we consider this type of field and thus consider $z=2$. This nonrelativistic field satisfies commutation or anticommutation relations
$$
[\psi_\alpha(\vec{x}),\psi_\beta^\dagger(\vec{y})]_\pm=\delta(\vec{x}-\vec{y})\delta_{\alpha\beta},
$$
depending on the spin of the field.
We can define the number and momentum densities as
$$
n(\vec{x}) = \psi^\dagger(\vec{x})\psi(\vec{x}), \;\;\;\;\;\;j_i(\vec{x}) = -\frac{i}{2} (\psi^\dagger(\vec{x})\partial_i\psi(\vec{x})-\partial_i\psi^\dagger(\vec{x})\psi(\vec{x})).
$$
(Notice that in our units, the number density is the same as the mass density.) Find all possible commutators between number and momentum densities. $\;\;\;\;\;\;\;\;\;\;\;\;\;\;\;\;\;\;\;\;\;\;$
\linebreak

(8) {\bf Nonrelativistic CFTs, Schrödinger algebra} \cite{ncft} : 
In this case, we have new symmetries in addition to the Galilean group transformations. Clearly we have invariance under dilatations
$$
t\rightarrow \lambda^2 t, \;\;\;\;\;\; x^i\rightarrow \lambda x^i
$$
generated by $D$. We also have invariance under the special transformations 
$$
t\rightarrow \frac{t}{1+bt}\;\;\;\;\;\;x^i\rightarrow \frac{x^i}{1+bt}
$$
generated by $K$. $\;\;\;\;\;\;\;\;\;\;\;\;\;\;\;\;\;\;\;\;\;\;\;\;\;\;\;\;\;\;\;\;\;\;\;\;\;\;\;\;\;\;\;\;\;\;$
\linebreak (a) Find the algebra for these generators (except ones involving $L$---feel free to omit these). One way to do this is to define the operators via
\begin{align*}
M &=\int d\vec{x} \;n(\vec{x}),  &  P_i &=\int d\vec{x} \;j_i(\vec{x}), & L_{ij} &=\int d\vec{x} \;( x_i j_j(\vec{x})-x_j j_i(\vec{x})), \\
G_i &=\int d\vec{x} \; x_i n(\vec{x}), & K &=\int d\vec{x} \;\frac{x^2}{2}n(\vec{x}), & D &=\int d\vec{x} \;x_i j_i(\vec{x}),
\end{align*}
as well as the Hamiltonian $H$.
This is the \emph{Schrödinger algebra}. $\;\;\;\;\;\;\;\;\;\;\;\;\;\;$
\linebreak (b) Verify that $M$ commutes with the entire algebra.  $\;\;\;\;\;\;\;\;\;\;\;\;\;\;\;\;\;\;\;\;\;\;\;\;$
\linebreak (c) Verify that  $H$, $K$, and $D$ again form an $SL(2,\mathbb{R})$ subalgebra, as expected from discussions in lecture. $\;\;\;\;\;\;\;\;\;\;\;\;\;\;\;\;\;\;\;\;\;\;\;\;\;\;\;\;\;\;\;\;\;\;\;\;\;\;$
\linebreak

(9) {\bf Nonrelativistic CFTs, Part 4} \cite{ncft}: As an aside, we briefly consider arbitrary $z$. We lose invariance under special transformations, but we still have invariance under Galilean transformations and dilatations. $\;\;\;\;\;\;\;\;\;\;\;\;\;\;\;\;$
\linebreak (a) Find the algebra for arbitrary $z$. Specifically, find the commutators involving $D$ (since they are the only ones that will change). In order to find the commutator of $D$ with $M$, use the Jacobi identity of $P$, $G$, and $D$. Notice that for general $z$, $M$ is no longer in the center. We will henceforth only consider $z=2$, such that $M$ is a good quantum number.
The Schrödinger group for $d$-dimensions can be embedded into the relativstic conformal group in $d+1$-dimensions. This is related to the fact that one can arrive at the Schrödinger equation from the massless Klein-Gordon equation through Kaluza-Klein compactification\footnote{The Schrödinger mass $M$ is the inverse of the compactification radius.}.$\;\;\;\;\;\;\;\;\;\;\;\;\;\;\;\;\;\;\;\;\;\;\;\;\;\;\;\;\;\;$
\linebreak (b) Find how to perform this embedding for $z=2$. Look back at the conformal algebra in higher dimensions for assistance. $\;\;\;\;\;\;\;\;\;\;\;\;\;\;\;\;\;\;\;\;\;\;\;\;$
\linebreak In a similar fashion to relativistic CFT, we say a local operator has scaling dimension $\Delta$ and mass $m$ if
$$
[D,\mathcal{O}(0)]=i\Delta \mathcal{O}(0), \;\;\;\;\;\; [M,\mathcal{O}(0)]=m \mathcal{O}(0).
$$
\linebreak (c) Use the algebra to show now there are four operators that we can use to construct states with larger or smaller $\Delta$.
Assuming the dimensions of operators are bounded below (as we proved was the case for unitary theories), we can again define a \emph{primary operator}. Now, a state is primary iff
$$
[G_i,\mathcal{O}(0)] = [C,\mathcal{O}(0)] = 0.
$$
Obviously we could push this theory much farther, but we leave additional exploration to the reader. $\;\;\;\;\;\;\;\;\;\;\;\;\;\;\;\;\;\;\;\;\;\;\;\;\;\;\;\;\;\;\;\;\;\;\;\;\;\;\;\;\;\;$
\linebreak

(10) {\bf Proof of Noether's Theorem} \cite{difran}: In this exercise, we ask you to derive the form of the conserved current used in the text. Consider an action functional 
$$
S = \int d^dx\;\mathcal{L}(\phi,\partial_\mu \phi).
$$
We will study the effect of a transformatiion $x\rightarrow x', \phi(x)\rightarrow \phi'(x') = F(\phi(x)).$ The change in the action functional is obtained by substituting $\phi'(x)$ for $\phi(x)$.
\\
(a) Show that this action can be expressed in the form
$$
S' = \int d^dx\; \left| \frac{\partial x'}{\partial x} \right| \mathcal{L} (F(\phi(x)), (\partial x^\nu / \partial x'^\mu)\partial_\nu F(\phi(x))).
$$
\\
(b) Consider some infinitesimal transformation, the effects of which are given by equation (\ref{eq:inftrans}). To first order in our small parameters, find the inverse Jacobian and the determinant of the Jacobian matrix.
\\
(c) Substitute these expressions into $S'$. The variation in the action, $\delta S = S'-S$, contains terms with no derivatives of the infinitesimal parameter. These will sum to zero for rigid transformations. By expanding the Lagrangian, we find that this variation in the action will depend only on terms going as a derivative of the parameter. Explicitly perform this expansion to find
$$
\delta S = - \int d^dx\; j^\mu \partial_\mu \epsilon_a.
$$
Show that this $j^\mu$ must be the expression quoted in the text. Integrating this variation by parts and demanding that the variation of the action vanishes, we have therefore proven that the current $j^\mu$ is conserved. $\;\;\;\;\;\;\;\;\;\;\;\;\;\;\;\;\;\;\;\;\;\;\;\;\;\;\;\;\;\;\;\;\;\;$
\linebreak

(11) {\bf Conserved currents for $d=2$ free fermion}: Consider the Lagrangian for a free fermion in two dimensions
$$
\mathcal{L} = \frac{i}{2} \Psi^\dagger \gamma^0\gamma^\mu\partial_\mu\Psi.
$$
Find the following quantities: $\;\;\;\;\;\;\;\;\;\;\;\;\;\;\;\;\;\;\;\;\;\;\;\;\;\;\;\;\;\;$
\linebreak
(a) The form of the spin generator $S_{\mu\nu}$ that ensures Lorentz invariance;  $\;\;\;\;\;$
\linebreak
(b) The canonical energy-momentum tensor; $\;\;\;\;\;\;\;\;\;\;\;\;\;\;\;\;\;\;\;\;\;\;\;\;\;\;\;\;\;\;$
\linebreak
(c) The Belinfante stress-energy tensor; $\;\;\;\;\;\;\;\;\;\;\;\;\;\;\;\;\;\;\;\;\;\;\;\;\;\;\;\;\;\;$
\linebreak
(d) The dilatation current; check that it is conserved. $\;\;\;\;\;\;\;\;\;\;\;\;\;\;\;\;\;\;\;\;\;\;\;\;\;\;\;\;\;\;$
\linebreak

(12) {\bf More examples of traceless stress-energy tensors}: These are some straightforward computations.
\linebreak
(a) Find the modification that will give a traceless stress-energy tensor for the free massless scalar field in $d>2$ dimensions. $\;\;\;\;\;\;\;\;\;\;\;\;\;\;\;\;\;\;\;\;\;\;\;\;\;\;\;\;\;\;$
\linebreak
(b) Find the modification that will give a traceless stress-energy tensor for massless $\phi^4$ theory in $d=4$ dimensions. $\;\;\;\;\;\;\;\;\;\;\;\;\;\;\;\;\;\;\;\;\;\;\;\;\;\;\;\;\;\;\;\;\;\;\;\;$
\linebreak

(13) {\bf Liouville field theory} \cite{difran}: Consider Liouville field theory in $d=2$ dimensions with Lagrangian density
$$
\mathcal{L} = \frac12 \partial_\mu\phi\partial^\mu\phi-\frac12m^2e^\phi.
$$
Find the canonical stress-energy tensor and add a term that makes it traceless while preserving its conservation laws. $\;\;\;\;\;\;\;\;\;\;\;\;\;\;\;\;\;\;\;\;\;\;\;\;\;\;\;\;\;\;$
\linebreak

(14) {\bf Finite transformation of the stress-energy tensor}: Recall the effect of an infinitesimal conformal transformation on the 2d stress-energy tensor
$$
\delta_\epsilon T(z) = \frac{c}{12}\partial^3_z \epsilon(z) + 2 T(z) \partial_z \epsilon(z) + \epsilon(z) \partial_z T(z).
$$
We claimed that under a finite conformal transformation, $T$ transforms as
$$
T(z)\rightarrow T'(z) = \left( \frac{\partial f}{\partial z}\right)^2  T(f(z))+\frac{c}{12} \frac{1}{(\partial_z f)^2} \left( (\partial_z f)(\partial_z^3 f)-\frac{3}{2} (\partial^2_z f)^2 \right).
$$
Rather than ``integrating up'' the infinitesimal transformation (which is a sensible way to proceed; we simply have not introduced the necessary tools here), we will argue via an alternate method that this statement makes sense. $\;\;\;\;\;\;\;\;\;\;\;\;$
\linebreak (a) Verify to first order in $\epsilon$ that this finite transformation reproduces the infinitesimal transformation.
\linebreak During the previous calculation, it became obvious that there are potentially many finite transformations that would reproduce the infinitesimal transformation. What else is required to show that this is truly the finite transformation rule? We require one additional property. $\;\;\;\;\;\;\;\;\;\;\;\;\;\;\;\;\;\;\;\;\;\;\;\;\;\;\;\;\;\;\;\;\;\;\;\;\;\;$
\linebreak (b) Verify the composition rule for finite conformal transformations: the result of two successive transformations $z\rightarrow w\rightarrow u$ should coincide with what is obtained from the single transformation $z\rightarrow u$. It can be shown that the Schwarzian derivative is the only possible addition to the tensor transformation law satisfying this group property that also vanishes for global conformal transformations (you already proved that the Schwarzian derivative of a global conformal map vanishes, as it must; $T(z)$ is a quasi-primary field). This must therefore be the transformation rule for the stress-energy under finite conformal transformations. $\;\;\;\;\;\;\;\;\;\;\;\;\;\;$
\linebreak

(15) {\bf Cluster property of the four-point function} \cite{difran}: Consider the expression for a generic four-point function. Assuming all scaling dimensions are positive, show that you recover a product of two-point functions when the four points are paired off in such a way that the two points in each part are much closer to one another than the distance between the pairs. $\;\;\;\;\;\;\;\;\;\;\;\;\;\;\;\;\;\;\;\;\;\;\;\;\;\;\;\;\;\;$
\linebreak

(16) {\bf Four-point function for the free boson} \cite{difran}: Calculate the four-point function $\langle \partial\phi\partial\phi\partial\phi\partial\phi \rangle$ using Wick's theorem. Compare it to the general expression and find $f(u,v)$.
\linebreak

(17) {\bf Bosonization, Part 1} \cite{bp}: Consider a system with two real chiral fermions in $d=2$ dimensions that we combine into a complex chiral fermion
$$
\psi(z) = \frac{1}{\sqrt{2}}\left(\psi_1(z)+i\psi_2(z)  \right).
$$
\linebreak (a) Expanding in a Laurent series, find the algebra satisfied by the modes $\psi_r, \psi^*_s$ using contour integral methods as we dicussed in lectures for stress-energy tensor modes.
\linebreak (b) Consider now the field $j(z) \equiv \,\,: \psi(z)\psi*(z):\,\, = -i:\psi_1(z)\psi_2(z):\,$. Verify the equality.
\linebreak (c) Now expand $j(z)$ as a Laurent series and find an expression for mode $j_m$ in terms of $\psi^{(a)}_r$ modes. $\;\;\;\;\;\;\;\;\;\;\;\;\;\;\;\;\;\;\;\;\;\;\;\;\;\;\;\;\;\;\;\;\;\;\;\;\;\;\;\;\;\;$
\linebreak

(18) {\bf Bosonization, Part 2} \cite{bp}: Now we can calculate interesting things.
\linebreak (a) Calculate the commutator $[L_m, j_n]$. No tricks, just calculations. $\;\;\;\;\;\;\;\;\;\;\;\;$
\linebreak (b) Find the commutator $[j_m, j_n]$. Because fermions are complicated, we cannot naively shift the summation index; we must also be wary of operator normal ordering. You will find that this current satisfies the $U(1)$ current algebra.
\linebreak (c) Determine the $U(1)$ charge of the complex fermion by calculating $[j_m, \psi_s]$.
This algebra is exactly the algebra realized by a free boson $\phi(z,\bar{z})$ compactified on a circle of radius $R=1$.$\;\;\;\;\;\;\;\;\;\;\;\;\;\;\;\;\;\;\;\;\;\;\;\;\;\;\;\;\;\;\;\;\;\;\;\;\;\;\;\;\;\;\;\;\;\;\;\;\;$ 
\linebreak

(19) {\bf The modular group $PSL(2,\mathbb{Z})$, Part 1} \cite{difran}: The aim of this exercise is to show that the $S-$ and $T$-transformations generate the modular group. This is a lengthy process.
\linebreak (a) Show that for two positive integers $a>c>0$, there is a unique pair of integers $a_1,c_1$ such that
$$
a=a_1c+c_1,\;\;\;\;\;\; 0 \leq c_1 < c.
$$
\linebreak (b) We denote the greatest common divisor of positive integers $a,c$ by $\mbox{gcd}(a,c)$. Show that $\mbox{gcd}(a,c)=\mbox{gcd}(c,c_1)$. $\;\;\;\;\;\;\;\;\;\;\;\;\;\;\;\;\;\;\;\;\;\;\;\;\;\;\;\;\;\;$
\linebreak (c) Show that there exist two integers $a_0$ and $c_0$ such that
$$
c_0 a-a_0 c = \mbox{gcd}(a,c).
$$
Do this by repeating the process $c=a_2c_1+c_2$, etc. The sequence $c>c_1>\cdots\geq 0$ is strictly decreasing, so there exists a finite $k$ such that $c_k = \mbox{gcd}(a,c)$ and $c_{k+1}=0$. $\;\;\;\;\;\;\;\;\;\;\;\;\;\;\;\;\;\;\;\;\;\;\;\;\;\;\;\;\;\;$
\linebreak (d) Deduce that the integers $a,c$ are coprime iff there exist two integers $a_0, c_0$ such that $c_0 a - a_0 c = 1$. $\;\;\;\;\;\;\;\;\;\;\;\;\;\;\;\;\;\;\;\;\;\;\;\;\;\;\;\;\;\;$
\linebreak

(20) {\bf The modular group $PSL(2,\mathbb{Z})$, Part 2} \cite{difran}: We turn to the modular group. 
\linebreak (a) Prove that any product of $S$'s and $T$'s is an element of $PSL(2,\mathbb{Z})$.
\linebreak (b) Argue\footnote{If the indices and arguments become overwhelming, refer to part (g).} for a generic element $x=(a\tau+b)/(c\tau+d)$ of $PSL(2,\mathbb{Z})$ that $a$ and $c$ are coprime.
\linebreak (c) For $a>c$, show that there exists an integer $\rho_0$ such that
$$
\frac{a\tau + b}{c\tau + d} = \rho_0 + \frac{a_1 \tau + b_1}{c_1 \tau + d_1},
$$
with $c_1 = c, d_1=d,$ and $0\leq a_1 < c$. $\;\;\;\;\;\;\;\;\;\;\;\;\;\;\;\;\;\;\;\;\;\;\;\;\;\;\;\;\;\;$
\linebreak (d) Case i: If $a_1=0$, show that one can take $-b_1=c_1=1$ and write $x$ as a composition of $S$- and $T$-transformations. $\;\;\;\;\;\;\;\;\;\;\;\;\;\;\;\;\;\;\;\;\;\;\;\;\;\;\;\;\;\;$
\linebreak (e) Case ii: If $a_1>0$, we can write
$$
\frac{a\tau + b}{c\tau + d} = \rho_0 - 1 \bigg/ \left(\frac{-c_1 \tau - d_1}{a_1 \tau + b_1}\right)
$$
and repeat the above procedure to get
$$
\frac{a\tau + b}{c\tau + d} = \rho_0 - 1 \bigg/ \left(\rho_1 + \frac{a_2 \tau + b_2}{c_2 \tau + d_2}\right),
$$
where $c_2=a_1, d_2=b_1,$ and $0\leq a_2 < a_1$. $\;\;\;\;\;\;\;\;\;\;\;\;\;\;\;\;\;\;\;\;\;\;\;\;\;\;\;\;\;\;$
\linebreak (f) Repeating this division procedure leds to five sequences, $\rho_i, a_i, b_i, c_i, d_i$. Argue that there exists a finite integer $k$ such that $a_k=0, a_k-1\neq 0$. Show that one can take $-b_k = c_k = 1$, and conclude that $x$ can be written as some composition of $S$- and $T-$transformations. $\;\;\;\;\;\;\;\;\;\;\;\;\;\;\;\;\;\;\;\;\;\;\;\;\;\;\;\;\;\;$
\linebreak (g) If at any point this exercise becomes confusing, try to do this procedure for the specific case
$$
x=\begin{pmatrix}
8 & 5 \\
3 & 2
\end{pmatrix}
$$
and express it as a product of $S$ and $T$ transformations. $\;\;\;\;\;\;\;\;\;\;\;\;\;\;\;\;\;\;\;\;\;\;\;\;\;\;\;\;\;\;$
\linebreak

(21) {\bf Virasoro descendant inner product} \cite{difran}: Show that the norm of the state $(L_{-1})^n|h\rangle$ is
$$
2^n n! \prod_{i=1}^n(h-(i-1)/2).
$$
\linebreak

(22) {\bf Unitarity of $SU(2)$ representations} : As a refresher on highest weight representations, consider the $SU(2)$ Lie algebra in the form
$$
[J_+,J_-] =2J_0 ,\;\;\;\;\;\;[J_0,J_\pm]=\pm J_\pm.
$$
Let $|j\rangle$ be a highest weight state ($J_+|j\rangle = 0$) labeled by the $J_0$ eigenvalue $j$. Prove that unitary highest weight representations force us to consider either positive half-integer $j$ or non-negative integer $j$. $\;\;\;\;\;\;\;\;\;\;\;\;\;\;\;\;\;\;\;\;\;\;\;\;\;\;\;\;\;\;$
\linebreak

(23) {\bf Correlation function of descendant fields}: In this exercise, we will calculate some correlation functions of intermediate difficulty in two spacetime dimensions. Define all conformal dimensions and the central charge as necessary.
\linebreak (a) Calculate the two-point correlation function of two descendant fields $\langle \phi^{-2}_{i}(w_1)\phi^{-3}_{i}(w_2) \rangle$.
\linebreak (b) Calculate the three-point correlation function $\langle T(z) \phi_1(w_1) \phi_2(w_2) \rangle$.
\linebreak (c) Calculate the three-point correlation function involving descendant fields 
\linebreak $\langle  \phi^{-m}_i(w_i) \phi^{-n}_j(w_j) \phi_k(w_k) \rangle$.$\;\;\;\;\;\;\;\;\;\;\;\;\;\;\;\;\;\;\;\;\;\;\;\;\;\;\;\;\;\;$
\linebreak

(24) {\bf Conformal Casimir operator}: Calculate the conformal group quadratic Casimir operator $C_2 \equiv \frac12 J^{\mu\nu}J_{\mu\nu}.$  $\;\;\;\;\;\;\;\;\;\;\;\;\;\;\;\;\;\;\;\;\;\;\;\;\;\;\;\;\;\;$
\linebreak

(25) {\bf Stress-energy tensor improvement gymnastics} \cite{difran}: We claimed that adding a term of the form $\frac12 \partial_\lambda\partial_\rho X^{\lambda\rho\mu\nu}$ would give us an improved stress-energy tensor with desirable properties. Now we will investigate these claims. We said that this improvement will be possible when the virial $V^\mu$ is the divergence of another tensor $\sigma^{\alpha\mu}$. First, define the symmetric part of $\sigma$ as $\sigma_+^{\alpha\mu}$. Then $X$ can be defined as
$$
X^{\lambda\rho\mu\nu}=\frac{2}{d-2}\left[ \eta^{\lambda\rho}\sigma_+^{\mu\nu} - \eta^{\lambda\mu}\sigma_+^{\rho\nu} - \eta^{\lambda\mu}\sigma_+^{\nu\rho} + \eta^{\mu\nu}\sigma_+^{\lambda\rho} + \frac{1}{d-1}(\eta^{\lambda\rho}\eta^{\mu\nu} - \eta^{\lambda\mu}\eta^{\rho\nu})\sigma_{+\alpha}^\alpha\right].
$$
\linebreak (a) Show that $\partial_\mu\partial_\lambda\partial_\rho X^{\lambda\rho\mu\nu}=0$. $\;\;\;\;\;\;\;\;\;\;\;\;\;\;\;\;\;\;\;\;\;\;\;\;\;\;\;\;\;\;$
\linebreak (b) Show that the trace of this term is
$$
\frac12\partial_\lambda\partial_\rho X^{\lambda\rho\mu}_\mu = \partial_\lambda\partial_\rho \sigma_+^{\lambda\rho} = \partial_\mu V^\mu.
$$
\linebreak

(26) {\bf General statements about unitarity bounds} \cite{rychkov}: In lecture, we discussed deriving unitarity bounds for scalars and other particles. Here we make general statements about unitarity bounds. $\;\;\;\;\;\;\;\;\;\;\;\;\;\;\;\;\;\;\;\;\;\;\;\;\;\;\;\;\;\;$
\linebreak (a) Consider the matrix
$$
N_{\nu\{t\},\mu\{s\}} ={}_{\{t\}}\langle \Delta, l | K_\nu P_\mu | \Delta, l\rangle_{\{s\}}.
$$
Use proof by contradiction to show that this matrix must have only positive eigenvalues in a unitary theory. $\;\;\;\;\;\;\;\;\;\;\;\;\;\;\;\;\;\;\;\;\;\;\;\;\;\;\;\;\;\;$
\linebreak (b) Use the conformal algebra to show that these eigenvalues gets contributions proportional to $\Delta$ and contributions that are eigenvalues of the Hermitian matrix
$$
C_{\nu\{t\},\mu\{s\}} = \langle \{t\} | i M_{\mu\nu} | \{s\} \rangle.
$$
Therefore our unitarity condition becomes
$$
\Delta \geq \lambda_{{max}}(C), 
$$
where $\lambda_{{max}}(C)$ is the maximum eigenvalue of $C$. $\;\;\;\;\;\;\;\;\;\;\;\;\;\;\;\;\;\;\;\;\;\;\;\;\;\;\;\;\;\;$
\linebreak (c) We express the action of the operator $M_{\mu\nu}$ as
$$
-iM_{\mu\nu} = -\frac12 (V^{\alpha\beta})_{\mu\nu} (M_{\alpha\beta})_{\{s\},\{t\}},
$$
where the generator $V$ in the vector representation is given by
$$
(V^{\alpha\beta})_{\mu\nu} = i( \delta^{\alpha}_{\mu} \delta^{\beta}_{\nu} - \delta^{\alpha}_{\nu} \delta^{\beta}_{\mu} ).
$$
We can compare this problem with the problem in quantum mechanics of finding the eigenvalues of $L\cdot S$---both $S$ and $M$ act in the space of spin indices, and the coordinate space in which $L$ acts is replaced by a vector space in which $V$ acts. Express $L\cdot S$  as a combination of quadratic operators. $\;\;\;\;\;\;\;\;\;\;\;\;\;\;\;\;\;\;\;\;\;\;\;\;\;\;\;\;\;\;$
\linebreak (d) In the result from part (c), we find two of the operators are Casimirs with obvious eigenvalues, and the third is the Casimir of the tensor product representation $L \otimes S$ with eigenvalues $j(j+1)/2,\,\, j=|l-s|,\cdots,l+s$. Guided by this example from quantum mechanics, write down an expression for the maximal eigenvalue in terms of the Casimirs of spin representation $l$, vector representation $V$, and a tensor representation $R' \in V \otimes l$. The expression can be shown equal to $d-2+l$ for $l\geq 1$, giving us the unitarity bound given in lecture. $\;\;\;\;\;\;\;\;\;\;\;\;\;\;\;\;\;\;\;\;\;\;\;\;\;\;\;\;\;\;$
\linebreak

(27) {\bf Physics is not always phun }: Prove equation (\ref{eq:modehw}) using the steps suggested.
\linebreak

(28) {\bf More explicit calculations} \cite{bp}: (a) Using equation (\ref{eq:modehw}), prove the norm of the state $|\phi\rangle = \phi_{-h}|0\rangle$ is equal to the structure constant of the two-point function $d_{\phi\phi}$. \linebreak (b) Similarly, show that the three-point function of $\phi_1,\phi_2, \phi_3$ gives the constant $C_{123}$.
\linebreak

(29) {\bf Poisson resummation formula} \cite{bp}: In this exericse, we will derive this resummation formula and use it to prove invariance under the modular $S$-transformation of the partition function for a free boson compactified on a circle.
\linebreak (a) Use the discrete Fourier transform of the periodic function
$$
\sum_{n\in\mathbb{Z}}{\delta(x-n)} = \sum_{k\in \mathbb{Z}} e^{2\pi i k x}
$$
to prove the Poisson resummation formula (\ref{eq:poisson}). $\;\;\;\;\;\;\;\;\;\;\;\;\;\;\;\;\;\;\;\;\;\;\;\;\;\;\;\;\;\;$
\linebreak (b) Use this formula to show that the partition function is invariant under a modular $S$-transformation (you will need to use the formula twice). $\;\;\;\;\;\;\;\;\;\;\;\;\;\;\;\;\;\;\;\;\;\;\;\;\;\;\;\;\;\;$
\linebreak

(30) {\bf Jacobi triple product identity} \cite{bp}: in this exericse, we will derive this identity in the following way. From exercise (17), we have seen that algebra generated by two fermions in the NS sector with $j_0$ eigenvalues $\pm1$ is equivalent to the algebra generated by the currents $j(z), j^\pm(z)$. From discussions following our analysis of the free boson compactified on a circle, we know that in addition to these currents we have primary fields given by vertex operators
$$
V_{\pm N}(z) = :e^{\pm i N \phi}:, \;\;\;\;\mbox{with}\;\;\;\; (h,\alpha) = \left(\frac{N^2}{2},N\right).
$$
Here $h$ is the conformal weight and $\alpha$ is the $j_0$ charge of the vertex operator. 
\linebreak (a) Consider a charged character $\chi(\tau,z)$; that is, consider
$$
\chi(\tau,z) \equiv \mbox{Tr}_\mathcal{H} \left( q^{L_0-\frac{c}{24}}w^{j_0} \right), \;\;\;\; w = \exp(2\pi i z). 
$$
Write down the charged character for the two complex fermions $\Psi(z)$ and $\bar{\Psi}(z)$ (having the same charges as before). $\;\;\;\;\;\;\;\;\;\;\;\;\;\;\;\;\;\;\;\;\;\;\;\;\;\;\;\;\;\;$
\linebreak (b) We have already found the character for the primary field $j(z)$. Find the characters corresponding to the additional vertex operators (Hint: states in the Hilbert space are written as $|\alpha,n_1,n_2,\cdots\rangle = \lim_{z,\bar{z}\rightarrow0} j_{-1}^{n_1} j_{-2}^{n_2}\cdots V_\alpha(z,\bar{z})|0\rangle$.)
\linebreak (c) Find the sum of all characters for this bosonic theory. Due to bosonization, for $R=1$ the expressions we have found must be equal. We have therefore established the Jacobi triple product identity. $\;\;\;\;\;\;\;\;\;\;\;\;\;\;\;\;\;\;\;\;\;\;\;\;\;\;\;\;\;\;$
\linebreak

(31) {\bf Modular partition functions} $Z_{orb.}\bigg|_{R=\sqrt{2}} = Z_{circ.}\bigg|_{R=2\sqrt{2}}$ \cite{bp}: Explicitly show that the moduli spaces of these partition functions intersect at this point. This will give you plenty of practice with modular functions. $\;\;\;\;\;\;\;\;\;\;\;\;\;\;\;\;\;\;\;\;\;\;\;\;\;\;\;\;\;\;$
\linebreak

(32) {\bf Computing fusion coefficients} \cite{bp}: in this exercise, we compute some fusion coefficients using the Verlinde formula. $\;\;\;\;\;\;\;\;\;\;\;\;\;\;\;\;\;\;\;\;\;\;\;\;\;\;\;\;\;\;$
\linebreak (a) Consider the free boson compactified on a circle of radius $R=\sqrt{2k}$---the $\hat{u}(1)_k$ theory. Compute the fusion coefficients for this theory. $\;\;\;\;\;\;\;\;\;\;\;\;\;\;\;\;\;\;\;\;\;\;\;\;\;\;\;\;\;\;$
\linebreak (b) The partition function of $\hat{u}(1)_1$ is the same as for $\hat{su}(2)_1$; that is, the free boson at the self-dual radius has vertex operators that combine with the current $j(z)$ to give an $su(2)$ Ka\u{c}-Moody algebra. Write the fusion rules for the two highest weight representations of this algebra. $\;\;\;\;\;\;\;\;\;\;\;\;\;\;\;\;\;\;\;\;\;\;\;\;\;\;\;\;\;\;$
\linebreak

(33)  {\bf The modular $T$-matrix} : In this exercise, we expand on this idea.
\linebreak (a) Consider the $\hat{u}(1)_k$ theory. Do a series expansion of the character $\chi_m^{(k)}$ to show the highest weight state corresponding to this character has conformal dimension $h=m^2/4k$. Use this to check that the modular matrix $T_{ij}$ is of the form quoted in lecture.
\linebreak (b) Consider the free fermion theory. Using $\chi_0, \chi_{\frac12}, \chi_{\frac{1}{16}}$, compute the matrix $T_{ij}$ for this theory. $\;\;\;\;\;\;\;\;\;\;\;\;\;\;\;\;\;\;\;\;\;\;\;\;\;\;\;\;\;\;$
\linebreak
\begin{figure}
\centering
\includegraphics[scale=.5]{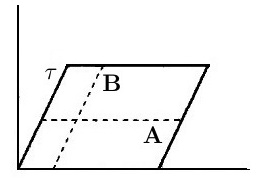}
\caption{Basis of homological one-cycles on the torus (based on \cite{bp})} \label{figure:toruscycles}
\end{figure}

(34) {\bf Proof/sketch of Verlinde formula} \cite{bp}: We will not give a full proof of the Verlinde formula. For further details, refer to the source references. The characters $\chi_j$ can be viewed as conformal blocks for the zero-point amplitude on the torus. The key idea is that this amplitude is identical to the one-point amplitude of the identity operator. This identity means that the character can also be written as a certain scaling limit of the conformal block of the two-point function $\langle \phi_i(z) \phi_i^*(z)\rangle_{\mathbb{T}^2}$ on the torus. We use $\mathcal{F}^{i,i*}(z-w)$ for the conformal block and $\phi_i^*$ denotes the conjugate operator of $\phi_i$. Then
$$
\chi_j \sim \lim_{z\rightarrow w} (z-w)^{2h_i} \mathcal{F}^{i,i*}_j(z-w).
$$
Next, one defines a monodromy operator $\Phi_i({\bf C})$ acting on the characters as
$$
\Phi_i({\bf C}) \chi_j = \lim_{z\rightarrow w} (z-w)^{2h_i} M_{\phi_i,{\bf C}}\left( \mathcal{F}^{i,i*}_j(z-w) \right).
$$
Here $M_{\phi_i,{\bf C}}$ is defined by taking $\phi_i$, moving it around the one-cycle ${\bf C}$ on the torus $\mathbb{T}^2$, and computing the effect of that monodromy on the conformal block. A basis of homological one-cycles is given by the fundamental cycles on $\mathbb{T}^2$, denoted by ${\bf A}$ and ${\bf B}$ (see Figure \ref{figure:toruscycles}). In our conventions, ${\bf A}$ is a space-like cycle with $0 \leq \mbox{Re} w \leq 2\pi$ and ${\bf B}$ is the time-like cycle in the $\tau$ direction. $\;\;\;\;\;\;\;\;\;\;\;\;\;\;\;\;\;\;\;\;\;\;\;\;\;\;\;\;\;\;$
\linebreak (a) Show that the modular $S$-transformation exchanges ${\bf A}$ and ${\bf B}$.
\linebreak Moving $\phi_i$ around the ${\bf A}$-cycle does not change the conformal family $\phi_j$ circulating along the time-like direction. Thus $\Phi_i({\bf A})$ acts diagonally on the characters
$$
\Phi_i({\bf A})\chi_j = \lambda_i^j \chi_j.
$$
The action $\Phi_i({\bf B})$ is more complicated. Without giving details, we use something called the \emph{pentagon identity}\footnote{Refer to \cite{bp} for details.} to get the result
$$
\Phi_i({\bf B})\chi_j = N^k_{ij}\chi_k.
$$
But because the $S$-transformation exchanges these cycles, we also know that $\Phi_i({\bf B}) = S\Phi_i({\bf A})S$. Then the $S$-transformation can be said to diagonalize the fusion rules and we can write
$$
N^k_{ij}=\sum_{m}S_{jm}\lambda_i^m\bar{S}_{mk}.
$$
\linebreak (b) Using this formula, derive the Verlinde formula. $\;\;\;\;\;\;\;\;\;\;\;\;\;\;\;\;\;\;\;\;\;\;\;\;\;\;\;\;\;\;$
\linebreak

(35) {\bf The trace anomaly, Part 1} \cite{difran,trace1,trace2}: In this exercise, we give an alternate derivation of the trace anomaly for a free boson. Consider the generating functional
$$
Z[g] = \int D\phi e^{-S[\phi,g]} = e^{-W[g]},
$$
where $S[\phi,g] = \int d^2 x \, \sqrt{g} g^{\mu\nu}\partial_\mu\phi\partial_\nu\phi =  - \int d^2x\, \sqrt{g} \phi \Delta\phi$. $\;\;\;\;\;\;\;\;\;\;\;\;\;\;\;\;\;\;\;\;\;\;\;\;\;\;\;\;\;\;$
\linebreak (a) Find the form of the Laplacian operator $\Delta$ acting on $\phi$.
Under a local scale transformation of the metric $\delta g_{\mu\nu} = \sigma(x)g_{\mu\nu}$, the action varies as
$$
\delta S[\phi,g] = -\frac12 \int d^2 \,x \sigma(x)T^\mu_\mu.
$$
\linebreak (b) Find the variation of the connected vacuum functional $\delta W[g]$. 
\linebreak
We intend to show that this variation no longer vanishes on an arbitrary manifold.
First, we define the functional measure $D\phi$ is a more convenient manner.  We introduce a complete set of orthonormal functions $\{ \phi_n \}$ such that
$$
\langle \phi_m, \phi_n \rangle = \int d^2x\, \sqrt{g}\phi_m^*\phi_n = \delta_{mn}.
$$
\linebreak (c) By expressing a general field configuration as $\phi=\sum c_n\phi_n$, find the line element $||\delta\phi||^2$.
This allows us to define the functional integration measure as 
$$
D\phi = \prod_n dc_n.
$$
The most convenient complete set is the set of normalized eigenfunctions of the Laplacian with eigenvalues $-\lambda_n$. $\;\;\;\;\;\;\;\;\;\;\;\;\;\;\;\;\;\;\;\;\;\;\;\;\;\;\;\;\;\;$
\linebreak (d) Find the action of a configuration specified by the expansion coefficients $c_n$. Using this, find the naive vacuum functional.
Of course, this is not completely accurate. We saw this problem when considering the free boson on the torus. The zero-mode $\phi_0$ has a vanishing eigenvalue that serves a source of divergence. To fix this issue, we compactify the field $\phi$ by identifying $\phi$ and $\phi + a$. Then the range of integration of $c_0$ is the segment $\{0,a\sqrt{A} \}$, where $A$ is the area of the manifold (this follows from the condition that $\langle \phi_0, \phi_0\rangle = A \phi_0^2 = 1$). Then the vacuum functional is replaced by
$$
Z[g] = a\sqrt{A}\prod_{n\neq 0}\sqrt{\frac{2\pi}{\lambda_n}}.
$$
\linebreak (e) Using $\mbox{Tr}^\prime$ to indicate a trace taken over nonzero modes, find the corresponding expression for the connected function $W[g]$. $\;\;\;\;\;\;\;\;\;\;\;\;\;\;\;\;\;\;\;\;\;\;\;\;\;\;\;\;\;\;$
\linebreak

(36) {\bf The trace anomaly, Part 2} \cite{difran,trace1,trace2}: Now we place some mathematical games. 
\linebreak (a) Using the representation
$$
\ln B = -\lim_{\epsilon\rightarrow 0} \int_\epsilon^\infty \frac{dt}{t} \left(e^{-Bt}-e^{-t}\right),
$$
show that
$$
W[g] = -\ln a -\frac12 \ln A - \frac12 \mbox{Tr'}\left[ \int_\epsilon^\infty \frac{dt}{t} \left( e^{t\Delta}-e^{-2\pi t} \right)  \right].
$$
For now we will keep $\epsilon$ finite and send it to zero at the end.
\linebreak (b) For the variation $\delta g_{\mu\nu} = \sigma g_{\mu\nu}$, find the variation in the second term. Using the fact that nonzero modes have negative eigenvalues, find the variation in the third term.
\linebreak (c) Show that we can combine the two variations into a single expression
$$
\delta W[g] = \frac12 \mbox{Tr}(\sigma e^{\epsilon \Delta}).
$$
Now to proceed, we introduce the \emph{heat kernel}
$$
G(x,y;t) = 
\begin{cases}
\langle x|e^{t\Delta}|y\rangle, & t\geq 0 \\
0, & t < 0.
\end{cases} 
$$
\linebreak (d) Write the variation in terms of this kernel. Assuming for the moment that
$$
G(x,x;\epsilon) = \frac{1}{4\pi \epsilon} + \frac{1}{24\pi} R(x) + O(\epsilon),
$$
find the variation of $W[g]$. $\;\;\;\;\;\;\;\;\;\;\;\;\;\;\;\;\;\;\;\;\;\;\;\;\;\;\;\;\;\;\;\;\;\;\;\;\;\;\;\;\;\;\;\;\;\;\;\;\;\;$
\linebreak

(37) {\bf The trace anomaly, Part 3} \cite{difran,trace1,trace2}: The first problem is that we have claimed a particular short-time behavior for the diagonal kernel without proof. In a later version of this course, we will prove this claim. For now, we refer the reader to the original reference. The second problem is that in the limit $\epsilon\rightarrow0$, the first term becomes infinite. This divergence results from the assumed finite size of the manifold. To fix it, we add can add a $\phi$-independent counterterm to the original action of the form 
$$
S_{ct}[g] = \alpha \int d^2 x \, \sqrt{g}.
$$
\linebreak (a) Find the variation of this term under the same local scale transformation. What value of $\alpha$ cancels the divergent term? (The other piece cannot be canceled with a counterterm. Can you see why a term of the form $\sim \int d^2 x\, \sqrt{g} R(x)$  cannot work?)
\linebreak (b) By comparing equations, one finds the trace anomaly. Note here that $c=1$ for the free boson and that this reference uses some different normalizations. What will the form of the trace anomaly be with this normalization?
To honestly relate the trace anomaly to the central charge, we must somehow introduce the $TT$ two-point function. We do this by using the conformal gauge, a coordinate system where
$$
g_{\mu\nu}=\delta_{\mu\nu}e^{2\phi(x)}
$$
\linebreak (c) Find $\sqrt{g}$ and $\sqrt{g}R$ for this metric tensor.
Since a local scale tranformations amounts to a local variation of the field $\phi$, the corresponding variation of the connected functional $W[g]$ is
$$
\delta W[g] = -\frac{C}{24\pi}\int d^2 x \, \partial^2 \phi \delta\phi.
$$
Here $C$ is just some constant, with $C=1$ for a free boson. $\;\;\;\;\;\;\;\;\;\;\;\;\;\;\;\;\;\;\;\;\;\;\;\;\;\;\;\;\;\;$
\linebreak (d) Find the expression for $W[g]$. $\;\;\;\;\;\;\;\;\;\;\;\;\;\;\;\;\;\;\;\;\;\;\;\;\;\;\;\;\;\;$
\linebreak (e) Using the defining properties of the Laplacian Green function $\partial_x^2 K(x,y) = \delta(x-y)$, write this expression in terms of the Green function $K(x,y)$ of the Laplacian. Keep in mind that the expression must be symmetric in $x$ and $y$.
\linebreak (f) This result can be extended to an arbitrary coordinate system. We pick up factors $\sqrt{g(x)}, \sqrt{g(y)}$ in the integrand, $\partial^2 \phi$ is replaced by $R$, and $K(x,y)$ now satisfies
$$
\sqrt{g(x)}\Delta_x K(x,y) = \delta(x-y).
$$
Using the fact that
$$
\langle T_{\mu\nu}(x) T_{\rho\sigma}(y) \rangle = \frac{\delta^2 W}{\delta g_{\mu\nu}(x) \delta g_{\rho\sigma}(y)},
$$
argue that the central charge and the coefficient $C$ are one and the same thing.
\linebreak

(38) {\bf  Lüscher term} \cite{luscher}: The Nambu-Goto action of string theory is given by
$$
S=\frac{1}{2\pi \alpha^\prime} \int d^2\sigma \sqrt{\mbox{det}\left( \frac{\partial X^\mu}{\partial \sigma^a} \frac{\partial X_\mu}{\partial \sigma^b}
 \right)}.
$$
Although inspired from hadronic physics/flux tubes, this theory is not an adequate fundamental theory of mesons. Nevertheless, we will view it as an effective theory of the QCD flux tube. Here we interpret $X^\mu(\sigma^1, \sigma^2)$ as coordinates of the worldsheet swept out by the line running between the quark and antiquark as it propagates through $d$-dimensional spacetime. $\;\;\;\;\;\;\;\;\;\;\;\;\;\;\;\;\;\;\;\;\;\;\;\;\;\;\;\;\;\;$
\linebreak (a) The first step is to introduce static quarks by demanding that the worldsheet is bounded by some rectangular loop with sides $R \times T$, $T\gg R$. Denote coordinates by $X^\mu = (X_0,X_1,\vec{X}_\perp)$, and use reparametrization of the Nambu-Goto action to set $\sigma^0=x_0,\sigma^1=x_1,$ with the $R \times T$ loop lying in the $x_0 - x_1$ plane at $\vec{X}_\perp=0$. Express the static potential
$$
e^{-V(R)T} = \int DX^\mu e^{-S}
$$
as an integral over the perpendicular directions, to fourth order in $\partial\vec{X}_\perp$.
\linebreak (b) Expand the action to second order in the transverse fluctuations and evaluate the integral. $\;\;\;\;\;\;\;\;\;\;\;\;\;\;\;\;\;\;\;\;\;\;\;\;\;\;\;\;\;\;$
\linebreak The result will be the determinant of a two-dimensional Laplacian operator subject to Dirichlet boundary conditions on the $R\times T$ loop. $\;\;\;\;\;\;\;\;\;\;\;\;\;\;\;\;\;\;\;\;\;\;\;\;\;\;\;\;\;\;$
\linebreak (c) Regulate and evaluate via standard zeta-function methods. The result for $T\gg R$ gives an exponential function. The argument of this function is precisely the Lüscher term discussed in lecture. $\;\;\;\;\;\;\;\;\;\;\;\;\;\;\;\;\;\;\;\;\;\;\;\;\;\;\;\;\;\;$
\linebreak

(39) {\bf Stress-energy tensor for non-abelian gauge theory}: Consider non-abelian gauge theory with fermion fields and ghost fields
$$
\mathcal{L} = \bar{\psi}(i\gamma^\mu D_\mu - m)\psi - \frac14 F^a_{\mu\nu}F^{a\,\mu\nu} + \frac{1}{2\xi}(\partial_\mu A^{a\,\mu})^2 - \bar{c}^a\partial_\mu (D^{ab\,\mu} )c^b.
$$
Argue for the form of $K_\mu$ satisfies the properties we require: (1) it must have the correct dimension; (2) it must be BRST invariant \cite{peskin}. Are there any other properties we expect for this current or its divergence? $\;\;\;\;\;\;\;\;\;\;\;\;\;\;\;\;\;\;\;\;\;\;\;\;\;\;\;\;\;\;$
\linebreak

(40) {\bf The theory of elasticity, Part 1} \cite{cardymodel}: This exercise focuses on confirming claims made about the theory of elasticity in two dimensions. Begin with the action in Cartesian coordinates.$\;\;\;\;\;\;\;\;\;\;\;\;\;\;\;\;\;\;\;\;\;\;\;\;\;\;\;\;\;\;$
\linebreak (a) Find the canonical stress-energy tensor by the Noether procedure. Check that it is traceless. $\;\;\;\;\;\;\;\;\;\;\;\;\;\;\;\;\;\;\;\;\;\;\;\;\;\;\;\;\;\;$
\linebreak (b) Calculate the correction $B^{\rho\mu\nu}$ to the stress-energy tensor. Check that the improved Belinfante tensor is symmetric and traceless. $\;\;\;\;\;\;\;\;\;\;\;\;\;\;\;\;\;\;\;\;\;\;\;\;\;\;\;\;\;\;$
\linebreak (c) Find the equations of motion for this theory. $\;\;\;\;\;\;\;\;\;\;\;\;\;\;\;\;\;\;\;\;\;\;\;\;\;\;\;\;\;\;$
\linebreak

(41) {\bf The theory of elasticity, Part 2} \cite{cardymodel}: Now consider the action in complex coordinates. 
\linebreak (a) Calculate the stress-energy tensor. Then find its trace. $\;\;\;\;\;\;\;\;\;\;\;\;\;\;\;\;\;\;\;\;\;\;\;\;\;\;\;\;\;\;$
\linebreak (b) Find the OPEs: $u(z)u(w), \bar{u}(z)\bar{u}(w), u(z)\bar{u}(w)$. $\;\;\;\;\;\;\;\;\;\;\;\;\;\;\;\;\;\;\;\;\;\;\;\;\;\;\;\;\;\;$
\linebreak (c) Use Wick's theorem to find the two-point correlator $\langle T^\mu_\mu T^\nu_\nu \rangle$. $\;\;\;\;\;\;\;\;\;\;\;\;\;\;\;\;\;\;\;\;\;\;\;\;\;\;\;\;\;\;$
\linebreak (d) Check the claims made in eqs. (\ref{eq:sixseventyone}) and (\ref{eq:sixseventytwo}). $\;\;\;\;\;\;\;\;\;\;\;\;\;\;\;\;\;\;\;\;\;\;\;\;\;\;\;\;\;\;$
\linebreak

(42) {\bf Scale invariance vs. conformal invariance, Part 1} \cite{3dsc1,3dsc2}: The \emph{free Maxwell theory} in $d\neq4$ dimensions gives a physical example of a unitary, scale invariant theory that is \emph{not} conformally invariant.  $\;\;\;\;\;\;\;\;\;\;\;\;\;\;\;\;\;\;\;\;\;\;\;\;\;\;\;\;\;\;$
\linebreak (a) Find the stress-energy tensor for this theory, and then take its trace. 
\linebreak (b) You will see that for $d\neq4$, the trace of the stress-energy tensor can be written as the trace of some virial current. We claim that it can not be written as the divergence of some other tensor $L_{\mu\nu}$. Show this by writing down the only possibility from dimensional grounds and showing that it can not generate the virial current.
\linebreak We can also see the issue in another way. In position space,
$$
\langle A_\mu(x) A_\nu(0) \rangle = \frac{\eta_{\mu\nu}}{|x|^{d-2}} + \mbox{gauge terms},
$$
where we omit writing the gauge-dependent terms. Instead we should consider the gauge invariant field $F_{\mu\nu}$ and its operator products.  $\;\;\;\;\;\;\;\;\;\;\;\;\;\;\;\;\;\;\;\;\;\;\;\;\;\;\;\;\;\;$
\linebreak (c) Find the two point function
$$
\langle F_{\mu\nu}(x) F_{\rho\sigma}(0) \rangle. 
$$
(d) Why can't $F_{\mu\nu}$ be a primary field? Why can't it be a descendant field? Since it can be neither when $d\neq4$, this scale invariant theory cannot be conformal invariant.
\linebreak

(43) {\bf Scale invariance vs. conformal invariance, Part 2} \cite{3dsc1,3dsc2}: By adding new local fields, we could recover conformal invariance. 
\linebreak (a) Consider $d=3$ dimensions. Let's add a free scalar field $B$. We postulate that 
$$
F_{\mu\nu} = \epsilon_{\mu\nu\rho}\partial^\rho B,
$$
so that $F_{\mu\nu}$ is not a descendant field. Check that this prescription gives the appropriate $F$ two-point function. In $d=3$, it seems that free scalar theory contains a subsector that is isomorphic to Maxwell theory; we saved conformality by changing the set of local operators. We can therefore successfully embed the non-conformal Maxwell theory in $d=3$ into a unitary CFT. $\;\;\;\;\;\;\;\;\;\;\;\;\;\;\;\;\;\;\;\;\;\;\;\;\;\;\;\;\;\;$
\linebreak (b) Now consider $d\geq 5$. Can we add a field such that we can construct $F_{\mu\nu}$ as its descendant? Write down all the possible descendant relations and determine the scaling dimension of this new field. The trouble is, this dimension is inconsistent with unitarity bounds! We thus conclude it is impossible to extend these Maxwell theories into unitary conformal field theories. $\;\;\;\;\;\;\;\;\;\;\;\;\;\;\;\;\;\;\;\;\;\;\;\;\;\;\;\;\;\;$
\linebreak

(44) {\bf Exercise about Weyl and Euler tensors} \cite{atheorem}: In Lecture 6, we introduced the Weyl and Euler tensors as possible anomaly terms for $\langle  T^\mu_\mu \rangle $ in $d=4$ dimensions. 
\linebreak (a) We could have instead expressed the RHS of eq. (\ref{eq:4danomaly}) in terms of $R^2_{\mu\nu\rho\sigma}$ and $R^2_{\mu\nu}$. Do this. $\;\;\;\;\;\;\;\;\;\;\;\;\;\;\;\;\;\;\;\;\;\;\;\;\;\;\;\;\;\;$ $\;\;\;\;\;\;\;\;\;\;\;\;\;\;\;\;\;\;\;\;\;\;\;\;\;\;\;\;\;\;$
\linebreak But they did not do this. Instead, they have chosen the anomaly terms according to their transformation properties as curvature invariants. $\;\;\;\;\;\;\;\;\;\;\;\;\;\;\;\;\;\;\;\;\;\;\;\;\;\;\;\;\;\;$
\linebreak (b) Consider a Weyl transformation $g_{\mu\nu}\rightarrow e^{2\omega(x)} g_{\mu\nu}$. Calculate the transformation of $C^2$ under this transformation.  The Weyl tensor is sometimes known as the \emph{conformal tensor}.$\;\;\;\;\;\;\;\;\;\;\;\;\;\;\;\;\;\;\;\;\;\;\;\;\;\;\;\;\;\;$
\linebreak

(45) {\bf Proof of the a theorem} \cite{atheorem}: In Lecture 6, we presented the anomaly action (\ref{eq:4deriv}) whose variation produces the desired trace anomaly terms. Perform the variation to reproduce this result. This is a lengthy calculation.$\;\;\;\;\;\;\;\;\;\;\;\;\;\;\;\;\;\;\;\;\;\;\;\;\;\;\;\;\;\;$
\linebreak

(46) {\bf Higher order OPE associativity}: Give a diagrammatic explanation for why six-point crossing symmetry follows naturally from OPE associativity. Feel free to restrict yourself to just one pair of OPE channels.$\;\;\;\;\;\;\;\;\;\;\;\;\;\;\;\;\;\;\;\;\;\;\;\;\;\;\;\;\;\;$
\linebreak

(47) {\bf Analytic bootstrapping with spin} \cite{rychkovrad}: For now, we refer the reader to the cited paper. Follow through all the arguments through eq. (4.14). Check what upper bounds on $\Delta_{0,\rm{min}}$ can be obtained using this method via your favorite computer algebra system. $\;\;\;\;\;\;\;\;\;\;\;\;\;\;\;\;\;\;\;\;\;\;\;\;\;\;\;\;\;\;$
\linebreak

(48) {\bf Modular bootstrapping} \cite{hell}: Partition function calculations for primary fields aren't always easy; they involve Dedekind functions, for one thing. In this exercise, we will consider the modular bootstrap applied to a simpler partition function.
\linebreak (a) For purely imaginary $\tau$, we can instead consider the thermodynamic partition function
$$
Z(\beta) = \sum_{n=0} e^{-\beta E_n}.
$$
Following Hellerman's proof by contradiction, derive an upper bound on the conformal dimension of the lowest state. How does this bound compare? For what values of the central charge is it useful? Is this state a primary or descendant state?
\linebreak (b) Generalize the argument from part (a) to bounding $\Delta_n$.$\;\;\;\;\;\;\;\;\;\;\;\;\;\;\;\;\;\;\;\;\;\;\;\;\;\;\;\;\;\;$
\linebreak

\end{list}

\break

\end{document}